\documentclass{osa-article}

\journal{osajournal}
\articletype{Research Article}
\usepackage{color,soul}
\usepackage{wrapfig}
\usepackage{lineno}
\begin{document}

%\title{Piezoelectric actuation of photonic systems}
\title{Piezoelectric actuation for integrated photonics}

\author{Hao Tian,\authormark{1} Junqiu Liu,\authormark{2,3} Alaina Attanasio,\authormark{1} Anat Siddharth,\authormark{4} Terence Blésin,\authormark{4} Rui Ning Wang,\authormark{4} Andrey Voloshin,\authormark{4} Grigory Lihachev,\authormark{4} Johann Riemensberger,\authormark{4} Scott E. Kenning,\authormark{1} Yu Tian,\authormark{1} Tzu Han Chang,\authormark{1} Andrea Bancora,\authormark{4} Viacheslav Snigirev,\authormark{4} Vladimir Shadymov,\authormark{4} Tobias J. Kippenberg,\authormark{4} and Sunil A. Bhave\authormark{1,*}}

\address{\authormark{1}OxideMEMS Lab, Purdue University, 47907 West Lafayette, IN, USA\\
\authormark{2}International Quantum Academy, Shenzhen 518048, China\\
\authormark{3}Hefei National Laboratory, University of Science and Technology of China, Hefei 230026, China\\
\authormark{4}Institute of Physics, Swiss Federal Institute of Technology Lausanne (EPFL), CH-1015 Lausanne, Switzerland\\
%\authormark{5}These authors contributed equally to this work
}
\email{\authormark{*}bhave@purdue.edu} %% email address is required



\begin{abstract}
Recent decades have seen significant advancements in integrated photonics, driven by improvements in nanofabrication technology. This field has developed from integrated semiconductor lasers and low-loss waveguides to optical modulators, enabling the creation of sophisticated optical systems on a chip-scale capable of performing complex functions like optical sensing, signal processing, and metrology. The tight confinement of optical modes in photonic waveguides further enhances the optical nonlinearity, leading to a variety of nonlinear optical phenomena such as optical frequency combs, second-harmonic generation, and supercontinuum generation. Active tuning of photonic circuits is crucial not only for offsetting variations caused by fabrication in large-scale integration, but also serves as a fundamental component in programmable photonic circuits. Piezoelectric actuation in photonic devices offers a low-power, high-speed solution and is essential in the design of future photonic circuits due to its compatibility with materials like Si and Si$_3$N$_4$, which do not exhibit electro-optic effects. Here, we provide a detailed review of the latest developments in piezoelectric tuning and modulation, by examining various piezoelectric materials, actuator designs tailored to specific applications, and the capabilities and limitations of current technologies. Additionally, we explore the extensive applications enabled by piezoelectric actuators, including tunable lasers, frequency combs, quantum transducers, and optical isolators. These innovative ways of managing photon propagation and frequency on-chip are expected to be highly sought after in the future advancements of advanced photonic chips for both classical and quantum optical information processing and computing. 
\end{abstract}

%%%%%%%%%%%%%%%%%%%%%%%%%%  body  %%%%%%%%%%%%%%%%%%%%%%%%%%
\section{Introduction}
Integrated photonics \cite{Thomson:16, Agrell:16} has helped bring table-top optical systems in the research lab to the chip-scale with compact sizes, portable weight, and low power consumption. 
Thanks to the advancement of nanofabrication technologies, significant achievements have been made in integrated semiconductor lasers \cite{HuangD:19, xiang:2020, Jin:21, McKinzie:21, Xiang:21, Liang:21}, modulators \cite{Melikyan:14, WangC:18, HeM:19, Tian:20, Chen:14, Tadesse:14, Shao:19, Tian:21}, photodetectors \cite{Kang:09, Beling:16}, as well as ultralow-loss photonic integrated circuits (PICs) \cite{Zhang:17, Yang:18, Chang:20, Xuan:16, Ji:17, Liu:20b, Ye:19b, Puckett:21}. 
Moreover, the high confinement of light in integrated optical waveguides allows the exploration of the optical nonlinearities \cite{Moss:13, Gaeta:19}, giving rise to nonlinear photonics such as optical frequency comb (OFC) generation \cite{Kippenberg:18}, supercontinua \cite{johnson2015}, and second harmonic generation \cite{lu:2019, liu2021}. 
For instance, dissipative Kerr soliton (DKS) microcombs formed in Kerr-nonlinear optical microring resonators \cite{Kippenberg:18, Gaeta:19, Herr:14} has provided a miniaturized, coherent, broadband OFC source with repetition rates in the millimeter-wave to microwave domain, which has shown promise in widespread application domains ranging from atomic clocks \cite{Spencer:18, Newman:19}, telecommunication \cite{Pfeifle:14}, spectroscopy \cite{dutt2018chip}, to Light Detection and Ranging (LiDAR) \cite{riemensberger:2020}, as well as the heterogeneous integration with III-V/Si lasers \cite{Stern:18, Raja:19, Shen:20, Xiang:21}. 

Various material platforms have been developed in the past decades \cite{Kovach:20}, including Si, Si$_3$N$_4$ \cite{Xuan:16, Ji:17, Liu:20b, Ye:19b}, LiNbO$_3$ \cite{Wang:19, He:19, Zhang:17, Gong:20}, AlN \cite{Jung:13, LiuX:18} and AlGaAs \cite{Pu:16, Chang:20}. 
Because of the maturity of Complementary Metal-Oxide-Semiconductor (CMOS) processes, Si has been the leading material for commercialized optical modulators in data center and optical telecommunication systems. 
In the last decade, Si$_3$N$_4$ has attracted much attention due to its wide transparency window, ultra-low linear losses, large Kerr nonlinearity ($\chi ^{(3)}$), and flexibility in dispersion engineering \cite{Gaeta:19}. 
It has been widely used for the generation of soliton microcombs \cite{Kippenberg:11}, supercontina \cite{Gaeta:19}, optical gyroscopes \cite{gundavarapu2018}, and optical interconnects \cite{yao2018}. 
Recently, high-$Q$ Si$_3$N$_4$ microresonators have been applied to suppress the phase noise of semiconductor lasers \cite{Jin:21, lihachev2022, Li:21} with linewidth comparable with commercial fiber lasers. 

Efficient and high-speed tuning of integrated photonic devices is becoming inevitable for many advanced applications. 
For example, fast actuation of microring resonators allows stabilizing the microcomb repetition rate with high bandwidth \cite{Spencer:18, Liu:20a}. 
Also, resonance tuning can be used for tunable filters, compensating fabrication variation, and aligning optical resonances between optical resonators \cite{tian2023, tian2023high}. 
Moreover, the programmability of photonic circuits enables more complex photonic circuits to be built, such as photonic neural network \cite{shen2017deep} and photonic quantum computing \cite{Arrazola:21}. 
On the other hand, microwave frequency optical modulation is highly demanded recently in platforms such as spatio-temporal modulation based optical non-reciprocity \cite{Shi:18} and topological optical band structures in synthetic frequency dimensions \cite{yuan2021synthetic, dutt2019experimental, dutt2020single}, which could open up new avenues for studying novel physics phenomenon. 

Conventionally, the optical waveguides can be modulated by incorporating electro-optic materials, such as Lithium Niobate (LiNbO$_3$) and Aluminum Nitride (AlN), whose refractive index can be changed upon external electric field, originating from their $\chi ^{(2)}$ nonlinearity. 
However, for materials such as Si and Si$_3$N$_4$, due to their inversion symmetry, and thus the lack of $\chi ^{(2)}$ nonlinearity, it is difficult to electrically modulate them. 
Hence, the thermo-optic tuning is utilized \cite{dong2010, Xue:16, Joshi:16, lee2017, dutt2018chip, shen2017deep}, which is limited by slow tuning speed ($\sim$1 ms), high power consumption ($\sim$1-100 mW), and large thermal cross-talk. 
Therefore, it is incompatible with both large-scale photonic integration and cryogenic applications \cite{Moille:2019} that are prevalent for classical and quantum photonic computation. 
Although electro-optic modulation of Si waveguides through the free-carrier injection \cite{xu2005, lu2016} and carrier depletion effect \cite{gardes2009, Lira:12} has made promising progress, it suffers from large propagation loss due to free carrier absorption. 
Alternatively, heterogeneous or hybrid integration with electro-optic materials, e.g., graphene \cite{yao2018gate, phare2015, akinwande2019, gan2013chip}, monolayer WS$_2$ \cite{datta2020low}, lead zirconate titanate (PZT)  \cite{alexander2018}, LiNbO$_3$ \cite{WangC:18, weigel2018, ahmed2019, ahmed2019high, HeM:19, chiles2014mid, chang2017}, barium titanate (BaTiO$_3$) \cite{abel2013,abel2019,abel2020,abel2022}, and organic electro-optic materials \cite{kieninger2018}, has been intensively studied recently. 
However, there are still open questions with regard to optical losses and dispersion engineering as well as the complexity of CMOS-complexity of these materials. 

\begin{figure}[b]
\centering
\includegraphics[width=13 cm]{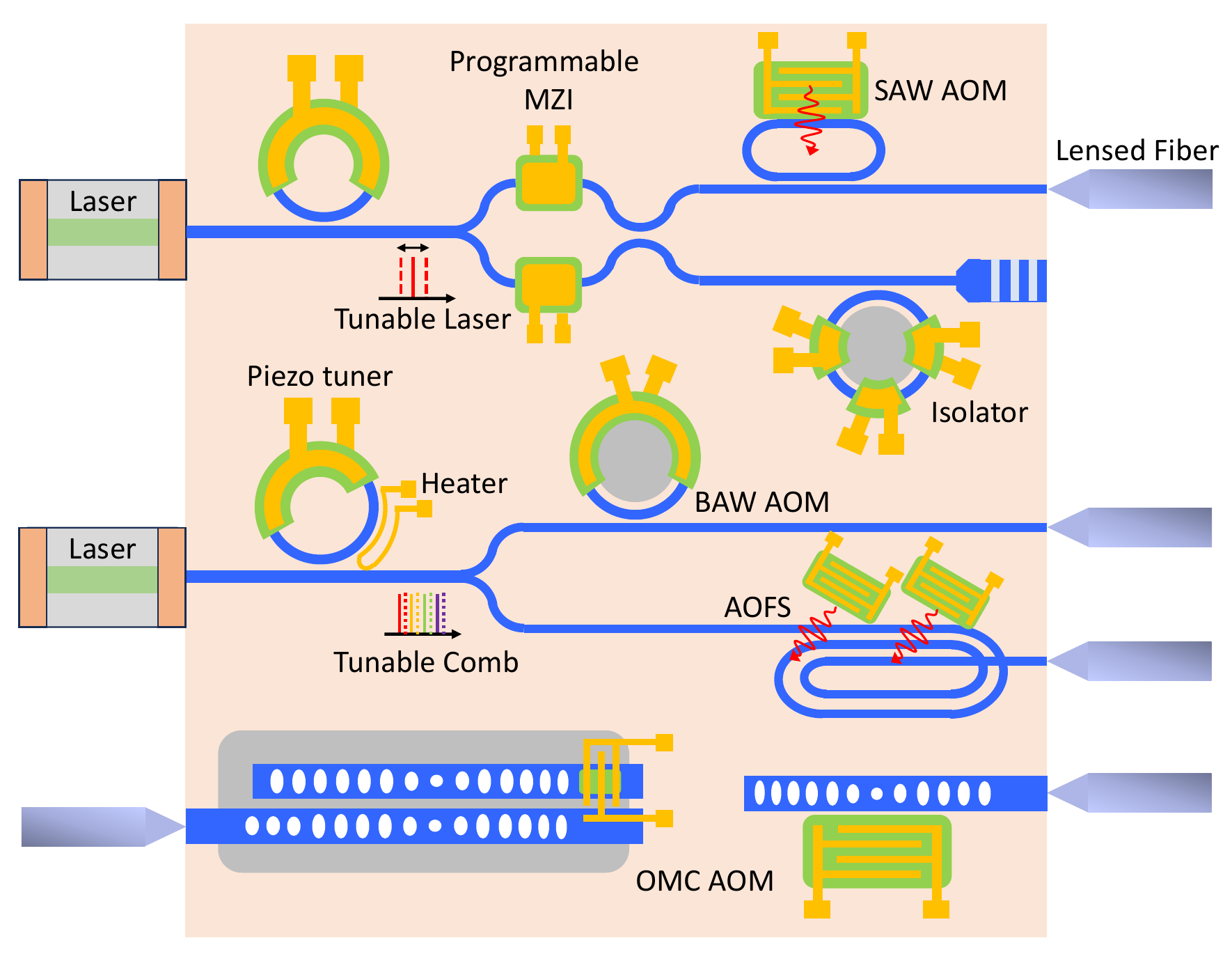}
\caption{
Schematic of a multi-functional integrated photonic chip enabled by piezoelectric actuators, from tunable light sources (laser and frequency comb), controllable light routing (Mach-Zehnder interferometer), acousto-optic modulation (using surface or bulk acoustic wave, or optomechanical crystal), to non-reciprocal optical devices (optical isolator) and frequency shifting (AOFS).The integration compatibility of these devices needs specific design and fabrication considerations, which will be discussed in each of the following chapters.    
}
\label{ch1_chip}
\end{figure}

The stress-optic effect has gained significant attention for the actuation of passive photonic waveguides and microring resonators due to the advances in Micro-Electro-Mechanical Systems (MEMS) technologies \cite{midolo2018nano, QuackMEMS}. 
By applying external forces either piezoelectrically \cite{Jin:18, stanfield2019cmos, Tian:20, everhardt12004, hosseini2015, wang2021} or electrostatically \cite{errando2015mems, errando2018, quack2019mems, bekker2018free, pruessner2019}, the deformation of the mechanical structure will generate stress and strain on the optical waveguides, and thus change the effective refractive index of the waveguide through photoelastic \cite{huang2003} and moving boundary effects \cite{johnson2002}. 
Although electrostatic actuation has been successfully applied to mechanical bending of photonic waveguides and resonators in various applications, e.g. directional coupler \cite{gyger2020}, optical switches \cite{haffner2019nano, seok2019silicon, han202132}, and phase shifters \cite{edinger2021silicon}, there remain challenges that need to be addressed in the future. 
For example, the optical waveguides have to be suspended, which may compromise their robustness, and the air-cladded waveguides present higher optical losses. 
In addition, the tuning is quadratic with applied voltage, and actuation bandwidth is limited by the fundamental mechanical resonance on the order of kHz-MHz \cite{jiang2020nanobenders}. 
On the contrary, the piezoelectric method circumvents these problems by fully cladding the waveguide in dielectric materials. 
Efficient, linear, bi-directional piezoelectric tuning of Si \cite{sebbag2012} and Si$_3$N$_4$ \cite{Jin:18, stanfield2019cmos, Tian:20, wang2022silicon} microring resonators has been experimentally demonstrated, with sub-$\mu$s tuning speed and nW on-hold power consumption. 
Moreover, acoustic waves can be excited piezoelectrically at several GHz, which enables chip-scale acousto-optic modulation. 
Therefore, stress-optic tuning can complement the traditional electro-optic and thermo-optic tuning in terms of material universality, CMOS compatibility, and power efficiency. 

This review focuses on the piezoelectric control of integrated photonic devices by giving comprehensive discussions on the aspects of piezoelectric materials, mechanical structures, and the operation frequency range. 
In Section 2, the fundamental physics will be introduced from piezoelectric effect to optomechanical interactions. 
Different mechanical modes and material platforms will be discussed. 
Section 3 reviews the state-of-the-art piezoelectric control of photonic integrated circuits with either released or unreleased mechanical structures, and compares the tuning performance of different piezoelectric materials. 
The applications of piezoelectric tuning will be elaborated in Section 4, from integrated semiconductor lasers to tunable optical frequency combs. 
In Section 5, acousto-optic modulation (AOM) at Radio-Frequency (RF) will be introduced regarding to different mechanical modes, with the applications of AOM discussed in Section 6. 
Section 7 will give our perspectives on current opportunities and challenges for the development of piezoelectric-controlled integrated photonic devices in the future. 
Finally, in Section 8, we will summarize the main features and conclude with implications for future applications. 
Figure \ref{ch1_chip} gives an outlook of an integrated photonic chip that is equipped with various piezoelectric actuators for different functionalities. The design of these devices will be discussed separately in each of the corresponding chapters. 

%%%%%%%%%%%%%%%%%%%%%%%%%%%%%%%%%%%%%%%%%%%%%%%%%%%
%%%%%%%%%%%%%%%%%%%%%%%%%%%%%%%%%%%%%%%%%%%%%%%%%%%
%%%%%%%%%%%%%%%%%%%%%%%%%%%%%%%%%%%%%%%%%%%%%%%%%%%
\section{Basics of Operation}
\begin{figure}[t]
\centering
\includegraphics[width=9 cm]{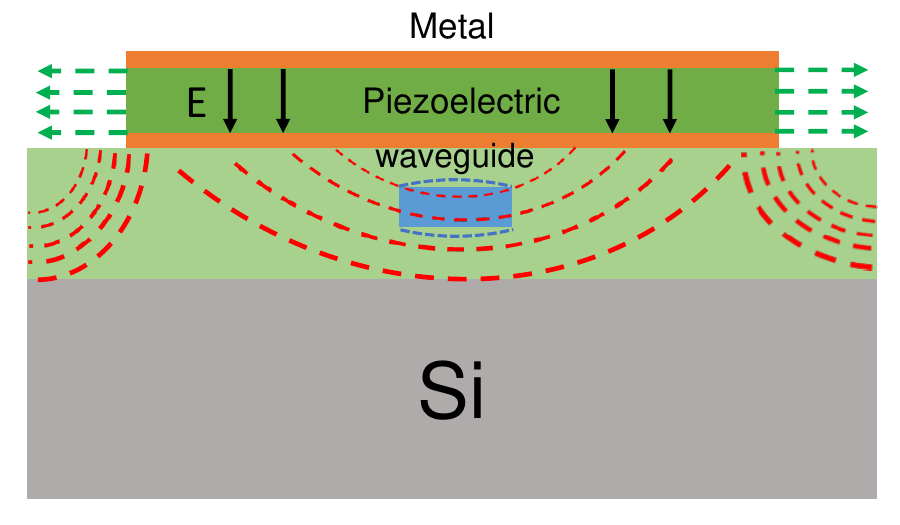}
\caption{
Schematic of a typical piezoelectric tuning of the optical waveguide. 
A piezoelectric actuator is placed on top of an optical waveguide (blue). 
Under external electrical signal, the piezoelectric film will deform and generate stress across the waveguide (red dashed lines), which modifies the effective refractive index of the waveguide through photoelastic effect. 
The boundary of the waveguide also displaces (dark blue dashed lines) which changes the shape of the waveguide.
}
\label{crossection}
\end{figure}

%\subsection{Optical waveguide and resonators}
%\cite{Liu:20b, Fan:18}
In this section, the basic knowledge behind the piezoelectric tuning of the integrated photonics devices will be given, covering from the generation of mechanical stress and strain via the piezoelectric effect, to the modulation of refractive index through the stress-optic effect. 
The cross-section of a typical device structure is as shown in Fig. \ref{crossection}, where a thin film of piezoelectric material is sandwitched between top and bottom metal electrodes and placed directly on top of the optical waveguide. 
By applying external electrical signal, the piezoelectric film will expend or squeeze, generating stress and strain around the optical waveguide, and thus changes the optical materials' refractive indices and waveguide shape. 
In the following, a full set of equations will be given to estimate the perturbation of the effective refractive index of the optical waveguides and the resonant frequency of the optical micro-resonators. 

%%%%%%%%%%%%%%%%%%%%%%%%%%%%%%%%%%%%%%%%%%%%%%%%%%%
%%%%%%%%%%%%%%%%%%%%%%%%%%%%%%%%%%%%%%%%%%%%%%%%%%%
\subsection{Piezoelectric Effect}
%As the key material property that the whole technique is based on, 
Piezoelectricity is an effect where charges can be generated and accumulated at the surface of the material under external pressure. 
It can happen only in crystalline materials with no inversion symmetry, such as PZT, AlN, and LiNbO$_3$. 
The inverse effect can also happen where stress and strain are built as external electric field is applied across the material. 
The direct and inverse piezoelectric effect can be mathematically described by relating the electric field to the strain of the material \cite{zgonik1994}:
\begin{eqnarray}
 \varepsilon _I &=& \sum ^6 _{J=1} s_{IJ} \sigma_J + \sum ^3 _{j=1} d_{Ij}^tE_{j}\\
 D _i &=& \sum ^6 _{J=1} d_{iJ} \sigma_J + \sum ^3 _{j=1} \epsilon_{ij} E_{j}
\end{eqnarray}
where $\varepsilon$ and $\sigma$ are the strain and stress, 
$D$ and $E$ are the electric displacement and electric field, 
$s_{IJ}$ is the elastic compliance tensor that describes the microscopic Hooke's law in linear elastic material, 
$\epsilon_{ij}$ is the permittivity of the dielectric, 
$d_{iJ}$ is the piezoelectric strain coefficient in the unit of $\text{m}/\text{V}$ (which is also equal to $\text{C}/\text{N}$). 
Note that the piezoelectric stress coefficient $e_{iJ}$ (in the unit of $\text{C}/\text{m}^2$) is also often quoted in the literature, which characterizes the surface charge density generated by the strain. 
$d_{iJ}$ and $e_{iJ}$ are generally related through the elastic stiffness tensor $c_{IJ}$ of the material through \cite{zgonik1994}:
\begin{equation}
    e_{iJ}=\sum ^6 _{K=1}d_{iK}c_{KJ}
\end{equation}

Subscript indices represent different components of each physical quantity. 
The capital indices $I,J$ are for stress and strain ranging from 1 to 6 to take into account the normal and shear components with the Voight notation that: 
$1\rightarrow xx, 2\rightarrow yy, 3\rightarrow zz, 4\rightarrow yz, 5\rightarrow xz, 6\rightarrow xy$. 
The indices $i,j$ are for electric field and displacement, and each has three components: 
$1\rightarrow x, 2\rightarrow y, 3\rightarrow z$. 
Note that due to the symmetry, $d_{Ij}^t=d_{jI}$. 
The piezoelectric coefficient is related to crystal structures of each piezoelectric material. 
Although $d_{iJ}$ has 18 components, due to the crystal symmetry, only few are non-zero.   

%%%%%%%%%%%%%%%%%%%%%%%%%%%%%%%%%%%%%%%%%%%%%%%%%%%
\begin{table}[t]
    \centering
    \caption{Piezoelectric and mechanical properties of typical piezoelectric materials. The scandium doping concentration in AlScN is indicated.  }
    \begin{tabular}{ p{1.2cm}|p{2.8cm}|p{1.7cm}|p{1.3cm}|p{1.3cm} |p{1.7cm} |p{2cm} }
 \hline
 Material  &  Piezoelectric stress coefficient ($\text{C}/\text{m}^2$) & Piezoelectric strain coefficient ($\text{pm}/\text{V}$) & Elastic modulus (GPa) & Acoustic Velocity (m/s) & Dielectric constant & Ref. \\
 \hline
 PZT & e$_{31}$=-6.5 e$_{33}$=23.3 & d$_{31}$=-150 d$_{33}$=60-130 & 68 & $V_l=$4600 $V_t=$2400 & 400-1400 & \cite{fraga2014, wu2009}\\
 
 AlN & e$_{31}$=-0.58 e$_{33}$=1.55 e$_{15}$=-0.48 & d$_{31}$=-2.0 d$_{33}$=3.9 & 308 & $V_l=$10169 $V_t=$6369 & $\epsilon_{r, 11}$=8 $\epsilon_{r, 33}$=9 & \cite{fraga2014, zhu2021, wang2020high, lueng2000, bernardini1997, liu2023, nie2023} \\
 
 AlScN & e$_{31}$=-2.5 (Sc 30\%) e$_{33}$=2.1 (Sc 30\%) & d$_{31}$=7 (Sc 30\%) d$_{33}$=15 (Sc 30\%) & 250 (Sc 30\%) & 9089 (Sc 20\%) & 14 (Sc 30\%) & \cite{pirro2021, kusano201836, ledesma20219, wang2022effects, teshigahara2012, akiyama2009, akiyama2013, fichtner2017, zou2022, chen2022, nie2023}\\
 
 ZnO & e$_{31}$=-0.57 e$_{33}$=1.32 & d$_{31}$=-5.0 d$_{33}$=5.9 & 201 & $V_l=$5790 $V_t=$2700 & $\epsilon_{r, 11}$=7.4 $\epsilon_{r, 33}$=7.8 & \cite{fraga2014, zhu2021, wang2020high, bernardini1997, nie2023} \\
 
 LiNbO$_3$ & e$_{31}$=0.23 e$_{33}$=1.33  e$_{22}$=2.43 e$_{15}$=3.83 & d$_{31}$=-1.0 d$_{33}$=6 d$_{22}$=20.8 d$_{15}$=69 & 203 & $V_l=$7316 $V_t=$4795 & $\epsilon_{r, 11}$=44 $\epsilon_{r, 33}$=27.9 & \cite{fraga2014, zhu2021, shur2010, wang2020high, kushibiki1999, liu2023}\\
 
 LiTaO$_3$ & e$_{31}$=-0.11 e$_{33}$=1.93 e$_{22}$=1.67 e$_{15}$=2.63 & d$_{31}$=-3.0 d$_{33}$=9.0 d$_{22}$=7.5 d$_{15}$=26 & 233 & $V_l=$5746 $V_t=$3536 & $\epsilon_{r, 11}$=38.3 $\epsilon_{r, 33}$=46.2 & \cite{zhu2021, shur2010, kushibiki1999} \\
 
 BaTiO$_3$ & e$_{31}$=-0.7 e$_{33}$=6.7 e$_{15}$=34.2 & d$_{31}$=-33.4 d$_{33}$=90 d$_{15}$=282 & 222 & $V_l=$6800 $V_t=$2500 & $\epsilon_{r, 11}$=2200 $\epsilon_{r, 33}$=56 &\cite{zgonik1994} \\
 
 GaAs & e$_{31}$=0.093 e$_{33}$=-0.185 e$_{15}$=0.093 e$_{14}$=-0.16 & d$_{14}$=1.345 & 118 & $V_l=$4730 $V_t=$3350 & 12.9 &\cite{zhu2021, bright1989, rampal2021optical, cottam1973, liu2023} \\
 
 GaP & e$_{31}$=0.03 e$_{33}$=-0.07 e$_{14}$=-0.1 & d$_{14}$=1.9 & 141 & $V_l=$5830 $V_t=$4120 & 11.1 & \cite{nelson1968electro, krempl1997gallium, weil1968elastic, stockill2022}\\
 
 GaN & e$_{31}$=-0.33 e$_{33}$=0.65 e$_{15}$=-0.33 e$_{14}$=0.56 & d$_{33}$=3.1 & 390 & $V_l=$7350 $V_t=$4578 & $\epsilon_{r, 11}$=9.5 $\epsilon_{r, 33}$=10.6 &\cite{shur1998, wang2020high, lueng2000, bernardini1997, polian1996, kane2011} \\

 \hline
\end{tabular}
    \label{tablepiezo}
\end{table}
%%%%%%%%%%%%%%%%%%%%%%%%%%%%%%%%%%%%%%%%%%%%%%%%%%%

The piezoelectric coefficients of typical piezoelectric materials are summarized in Table \ref{tablepiezo}, together with their mechanical properties. 
Note the specific values may vary, depending on the composition and growth conditions of the material. 
However, it is a good starting point for numerical simulations. 
For most demonstrated piezoelectric tuning of photonic circuits to date, PZT and AlN are the two main piezoelectric materials that are widely employed, because of their high e$_{33}$ and maturity of foundry processes. 
Although PZT shows a piezoelectric coefficient nearly 20 times larger, AlN advances in better linearity, smaller dielectric constant (thus smaller capacitance and shorter charge-discharge time), and zero hysteresis (since PZT is ferroelectric). 
Also, due to the high loss tangent at microwave frequencies, the actuation frequency of PZT is limited to below 1 GHz. 
On the other hand, AlN is excellent at microwave frequencies, and efficient gigahertz AOM has been demonstrated \cite{ghosh2016, Tadesse:14, li2015, Sohn:18, Kittlaus:21}. 
Therefore, depending on the specific requirement of applications (high efficiency, linearity, or speed), the right piezoelectric material should be considered via careful design and simulation. 

Scandium-doped AlN (AlScN) has attracted much attention due to its much improved piezoelectric coefficient e$_{33}$ compared to AlN. 
As for the LiNbO$_3$, while itself is widely adopted as an electro-optic material, its application in piezoelectric tuning is less explored. 
However, its application at microwave frequencies by exciting acoustic waves for the AOM has been extensively studied due to its much higher electromechanical conversion efficiency \cite{Shao:19, hassanien2021, shao:2020, yu2020, cai2019, sohn2021}. 
For the other piezoelectric materials Zinc Oxide (ZnO) and Lithium Tantalate (LiTaO$_3$), which show similar properties as AlN and LiNbO$_3$, they are still new in the field of piezoelectric tuning and AOM of photonic circuits, and further research in this direction could potentially open up new possibilities from the perspective of integration, efficiency, and cost. 

%%%%%%%%%%%%%%%%%%%%%%%%%%%%%%%%%%%%%%%%%%%%%%%%%%%
%%%%%%%%%%%%%%%%%%%%%%%%%%%%%%%%%%%%%%%%%%%%%%%%%%%
\subsection{Photoelastic and Moving Boundary Effect}

The generated stress and strain will modify the effective refractive index of the optical waveguide through the stress-optic effect, which consists mainly of photoelastic and moving boundary effects as will be elaborated in the following. 
These effects describe the interaction between light and mechanical structures (acoustic waves), which are widely adopted in fields including cavity optomechanics \cite{Tobias:2014}, stimulated Brillouin scattering \cite{wiederhecker2019}, and AOM \cite{Tadesse:14, Shao:19}. 
They offer operation in a wide frequency range from direct current (DC) to radio-frequency (RF). 
While the photoelastic and moving boundary effects describe the action of mechanics on the light, the inverse effects where the photon generates optical forces on the mechanical structures will also take place, which correspond to the electrostrictive \cite{rakich2010} and radiation-pressure forces \cite{rosenberg2009}, respectively.

The photoelastic effect is the change in refractive index $n$ (and the relative permittivity $\epsilon_r$) of an optical material under stress and strain \cite{xu1992acousto, huang2003}. 
In general, the relative permittivity is a rank-2 tensor which contains 6 independent components due to the symmetry \cite{huang2003}:
\begin{equation}
    \epsilon_r = \begin{pmatrix}
n_1^2 & n_6^2 & n_5^2\\
n_6^2 & n_2^2 & n_4^2\\
n_5^2 & n_4^2 & n_3^2
\end{pmatrix}
\end{equation}
where $n_{1-6}$ is the refractive index that relates electric fields in different directions, and the subscripts follow the same convention as the previous section. 
The relative change of refractive index under strain is conventionally described by the strain-optic coefficient $p_{ij}$ as:
\begin{equation}
    \Delta \frac{1}{n_I^2}=\sum ^6_{J=1} p_{IJ} \varepsilon _J
    \label{eq4}
\end{equation}

For most of the optical materials used in integrated photonic circuits, such as amorphous SiO$_2$ and Si$_3$N$_4$, and Si with cubic lattice, the strain-optic coefficients have only two independent components due to the symmetry of the material \cite{huang2003}:
\begin{equation}
   \Delta  \begin{pmatrix}
     1/n_1^2\\1/n_2^2\\ 1/n_3^2\\ 1/n_4^2\\ 1/n_5^2\\ 1/n_6^2
    \end{pmatrix}
    = \begin{pmatrix}
    p_{11} & p_{12} & p_{12} & 0 & 0 & 0\\
    p_{12} & p_{11} & p_{12} & 0 & 0 & 0\\
    p_{12} & p_{12} & p_{11} & 0 & 0 & 0\\
    0 & 0 & 0 & p_{44} & 0 & 0\\
    0 & 0 & 0 & 0 & p_{44} & 0\\
    0 & 0 & 0 & 0 & 0 & p_{44}\\
    \end{pmatrix}
    \begin{pmatrix}
    \varepsilon_1 \\\varepsilon_2 \\\varepsilon_3 \\\varepsilon_4 \\\varepsilon_5 \\\varepsilon_6
    \end{pmatrix}
    \label{eq1.13}
\end{equation}
where $p_{44}=(p_{11}-p_{12})/2$. 
The stress-optic coefficient $C_i$ of these materials is also often referred in the literature which directly relates the change of index with the stress $\sigma$ as \cite{huang2003}:
\begin{equation}
     \begin{pmatrix}
     n_1\\n_2\\ n_3\\ n_4\\ n_5\\ n_6
    \end{pmatrix}
    =   \begin{pmatrix}
     n_0\\n_0\\ n_0\\ 0\\ 0\\ 0
    \end{pmatrix}
    -\begin{pmatrix}
    C_1 & C_{2} & C_{2} & 0 & 0 & 0\\
    C_{2} & C_{1} & C_{2} & 0 & 0 & 0\\
    C_{2} & C_{2} & C_{1} & 0 & 0 & 0\\
    0 & 0 & 0 & C_{3} & 0 & 0\\
    0 & 0 & 0 & 0 & C_{3} & 0\\
    0 & 0 & 0 & 0 & 0 & C_{3}\\
    \end{pmatrix}
    \begin{pmatrix}
    \sigma_1 \\\sigma_2 \\\sigma_3 \\\sigma_4 \\\sigma_5 \\\sigma_6
    \end{pmatrix}
\end{equation}
where $n_0$ is the original refractive index of the optical materials. 
It can be intuitively interpreted that $C_1$ relates the index and stress in the same direction, while $C_2$ relates those in the orthogonal direction. 
The stress-optic coefficient $C_i$ are related with the strain-optic coefficient $p_{IJ}$ through the mechanical properties of the material \cite{huang2003}:
\begin{eqnarray}
C_1 &=& n_0^3\frac{p_{11}-2\nu p_{12}}{2E_{\text{mod}}}\\
C_2 &=& n_0^3\frac{p_{12}-\nu (p_{11}+p_{12})}{2E_\text{{mod}}}\\
C_3 &=& n_0^3\frac{p_{44}}{2G}
\end{eqnarray}
where $E_{\text{mod}}, ~G, ~\nu$ are Young's modulus, shear modulus, and Poisson ratio, respectively. 

Now that we know how the stress modifies the refractive index, the change of effective refractive of the optical waveguide due to the photoelastic effect can be calculated by the Bethe-Schwinger perturbation theory \cite{Shao:19, wiederhecker2019, chan2012}:
\begin{equation}
    \Delta n_{\text{eff,PE}} = \frac{n_{\text{eff,0}}}{2} \frac{\iiint \textbf{E}^*\cdot \Delta \epsilon \cdot \textbf{E} ~dV}{\iiint \textbf{E}^* \epsilon \textbf{E} ~dV}
    \label{eq1.18}
\end{equation}
where the numerator is the perturbation of the electromagnetic energy in the optical waveguide, and the denominator is the total energy of the optical mode. 
$n_{\text{eff,0}}$ is the effective refractive index of the waveguide. 
The volume integral is taken over the entire region where the electromagnetic field resides. 
$\Delta \epsilon$ is a rank-2 tensor and the integrand in the numerator can be expanded as \cite{Shao:19}:
\begin{equation}
    \textbf{E}^*\cdot\Delta \epsilon \cdot\textbf{E} = -\epsilon_0n^4\begin{pmatrix}
    E_1^* & E_2^* & E_3^*
    \end{pmatrix} 
    \begin{pmatrix}
    \Delta \frac{1}{n_1^2} & \Delta \frac{1}{n_6^2} & \Delta \frac{1}{n_5^2}\\
    \Delta \frac{1}{n_6^2} & \Delta \frac{1}{n_2^2} & \Delta \frac{1}{n_4^2}\\
    \Delta \frac{1}{n_5^2} & \Delta \frac{1}{n_4^2} & \Delta \frac{1}{n_3^2}
    \end{pmatrix}
    \begin{pmatrix}
    E_1 \\ E_2 \\ E_3
    \end{pmatrix}
    \label{eq1.19}
\end{equation}
where $\epsilon_0$ is the vacuum permittivity. 
The $\Delta (1/n_i^2)$ can be directly calculated from the strain-optic relation in Eq. \ref{eq4}. 
From Eqs. \ref{eq1.18}-\ref{eq1.19}, we can estimate the change of the refractive index of the optical waveguide by simulating the optical and strain field distribution using Finite-Element Method (FEM). 

%%%%%%%%%%%%%%%%%%%%%%%%%%%%%%%%%%%%%%%%%%%%%%%%%%%
\begin{table}[t]
    \centering
    \caption{Summary of the optical and photoelastic properties of typical optical materials used in integrated photonic circuits. The wavelength at which the linear and nonlinear refractive index are measured is at 1550 nm if not specified. o and e indicate ordinary and extraordinary optical axis.  }
    \begin{tabular}
    { p{1.5cm}|p{1.8cm}|p{2.4cm}|p{3.2cm}|p{1.6cm}|p{1.5cm} }
 \hline
 Material  &  Optical refractive index n & Nonlinear refractive index n$_2$ (m$^2$/W) & Photoelastic coefficient & Electro-optic coefficient (pm/V) &Ref.  \\
 \hline
 Si  &  3.48 & 5$\times$10$^{-18}$ & p$_{11}$=-0.094 p$_{12}$=0.017 p$_{44}$=-0.051 & 0 &\cite{huang2003, zhu2021, chan2012, xu2011polarization} \\
 
 SiO$_2$  &  1.44 & 3$\times$10$^{-20}$ & p$_{11}$=0.121 p$_{12}$=0.270 & 0 &\cite{huang2003, zhu2021} \\
 
 Si$_3$N$_4$  &  2.0 & 2.5$\times$10$^{-19}$ & p$_{11}$=-0.125 p$_{12}$=0.047 & 0 &\cite{zhu2021, Gyger:20, capelle2017} \\
 
 AlN  &  2.12 (o)  2.16 (e) & 2.3$\times$10$^{-19}$ & p$_{11}$=-0.1 p$_{12}$=-0.027 p$_{13}$=-0.019 p$_{33}$=-0.107 p$_{44}$=-0.032 p$_{66}$=-0.037 & r$_{13}$=0.67 r$_{33}$=-0.59 &\cite{zhu2021, graupner1992, davydov1997} \\
 
 LiNbO$_3$  &  2.21 (o)  2.14 (e)& 1.8$\times$10$^{-19}$ & p$_{11}$=-0.026 p$_{12}$=0.09 p$_{13}$=0.133 p$_{14}$=-0.075 p$_{31}$=0.179 p$_{33}$=0.071 p$_{41}$=-0.151 p$_{44}$=0.146 & r$_{13}$=9.6 r$_{22}$=6.8 r$_{33}$=30.9 r$_{51}$=32.6 &\cite{huang2003, zhu2021, dixon1967, liu2023}\\
 
 LiTaO$_3$  &  2.119 (o)  2.123 (e)& 1.46$\times$10$^{-19}$ (800 nm) & p$_{11}$=-0.081 p$_{12}$=0.081 p$_{13}$=0.093 p$_{14}$=-0.026 p$_{31}$=0.089 p$_{33}$=-0.044 p$_{41}$=-0.085 p$_{44}$=0.028 & r$_{13}$=8.4 r$_{22}$=-0.2 r$_{33}$=30.5 r$_{51}$=20 &\cite{huang2003, zhu2021, dixon1967}\\

BaTiO$_3$  &  2.297 (o)  2.268 (e)& - & p$_{11}$=0.5 p$_{12}$=0.106 p$_{13}$=0.2 p$_{31}$=0.07 p$_{33}$=0.77 p$_{44}$=1.0 & r$_{13}$=8 r$_{33}$=40.6 r$_{51}$=730 &\cite{zgonik1994, karvounis2020, abel2012electro}\\
  
GaAs  &  3.38  & 2.6$\times$10$^{-17}$ & p$_{11}$=-0.165 p$_{12}$=-0.140 p$_{44}$=-0.072 &  r$_{41}$=1.43 &\cite{zhu2021, Balram:14, dixon1967, liu2023}\\
 
GaP  &  3.05  & 1.1$\times$10$^{-17}$ & p$_{11}$=-0.23 p$_{12}$=-0.13 p$_{44}$=-0.10 &  r$_{41}$=-1.1 &\cite{huang2003, nelson1968electro, schneider2019, wilson2020, dixon1967}\\

GaN  &  2.3  & 1.4$\times$10$^{-18}$ & p$_{11}$=0.031 p$_{12}$=0.008 p$_{13}$=0.006 p$_{33}$=0.033 p$_{44}$=0.01 p$_{66}$=0.012 &  r$_{13}$=1.00 r$_{33}$=1.60 &\cite{pezzagna2008, zheng2022integrated, cuniot2014, davydov1997}\\
 
 \hline
 
 \hline
\end{tabular}
    \label{tableopt}
\end{table}
%%%%%%%%%%%%%%%%%%%%%%%%%%%%%%%%%%%%%%%%%%%%%%%%%%%

Due to the large contrast between the refractive index of the waveguide core and the cladding material, the abrupt displacement of the waveguide's boundary will modify the overall effective refractive index of the waveguide, which is the so-called moving boundary effect. 
It will play an important role when the shape of the waveguide deforms under external mechanical strain. 
Rigorous derivation of the relation between the index and the boundary displacement can be found in Ref. \cite{johnson2002}, and the main result is summarized as below \cite{chan2012, Shao:19}:
\begin{equation}
    \Delta n_{\text{eff,MB}}= \frac{n_{\text{eff,0}}}{2}\frac{\iint (\textbf{Q}\cdot \hat{\textbf{n}})(\Delta \epsilon \text{E}_{\parallel}^2-\Delta \epsilon^{-1} \text{D}_{\perp}^2) ~dS}{\iiint \textbf{E}^* \epsilon \textbf{E} ~dV}
    \label{eq1.20}
\end{equation}
where $\textbf{Q}$ is the mechanical displacement of the boundary, and $\hat{\textbf{n}}$ is the normal vector of the boundary facing outward. 
The dot product of $\textbf{Q}\cdot \hat{\textbf{n}}$ picks the component of $\textbf{Q}$ normal to the boundary. 
$\text{E}_{\parallel}$ only takes the component parallel to the boundary, and $\text{D}_{\perp}$ is the perpendicular component. 
$\Delta \epsilon= \epsilon_{\text{core}}-\epsilon_{\text{cladding}}$ and $\Delta \epsilon^{-1}= \epsilon^{-1}_{\text{core}}-\epsilon^{-1}_{\text{cladding}}$ are the difference of permittivity and inverse of permittivity between the waveguide core and cladding materials. 
The surface integral in the numerator is taken over all the surfaces of the waveguide. 

The resonant frequency tuning of an optical resonator can be inferred from the perturbation of the effective refractive index of the waveguide as \cite{Shao:19}:
\begin{equation}
    \frac{\Delta \omega _0}{\omega _0} = -\frac{\Delta n_{\text{eff,PE}}+\Delta n_{ \text{eff,MB}}}{n_{\text{eff,0}}}-\frac{\Delta r}{r}
    \label{eq13}
\end{equation}
where $\omega _0$ and $n_{\text{eff,0}}$ are the original optical resonant angular frequency and the effective refractive index of the waveguide. 
The second term on the right-hand side is the relative change of the radius of the optical microring resonator due to the mechanical deformation, which can be replaced by the relative change of the cavity length for the race-track resonator. 
Note this is the change of the optical path along the longitudinal direction which should not be confused with the moving boundary effect in the transverse direction.  

The above equations will be useful for simulating the tuning of the optical resonance upon quasi-DC voltages applied to the actuator. 
For AOM at microwave frequencies, it is widely adopted to describe the interaction between photon and phonon by using the single photon-phonon optomechanical coupling rate $g_0$, which can be calculated by normalizing Eq. \ref{eq13} using the Zero-Point Fluctuation (ZPF) of the quantum ground state of the mechanical motion as:
\begin{eqnarray}
\label{eqg0}
     g_0 &=&  \Delta \omega_0\frac{x_{\text{ZPF}}}{\text{max}\{Q\}}\\
     x_{\text{ZPF}} &=& \sqrt{\frac{\hbar}{2m_{\text{eff}}\Omega_{\text{m}}}}\\
     m_{\text{eff}} &=& \frac{\iiint \rho Q^2 ~dV}{\text{max}\{Q^2\}}
\end{eqnarray}
where $x_{\text{ZPF}}$, $m_{\text{eff}}$ and $\Omega_{\text{m}}$ are the amplitude of zero-point motion, and the effective mass and resonant frequency of the mechanical resonator, respectively. 
$\rho$ is the material's mass density, and $\text{max}\{Q\}$ and $\text{max}\{Q^2\}$ are the maximum displacement and its square over the whole mechanical mode. 
Intuitively, $g_0$ measures the scattering rate between a single photon and phonon in an optomechanical cavity, and is related with the overlap between the optical and mechanical modes. 
It can be experimentally calibrated by optical readout of the thermal Brownian noise of the mechanical resonator \cite{gorodetksy2010}, typically in a homodyne interferometer. 
From the expressions, we can also see the larger the mechanical mass $m_{\text{eff}}$ and frequency $\Omega_{\text{m}}$, the smaller the $x_{\text{ZPF}}$, and thus smaller coupling rate. 
In the case where large $g_0$ is desired, it is preferred to tightly confine the optical and mechanical modes in a small volume. 

The photoelastic coefficients and the optical properties of typical optical materials for integrated photonics are summarized in Table \ref{tableopt}. 
The first three rows are materials with inversion symmetry, which have neither piezoelectric nor electro-optic effects. 
However, they are the most widely used optical materials for optical waveguides, showing ultra-low optical losses, high third-order nonlinearity, and large-scale production. 
On the other hand, they exhibit reasonable amount of photoelastic coefficient, which is one of the main reasons for piezoelectric tuning. 
For the piezoelectric materials presented in Table \ref{tablepiezo}, they are also good electro-optic materials and have been applied in electro-optic modulators \cite{xiong2012low, WangC:18}. 
This `coincidence' is because both piezoelectric and electro-optic effects have the same microscopic origin, which is due to the non-centrosymmetry of the crystal structure of these materials, such that they support spontaneous polarization that is dependent on external stress and electric field. 
On the other hand, they can be modulated by acoustic waves excited through their piezoelectricity by fabricating interdigital transducers (IDT) on the same layer (see next section). 
Most recently, efficient AOMs have been demonstrated on these material platforms \cite{Tadesse:14, Shao:19}. 
GaAs and GaP have recently attracted much attention in piezo-optomechanics by working as the optomechanical crystals due to their high refractive index and photoelastic coefficients \cite{schneider2019, stockill2019, stockill2022, honl2022, Balram:14, forsch2020}. 
Sub-wavelength confinement of both optical and mechanical modes strongly enhances the optomechanical coupling rate g$_0$. 
Meanwhile, the high third-order nonlinearity makes them promising for nonlinear photonics, having recently crossed the elusive threshold for continous-travelling wave Kerr parametric gain [cite new arxiv].
Therefore, these different materials show their own advantages in one or more aspects, and a hybrid combination is usually preferred to take full advantage. 

%%%%%%%%%%%%%%%%%%%%%%%%%%%%%%%%%%%%%%%%%%%%%%%%%%%
\begin{figure}[t]
\centering
\includegraphics[width=13.5 cm]{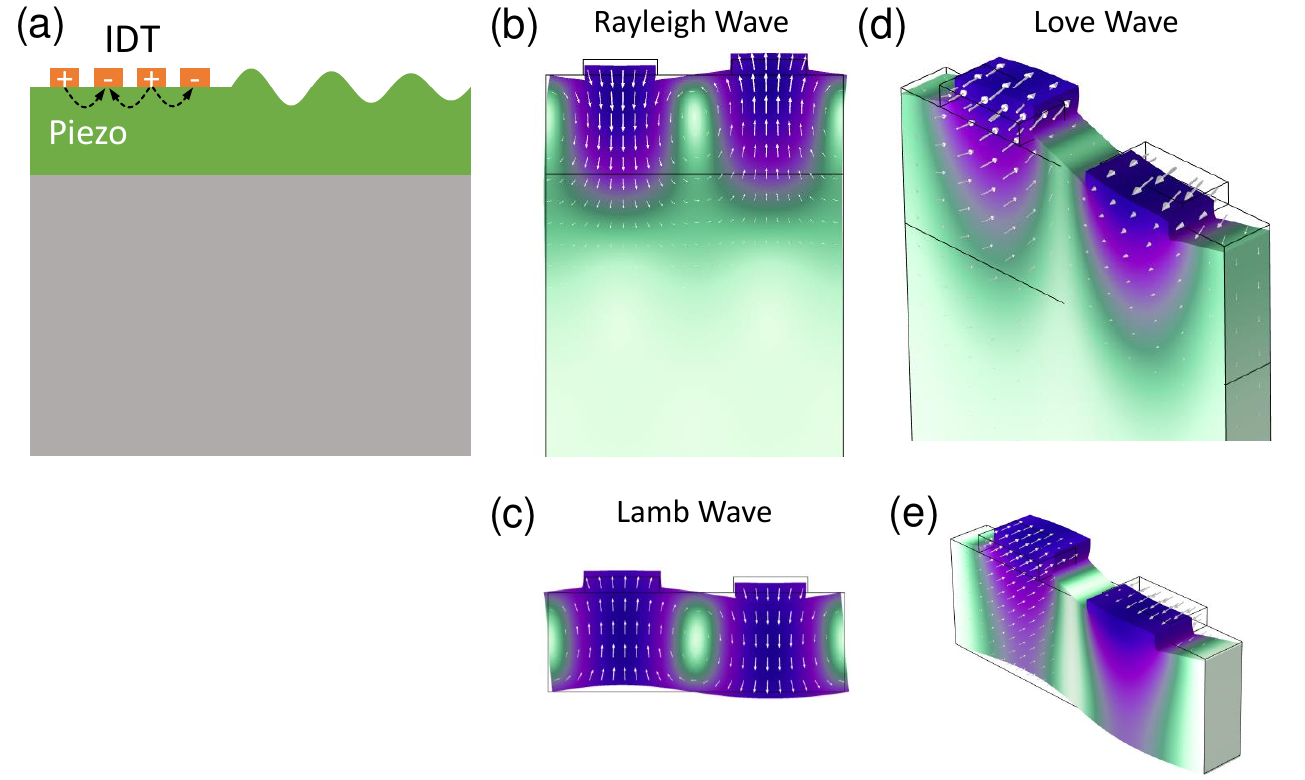}
\caption{
Typical surface acoustic waves excited by an IDT. 
(a) Schematic showing the working principle of IDT. 
The black dashed lines illustrate the alternative electric field. 
(b) Rayleigh wave of a piezoelectric thin film on a substrate. 
(c) Lamb wave of a released piezoelectric film. 
The white arrows show the particle displacement. 
Love wave of an (d) unreleased and (e) released piezoelectric film. 
Only one period is shown in these mode illustrations. 
}
\label{SAW}
\end{figure}

%%%%%%%%%%%%%%%%%%%%%%%%%%%%%%%%%%%%%%%%%%%%%%%%%%%
%%%%%%%%%%%%%%%%%%%%%%%%%%%%%%%%%%%%%%%%%%%%%%%%%%%
\subsection{Surface and Bulk Acoustic Waves}

Besides the quasi-DC (DC-MHz) actuation, when driving at microwave frequencies (GHz), the vibration of the piezoelectric actuator will launch micron to sub-micron wavelength acoustic waves.
Depending on the propagation direction of the acoustic waves, they can be classified mainly into surface and bulk acoustic waves. 
An acoustic resonator functions as a container that confines vibrational waves for either amplification or absorption. The behavior of acoustic waves within such a resonator is governed by the acoustic wave equation and appropriate boundary conditions, which can either be free or fixed. A free boundary condition implies that displacement is at its maximum while stress is zero, typically occurring at the interface with air or vacuum. Conversely, a fixed boundary condition means that displacement is zero and stress is at its peak. At the boundaries between different materials, acoustic waves may be partially or fully reflected due to mismatched acoustic impedance. Acoustic resonance arises when these reflected waves constructively interfere with each other, with the resonator's cavity length corresponding to an integer multiple of half-wavelengths. The piezoelectric actuator acts as an acoustic source, facilitating the entry of more vibrational waves into the cavity, generally by exerting pressure on one wall or gap within the cavity. \cite{staelin2011}.
With the acoustic energy enhanced by the acoustic resonances, efficient AOM can be realized. 
AOM has been conventionally adopted in laser systems, spectroscopy, and atomic physics because of its flexibility in controlling the light direction, intensity, and frequency precisely.

The traditional AOMs are mostly realized by exciting surface acoustic waves (SAW) through IDT fabricated on the surface of a bulk piezoelectric crystal (e.g., fused silica, LiNbO$_3$, chalcogenide glasses). 
They mainly work for free-space light with an operating frequency on the order of 100 MHz and consume a substantial electrical power (a few Watts). 
Thanks to the advances in nano-fabrication and wafer-level deposition of high-quality piezoelectric thin films, it is only within the last decade that the integration of AOM on chip has been successfully demonstrated, extending the modulation frequency beyond 10 GHz \cite{Tadesse:14, ghosh2016, li2015, cai2019, yu2020, sarabalis2020, shao:2020, hassanien2021}. 
Most of these demonstrations excite SAW by fabricating co-planar IDT electrodes as shown in Fig. \ref{SAW}(a), where an alternating electric field is generated which in turn excites periodically oscillating strain field. 
Therefore, the wavelength of the acoustic wave is dictated by the period of the IDT. 
At microwave frequencies, due to the slow acoustic velocity (five orders of magnitude smaller than the speed of light), the acoustic waves have wavelength on the similar order with the optical wavelength, which improves mode overlap and interaction strength \cite{Tadesse:14}. 

Depending on the polarization direction of the propagating soundwave, different acoustic modes can be identified, such as the Rayleigh wave \cite{Tadesse:14}, Lamb wave \cite{park2020} and Love wave \cite{zhang2020}, as illustrated in Fig. \ref{SAW}(b-e). 
While the first two oscillate out of plane, the last one is in plane. 
In the Rayleigh wave, the particles follow an ellipse trajectory with a long axis in the out of plane direction. 
It is primarily a shear (transverse) wave but also has a longitudinal component. 
The Lamb wave is similar as the Rayleigh wave but mainly refers to a plate with finite thickness, such as the released piezoelectric thin film. 
The Love wave, also known as the Shear Horizontal (SH) wave, can be supported in both released and unreleased configurations. 
The excitation of these acoustic modes is determined by the alignment of the electric field with the corresponding component in the piezoelectric coefficient tensor, which should be chosen carefully for specific crystal orientations in order to achieve the highest efficiency.

%Since the acoustic waves are confined along the surface of the substrate, to have efficient optical modulation, the optical waveguides are usually brought as close as possible to the surface of the substrate by either directly etching trenches on the surface of the piezoelectric film \cite{Tadesse:14}, or burying the waveguide in dielectric cladding and within one acoustic wavelength (sub-micron) away from the piezoelectric film \cite{Kittlaus:21}, or fabricating the waveguide directly on top of the piezoelectric layer \cite{yu2020, marinkov2021} (see Fig. \ref{acoustics}(a-c)). These waveguide arrangements are at the expense of scattering light into the environment and increasing the optical loss.

%%%%%%%%%%%%%%%%%%%%%%%%%%%%%%%%%%%%%%%%%%%%%%%%%%%
\begin{figure}[t]
\centering
\includegraphics[width=10 cm]{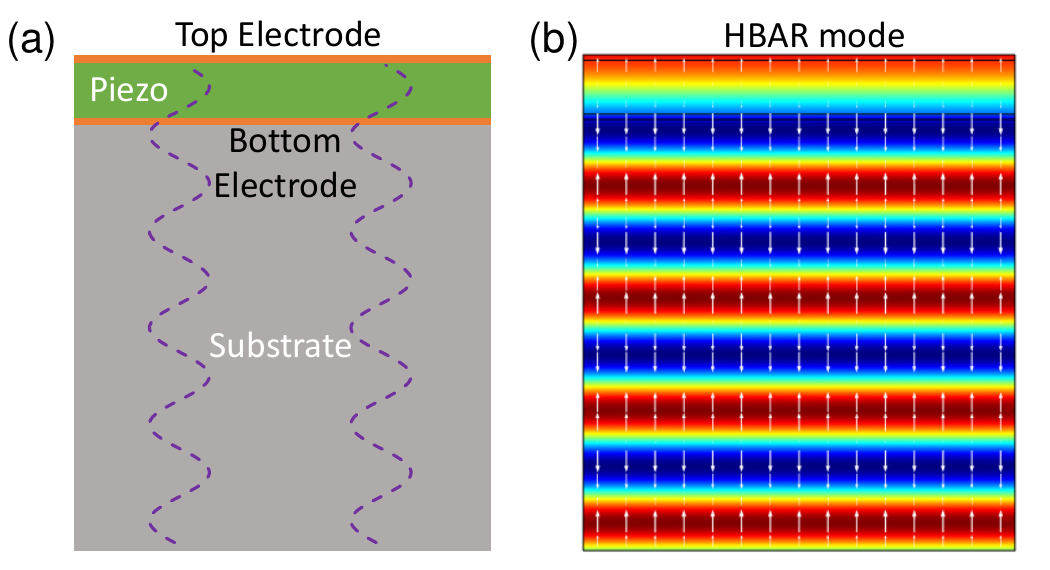}
\caption{
Bulk acoustic wave resonator. 
(a) Cross-section and (b) longitudinal displacement distribution of a typical HBAR mode. }
\label{BAW}
\end{figure}

Apart from SAW, by placing electrodes at the top and bottom of the piezoelectric layer, the acoustic waves can be directed vertically towards the substrate, exciting so-called High-overtone Bulk Acoustic Waves Resonances (HBAR). 
As shown in Fig. \ref{BAW}, an acoustic Fabry-Pérot cavity is naturally formed by the top and bottom surfaces of the substrate, which confines tightly the acoustic energy and forms a rich family of acoustic resonant modes. 
As illustrated in the simulation of Fig. \ref{BAW}(b), an oscillating mechanical displacement wave is excited and occupies uniformly the entire substrate. 
Depending on the piezoelectric material and the placement of the electrodes, the excitation of both longitudinal \cite{Tian:20} and transverse (shear) \cite{mckenna2020} acoustic waves has been demonstrated. 
There is no restriction on the thickness of the substrate, which can be either a thick unreleased substrate (e.g., 200-500 $\mu$m) or a suspended thin film (which is also referred as the Film Bulk Acoustic Waves Resonator, FBAR). 
HBAR has been ubiquitously deployed in modern wireless systems for high-quality RF filters \cite{bi2008bulk, gokhale2021} and oscillators \cite{boudot2016, daugey2015}, and also coupled with quantum systems such as the superconducting (SC) qubits \cite{chu2017, chu2018, chu2021} and the Nitrogen Vacancy centers \cite{macquarrie2013, macquarrie2015}. 
Most recently, efficient AOM by coupling HBAR with photonic microring resonators has been demonstrated \cite{Tian:20, Liu:20}. 

%Different from SAW, HBAR transmits vertically and perpendicular to optical paths. The coupling between vertical acoustic waves and in-plane optical circuits makes it possible for independent optimization of the actuator and optical components. Thus, waveguides can be fully cladded for preserving high quality factor and for desirable dispersion engineering. Furthermore, the high lateral acoustic mode confinement enables low cross-talk and compact integration. However, SAW AOM usually shows higher electromechanical coupling efficiency than BAW AOM, because multiple pairs of electrode fingers can be fabricated in SAW, while BAW only has one pair of electrodes (the top and bottom electrodes). Also, since SAW is tightly confined at the surface, it exhibits smaller mechanical mode volume than BAW, which would also help with the higher electromechanical and optomechanical coupling efficiency. Therefore, depending on the application, the acoustic waves, piezoelectric material, and electrode structure should be selected and designed carefully.

%%%%%%%%%%%%%%%%%%%%%%%%%%%%%%%%%%%%%%%%%%%%%%%%%%%
%%%%%%%%%%%%%%%%%%%%%%%%%%%%%%%%%%%%%%%%%%%%%%%%%%%
\subsubsection{Piezo-Optomechanical Model and Coupling Efficiency}

The excitation of the SAW and bulk acoustic waves (BAW) via the piezoelectric actuator can be described either by the Mason model \cite{tirado2010bulk, Tian:20, ahmed2023} or the Butterworth–Van Dyke (BVD) model \cite{larson:2000, Blesin:21}. 
The Mason model describes the acoustic wave based on the transmission line model which captures the physical nature of the acoustic wave propagation and can be used for predicting the electromechanical response based on the structure. 
Although the Mason model is more general, it is complicated, especially when it is embedded in a complex electric circuit. 
On the other hand, the BVD model is built on a lumped element circuit, which directly relates with the properties of the specific mechanical mode. 
It is an empirical model, which needs to be fitted with the electromechanical response obtained from either the simulation or the experiment. 
Nevertheless, the BVD model is one of the most widely used models in the MEMS community. 

\begin{figure}[b]
\centering
\includegraphics[width=11 cm]{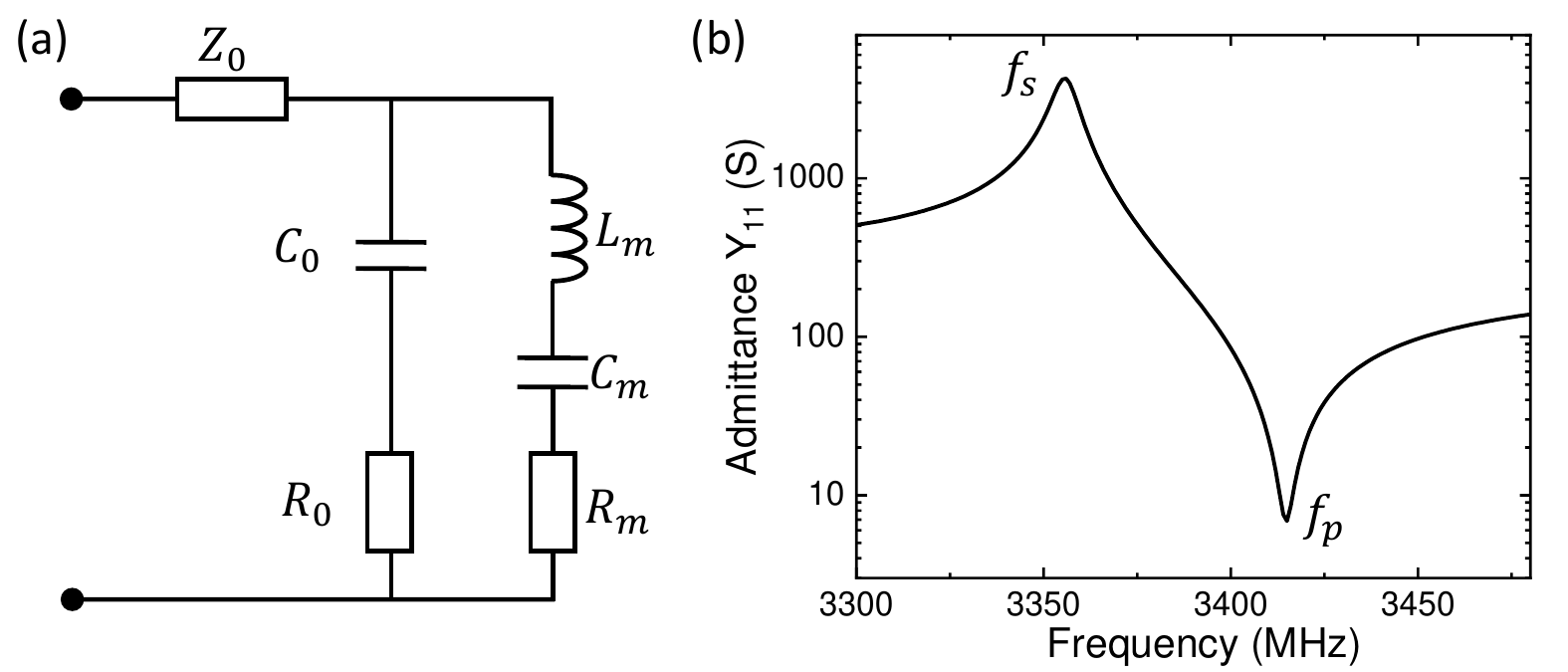}
\caption{
Modified Butterworth–Van Dyke (BVD) model. 
(a) Lumped element circuit of the BVD model. 
(b) Electrical admittance of a typical SAW resonator excited by an IDT. 
The resonance and anti-resonance are labeled as $f_s$ and $f_p$, respectively.
}
\label{BVD}
\end{figure}

The circuit of the BVD model is shown in Fig. \ref{BVD}(a), where an $L_mC_m$ resonator is parallelly connected with a capacitor $C_0$. 
While $C_0$ represents the capacitance of the piezoelectric actuator itself, $L_mC_m$ reflects the mechanical resonance with $L_m$ and $C_m$ being related with the mechanical kinetic and potential energy, respectively. 
Although the original BVD model did not include losses, it was later modified to include losses of the capacitor and mechanical mode as resistors, which is referred to as the modified BVD (mBVD) model \cite{larson:2000}. 
The $R_m$ is related with the intrinsic loss rate $\gamma _{m, i}$ of the mechanical mode as \cite{Blesin:21}:
\begin{equation}
    R_m = L_m\gamma_{m, i}
\end{equation}
The admittance of one typical mechanical mode is illustrated in Fig. \ref{BVD}(b) which consists of one resonance and an anti-resonance. 
The resonance is often referred to as the series resonant frequency $f_s$ that corresponds to the resonant frequency of the series resonator $L_mC_m$ (which is also the mechanical resonance). 
The anti-resonance comes from the interplay between the two parallel branches where their admittance is added to be near zero. 
Thus, this is called the parallel resonant frequency $f_p$. 

The piezoelectric coupling efficiency is conventionally characterized by the figure of merit called $k^2_{t,\text{eff}}$, which is related with the difference between $f_s$ and $f_p$ as:
\begin{equation}
    k^2_{t,\text{eff}}= \frac{\pi ^2}{8}\left(1-\frac{f_s^2}{f_p^2} \right)
\end{equation}
While there are other definitions under different situations, the above is one of the most widely adopted expression \cite{chen2005}. 
From it we can see the larger the spacing between $f_s$ and $f_p$, the higher the $k^2_{t,\text{eff}}$ and thus the stronger the piezoelectric transduction. 
Also, by analyzing the circuit of the BVD model, $f_s$ and $f_p$ can be further related with the value of the lumped elements, such that the $k^2_{t,\text{eff}}$ can be expressed as:
\begin{equation}
    k^2_{t,\text{eff}}= \frac{\pi ^2}{8}\left(\frac{C_m}{C_0+C_m} \right)
\end{equation}
which is approximately proportional to the ratio of the motional capacitance $C_m$ to the actuator's electrical capacitance $C_0$. 
The $k^2_{t,\text{eff}}$ of an acoustic resonator is ultimately limited by the piezoelectric material and is related to its piezoelectric coefficient, elasticity, and permittivity \cite{Blesin:21, arrangoiz2016}. 
The highest coupling efficiency demonstrated to date has been realized on LiNbO$_3$ SAW resonators with efficiency approaching 30$\%$ \cite{zhang2020, hsu2021}. 

\begin{figure}[b]
\centering
\includegraphics[width=11 cm]{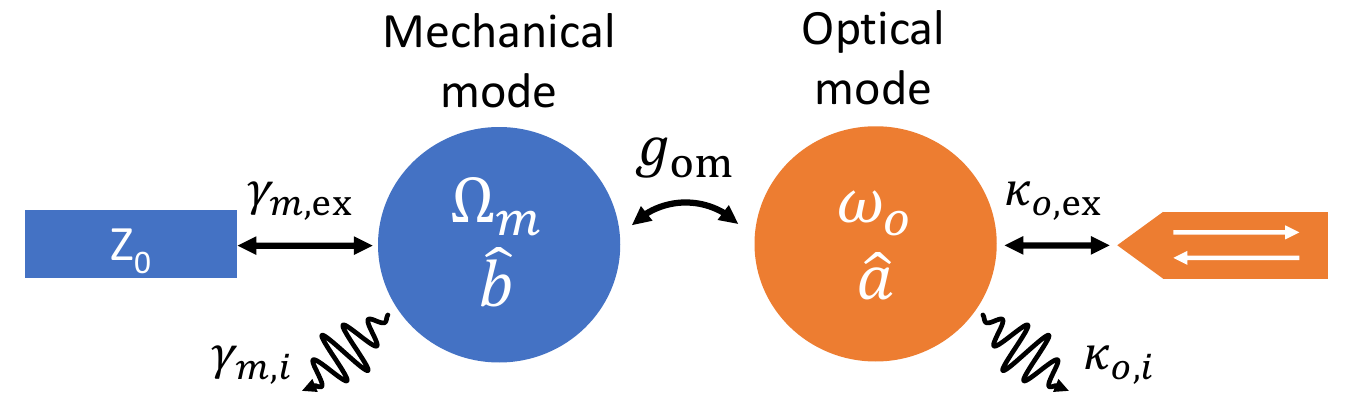}
\caption{
Schematic of the transduction between the mechanical and optical modes, where they are coupled through the optomechanical coupling rate $g_{\text{om}}$. 
While the mechanical mode is driven via a microwave transmission line (with impedance of $Z_0$) at the rate of $\gamma_{m,\text{ex}}$, the optical mode is coupled through an optical waveguide at the rate of $\kappa_{o,\text{ex}}$. 
The resonant angular frequency of the optical (mechanical) mode is $\omega_o$ ($\Omega_m$). 
}
\label{mode}
\end{figure}

In addition to the BVD model, the mechanical resonator can also be described by input-output theory, where the mechanical mode is parameterized by its frequency and amplitude of the intra-cavity field \cite{wollack2021loss}. 
It is more convenient for modeling the coupling between the mechanical and optical modes, which is not only applicable for classical AOM but is also compatible with the quantum transduction. 
As shown in Fig. \ref{mode}, the mechanical and optical modes are coupled at the rate determined by the optomechanical coupling strength enhanced by the intra-cavity optical pump photon number $n_{o,p}$, such that $g_{\text{om}}=g_0\sqrt{n_{o,p}}$, where $g_0$ can be found from Eq. \ref{eqg0}. 
The dynamics of the mechanical (optical) mode can be described by its field amplitude $\hat{b}$ ($\hat{a}$) governed by the Heisenberg-Langevin equations of motion \cite{Shao:19, aspelmeyer2014, Fan:18, safavi2013, safavi2019}:
\begin{eqnarray}
\label{eqmotion1}
     \Dot{\hat{a}} &=&  -\left(i\Delta +\frac{\kappa_o}{2} \right)\hat{a}-ig_0\hat{a}\left(\hat{b}+\hat{b}^{\dagger} \right)+\sqrt{\kappa_{o,\text{ex}}}\hat{a}_{\text{in}}\\
     \Dot{\hat{b}} &=&  -\left(i\Omega_m +\frac{\gamma_m}{2} \right)\hat{b}-ig_0\hat{a}^{\dagger}\hat{a}+\sqrt{\gamma_{m,\text{ex}}}\hat{b}_{\text{in}}
     \label{eqmotion2}
\end{eqnarray}
where $\kappa_o$ (= $\kappa_{o,i}+\kappa_{o,\text{ex}}$) and $\gamma_m$ (= $\gamma_{m,i}+\gamma_{m,\text{ex}}$) are the total loss of the optical and mechanical modes, respectively. 
A rotating frame around the frequency of the pump light is adopted such that $\Delta=\omega_{\text{cav}}-\omega_p$ is the relative detuning between the pump laser and the optical cavity. 
$\hat{a}_{\text{in}}$ ($\hat{b}_{\text{in}}$) represents the input field in the optical waveguide (microwave transmission line). 
They are normalized such that $\left<\hat{a}_{\text{in}}^{\dagger}\hat{a}_{\text{in}}\right>=P_o/\hbar\omega_p$ ($\left<\hat{b}_{\text{in}}^{\dagger}\hat{b}_{\text{in}}\right>=P_m/\hbar\Omega_p$) is the photon flux of the pump light (microwave signal) \cite{aspelmeyer2014}. 
With the same normalization, $\left<\hat{a}^{\dagger}\hat{a}\right>$ ($\left<\hat{b}^{\dagger}\hat{b}\right>$) is the average intra-cavity photon number (phonon number). 

The above model can be used to describe quantum transduction between optical and mechanical modes by treating $\hat{a}$ and $\hat{b}$ as the annihilation operator. 
It can also be easily incorporated into other quantum systems, such as the SC qubit \cite{arrangoiz2016}. 
On the other hand, the classical AOM can be modeled simply by taking the time average on the field amplitude and removing the hat: $a=\left<\hat{a}\right>$, $b=\left<\hat{b}\right>$ \cite{Shao:19, aspelmeyer2014}. 
Moreover, the model is generic to any optomechanical system, regardless of the structure, size, and shape of the optical and mechanical modes, such as the Fabry-Pérot cavity, the microring resonator, or optomechanical crystal. 
All we need to do is extracting the parameters of the system through simulation and/or experiments. 
One of the important parameters is the external coupling rate from the mechanical mode to the microwave transmission line with impedance of $Z_0$. 
Its relation with the electromechanical BVD model has been derived in Refs. \cite{Blesin:21, wollack2021loss}:
\begin{equation}
    \gamma_{m,\text{ex}}=\frac{Z_0}{L_m}=\Omega_m^2C_mZ_0\approx\Omega_m^2\frac{8}{\pi^2}k^2_{t,\text{eff}}C_0Z_0
\end{equation}
From the above equation we can see that, larger electromechanical coupling $k^2_{t,\text{eff}}$ leads to stronger external coupling rate. 

These differently derived piezo-optomechanical models can find uses under different circumstances. 
The BVD model can be easily incorporated into other circuits and simplify the problem as the circuit design. 
For example, acoustic devices have been successfully introduced into superconducting circuits, emerging as a new field of circuit quantum acousto-dynamics (circuit QAD) \cite{o2010quantum, chu2017, bienfait2019, arrangoiz2019}. 
Moreover, the circuit model that describes the optomechanical interaction has been established \cite{zeuthen2018, wu2020}. 
On the other hand, the input-output formalism provides a physical intuition and is more closely related with the experimental parameters. 
Therefore, depending on the specific problem, the right model should be chosen, and sometimes the combination of them will be beneficial. 

%%%%%%%%%%%%%%%%%%%%%%%%%%%%%%%%%%%%%%%%%%%%%%%%%%%
%%%%%%%%%%%%%%%%%%%%%%%%%%%%%%%%%%%%%%%%%%%%%%%%%%%
%%%%%%%%%%%%%%%%%%%%%%%%%%%%%%%%%%%%%%%%%%%%%%%%%%%
\section{Ultra-fast Piezoelectric Tuning of Integrated Photonic Devices}
%JL: this section should contain or better be started with an introduction on integrated photonics, particularly silicon nitride

In the last decades, integrated photonics has evolved into a mature technology enabling the synthesis, processing and detection of optical signals on-chip.
Currently developed PICs have enabled nearly all core functionalities via heterogeneous or hybrid integration of different material platforms \cite{Smit:14,  Horikawa:16, Park:20, Xiang:22a}, and is now utilized at a commercial level for telecommunications and datacenters. 
A second wave is currently under way in which the nonlinearity of PICs becomes relevant for applications, i.e. \emph{integrated nonlinear photonics}. 
The Kerr, Pockels and Brillouin nonlinearities inherent in these waveguides enable new capability for nonlinear optical signal generation and processing \cite{Moss:13, Eggleton:19, Gaeta:19, zhu2021}.

Among all material platforms developed for integrated photonics so far, silicon, indium phosphide (InP) and Si$_3$N$_4$ are the three leading platforms. 
Compared with silicon and InP which is commonly used to build active elements such as lasers, modulators and photodetectors, Si$_3$N$_4$ has limited capability due to its insulator and amorphous nature.
Nevertheless, Si$_3$N$_4$ is extremely excellent in optical losses -- it has an ultralow linear optical loss based on high-quality films grown via chemical vapor deposition (CVD), and a negligible nonlinear loss due to its 5 eV bandgap that makes Si$_3$N$_4$ immune from two-photon absorption (TPA) in the telecommunication wavelength around 1550 nm.
The exceptional low losses in Si$_3$N$_4$ are currently not obtainable in any other material including silicon and InP. 
In addition, the refractive index of Si$_3$N$_4$ ($n(\text{Si}_3\text{N}_4)=2.0$) is lower than the refractive indices of silicon and InP ($n(\text{Si})=3.5$, $n(\text{InP})=3.1$). 
The lower refractive index enables lower optical confinement as a con, but it meanwhile reduces the mode mismatch to an optical fiber mode and can facilitate the fiber-chip edge coupling.

Yet, Si$_3$N$_4$ is all passive, however few active function can still be realized on Si$_3$N$_4$ in combination of other materials, thanks to the maturity and versatility of integrated photonics. 
For example, modulators can be realized on Si$_3$N$_4$ with integrated 2D-materials (such as graphene) \cite{Gruhler:13, Phare:15, Datta:20, He:21}, ferroelectric PZT \cite{hosseini2015, alexander2018, Jin:18, lihachev2022}, piezoelectric aluminium nitride \cite{stanfield2019cmos, Tian:20, Liu:20a} and electro-optic LiNbO$_3$ \cite{chang2017, Churaev:23, Snigirev:23}.
Multiple approaches have been demonstrated to combine Si$_3$N$_4$ photonic circuits with lasers via hybrid \cite{Stern:18, Raja:19, Voloshin:21, Shen:20, Fan:20, xiang:2020} or heterogeneous integration \cite{Xiang:21, Xiang:22a}. 
Today, Si$_3$N$_4$ has emerged as a leading material for integrated linear and nonlinear \cite{Moss:13, Xiang:22a} photonics due to its low loss and superior optical power handling capability \cite{Gyger:20}. 
To date, among all integrated platforms \cite{Kovach:20}, optical losses below 1 dB/m have been demonstrated in Si$_3$N$_4$ PIC \cite{Spencer:14, Gundavarapu:19, bauters2011, Jin:21, Xuan:16, Ji:17, Liu:20b, Ye:21a}.

Although bulk piezoelectric actuators have been widely used in today's commercial laser systems for tuning laser frequencies, it is only most recently that they are integrated on chip for tuning PIC, due to the maturity of depositing high-quality piezoelectric thin films on Si substrates. 
This chapter reviews the progress made in piezoelectric tuning of PIC made from Si and Si$_3$N$_4$. 
AlN and PZT are the two main piezoelectric materials that have been extensively explored to date, and will be discussed separately in the following sections.  

%%%%%%%%%%%%%%%%%%%%%%%%%%%%%%%%%%%%%%%%%%%%%%%%%%%
%%%%%%%%%%%%%%%%%%%%%%%%%%%%%%%%%%%%%%%%%%%%%%%%%%%
\subsection{Unreleased AlN Actuator}

%%%%%%%%%%%%%%%%%%%%%%%%%%%%%%%%%%%%%%%%%%%%%%%%%%%
\begin{figure}[b]
\centering
\includegraphics[width=13 cm]{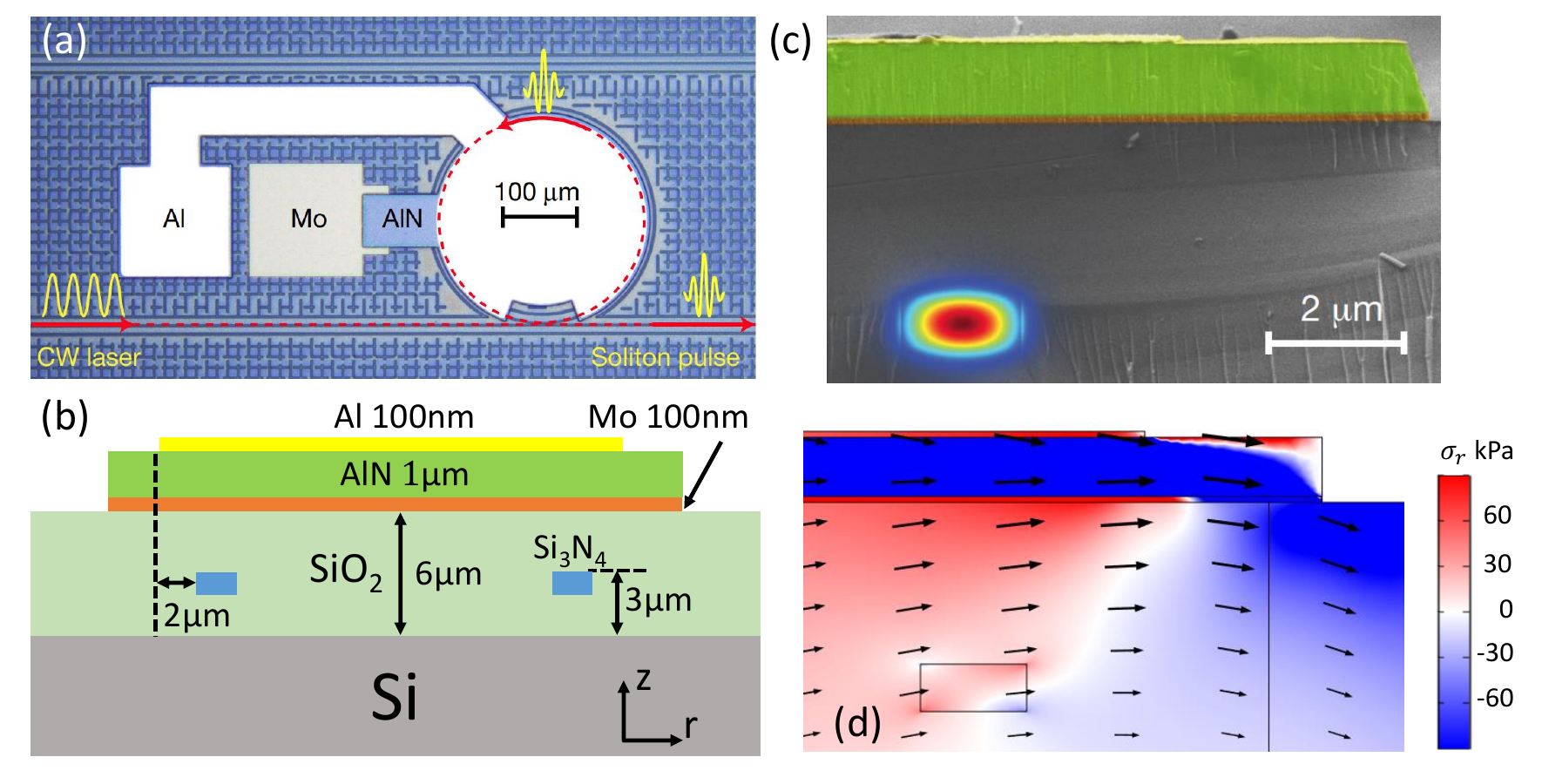}
\caption{
Design of an AlN piezoelectric actuator on Si$_3$N$_4$. 
(a) Optical microscope image of a disk AlN actuator covering a Si$_3$N$_4$ microring resonator (red dashed line). 
(b) Cross-section of the device with marked scale. 
(c) False-color SEM image of the cross section of the fabricated device with the simulated TE$_{00}$ optical mode overlapping on the optical waveguide. 
(d) Numerical simulation of the horizontal stress distribution around the optical waveguide under +60 V DC biasing. 
The overlaid black arrows denote the local mechanical displacement.
Panel (a, c) are reprinted from Ref. \cite{Liu:20a}.
}
\label{Ch3_device}
\end{figure}
%%%%%%%%%%%%%%%%%%%%%%%%%%%%%%%%%%%%%%%%%%%%%%%%%%%

%%%%%%%%%%%%%%%%%%%%%%%%%%%%%%%%%%%%%%%%%%%%%%%%%%%
\subsubsection{Actuator Design and DC Tuning}
As the well-known piezoelectric material, AlN has been widely deployed in modern acoustic RF filters due to its mature foundry-level process. 
We demonstrated for the first time the tuning of a Si$_3$N$_4$ microring resonator by integrating an AlN piezoelectric actuator on top \cite{Tian:20, Liu:20a}. 
One typical fabricated device is shown in Fig. \ref{Ch3_device}(a), where a disk shaped actuator overlaps on top of the microring resonator. 
The bus-waveguide-to-microresonator coupling region is opened in order to prevent any actuation-induced perturbation of light coupling (however, the impact might be very negligible). 
From the cross-section in Fig. \ref{Ch3_device}(b), 1-$\mu$m-thick AlN film is sandwiched between between top Aluminum (Al, 100 nm thick) and bottom Molybdenum (Mo, 100 nm thick) metal layers. 
Molybdenum is chosen to minimize the acoustic impedance with AlN compared to gold or titanium. 
Ultraow-loss Si$_3$N$_4$ waveguides are fabricated using the photonic Damascene process \cite{Pfeiffer:18b, Pfeiffer:18, Liu:20b}. 
The optical waveguide has a dimension of 0.8$\times$1.8 $\mu$m$^2$, which is fully buried in a 6-$\mu$m-thick SiO$_2$ cladding and placed 2 $\mu$m within the edge of the top metal. 
The 3-$\mu$m-thick SiO$_2$ above the waveguide helps to prevent metal absorption from the Mo layer. 
The entire device sits on a solid 230-$\mu$m-thick Si substrate, which is called an ``\textbf{unreleased}'' structure in the MEMS community. 
The false-colour SEM image of the cross-section of the device is shown in Fig. \ref{Ch3_device}(c), including the optical mode in the Si$_3$N$_4$ waveguide.

%%%%%%%%%%%%%%%%%%%%%%%%%%%%%%%%%%%%%%%%%%%%%%%%%%%
\begin{figure}[b]
\centering
\includegraphics[width=12 cm]{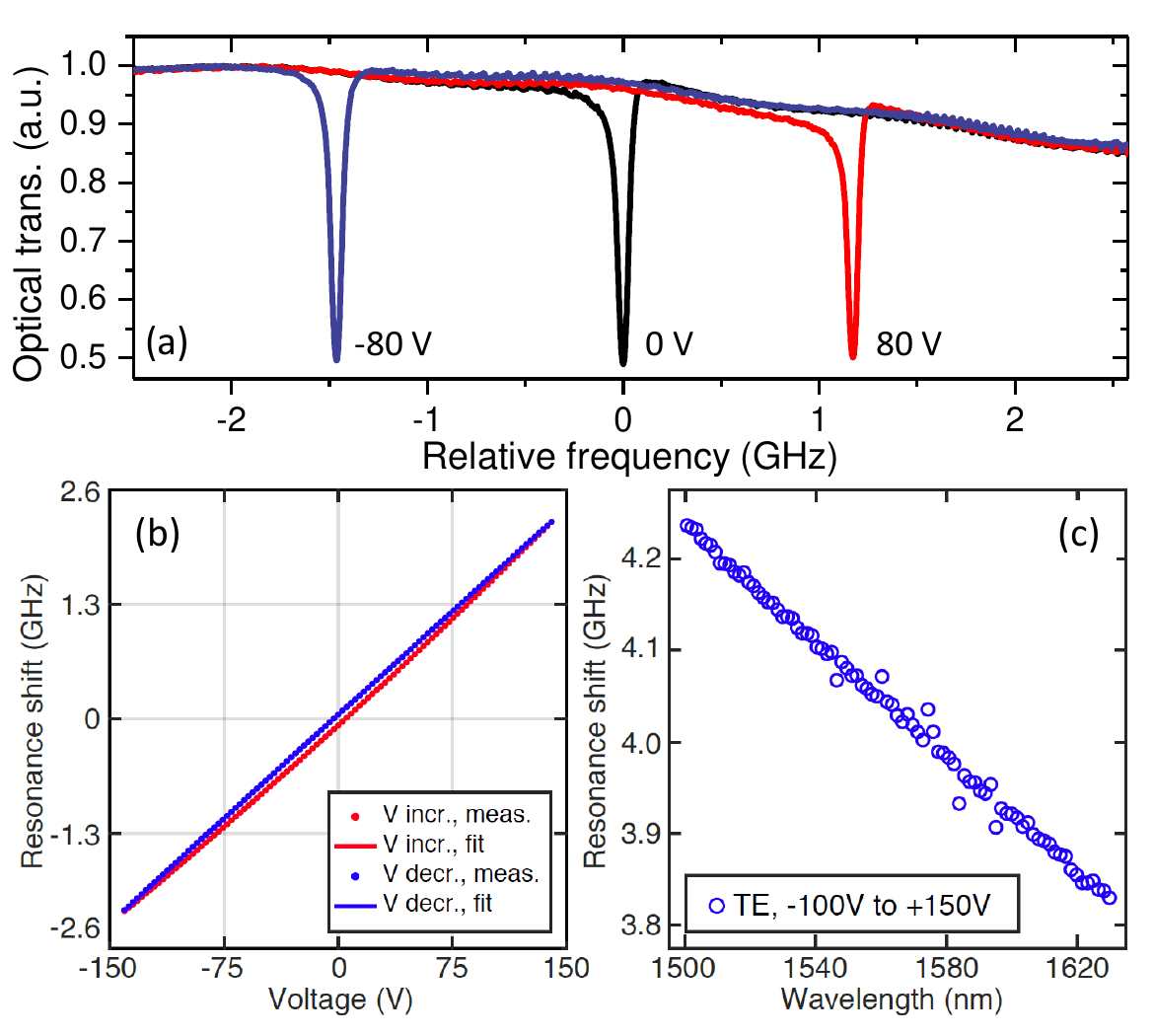}
\caption{
Piezoelectric tuning of the optical resonance. 
(a) Optical transmission of one optical TE$_{00}$ resonance under various voltages. 
(b) Resonance shift versus applied voltages when voltage increases (red) and then decreases (blue), showing a small hysteresis. 
(c) Total resonance shift for resonances with different center wavelengths when the voltage increases from $-100$ V to $+150$ V. 
The wavelength-dependent tuning is due to a 5.3 MHz change of the FSR under the piezoelectric actuation.
Panel (b, c) are reprinted from Ref. \cite{Liu:20a}.
}
\label{Ch3_tuning}
\end{figure}
%%%%%%%%%%%%%%%%%%%%%%%%%%%%%%%%%%%%%%%%%%%%%%%%%%%

Polycrystalline AlN thin film and Mo are sputtered on the SiO$_2$ cladding which has been polished via chemical-mechanical planarization (CMP) before the sputtering. 
The sub-nanometer roughness of the SiO$_2$'s surface helps maintain good crystal orientation of the AlN film, which usually has an X-Ray diffraction (XRD) rocking curve peak width around 2 degrees. 
The AlN has a wurtzite structure with $c$-axis pointing vertically in the positive $z$-axis. 
Upon applying a positive electrical voltage on the top metal (while bottom metal is grounded), an electric field along the negative $z$-direction is formed. 
Since AlN has a positive piezoelectric coefficient $d_{33}$, it will be squeezed in the $z$-direction and expand horizontally (AlN has a positive Poisson ratio). 
From the FEM numerical simulation in Fig. \ref{Ch3_device}(d), extensive (positive) stress and stress gradient are generated around the optical waveguide, and a horizontal mechanical displacement pushes the microring outwards. 
On the other hand, when applying negative voltages, AlN actuator changes from expanding to shrinking, such that the stress in the above analysis changes sign. 
In this way, bi-directional tuning can be achieved by reversing the applied voltage's sign. 
The optical mode distribution can be similarly simulated, and the fundamental transverse electric (TE$_{00}$) mode is shown in Fig. \ref{Ch3_device}(c). 
By combining the mechanical and optical simulations, the tuning of the resonant frequency of the optical microring can be estimated by Eq. \ref{eq1.20}, taking into account the photoelastic and moving boundary effects.  

The piezoelectric tuning of the TE$_{00}$ mode of the device in Fig. \ref{Ch3_device} is shown in Fig. \ref{Ch3_tuning}(a). 
When applying positive 80 V, the resonant frequency increases by 1.2 GHz, which corresponds to the blue shift. 
With negative 80 V, the resonance shifts to a lower frequency, showing bi-directional tuning. 
It can be seen the linewidth of the resonance is maintained during the tuning which means the high optical $Q$ is unaffected. The estimated intrinsic quality factor, $Q_0>15\times10^6$, with integrated AlN actuators is identical to bare microresonators without AlN \cite{Liu:18a}, demonstrating that the monolithically integrated AlN actuators are compatible with ultralow-loss Si$_3$N$_4$ waveguide platforms. 
Moreover, the DC current draw is maintained below 1 nA, with power consumption at the level of tens of nano-Watts, allowing large-scale integration and cryogenic applications. 
The dependence of resonance shift on the applied voltages is plotted in Fig. \ref{Ch3_tuning}(b), where a highly linear tuning is obtained up to 140 V with tuning efficiency of 15.7 MHz/V. 
However, a small hysteresis is observed when sweeping the voltage back and forth. 
While AlN is non-ferroelectric, the hysteresis can be caused by accumulated charges at the interfaces between AlN and metals and the grain boundaries of the polycrystalline AlN. 
The tuning range at different wavelengths is shown in Fig. \ref{Ch3_tuning}(c), which linearly decreases at longer wavelength. 
This results from a 5.3 MHz change of the microresonator's FSR (191.0 GHz). 
The linear dependence of the tuning on wavelength indicates no observable change of the dispersion (where the $\Delta$FSR varies with wavelength). 
While the capability of controlling the dispersion could find application in versatile programming of the frequency comb's spectrum \cite{yao2018gate}, the linear tuning of the FSR is preferred for stabilization of the repetition rate of the frequency comb \cite{Newman:19}. 

%%%%%%%%%%%%%%%%%%%%%%%%%%%%%%%%%%%%%%%%%%%%%%%%%%%
\begin{figure}[b]
\centering
\includegraphics[width=9 cm]{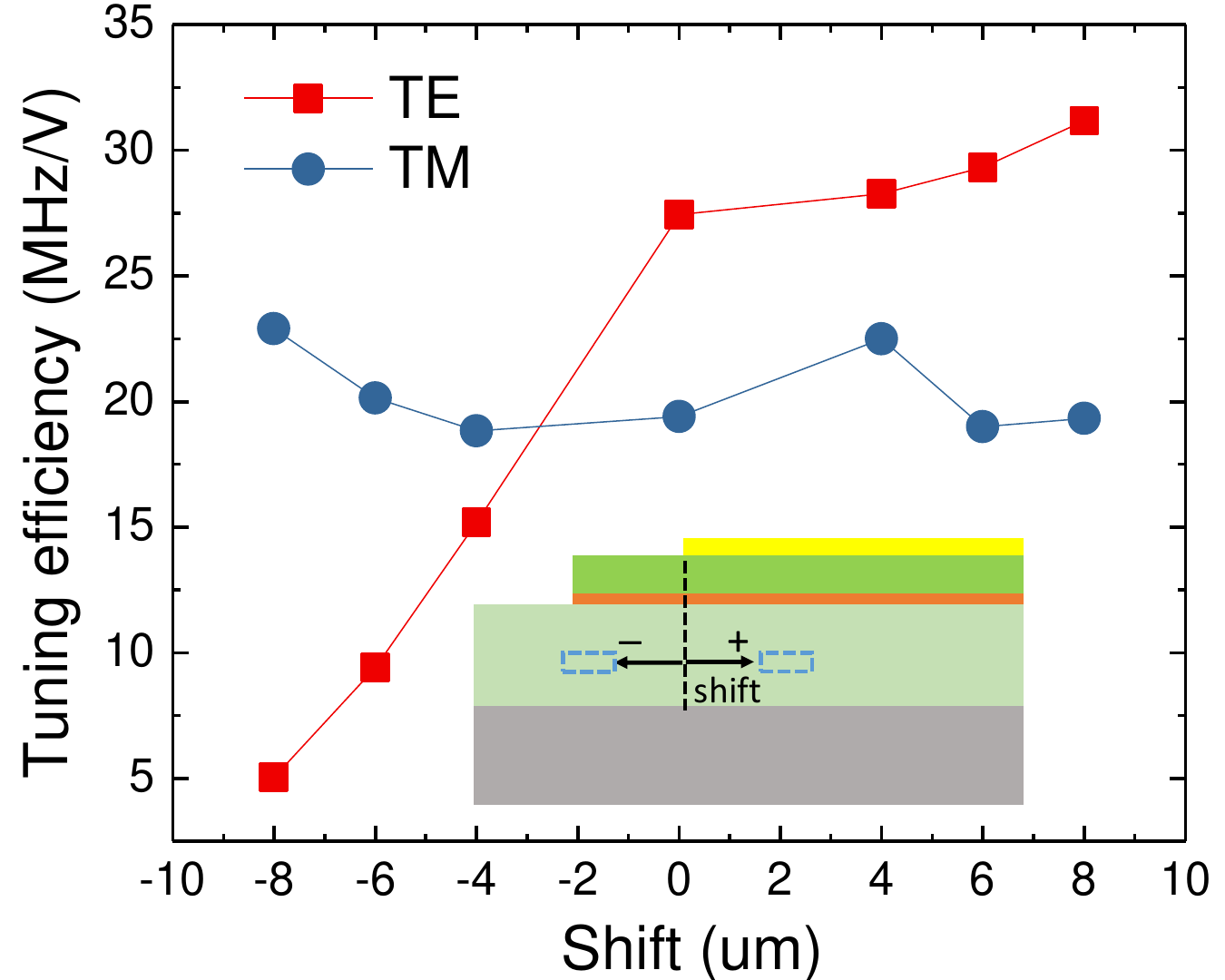}
\caption{
Tuning efficiency under various relative positions between the optical waveguide and the actuator for TE$_{00}$ (red) and TM$_{00}$ (blue) modes. 
The inset shows the definition of the relative shift of the optical waveguide (blue dashed box).
}
\label{Ch3_position}
\end{figure}
%%%%%%%%%%%%%%%%%%%%%%%%%%%%%%%%%%%%%%%%%%%%%%%%%%%

It can be seen from the stress simulation in Fig. \ref{Ch3_device}(d) that the stress varies dramatically around the edge of the actuator. 
It is thus necessary to optimize the position of the optical waveguide relative to the actuator.
We experimentally fabricated and measured devices with different waveguide positions as shown in Fig. \ref{Ch3_position}. 
As the waveguide moves from outside of the actuator to inside (negative shift to positive), the tuning efficiency for TE$_{00}$ mode increases and saturates as it move more inside, where the stress becomes more uniform. 
A maximum of 32 MHz/V tuning of TE$_{00}$ mode is achieved for +8 $\mu$m shift. 
Differently, the tuning maintains nearly constant around 20 MHz/V for TM$_{00}$ mode due to the anisotropy of the photoelastic effect. 
Therefore, it is preferred to place the waveguide well inside the actuator for better tuning efficiency.  
\begin{figure}[b]
\centering
\includegraphics[width=11 cm]{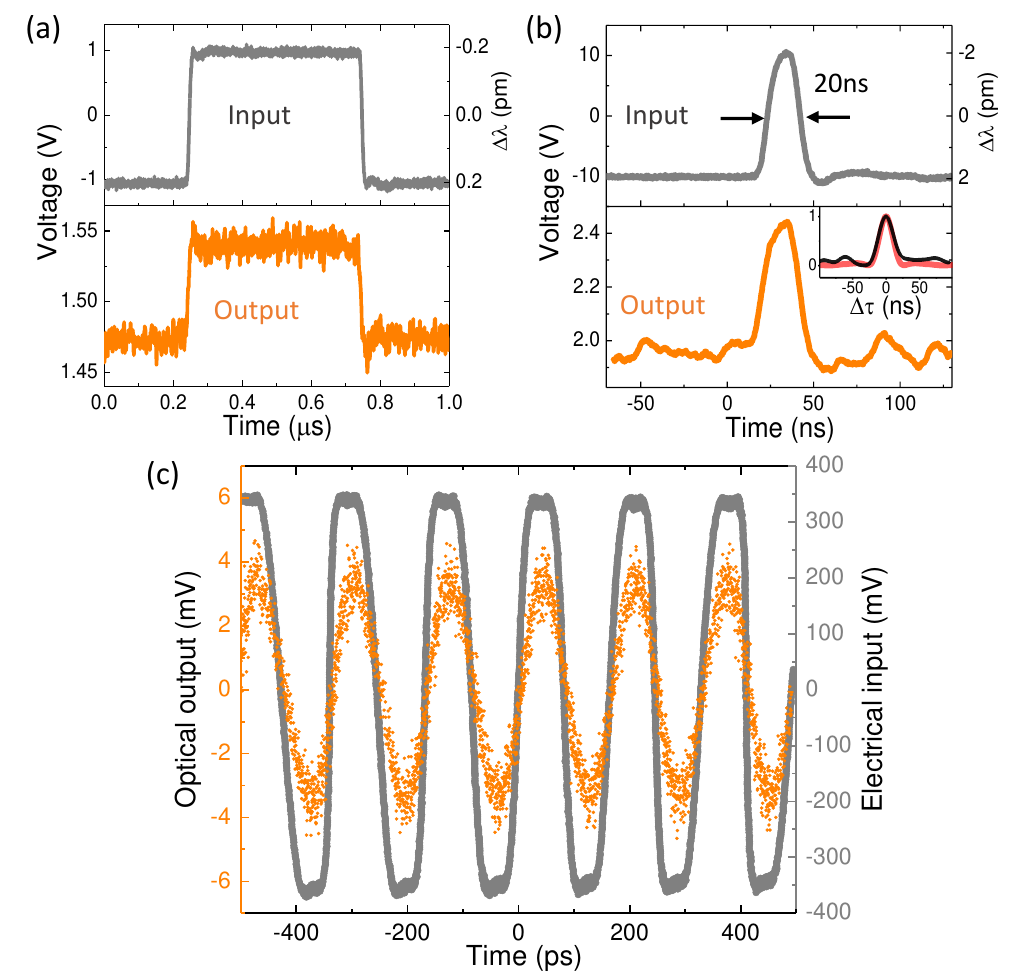}
\caption{
Time-domain response of the actuator. 
Modulation of the output light intensity (orange) under the electrical driving of (a) 1 MHz, 2 V V${_\text{pp}}$ square wave (gray), and (b) 20 ns short pulses with 20 V V${_\text{pp}}$ and 5 MHz repetition rate. 
The inset in (b) shows the normalized cross-correlation (black) between input and output signals and the auto-correlation (red) of the input signal. 
The right $Y$ axis in (a), (b) denotes the relative shift of the resonant wavelength. 
(c) Modulation under 6 GHz square wave. 
The electrical signal (gray) is measured after 100 times attenuation of the original signal. 
Reprinted from Ref. \cite{Tian:20}
}
\label{Ch3_time}
\end{figure}
%%%%%%%%%%%%%%%%%%%%%%%%%%%%%%%%%%%%%%%%%%%%%%%%%%%
\subsubsection{Tuning Speed}
In addition to low power, another key advantage of piezoelectric tuning over thermal tuning is the high tuning speed (i.e. wide actuation bandwidth). 
The speed mainly relies on three factors including RC constant time, photon lifetime, and mechanical transduction time. 
Due to the small area of the device and small dielectric constant of AlN ($\epsilon _r =9$), the capacitor is estimated to be 10 pF which guarantees fast electrical response of $\sim$10 ps. 
The photon lifetime is specific for optical resonators and can be estimated by $\tau _{ph}=\lambda Q/2\pi c$ \cite{xiong2012}, which is $\sim$0.1 ns with optical $Q\sim 10^5$. 
This is not a limitation for wide-bandwidth photonic devices such as Mach-Zehnder interferometers (MZI). 
The high acoustic wave velocity in SiO$_2$ (5900 m/s) guarantees fast ($\sim$0.5 ns) building up of the stress field around the waveguide, theoretically allowing sub-ps response. 
Based on these observations, an overall sub-nanosecond dynamic response can be expected.

The actuation speed can be characterized by studying the time domain dynamic response of the device as shown in Fig. \ref{Ch3_time}. 
Under the driving of a small signal (V${_\text{pp}}$=2 V) 1 MHz square wave while biasing the laser at the slope of the optical resonance, the output light intensity is modulated [see Fig. \ref{Ch3_time}(a)]. 
The output follows the input square wave and shows sharp switching between the two voltage levels with high signal to noise ratio (SNR). 
Furthermore, 5 MHz electrical pulses with 20 ns width are applied in Fig. \ref{Ch3_time}(b). 
To quantitatively characterize how well the actuator responds to the input electrical signal, the normalized cross-correlation between the input and output signals is calculated as shown in Fig. \ref{Ch3_time}(b) inset. 
The similarity between the auto-correlation (red) of the input signal itself and the cross-correlation (black) demonstrates ultra-fast actuation. 
The ultimate tuning speed of the actuator is tested by applying 6 GHz square wave (V${_\text{pp}}$=7 V) as in Fig. \ref{Ch3_time}(c). 
The fact that the output optical modulation (orange) shows oscillation at the same repetition rate of 6 GHz suggests that the piezoelectric actuator is capable of high-speed signal transduction. 

\begin{figure}[b]
\centering
\includegraphics[width=13 cm]{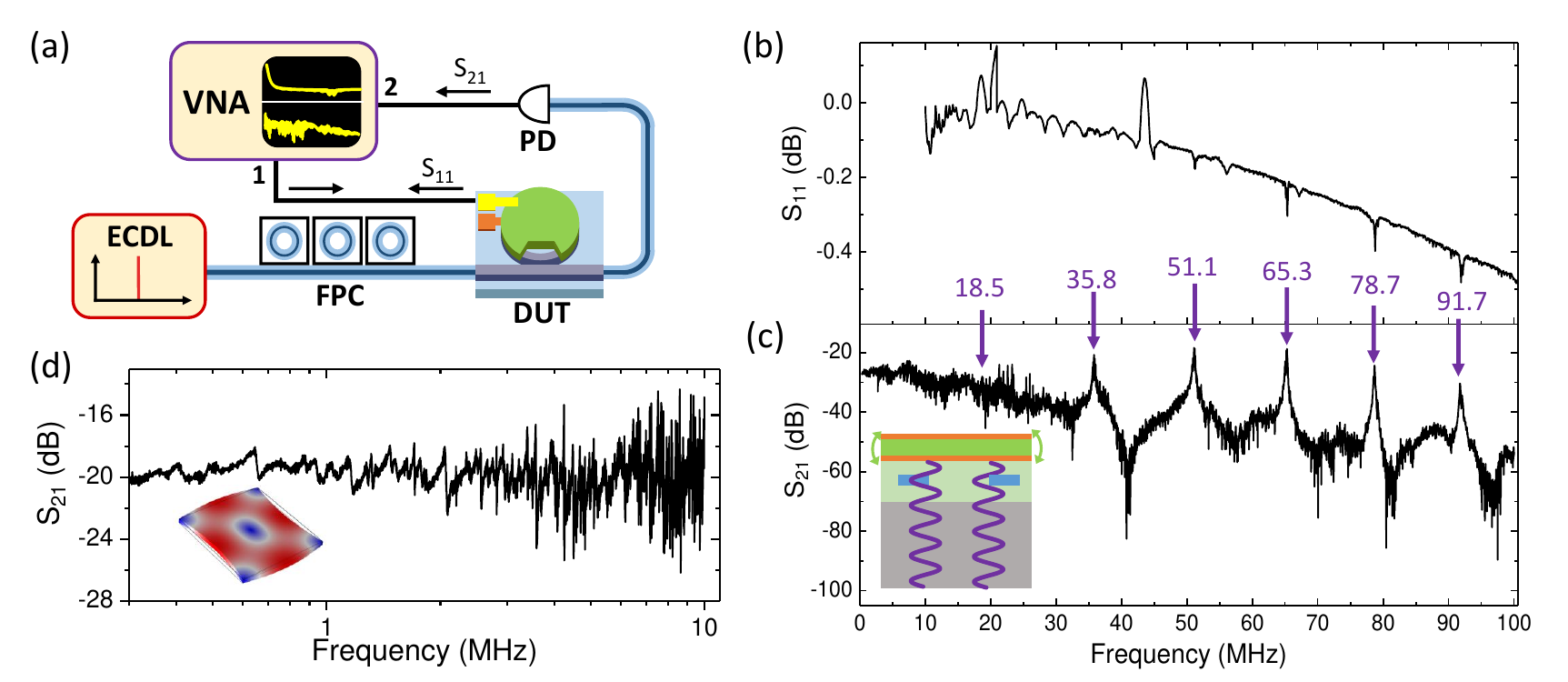}
\caption{
Frequency-domain response of the piezoelectric actuator. 
(a) Schematic of the setup for frequency-response measurement. 
ECDL: external-cavity diode laser. 
PC: polarization controller. 
DUT: device under test. 
PD: photo-diode. 
VNA: vector network analyzer. 
(b) Electromechanical S$_{11}$ and (c) optomechanical S$_{21}$ spectrum response of the actuator. 
The positions and frequencies of each HBAR modes are as labeled. 
Inset in (c) is the schematic of the HBAR mode formed in the Si substrate. 
(d) Zoom-in of S$_{21}$ below 10 MHz with frequency plotted in log scale. 
Inset shows one typical mechanical vibration mode of the photonic chip at 755 kHz. 
(b-d) reprinted from Ref. \cite{tianthesis}. 
}
\label{Ch3_frequency}
\end{figure}

In practical applications, linear transduction is often required, which not only maintains high purity of input electrical signals, but also eases the design of driving electric circuits and signal post-processing. 
This corresponds to a flat response in frequency domain. 
Despite of the high transduction speed of the piezoelectric actuator, it is observed that there exist sharp spikes at tuning edges when switching from one voltage state to the other [see Fig. \ref{Ch3_time}(a)]. 
They come from the ringing of mechanical resonances excited in the bulk Si substrate and the vibration of the entire photonic chip (with size 5$\times$5 mm$^2$) itself. 

The frequency response can be measured using the setup in Fig. \ref{Ch3_frequency}(a), where the scattering parameters of the actuator are characterized by a vector network analyzer (VNA). 
The electromechanical S$_{11}$ measures the reflection of the electrical energy from the actuator while driving with port 1 of the VNA. 
As shown in Fig. \ref{Ch3_frequency}(b), mechanical resonances are found from the dips where the electrical energy is converted to the acoustic energy into the device. 
These resonances will manifest themselves in the modulation of light as measured by the optomechanical S$_{21}$ response in Fig. \ref{Ch3_frequency}(c). 
A series of equally spaced mechanical resonances can be found which align with the dips in S$_{11}$. 
These resonances are the HBAR modes formed inside the thick Si substrate which works naturally as an acoustic Fabry-Pérot cavity. 
As in optics, the spacing between them (also know as the free spectral range, FSR) is inversely proportional to the Si thickness. 
The first harmonic of the HBAR mode locates at 18.5 MHz. 
Due to the high quality of the single crystalline Si substrate, a mechanical $Q\sim100$ is obtained, and the high signal contrast between the resonances and anti-resonances severely limits the flatness of the frequency response and thus the linear transduction bandwidth.  
The mechanical Q is mainly influenced by the scattering losses occurring at the interfaces of the different layers, which depend on the smoothness of each layer's surface. Additionally, the damping of the HBAR mode by the bottom chip holder affects the Q, which is related with how the chip is mounted. These loss mechanisms can be exploited to further suppress the mechanical resonances. 
Despite of being undesired at low frequencies, these HBAR modes extend up to several GHz which opens a new avenue for efficient acousto-optic modulation as will be discussed later.  

Bulk mechanical modes of the entire chip at lower frequencies (100s kHz) will also be excited by the piezoelectric actuator as shown in Fig. \ref{Ch3_frequency}(d). 
Multiple mechanical modes can be supported including the flexural mode, Lamé mode, and face-shear mode \cite{wu2020mems}. 
While the vibration of the flexural mode is mainly out of plane, that of the Lamé and face-shear mode is in plane. 
The profile of one of the typical Lamé modes is shown in the inset of Fig. \ref{Ch3_frequency}(d). 
Both the HBAR and chip modes will distort the transduced signal, in particular in cases where broadband signals such as square and triangle wave modulation is required, and need to be mitigated carefully in practical applications. 

\begin{figure}[b]
\centering
\includegraphics[width=13 cm]{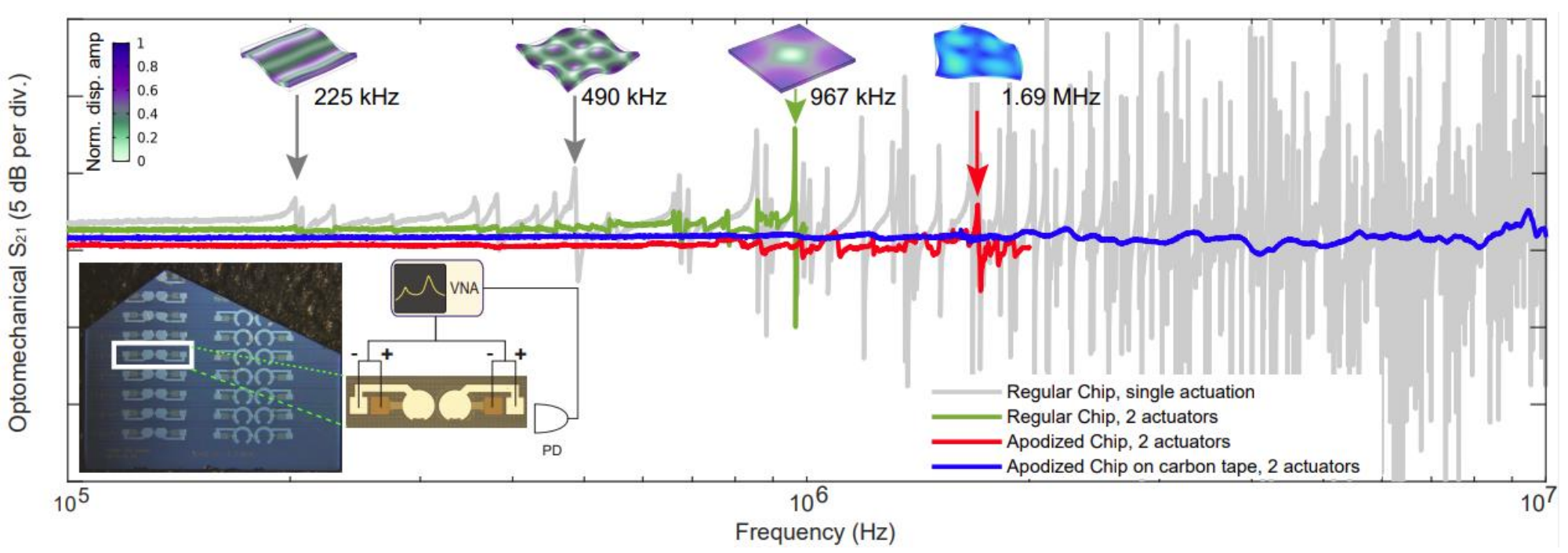}
\caption{
Strategies for removing chip modes. 
Optomechanical S$_{21}$ responses with single actuation (gray) and dual actuators with differential drive on a regular square chip (green), an apodized chip (red), and an apodized chip on a carbon tape (blue). 
Inset: numerical simulation of three mechanical modes of the square chip and the apodized chip (1.69 MHz); 
photo of the apodized chip with the dual-actuator configuration; 
experimental schematic for differential driving of dual actuators.
Reprinted from Ref. \cite{lihachev2022}.
}
\label{Ch3_strategy}
\end{figure}

%%%%%%%%%%%%%%%%%%%%%%%%%%%%%%%%%%%%%%%%%%%%%%%%%%%
\subsubsection{Strategies to Improve the Tuning Speed}

As mentioned in the last section, the high density of mechanical modes of the chip at MHz frequencies limits the actuation optomechanical bandwidth. 
In Ref. \cite{lihachev2022}, we investigated a combination of strategies to suppress these chip mechanical modes which resulted in flat response up to 10 MHz. 
At low frequency, the actuator is relatively small compared with the wavelength of the acoustic wave, so it can be treated as a point source that locally emits acoustic waves. 
As inspired by Ref. \cite{wilson2011} where the flexural modes of a nanomechanical membrane are suppressed by the far-field destructive interference between elastic waves, the flexural mode can be mitigated by differentially driving of two adjacent actuators. 
As shown in the optical image of the insets in Fig. \ref{Ch3_strategy}, an auxiliary disk actuator is placed adjacent to the actuator that is on the optical microring resonator. 
By driving them with the same signal but with $\pi$ phase difference, the excitation of mechanical resonances can be effectively suppressed as shown in the green curve of Fig. \ref{Ch3_strategy}. 
However, it is mostly effective for modes below 1 MHz where most flexural modes reside (e.g., modes at 225 and 490 kHz). 
The in plane Lamé and shear modes are still significant such as the 967 KHz resonance. 

\begin{figure}[b]
\centering
\includegraphics[width=12 cm]{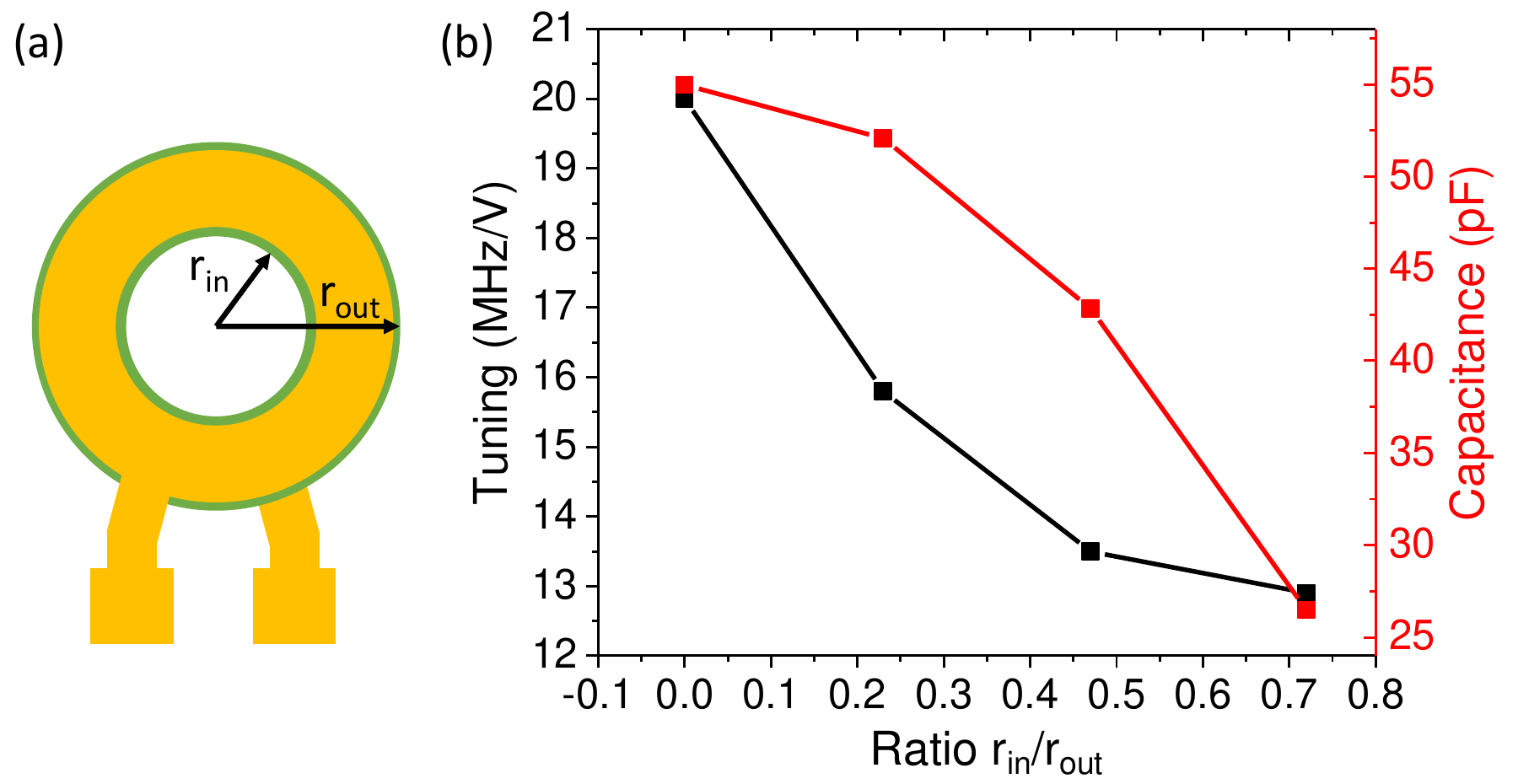}
\caption{
Strategy for reducing the RC constant time of the actuator. 
(a) Schematic of the donut-shaped actuator with inner and outer radius of $r_{\text{in}}$ and $r_{\text{out}}$. 
(b) Experimentally measured tuning efficiency (black curve) for different ratios between the inner and outer radius. 
The corresponding capacitance of the actuator is calculated as the red curve.
}
\label{Ch3_donut}
\end{figure}

To further suppress the in-plane acoustic waves, the entire chip is apodized into a polygon with non-parallel edges to deflect the reflection such that the in-plane acoustic waves cannot resonate. 
The lower left optical image in Fig. \ref{Ch3_strategy} shows one of the apodized photonic chip where the top two corners are intentionally cut to reduce parallel faces. 
This strategy extends the flatness of the response further to beyond 1 MHz as shown by the red curve in Fig. \ref{Ch3_strategy}. 
However, since the light has to be coupled in and out through the left and right edges of the chip, they are still parallel. 
One of the mechanical mode of the apodized chip at 1.69 MHz is shown in Fig. \ref{Ch3_strategy}. 
Finally, by attaching the apodized chip on a piece of carbon tape and then differentially driving the actuators, the flat response is extended up to 10 MHz with variation smaller than 1 dB (blue curve in Fig. \ref{Ch3_strategy}). 
This flat response will improve the linearity of piezoelectric actuation for integrated tunable laser, and frequency-modulated continuous-wave (FMCW) LiDAR where any nonlinearity will reduce the SNR as light travels over long distance ($>$100 m) \cite{lihachev2022}. 
While these strategies are effective for mechanical modes of the chip, the HBAR modes at tens of MHz that reside vertically in the substrate will require extra solutions. 
In Ref. \cite{Tian:20}, we proposed to eliminate the HBAR modes by roughing the backside surface of the Si substrate via isotropic XeF$_2$ etch, and attaching a layer of polyurethane epoxy for absorbing the acoustic waves \cite{pinrod2018}. 
However, this method is only effective for HBAR modes at GHz frequencies, where the acoustic wavelength has similar length scale with the backside roughness. 
On the other hand, since the first harmonic of the HBAR is inversely proportional to the thickness of the substrate, by thinning the Si substrate, the HBAR modes' frequencies can be increased which can broaden linear actuation bandwidth.

As the RC time constant ultimately limits the speed of the capacitive actuator, we further study a new design of a donut-shaped actuator which has smaller area as shown in Fig. \ref{Ch3_donut}. 
A series of actuators with different inner radius is designed while keeping the outer radius as 470 $\mu$m and the optical microring at the outer edge of the donut. 
Figure \ref{Ch3_donut}(b) plots the tuning efficiency and capacitance as a function of the ratio between the inner and outer radius. 
Zero inner radius corresponds to the disk-shaped actuator as discussed above. 
As the inner radius increases, both of the tuning efficiency and capacitance decrease. 
This is because the donut actuator has more free boundaries which compromises the stress confinement. 
However, the slope of the curve decreases for the tuning efficiency but increases for capacitance. 
Therefore, it is possible to design a donut actuator that has largely reduced capacitance while maintaining reasonable tuning efficiency. 
Although the AlN actuator is not limited by its RC time, this strategy is promising for PZT actuators which has a large capacitance on the order of nF, due to its large dielectric constant. 

\begin{figure}[b]
\centering
\includegraphics[width=13 cm]{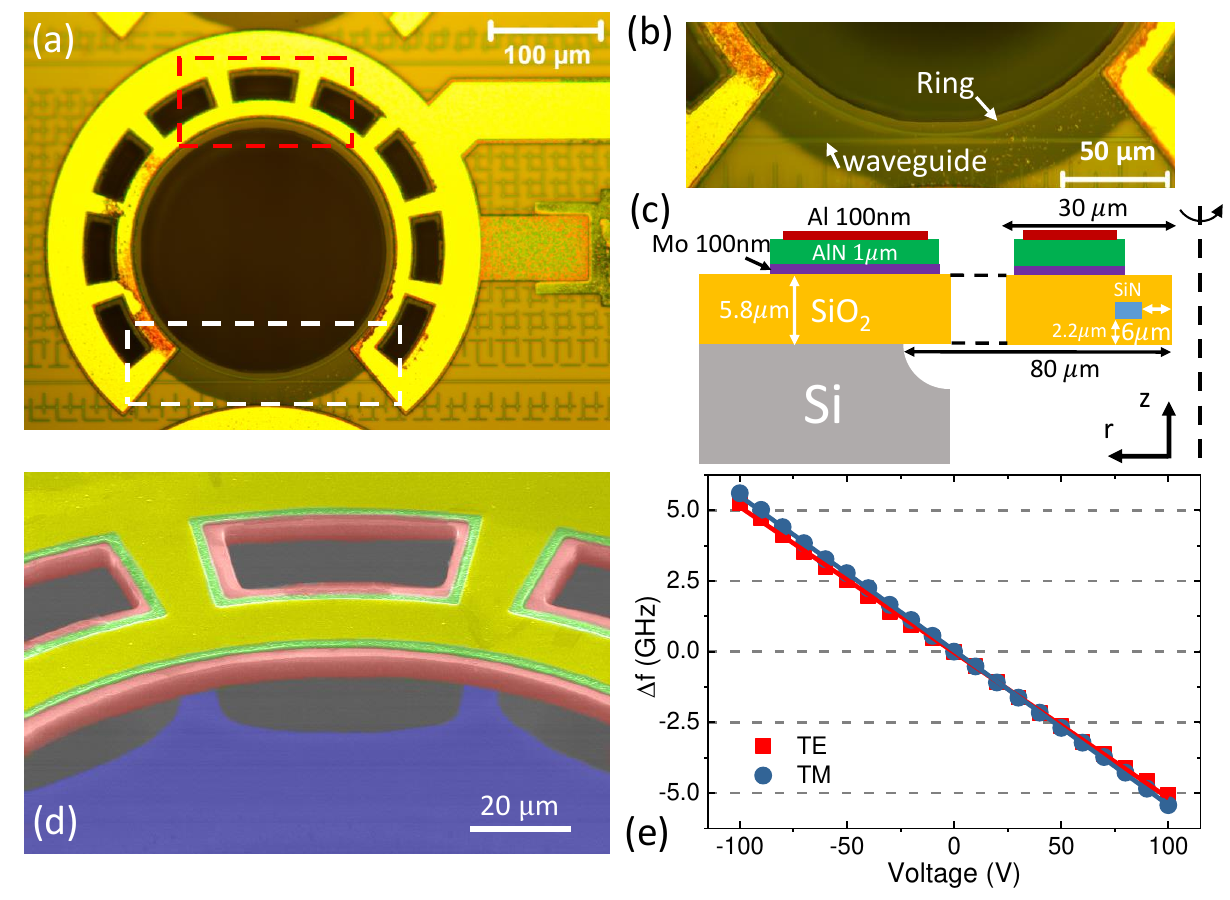}
\caption{
DC tuning of a Si$_3$N$_4$ microring resonator with released AlN piezoelectric actuator. 
(a) Optical microscope image of a released device. 
(b) Zoom-in around bus waveguide-ring coupling region in the white box in (a). 
(c) Cross-section in the radial direction with dimensions as labeled. 
(d) False colored SEM in the red box in (a). 
(a-d) reprinted from Ref. \cite{Tian:21}. 
(e) DC tuning of the optical resonance of TE$_{00}$ and TM$_{00}$ modes. 
}
\label{Ch3_release}
\end{figure}

%%%%%%%%%%%%%%%%%%%%%%%%%%%%%%%%%%%%%%%%%%%%%%%%%%%
%%%%%%%%%%%%%%%%%%%%%%%%%%%%%%%%%%%%%%%%%%%%%%%%%%%
\subsection{Released AlN Actuator}

The devices presented in the previous section is solidly mounted on a Si substrate such that the mechanical deformation is limited and the tuning mainly relies on the photoelastic effect. 
By partially removing the Si beneath the piezoelectric actuator, the actuator gains more degree of freedom to generate mechanical displacement that deforms the shape of the photonic devices. 
This technique is called ``releasing'' which has been widely applied in conventional MEMS technologies. 
The mechanical deformation will modify the photonic devices' properties in different ways depending on the device design \cite{midolo2018nano}. 
For example, the gap in an optical directional coupler can be mechanically controlled for tuning the output power ratio in an optical switch \cite{han202132}. 
The lengths of the arms in a Mach-Zehnder interferometer can be mechanically stretched for modulating the relative phase \cite{dong2022high, stanfield2019cmos}. 
For the optical microring resonators that will be discussed in the following sections, the size of the ring can be expended or shrunk by the piezoelectric actuators which changes the effective cavity length. 
This pure mechanical effect is usually stronger than the photoelastic effect and shows larger tuning, and it is more prominent for smaller devices since it's proportional to the relative change of the radius $\Delta r/r$. 

\begin{figure}[b]
\centering
\includegraphics[width=13 cm]{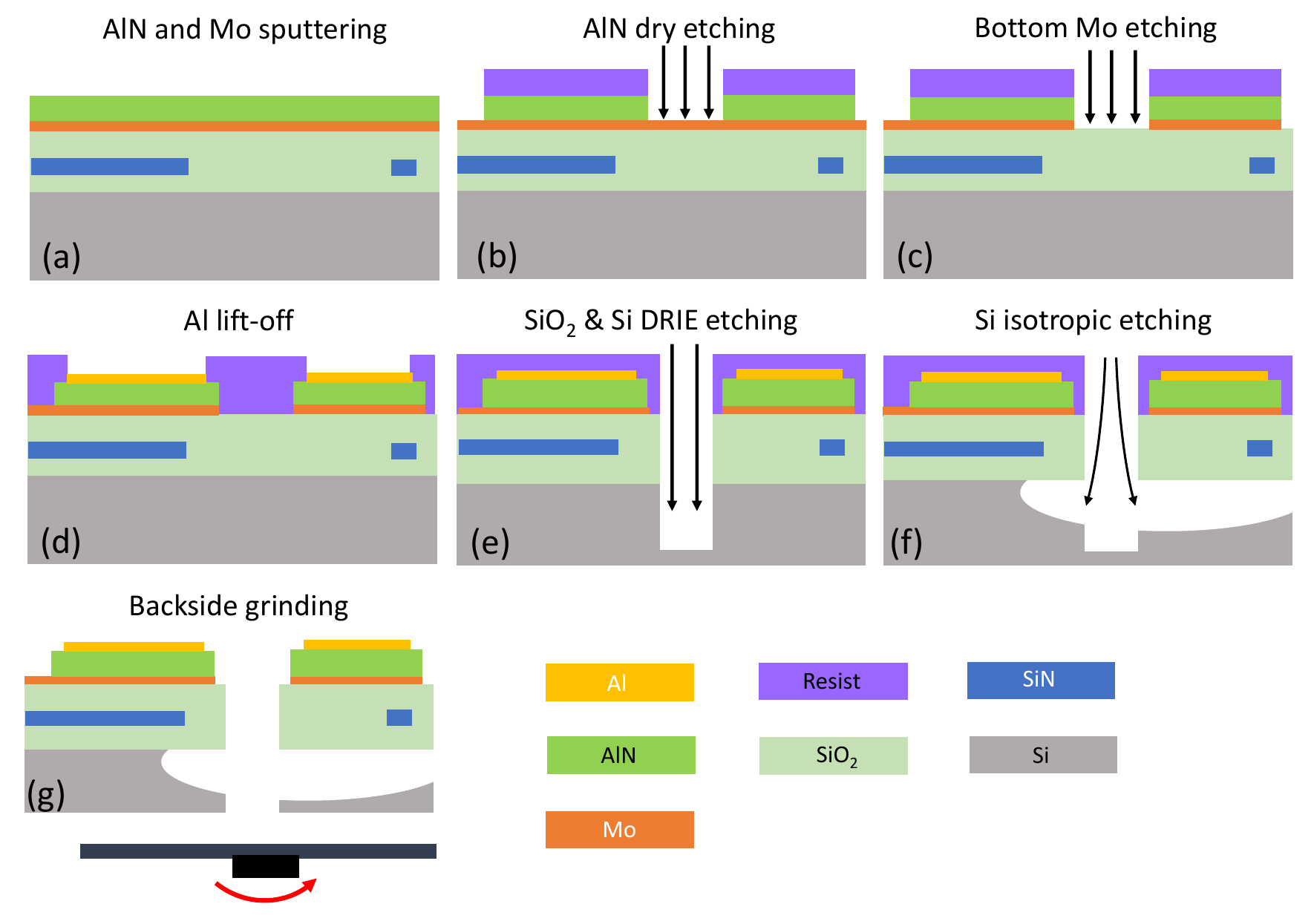}
\caption{
Fabrication flow for making released AlN piezoelectric actuator in Fig. \ref{Ch3_release}. }
\label{Ch3_fab}
\end{figure}
%%%%%%%%%%%%%%%%%%%%%%%%%%%%%%%%%%%%%%%%%%%%%%%%%%%
\subsubsection{DC Tuning}

One of the typical released device with AlN piezoelectric actuator is shown in Fig. \ref{Ch3_release} \cite{dong2018port}. 
A donut shaped actuator is fabricated on the SiO$_2$ cladding which is released by Si isotropic etch under the SiO$_2$ layer. 
A big center hole together with nine outside holes are opened for the etching gas to react with the Si underneath and form a uniformly released oxide membrane, as can be seen in Fig. \ref{Ch3_release}(b). 
The free-standing membrane can be seen more clearly from the skewed false-color SEM in Fig. \ref{Ch3_release}(d). 
The Si$_3$N$_4$ microring is embedded in the SiO$_2$ cladding and located at the inner periphery of the actuator. 
The waveguide is 6 $\mu$m away from the inner edge of the oxide cladding [see Fig. \ref{Ch3_release}(c)] which prevents the scattering loss from the sidewall roughness due to the oxide etching. 
The optical $Q$ is thus maintained intact after releasing. 
The DC tuning of the TE$_{00}$ and TM$_{00}$ modes is shown in Fig. \ref{Ch3_release}(e), where linear and bi-directional tuning is achieved. 
The tuning efficiency is around 50 MHz/V for both TE$_{00}$ and TM$_{00}$ modes respectively, which is over 2 times larger than unreleased device with similar size \cite{Tian:20}.

The fabrication flow for the released device is shown in Fig. \ref{Ch3_fab}. 
The Si$_3$N$_4$ photonic wafer is firstly fabricated with either the subtractive process \cite{luke2013} or the photonic Damascene process \cite{pfeiffer2016, Pfeiffer:18, Liu:20b}. 
The top surface of the SiO$_2$ cladding undergoes chemical-mechanical polishing (CMP) before the AlN deposition in order to have sub-nanometer surface roughness for better crystalline orientation of the AlN film. 
Polycrystalline AlN and bottom Mo metal are sputtered on the photonic wafer sequentially in the same vacuum chamber. 
The AlN is first dry-etched by the reactive-ion-etching (RIE) with gases Cl$_2$ and BCl$_3$ for defining the shape of the AlN actuator. 
The bottom Mo is dry-etched for patterning the bottom electrode contact pad. 
The top Al metal is patterned with the standard lift-off process utilizing electron-beam evaporation. 
In the next step, the release holes are fabricated by deep RIE (DRIE) of the SiO$_2$ and Si, which is anisotropic and has good selectivity in the vertical direction as illustrated in Fig. \ref{Ch3_fab}(e). 
Note that the dicing lines between chips are also etched in the same step, which produces smooth chip edge for good lensed fiber to chip edge coupling. 
Subsequently, Si isotropic dry etching using SF$_6$ is performed to undercut and suspend the SiO$_2$ membrane. 
The mechanical grinding in the final step serves to singularize the wafer into chips. 

\begin{figure}[b]
\centering
\includegraphics[width=13 cm]{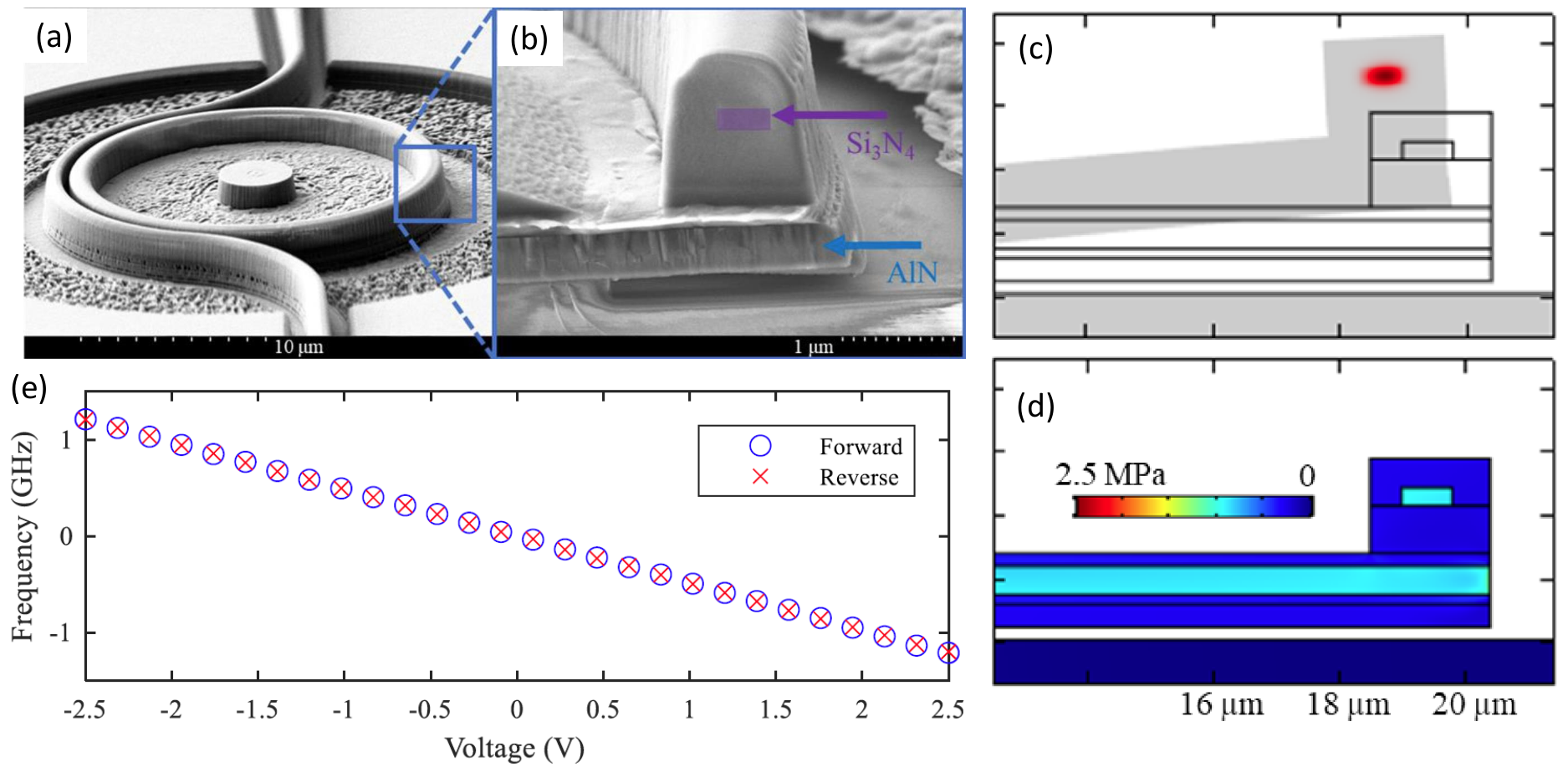}
\caption{
Released AlN actuator with a buried design. 
(a) SEM of a typical released optical microring resonator with radius of 20 $\mu$m. 
(b) Zoom-in SEM of the cross-section of the AlN actuator and the Si$_3$N$_4$ waveguide (purple). 
(c) Finite-element simulation of the mechanical deformation (scaled by 10000x) under -1 V. 
The simulated TE$_{00}$ optical mode is shown in red. 
(d) Von Mises stress distribution at -1 V. 
(e) Tuning of the TE$_{00}$ resonance at room temperature in forward (blue circle) and reverse (red cross) sweeping direction, showing a 480 MHz/V tuning efficiency.
Reprinted from Ref. \cite{stanfield2019cmos}. 
}
\label{Ch3_Mattdevice}
\end{figure}

For the devices in Fig. \ref{Ch3_release}, the AlN has to be fabricated on flat SiO$_2$ cladding layer for maintaining high quality crystal alignment in the vertical direction of the AlN. 
This limits the design degree of freedom in the horizontal direction on the SiO$_2$ layer.
As shown in Fig. \ref{Ch3_Mattdevice}, Eichenfield \textit{et al}. \cite{stanfield2019cmos} demonstrated a buried actuator design where AlN is firstly fabricated before the deposition and etching of the Si$_3$N$_4$ microring resonator. 
This allows to etch the SiO$_2$ layer cladding to optimize the mechanical deformation. 
Here, they removed most of the SiO$_2$ in the center and only left a thin wall at the periphery for the optical waveguide [see Fig. \ref{Ch3_Mattdevice}(b)]. 
In this way the waveguide has more freedom to move radially for larger relative change of the radius of the microring, as can be seen in Fig. \ref{Ch3_Mattdevice}(c). 
On the other hand, the stress is released and is small around the waveguide [see Fig. \ref{Ch3_Mattdevice}(d)]. 
Thus, the tuning is mainly from the mechanical deformation of the optical ring. 

The tuning at room temperature is experimentally measured as shown in Fig. \ref{Ch3_Mattdevice}(e). 
A linear tuning with much higher efficiency of 480 MHz/V is obtained. 
The efficiency is 2.4 times larger than the unreleased device with similar design which is also studied in Ref. \cite{stanfield2019cmos}. 
No hysteresis is observed, illustrating high quality of the AlN film. 
An ultra-low static power consumption is reported with 8 pW at 2 V, corresponding to power efficiency of 8 fW/MHz. 
With this low power, they further demonstrated tuning at the cryogenic temperature (7 K) with efficiency of 260 MHz/V. 
The resistance of the AlN actuator is measured to increase by 40 times at cryogenic temperature which further reduces the leakage current and electrical power dissipation to sub-picowatt levels. 
These results prove the suitability of AlN piezoelectric actuators in cryogenic applications which have stringent requirement on power dissipation. 

\begin{figure}[b]
\centering
\includegraphics[width=12 cm]{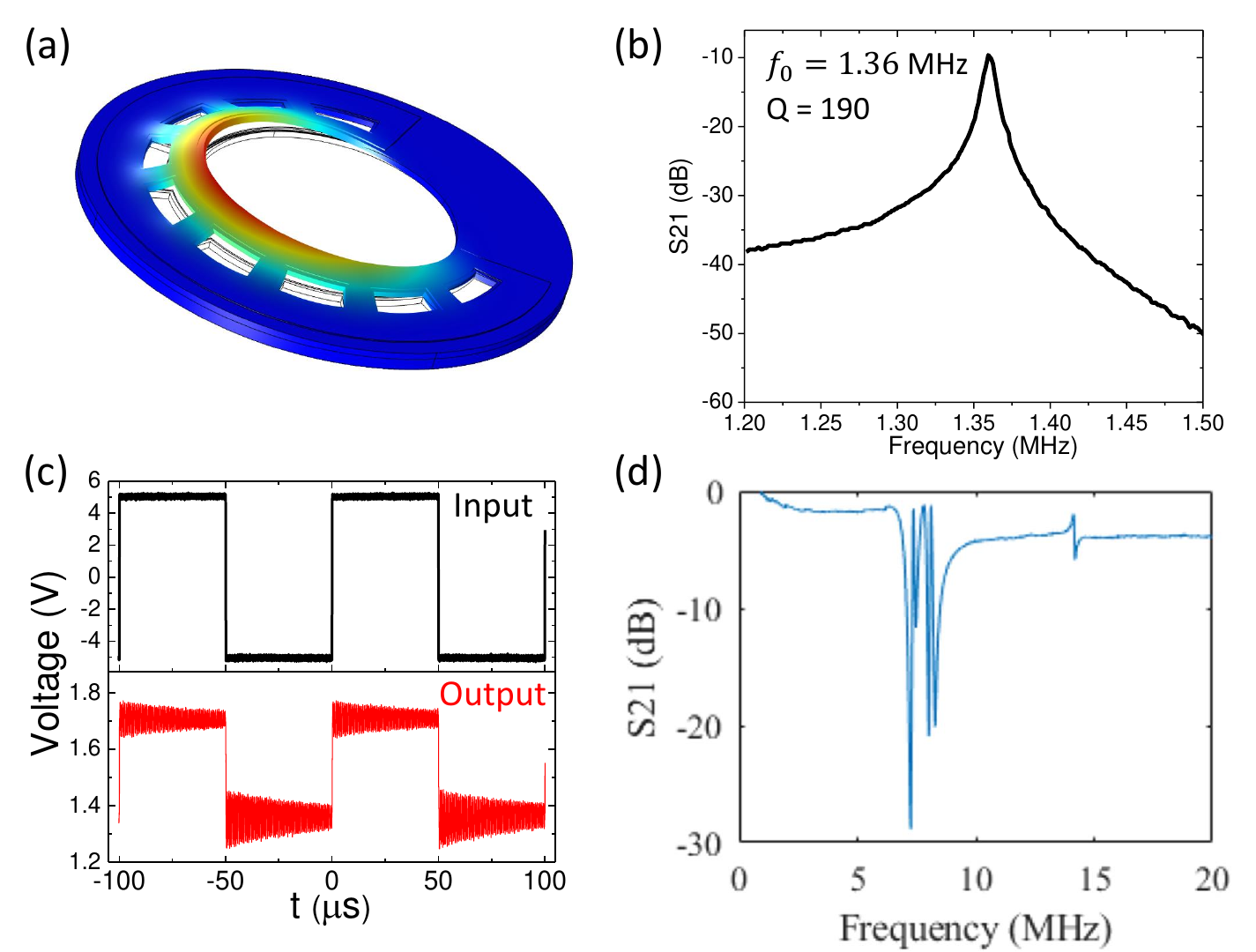}
\caption{
Dynamic tuning and mechanical resonances of the released actuator. 
(a) Simulation of the displacement of the fundamental mechanical mode. 
(b) Experimentally measured optomechanical S$_{21}$ response showing the mechanical mode at 1.36 MHz. 
(c) Time dynamic response of the device under 10 kHz square wave with 10 V V$_{\text{pp}}$. 
Results in (a-c) are from the device in Fig. \ref{Ch3_release}.
(a-c) reprinted from Ref. \cite{tianthesis}.
(d) Optomechanical S$_{21}$ response of the device in Fig. \ref{Ch3_Mattdevice} with first mechanical mode at 7.26 MHz. Reprinted from Ref. \cite{stanfield2019cmos}. 
}
\label{Ch3_mechanics}
\end{figure}

%%%%%%%%%%%%%%%%%%%%%%%%%%%%%%%%%%%%%%%%%%%%%%%%%%%
\subsubsection{Mechanical Resonances of Released Actuators}

Although the actuator will have larger deformation and tuning effect due to the smaller stiffness after releasing, this will inevitably lower the fundamental mechanical frequency which then limits the tuning bandwidth and linearity. 
Therefore, the trade-off between the tuning bandwidth and the range has to be taken into account in the design for specific applications. 
Figure \ref{Ch3_mechanics}(a) shows the fundamental mechanical vibration mode of the device in Fig. \ref{Ch3_release} with the oxide membrane bending up and down. 
Its frequency is experimentally measured at 1.36 MHz with a mechanical $Q\approx190$ as shown in Fig. \ref{Ch3_mechanics}(b). 
Its influence on the tuning is manifested as the mechanical ringing when switching from one state to the other. 
In Fig. \ref{Ch3_mechanics}(c), a 10 kHz square wave is applied to the released actuator. 
However, due to the high mechanical $Q$, it takes tens of microsecond for the mechanical vibration to ring down, which largely limits the tuning speed. 
Similarly, for the device shown in Fig. \ref{Ch3_Mattdevice}, the first mechanical mode appears to be at 7.26 MHz [see Fig. \ref{Ch3_mechanics}(d)]. 
These mechanical modes can be mitigated by advanced packaging techniques, where the devices are encapsulated in inertial damping gases \cite{kim2011}. 
Both of the mechanical vibration amplitude and $Q$ can be largely suppressed. 
We can also precondition the input signal by filtering the frequency components at the mechanical resonant frequencies. 
These advanced strategies can reduce the influence of the mechanical resonance and worth more investigation in the future.

\begin{figure}[b]
\centering
\includegraphics[width=12 cm]{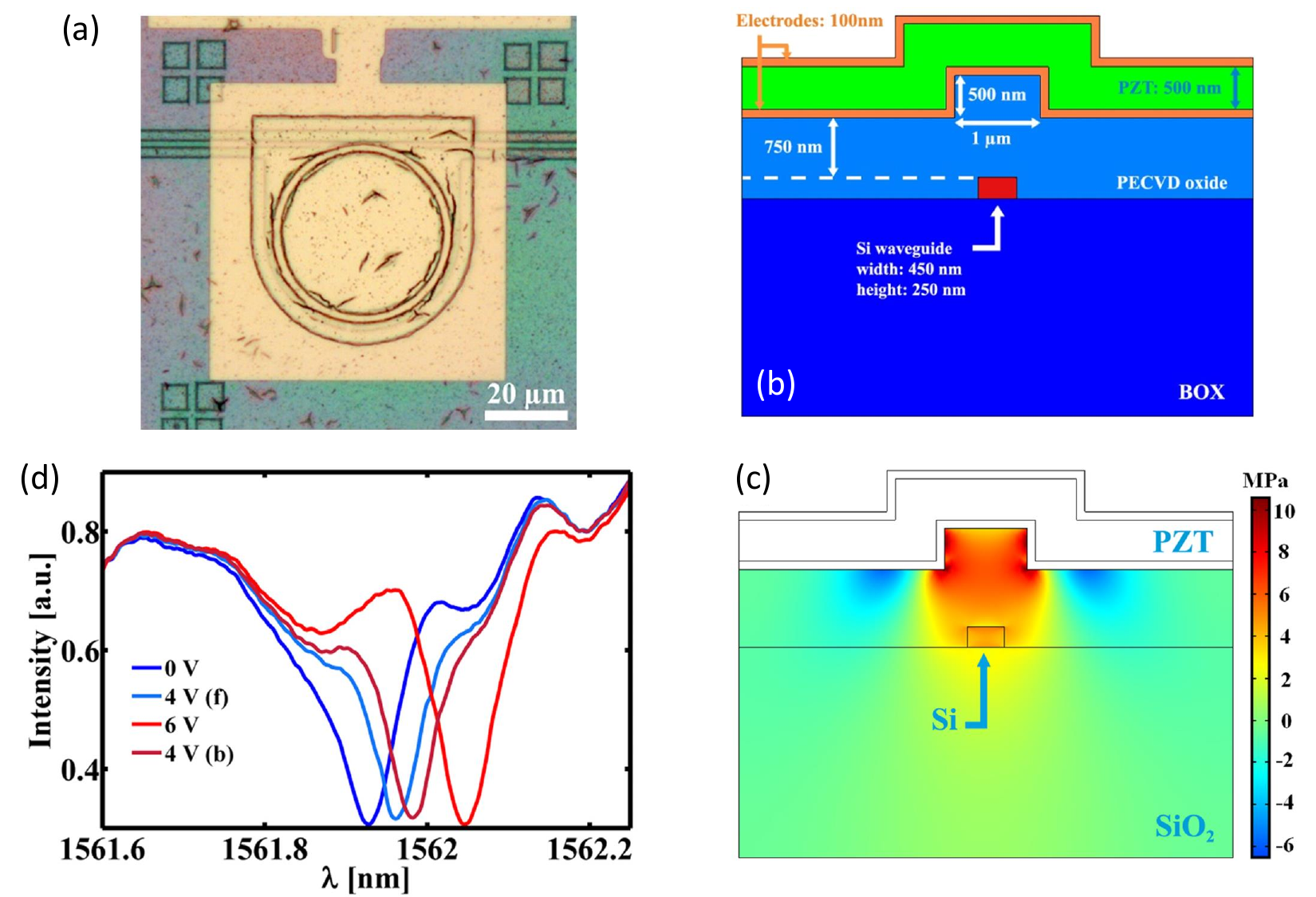}
\caption{
Tuning of the Si microring resonator using PZT actuator. 
(a) Optical image of the fabricated PZT actuator and microring resonator. 
(b) Cross-section of the actuator and optical waveguide, with critical dimensions as labeled. 
(c) Stress $\sigma _{\text{x}}$ distribution under 5 V applied DC voltage. 
(d) Optical transmission spectra under different voltages. 
The spectra at 4 V in the forward (blue) and backward (dark red) sweep direction are measured. 
Reprinted from Ref. \cite{sebbag2012}.
}
\label{Ch3_PZTSi}
\end{figure}

%%%%%%%%%%%%%%%%%%%%%%%%%%%%%%%%%%%%%%%%%%%%%%%%%%%
%%%%%%%%%%%%%%%%%%%%%%%%%%%%%%%%%%%%%%%%%%%%%%%%%%%
\subsection{Piezoelectric Tuning Using PZT Actuator} 

Despite of the successful demonstration of piezoelectric tuning using AlN released and unreleased actuator, AlN has a relatively small piezoelectric coefficient which requires high voltage up to 100 V. 
Although this can be achieved on-chip using charge-discharge pump circuits, they are usually very slow due to their large capacitance. 
PZT, a traditional piezoelectric ceramic that has nearly 20 times larger piezoelectric response, has been widely used in sensors, actuators, and transducers in MEMS community \cite{smith2012pzt}. 
The successful deposition of PZT thin films and fabrication of PZT actuators on photonic wafers have only been demonstrated in the most recent decade. 
This section reviews the application of both unreleased and released PZT actuators in the tuning of photonic devices.

%%%%%%%%%%%%%%%%%%%%%%%%%%%%%%%%%%%%%%%%%%%%%%%%%%%
\subsubsection{Unreleased PZT tuning of Si Microring Resonator}

The PZT tuning of photonic devices is pioneered by U. Levy et al. \cite{sebbag2012} as shown in Fig. \ref{Ch3_PZTSi}. 
A Si microring resonator with 20 $\mu$m radius was fabricated on an SOI wafer with 250 nm device layer on a 2-$\mu$m-thick buried oxide layer (BOX). 
A 500 nm sol-gel morphotropic PZT (Zr/Ti = 52/48) thin film was spin coated on the photonic wafer, which was then annealed at 650 $^{\circ}$C for it to crystallize. 
The PZT is sandwiched between top (100 nm Au) and bottom (100 nm Pt) metal electrodes for applying voltage. 
Different from previous AlN actuators, the PZT does not require flat top surface to grow polycrystalline, allowing to pattern the top oxide cladding layer and engineer the distribution of stress. 
They found that the formation of an oxide cap above the waveguide could concentrate the stress around the oxide corners for enhanced stress-optic tuning, as shown in Fig. \ref{Ch3_PZTSi}(c). 

Since PZT is also a ferroelectric material, it was electrically poled at 50$^{\circ}$C for 10 min to align the polarization in different domains in the same direction for maximizing its piezoelectric response. 
The tuning of one optical resonance under different voltages is presented in Fig. \ref{Ch3_PZTSi}(d), which shifts to longer wavelength (red tuning) for larger voltage. 
The resonance is shifted by 34 pm (corresponding to 4.2 GHz change in frequency) at 4 V and by 115 pm (14.4 GHz) at 6 V, which is much larger than previous demonstrations with AlN actuators. 
Besides the larger piezoelectric response, the larger photoelastic effect of Si also helps to improve the tuning efficiency. 
Notably, the difference between the spectra at 4 V measured when increasing and decreasing the voltage illustrates the hysteresis behaviour of the PZT actuator. 
It comes from the ferroelectricity where the polarization can be reoriented by the electric field. 
Despite of being undesired in applications where linear tuning is required, the hysteresis can be used to realize an optically detected memory system \cite{sebbag2012}. 
The tuning speed was tested by applying a 1 kHz square wave to the actuator, and a rise time of 100 $\mu$s was observed from the optical modulation, which was limited by the photodetector.  

\begin{figure}[b]
\centering
\includegraphics[width=13.3 cm]{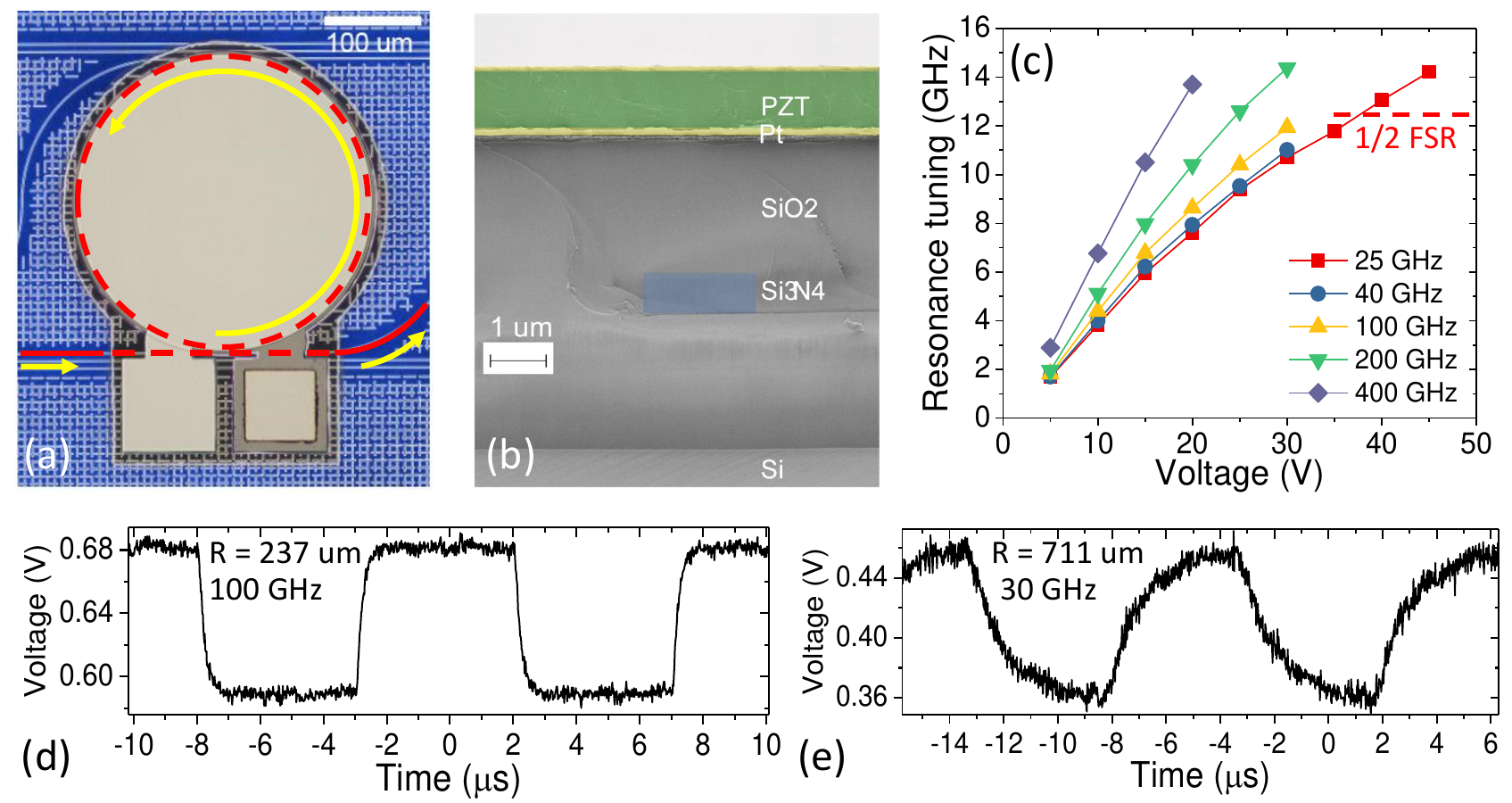}
\caption{
Tuning of thick Si$_3$N$_4$ microring resonator with PZT actuators. 
(a) Optical image of a PZT actuator on an optical microring resonator (red lines). 
(b) False-colored SEM of the cross-section of the device in (a). 
(a, b) reprinted from Ref. \cite{lihachev2022}. 
(c) Tuning of the optical resonance for devices with different radius and FSR (as labeled). 
Tuning dynamics under 100 kHz square wave for (d) 100 GHz FSR and (e) 30 GHz FSR devices. 
(c-e) reprinted from Ref. \cite{tianthesis}.
}
\label{Ch3_PZTSiN}
\end{figure}

%%%%%%%%%%%%%%%%%%%%%%%%%%%%%%%%%%%%%%%%%%%%%%%%%%%
\subsubsection{Unreleased PZT Tuning of Si$_3$N$_4$ Microring Resonator}

The PZT tuning of Si$_3$N$_4$ photonic devices has been explored in the past decade, as the fabrication of high-quality, low-loss Si$_3$N$_4$ waveguide matures \cite{pfeiffer2016, Pfeiffer:18b, Pfeiffer:18, Liu:20b, luke2013}. 
Two kinds of Si$_3$N$_4$ waveguide design are widely explored. 
One is the thick Si$_3$N$_4$ waveguide, which has tight mode confinement and is fabricated by the photonic Damascene process \cite{pfeiffer2016, Pfeiffer:18b, Liu:18a, Liu:20b} or the thermal cycling process to address the high stress built in thick Si$_3$N$_4$ film \cite{levy2010, ye2019high}. 
The other design uses a thin Si$_3$N$_4$ waveguide fabricated by the subtractive process where the optical mode mainly resides in the SiO$_2$ cladding \cite{bauters2011, Gundavarapu:19, Jin:21}. 
Since the mode is less confined, the radius of the thin Si$_3$N$_4$ optical ring resonator is on the order of millimeter for maintaining low bending loss, while the radius of thick Si$_3$N$_4$ ring resonator can be as low as 20 $\mu$m \cite{karpov2018}.

Figure \ref{Ch3_PZTSiN}(a-c) show the piezoelectric tuning of thick Si$_3$N$_4$ microring resonator with PZT disk actuator \cite{lihachev2022, tian2023}. 
Following the findings in Fig. \ref{Ch3_position}, the microring is placed 8 $\mu$m inside the actuator. 
From the false-color SEM in Fig. \ref{Ch3_PZTSiN}(b), it can be seen 1 $\mu$m PZT film is sandwiched between top and bottom 100 nm platinum (Pt) electrodes. 
The key difference from the AlN actuator is that the PZT process allows a patterned bottom electrode to eliminate the bond pad capacitance. 
This is especially important for PZT since it has a large dielectric constant on the order of 1000 \cite{wu2009}. 
The Si$_3$N$_4$ waveguide (0.9$\times$2.2 $\mu$m$^2$) is embedded in 6 $\mu$m SiO$_2$ cladding and is 2.5 $\mu$m below the bottom Pt electrode. 
Similarly as before, the PZT actuator was poled under 25 V at room temperature to align the polarization for maximal tuning efficiency. 
The DC tuning of microring resonators with different radii and FSRs is plotted in Fig. \ref{Ch3_PZTSiN}(c). 
Over 10 GHz tuning is achieved for them under applied 30 V. 
However, the tuning gradually saturates for higher voltages, due to the saturation of the polarization in the domains of PZT. 
The current draw from the actuator is kept below 1 nA at 20 V, indicating a low power consumption on the order of nanowatt. 
Noticeably, up to 50 V is applied without breaking down the PZT film, showing a high breakdown field beyond 50 MV/m. 

The tuning efficiency increases as the device's size decreases (FSR increases). 
This is due to the contribution from the relative change of the radius $\Delta r/r$ to the resonant frequency (see Eq.\ref{eq13}), which is larger for smaller radius $r$. 
For the 25-GHz-FSR device (red curve), over half FSR tuning is obtained under 40 V, leading to $\pi$ phase shift of the light in the microring. 
The figure of merit (modulation length product) $V_{\pi} L$ is calculated to be 22 V$\cdot$cm, which is an order of magnitude larger than the state of the electro-optic modulation using LiNbO$_3$ \cite{WangC:18}. 
However, the modulation loss product $V_{\pi} L  \alpha =1.87$ V$\cdot$dB is comparable with LiNbO$_3$ (1 V$\cdot$dB), thanks to the ultra-low loss of the Si$_3$N$_4$ waveguide. 
The tuning speed can be inferred from the time response as shown in Fig. \ref{Ch3_PZTSiN}(c-d). 
For a smaller device (FSR = 100 GHz), it takes 0.2 $\mu$s to switch from one state to the other. 
Due to the large permittivity of PZT, the speed is ultimately limited by the large RC time constant for the large devices. 
Thus, the 30 GHz FSR device takes 1.4 $\mu$s to respond, which is the time to charge the large PZT plate capacitor. 

\begin{figure}[t]
\centering
\includegraphics[width=12 cm]{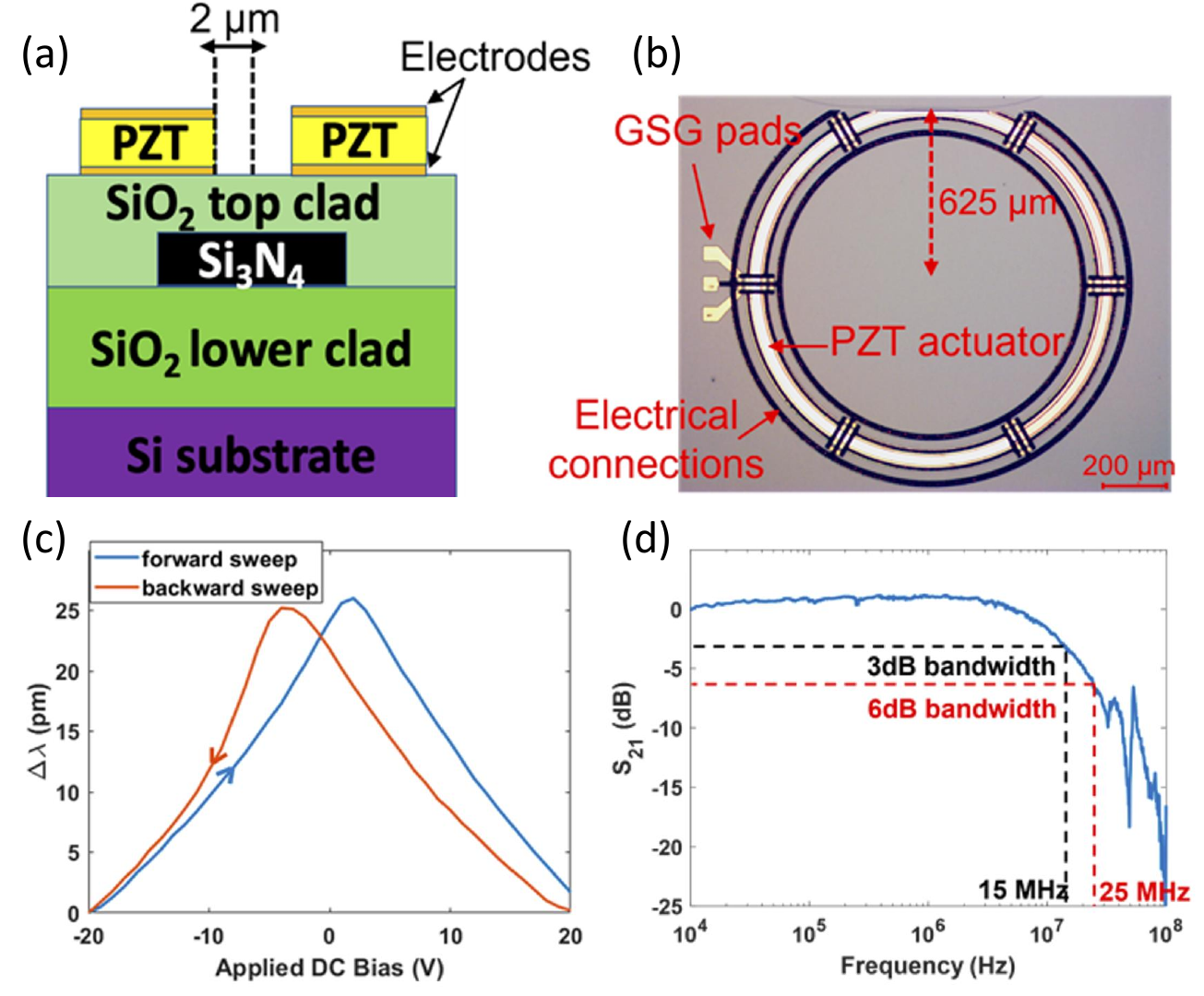}
\caption{
Tuning of thin Si$_3$N$_4$ microring resonator using PZT actuators. 
(a) Cross-section of the PZT actuator on a thin Si$_3$N$_4$ waveguide. 
(b) Optical image of the fabricated device. 
(c) Tuning of the resonant wavelength in forward (blue) and backward (red) sweeping directions, showing hysteresis and nonlinear tuning. 
(d) Frequency response of the actuator showing 3-dB and 6-dB bandwidth of 15 MHz and 25 MHz respectively. 
Reprinted from Ref. \cite{wang2022silicon}. 
}
\label{Ch3_PZTSiN2}
\end{figure}

The piezoelectric tuning of thin Si$_3$N$_4$ waveguide (175 nm thickness) is demonstrated by Wang \textit{et al}. \cite{wang2022silicon} as shown in Fig. \ref{Ch3_PZTSiN2}. 
Due to the low confinement of the thin Si$_3$N$_4$ waveguide, to maintain the high intrinsic optical $Q_0\sim7\times10^6$, a gap is opened in the PZT actuator right above the waveguide, and the PZT is offset by 2 $\mu$m relative to the center of the waveguide [see Fig. \ref{Ch3_PZTSiN2}(a)]. 
The optical image of the fabricated device with 625 $\mu$m radius is shown in Fig. \ref{Ch3_PZTSiN2}(b). 
The DC tuning of the resonant wavelength under forward and backward voltage sweep is plotted in Fig. \ref{Ch3_PZTSiN2}(c). 
A nonlinear and hysteresis tuning can be observed with tuning efficiency of 162 MHz/V, which leads a $V_{\pi} L$ of 43 V$\cdot$cm and $V_{\pi} L  \alpha$ of 1.3 V$\cdot$dB. 
Similarly, a 20 nW power consumption at 20 V was reported. 

Interestingly, a butterfly-like curve of the tuning-voltage is obtained which agrees with the typical strain-electric field curve for ferroelectric materials. 
It is related with the switching of polarization domains under the electric field. 
For example, as the voltage increases from negative to zero (blue curve), the population of domains that align in the negative direction gradually decreases but still with some remaining at $V=0$. 
As the voltage further changes to positive sign but at small value, the electric field and the polarization have the opposite sign and the wavelength increases (relative to $V=0$). 
As the voltage further increases, all the domains align with the electric field and now in the positive direction, and then the wavelength changes from increasing to decreasing. 
The coercive field of the PZT can be found at the turning point where all the domains are switched by the field. 
Similar effect can be observed when sweeping the voltage backward. 
Therefore, different from the AlN, the PZT shows the same tuning direction at high positive and negative voltages since the domains are kept aligned with the field. 
To achieved bi-directional tuning, a DC bias needs to be applied and maintained. 
A 3-dB tuning bandwidth of 15 MHz (corresponding to 0.067 $\mu$s) is achieved, due to the donut-shaped actuator which has smaller area than the disk. 
Therefore, depending on the specific application, the shape and area of the actuator should be designed carefully, trading-off between tuning efficiency and speed. 

\begin{figure}[t]
\centering
\includegraphics[width=13.3 cm]{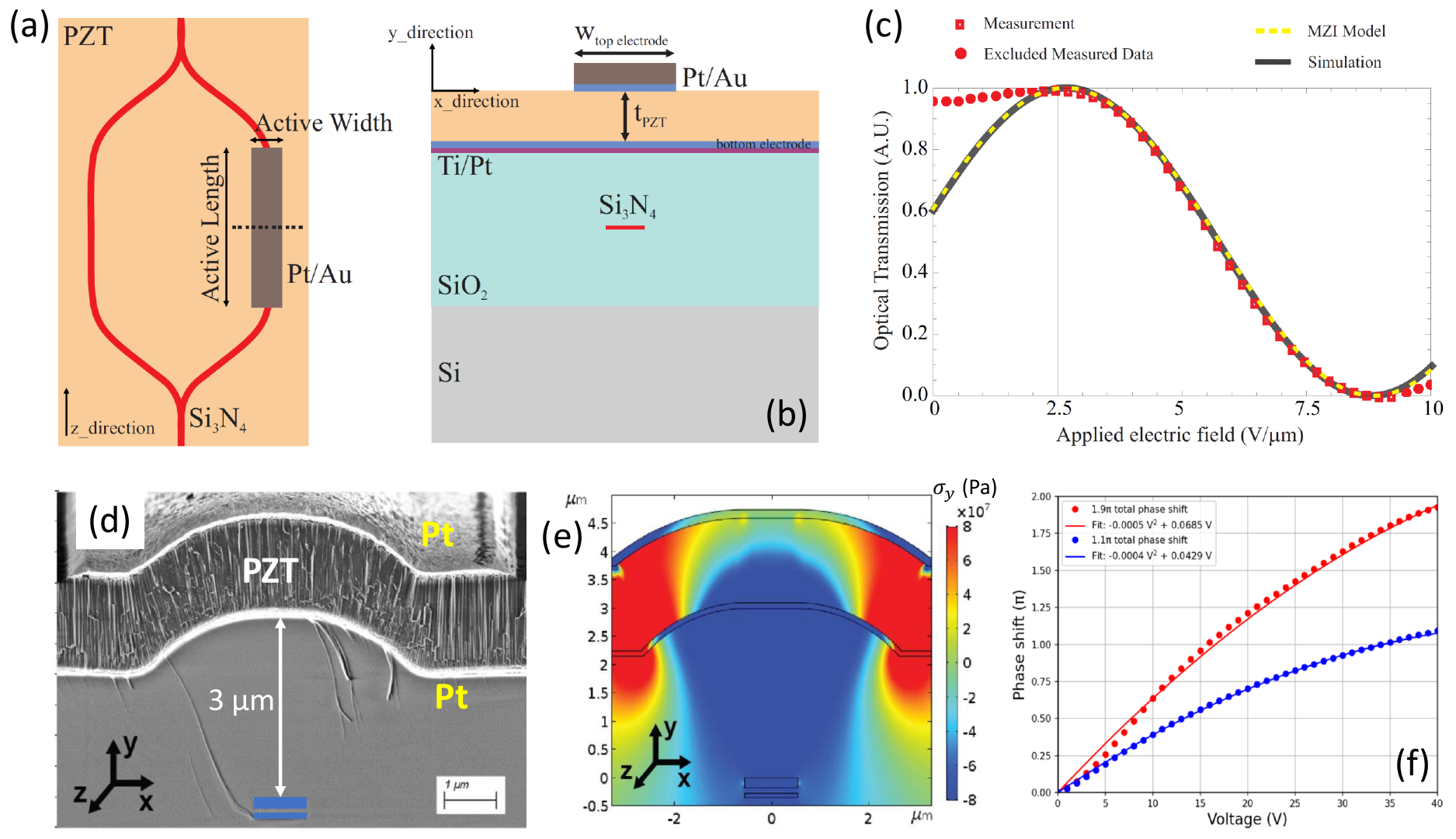}
\caption{
PZT Tuning of Si$_3$N$_4$ Mach-Zehnder Interferometer. 
(a) Top view of the Si$_3$N$_4$ MZI with PZT actuator on one arm. 
(b) Cross-section of the actuator and thin Si$_3$N$_4$ waveguide. 
(c) Measured (red dots) and simulated (gray solid line) optical transmission of the MZI under increasing electric field to the PZT. 
(a-c) reprinted from Ref. \cite{hosseini2015}. 
(d) SEM of the cross-section of a dome-shape PZT actuator on top of an asymmetric double stripe Si$_3$N$_4$ waveguide (which is false colored as blue). 
(e) Numerical simulation of the vertical stress ($\sigma_{y}$) under 40 V applied on the top electrode. 
(f) Phase shift of one arm of the MZI induced by the dome-shape PZT actuator under different voltages. 
Phase shift between 1.1$\pi$ and 1.9$\pi$ is obtained under 40 V, depending on the quality of PZT. 
(d-f) reprinted from Ref. \cite{everhardt12004}.
}
\label{Ch3_PZTMZI}
\end{figure}

%%%%%%%%%%%%%%%%%%%%%%%%%%%%%%%%%%%%%%%%%%%%%%%%%%%
\subsubsection{Unreleased PZT Tuning of Si$_3$N$_4$ Mach-Zehnder Interferometer}

Besides the microring resonator, the optical MZI has been another important building block in integrated optical circuits. 
Different from the microring which can only operate at discrete optical resonant frequencies with narrow linewidth, the MZI can support a wide optical bandwidth at the expense of a larger footprint. 
This feature helps it to stand out in electro-optic modulators where it usually shows higher modulation bandwidth than the microring \cite{WangC:18} and is easier to implement wavelength-division multiplexing (WDM) in optical telecommunication. 
Moreover, MZI can conduct beam-splitter operation on chip, which is the basic element in optical neural networks \cite{shen2017deep} and photonic quantum computing \cite{Arrazola:21, harris2016large}. 
Active tuning of the MZI is highly desired for programmable photonic circuits to be reconfigured to conduct complex computing tasks, such as unitary transformations \cite{bogaerts2020, harris2018, annoni2017}. 
However, most of the demonstrations rely on thermo-optic tuning, which is practically limited in scaling, power consumption, and cross-talk \cite{shen2017deep, Arrazola:21, annoni2017}. 
In that sense, piezoelectric tuning is gaining much attention and will play an important role in the future. 

The tuning of Si$_3$N$_4$ Mach-Zehnder interferometer based on PZT actuators was pioneered by Heideman \textit{et al}. \cite{hosseini2015} as shown in Fig. \ref{Ch3_PZTMZI}(a), where a rectangular PZT actuator is placed on top of one arm of the MZI. 
The stress induced by the PZT will change the effective refractive index of the waveguide and thus the relative phases between the two arms and the output light intensity after interference. 
A 25 nm thin Si$_3$N$_4$ waveguide (4 $\mu$m wide) is designed to support the fundamental TM$_{00}$ mode at an optical wavelength of 640 nm. 
Due to the weak confinement of the optical mode, the waveguide is placed 8 $\mu$m below the bottom metal to maintain a low optical loss. 
Unlike the works shown in previous sections, 2 $\mu$m PZT is deposited by pulsed-laser deposition (PLD). 
Note that the PZT uniformly covers the whole device, and only the top metal is patterned to generate the local stress. 

The actuator parameters need to be designed carefully to have a maximal phase shift. 
First, the phase accumulated in the arm is proportional to the actuator's length, which is usually captured in the figure of merit of half-wave voltage-length product $V_{\pi}L$. 
Second, the thicker the PZT film, the larger the stress that can be generated under the same amount of electric field. 
The thickness of the PZT is mainly limited by the quality of the PZT that can be maintained for thicker film and depends on the deposition method. 
Finally, the width of the actuator needs to be optimized via FEM simulations targeting at the highest refractive index change. 
The optimal value of 30 $\mu$m is found in Ref. \cite{hosseini2015}. 
The optical transmission of the MZI under various electric fields is measured in Fig. \ref{Ch3_PZTMZI}(c), where a typical cosine function can be found. 
A $\pi$ phase shift can be obtained from 2.5 V/$\mu$m to 8.5 V/$\mu$m, corresponding to $V_{\pi}=12$ V and $V_{\pi}L=12$ $V\cdot $cm (length of the actuator is 1 cm). 
Note that the phase shift is inversely proportional to the wavelength, so the figure of merit at 640 nm needs to be scaled to compare with the results at typical telecommunication C-band around 1550 nm. 

The phase shift of the MZI was further improved by the same group recently by optimizing the design of both the PZT actuator and the Si$_3$N$_4$ waveguide \cite{epping2017, everhardt12004}. 
As shown in Fig. \ref{Ch3_PZTMZI}(d), a dome-shape actuator was introduced which can concentrate the stress efficiently on the waveguide as can be seen from the numerical simulation in Fig. \ref{Ch3_PZTMZI}(e). 
This result agrees with the previous finding of the work on PZT tuning of the Si microring resonator \cite{sebbag2012} (see Fig. \ref{Ch3_PZTSi}). 
Moreover, the dome can be naturally formed during the subtractive fabrication of the Si$_3$N$_4$ waveguide, where etching the Si$_3$N$_4$ leaves a surface topology with a height of 1 $\mu$m and the dome is formed by uniformly depositing SiO$_2$ top cladding covering the whole waveguide. 
On the other hand, an asymmetric double-strip waveguide was designed to tightly confine the optical mode so that the thickness of the top cladding was reduced from 8 $\mu$m to 3 $\mu$m. 
This can largely increase the overlap between the optical mode and the stress distribution. 
With these improvements being implemented, the phase shift between 1$\pi$ to 2$\pi$ can be achieved under 40 V, depending on the structural properties of the deposited PZT films, as shown in Fig. \ref{Ch3_PZTMZI}(f). 
The maximum $V_{\pi}L$ of 16 V$\cdot$ cm is nearly twice better than the PZT actuator with flat structure. 
The tuning speed was further inferred from the time domain step response with a fitted time constant of 0.85 $\mu$s. 
The high speed benefits from the dome-shape design which allows for a narrower actuator (10 $\mu$m in this case) with smaller area and capacitance \cite{epping2017}. 

\begin{figure}[t]
\centering
\includegraphics[width=13.3 cm]{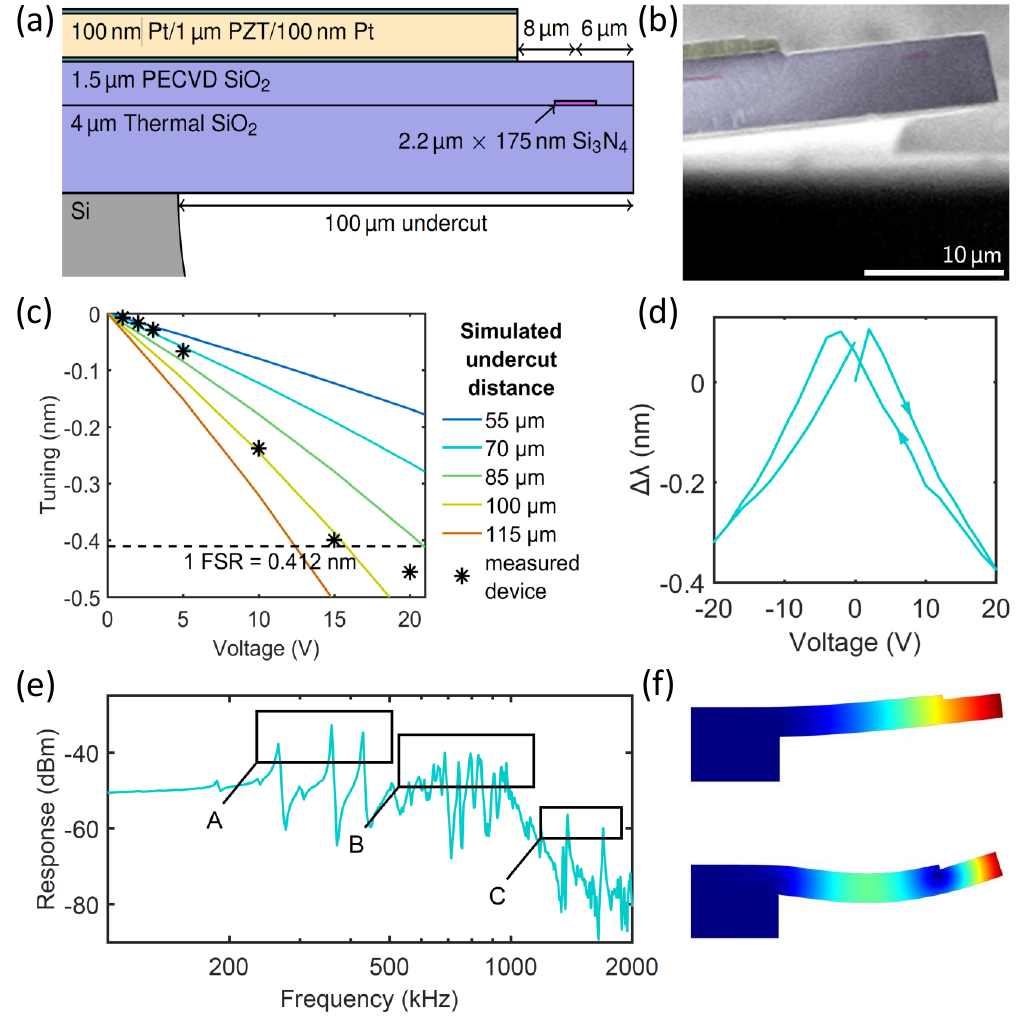}
\caption{
Released PZT tuning of thin Si$_3$N$_4$ microring resonator. 
(a) Cross-section of the device with critical dimensions as labeled. 
(b) False colored SEM of the fabricated device, showing the buckling of the released SiO$_2$ membrane.
(c) Measured (star dot) and simulated (solid lines) tuning of the resonant wavelength. 
The dashed black line denotes the FSR of the microring ring. 
(d) Wavelength tuning under forward and backward sweeping voltages. 
(e) Frequency response of the actuator showing mechanical resonances of the released actuator. 
(f) Numerical simulation of the fundamental and the first higher order vibrational mode of the cantilever. 
Reprinted from Ref. \cite{Jin:18}. 
}
\label{Ch3_PZTrelease}
\end{figure}

\begin{table}[t]
%\centering
\caption{
Comparison among the-state-of-the-art piezoelectric tuning of photonic devices. 
MZI: Mach-Zehnder interferometer.
WG: waveguide. 
WL: wavelength. 
BW: bandwidth. 
The sizes of ring devices are denoted by their radius $R$, while the sizes of MZI devices are by their lengths $L$. 
The power for ring devices is the power consumption per resonance tuning, and that for MZI device is power required for $\pi$ phase shift. 
}
    
\begin{tabular}{ p{0.8cm}|p{0.8cm}|p{0.8cm}|p{0.8cm}|p{1cm}|p{1.2cm}|p{1.2cm}|p{1.2cm}|p{1cm} |p{1cm}|p{1cm} |p{1.2cm} }
 \hline
 Ref.  &  Piezo &WG &WL (nm) & Structure&Released &Size& Tuning (MHz/V) & $V_{\pi} L$ (V$\cdot$cm) &Loss (dB/cm) &BW (MHz)& Power \\
 \hline
\cite{Liu:20a}   & AlN &Si$_3$N$_4$&1550 & Ring & No & R=118$\mu$m  & 15.7  &   450  &0.025 &  10  & 32 pW/MHz\\
\cite{dong2018port}   & AlN &Si$_3$N$_4$&1550 & Ring & Yes & R=118$\mu$m  & 50  &   141 & 1.24 &  1.3  & 20 pW/MHz\\
\cite{stanfield2019cmos}   & AlN &Si$_3$N$_4$& 770& Ring & Yes & R=19.8$\mu$m  & 480  &   16 &0.4 &  7.26  & 8 fW/MHz\\
\cite{dong2022high}   & AlN &Si$_3$N$_4$& 740& MZI & Yes & L=1cm  &   &   50 &0.39 &  120  & 6 nW\\
\cite{sebbag2012}   & PZT &Si& 1550& Ring & No & R=20$\mu$m  & 2000  &   3 & $\sim$10 &  0.01  & \\
\cite{tian2023}   & PZT &Si$_3$N$_4$& 1550& Ring & No & R=57$\mu$m  & 578  &   12.5 & 0.085 &  10  & 1.7 pW/MHz\\
\cite{wang2022silicon}   & PZT &Si$_3$N$_4$& 1550& Ring & No & R=625$\mu$m  & 162  &   43 & 0.03 &  15  & 6 pW/MHz\\
\cite{Jin:18}   & PZT &Si$_3$N$_4$& 1550& Ring & Yes & R=580$\mu$m  & 3250  &   3.6 & 0.3 &  0.27  & 3 pW/MHz\\
\cite{hosseini2015}   & PZT &Si$_3$N$_4$& 640& MZI & No & L=1cm  & -  &   12 & - &  0.6  & 300nW\\
\cite{everhardt12004}   & PZT &Si$_3$N$_4$& 1550& MZI & No & L=1cm  & -  &   16 &0.1  &  1  & 1$\mu$W\\
\cite{shen2023}   & HZO &Si-HZO & 1550& Ring & No & R=70$\mu$m  & 1050  &   3.24 & 48  &  -  & 0.96 pW/MHz\\

\hline
\end{tabular}
\label{table_compare}
\end{table}

%%%%%%%%%%%%%%%%%%%%%%%%%%%%%%%%%%%%%%%%%%%%%%%%%%%
\subsubsection{Released PZT Actuator}

The tuning efficiency can be largely improved by releasing the PZT actuator, as has been demonstrated by Jin \textit{et al}. \cite{Jin:18} and shown in Fig. \ref{Ch3_PZTrelease}. 
The cross-section of the designed device is illustrated in Fig. \ref{Ch3_PZTrelease}(a) where a 1-$\mu$m-thick PZT actuator is placed on a flat SiO$_2$ cladding that is undercut. 
The resonance tuning is primarily realized via bending the SiO$_2$ membrane upon actuation and thus changing the radius. 
For this reason, the Si$_3$N$_4$ waveguide does not need to be underneath the PZT actuator (where the stress is maximum), but instead should be placed close to the edge. 
From the simulation in Fig. \ref{Ch3_PZTrelease}(c), it can be seen that larger tuning can be obtained with wider undercut, since the mechanical structure is softer and easier to bend. 
However, larger undercut leads to lower mechanical vibration frequency and becomes fragile upon release. 
Simulations also show that when the radius of the microring is much larger than the undercut, the change in the radius of the ring $\Delta R$ is nearly independent of the radius. 
Since the wavelength tuning $\Delta \lambda$ is proportional to the relative change of the radius $\Delta R/R$, larger tuning can be obtained for a smaller ring, which is consistent with the results in Fig. \ref{Ch3_PZTSiN}. 
On the other hand, the ratio of the tuning relative to the FSR ($\Delta \lambda/$FSR) is kept constant with respect to the radius. 

The SEM of the fabricated device is shown in Fig. \ref{Ch3_PZTrelease}(b). 
Note that the bending of the cantilever is not due to the piezoelectric actuation, but due to the buckling of the SiO$_2$ membrane after releasing as a result of the build-in stress in SiO$_2$. 
Over one FSR tuning of the resonance is achieved with applied voltage as low as 16 V [see Fig. \ref{Ch3_PZTrelease}(c)], which is equivalent to the tuning efficiency of 3250 MHz/V. 
The tuning becomes highly nonlinear for larger tuning which is due to the ferroelectric nature of the PZT and the rotational motion of the cantilever \cite{Jin:18}. 
Since tuning one FSR is to shift the light by 2$\pi$ in the ring, the $V_{\pi}L$ can be inferred for the 580-$\mu$m-radius device to be 3.6 V$\cdot$cm. 
The tuning hysteresis is reported and characterized in Fig. \ref{Ch3_PZTrelease}(d), where a butterfly-shape curve is also observed. 
As for the power consumption, while a low leakage current ($<10$ nA) is maintained below 5 V, the current keeps increasing and finally leads to the breakdown of the PZT beyond 25 V. 
This is probably due to the low quality of the deposited PZT film, and nA level leakage current can be generally obtained for well prepared PZT \cite{melnick1991}. 
As shown in Fig. \ref{Ch3_PZTrelease}(e), the tuning bandwidth is primarily limited by the mechanical vibrational mode of the cantilever, with the fundamental mode at 265 kHz. 
The simulated mode shapes of the fundamental and first order modes are illustrated in Fig. \ref{Ch3_PZTrelease}(f). 
Note the cut-off of the frequency response beyond 1 MHz in Fig. \ref{Ch3_PZTrelease}(e) is due to the RC time constant of the actuator with 10 nF capacitance. 

%%%%%%%%%%%%%%%%%%%%%%%%%%%%%%%%%%%%%%%%%%%%%%%%%%%
%%%%%%%%%%%%%%%%%%%%%%%%%%%%%%%%%%%%%%%%%%%%%%%%%%%
\subsection{Comparison of the State-of-the-Arts}

In this section, we summarize and compare the performance of piezoelectric tuning of integrated photonic devices demonstrated with AlN and PZT actuators, providing a guidance for future device design, as shown in Table \ref{table_compare}. 
It can be seen that PZT actuators generally show an order of magnitude higher tuning efficiency than AlN, because of its larger piezoelectric response. 
However, the AlN actuator shows much more linear response, negligible hysteresis, low leakage current, bi-directional tuning (which means zero biasing), and compatibility with CMOS processes. 
These features are highly desired in applications that require fast and repeatable transduction of input signals, such as FMCW LiDAR \cite{riemensberger:2020, lihachev2022}. 
On the other hand, released actuator has larger tuning range than unreleased device, but at the expense of narrower tuning bandwidth. 
The currently most efficient tuning is achieved by combing the PZT and releasing process, as demonstrated by Ref. \cite{Jin:18}. 
Notably, Si waveguide shows several times larger tuning than Si$_3$N$_4$ because of its higher photoelastic coefficient and refractive index. 
However, Si waveguide demonstrated thus far has propagation loss much higher than the Si$_3$N$_4$. 
With further optimization of the fabrication to suppress the losses, piezoelectric tunable Si photonics could become a promising platform for programmable photonic networks \cite{dong2022high, harris2016large, shen2017deep} and reconfigurable optical signal processors. 

%%%%%%%%%%%%%%%%%%%%%%%%%%%%%%%%%%%%%%%%%%%%%%
%%%%%%%%%%%%%%%%%%%%%%%%%%%%%%%%%%%%%%%%%%%%%%
%%%%%%%%%%%%%%%%%%%%%%%%%%%%%%%%%%%%%%%%%%%%%%
\section{Piezoelectric Tuning in Laser Applications}

For many metrology and sensing applications of lasers and optical frequency combs, the ability to achieve high-speed actuation of laser frequency, comb teeth, and repetition rate is critical \cite{Cundiff2003}. 
For example, locking combs to microwave standards is key in optical frequency synthesis \cite{jones2001}. 
The phase-locking of combs to stable reference cavities is also central for low-noise microwave generation via optical frequency division \cite{fortier2011}. 
Optical clocks require frequency combs to be locked to atomic transitions \cite{takamoto2005}, and higher-bandwidth actuators are needed to keep pace with the improvement in clock fractional frequency uncertainty \cite{fortier2019}. 
Frequency combs based on mode-locked lasers have established elaborate techniques for wideband frequency actuation, typically realized using bulk phase or amplitude modulators within the laser cavity. 
In contrast, measurement-based high-speed feedback stabilization of chip-based microcombs is currently only achieved with off-chip bulk modulators that actuate on the pump laser \cite{Liu:20}. 
On the other hand, frequency agility of the laser is often a further key requirement to lock it to fibre gratings or achieve cycle-slip-free phase locking \cite{adler2004}. 
Also, the tuning speed is one of the limitations to the resolution of FMCW LiDAR \cite{riemensberger:2020}. 
Therefore, on-chip high-speed actuators are highly desirable for integrated laser and soliton microcomb sources.   

Due to its low optical losses and high Kerr nonlinearity, Si$_3$N$_4$ has become one of the main material platforms in hybrid integrated lasers with narrow linewidth \cite{Jin:21} and soliton microcombs with low pump threshold \cite{Kippenberg:11}. 
While modulators integrated with Si$_3$N$_4$, e.g. based on the electro-absorption of graphene \cite{Gruhler:13, phare2015, yao2018gate, Datta:20} and ferroelectric PZT \cite{alexander2018} have been demonstrated, they have not shown compatibility with soliton generation which requires maintaining ultralow optical losses. 
As a result, current integrated actuation techniques for soliton microcombs rely primarily on metallic heaters \cite{Joshi:16, Xue:16}, which have limited actuation bandwidth up to 10 kHz. 
Furthermore, heaters exhibit unidirectional tuning and typically consume electrical power exceeding 30 mW \cite{Stern:18}, higher than the optical power threshold for soliton formation in state-of-the-art integrated devices \cite{Stern:18, Raja:19}, and are not compatible with cryogenic operation \cite{Moille:2019, Stanfield:19}. 
Alternatively, the direct generation of solitons in Pockels materials such as AlN \cite{xiong2012, Jung:13, LiuX:18, gong2018} and LiNbO$_3$ \cite{Gong:20, He:19, Wang:19} can allow high-speed actuation, however these platforms are not yet as mature as Si$_3$N$_4$.

In this chapter, we will review current progress on the piezoelectric tuning of chip-scale integrated lasers and soliton microcombs. 
We will elaborate on the working principle, main characteristics, and design considerations. 
Finally, we will compare with other schemes and discuss future applications and improvements. 

%%%%%%%%%%%%%%%%%%%%%%%%%%%%%%%%%%%%%%%%%%%%%%
%%%%%%%%%%%%%%%%%%%%%%%%%%%%%%%%%%%%%%%%%%%%%%
\subsection{Narrow-Linewidth, Frequency-Agile Integrated Laser} 

Over the past decades, the heterogeneous integrated semiconductor laser has revolutionized the application of laser in our daily life, from a laser pointer to a data center, due to its compact size and low cost \cite{komljenovic2015}.
However, its frequency noise is still higher than the fiber laser. 
It is only most recently that the coupling to a low-loss Si$_3$N$_4$ photonic circuit has made it possible for a chip-scale, low-noise laser, thanks to the maturity of hybrid and heterogeneous integration \cite{porter2023}. 
A variety of coupling schemes has been invented, such as wafer bonding \cite{Xiang:21, xiang:2020, xiang2021, xiang2023}, edge coupling \cite{Shen:20, Jin:21, briles2021}, and photonic wire bonding \cite{xu2021hybrid, lindenmann2012, lindenmann2014}. 
A Hertz-linewidth laser has been achieved by properly tuning the laser injection current such that the CW laser's frequency coincides with the resonance of an external Si$_3$N$_4$ microresonator \cite{Jin:21, guo2022chip}.
The laser emission frequency is then self-injection-locked to the Si$_3$N$_4$ microresonator, a phenomenon in which the laser dynamics is impacted by external feedback light in the case without an optical isolator \cite{kondratiev2017}.
When light is coupled into the microresonator, bulk and surface Rayleigh scattering inside the microresonator can backscatter some portion of the light and inject it into the laser, which triggers laser self-injection locking and impacts the laser dynamics. 
The consequence is that the new laser emission frequency is tightly locked to the microresonator resonance whose frequency is close to the initial free-running laser's emission frequency. 
The laser linewidth and noise are significantly reduced in the presence of this passive optical feedback (in contrast to active locking of laser frequency such as the Pound–Drever–Hall (PDH) lock), and the linewidth is ultimately limited by the thermal-refractive noise (TRN) of the Si$_3$N$_4$ microresonator \cite{huang2019}. 

\begin{figure}[t]
\centering
\includegraphics[width=13.3 cm]{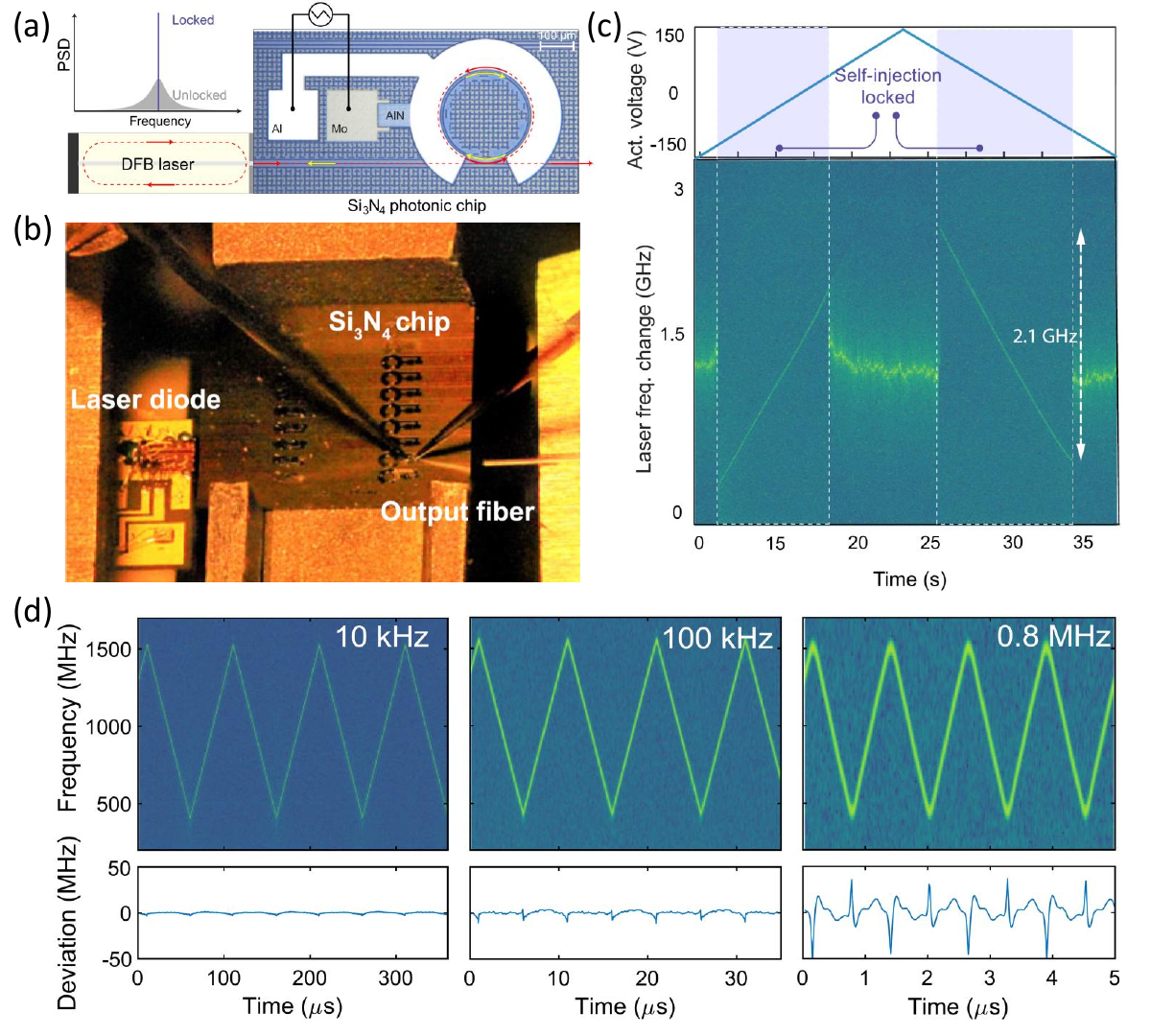}
\caption{
Tunable hybrid DFB laser system integrated with AlN actuator. 
(a) Schematic of the DFB laser self-injection locked to a high-Q Si$_3$N$_4$ microring resonator, which is tuned by the AlN actuator. 
(b) Photo of the experimental setup with DFB laser edge coupled to the Si$_3$N$_4$ chip. 
(c) Spectrogram of laser frequency change by sweeping the voltage applied to the AlN actuator from -150 to 150 V. 
The purple shaded areas correspond to the self-injection locking range of 2.1 GHz. 
(d) Time-frequency spectrogram of the heterodyne beat-notes between the SIL laser and a reference diode laser under triangular chirp at different repetition rates.
Bottom insets are the deviation from the ideal triangular wave.
Reprinted from Ref. \cite{lihachev2022}. 
}
\label{Ch4_DFB_AlN}
\end{figure}

\subsubsection{Piezoelectric Tunable Self-Injection Locked DFB Laser}
In the self-injection locking (SIL) state, the laser frequency is predominantly determined by the resonant frequency of the Si$_3$N$_4$ microresonator, which is fixed upon fabrication. 
The fast tuning of the laser frequency is made possible by monolithic integration of the piezoelectric actuator on top of the Si$_3$N$_4$ microresonator, which was first demonstrated in Ref. \cite{lihachev2022}.
As illustrated in Fig. \ref{Ch4_DFB_AlN}(a,b), a Distributed Feedback (DFB) laser is edge coupled to a Si$_3$N$_4$ microring resonator which is actuated by a donut-shaped AlN actuator. 
Upon applying electrical signal, the frequency tuning is transferred from the optical resonance to the locked laser frequency, as shown in Fig. \ref{Ch4_DFB_AlN}(c). 
Initially, the frequencies of the resonator and laser are brought close and set into SIL state by adjusting the current of the laser. 
The laser frequency can be tuned by the AlN actuator as long as it is maintained in the SIL state.
Therefore, the tuning range is ultimately limited by the SIL range which is 2.1 GHz in Ref. \cite{lihachev2022}. 
This range can be maximized by adjusting the gap between the laser chip and Si$_3$N$_4$ chip for optimal feedback phase, and can be further extended by increasing the backscattering \cite{Voloshin:21, kondratiev2017}, or tuning the laser and resonator concurrently.
It can be seen from Fig. \ref{Ch4_DFB_AlN}(c) that the laser linewidth narrowing in the SIL state is maintained during the tuning.
The frequency noise is characterized by the optical cross-correlation-based noise spectrum \cite{xie2017}, and the intrinsic laser linewidth is inferred from the white noise floor of the spectrum \cite{mercer19911}.
An intrinsic linewidth of 900 Hz is obtained for the device shown in Fig. \ref{Ch4_DFB_AlN} with a maximum output power of 1.5 mW. 

Figure \ref{Ch4_DFB_AlN}(d) shows the laser frequency tuning under a triangularly modulated voltage (150 V$_{\text{pp}}$) applied on the AlN actuator. 
The time-varying laser frequency is generally characterised by measuring a heterodyne beatnote with a reference ECDL on a fast photodetector, which produces a time-frequency spectrogram. 
Consequently, the laser output frequency is triangularly chirped with frequency excursion of 1.1 GHz and speed up to 0.8 MHz.
Due to the linearity of the AlN actuator, the frequency chirping nonlinearity is below 10 MHz (defined as the Root Mean Square (RMS) deviation of the measured frequency from an ideal triangular wave calculated by least-squares fitting), and no tuning hysteresis is observed during continuous operation. 
The degradation of the deviation is primarily due to the nonlinearity of the voltage amplifier at high frequency. 
This highly linear tuning behavior facilitates the generation of chirps without the need for active or passive linearization, which is ideal for direct implementation in long-range FMCW LiDAR systems \cite{feneyrou2017, feneyrou2017frequency, riemensberger:2020}. 
As discussed in the previous chapter, the modulation speed is limited by the chip's mechanical modes, such as the contour modes \cite{piazza2006} and HBAR modes \cite{Tian:20}. 
In Ref. \cite{lihachev2022}, a flat modulation response is extended to 10 MHz by implementing a differential drive on an apodized chip with carbon tape.
The linear frequency tuning speed of 1.6 PHz/s is achieved. The tuning speed of the hybrid laser is ultimately limited by the photon lifetime of the cavity and the carrier lifetime of the laser and EHz/s tuning speeds should be possible similar to recent work in electro-optical tuning \cite{Snigirev:23}.

\begin{figure}[t]
\centering
\includegraphics[width=13.3 cm]{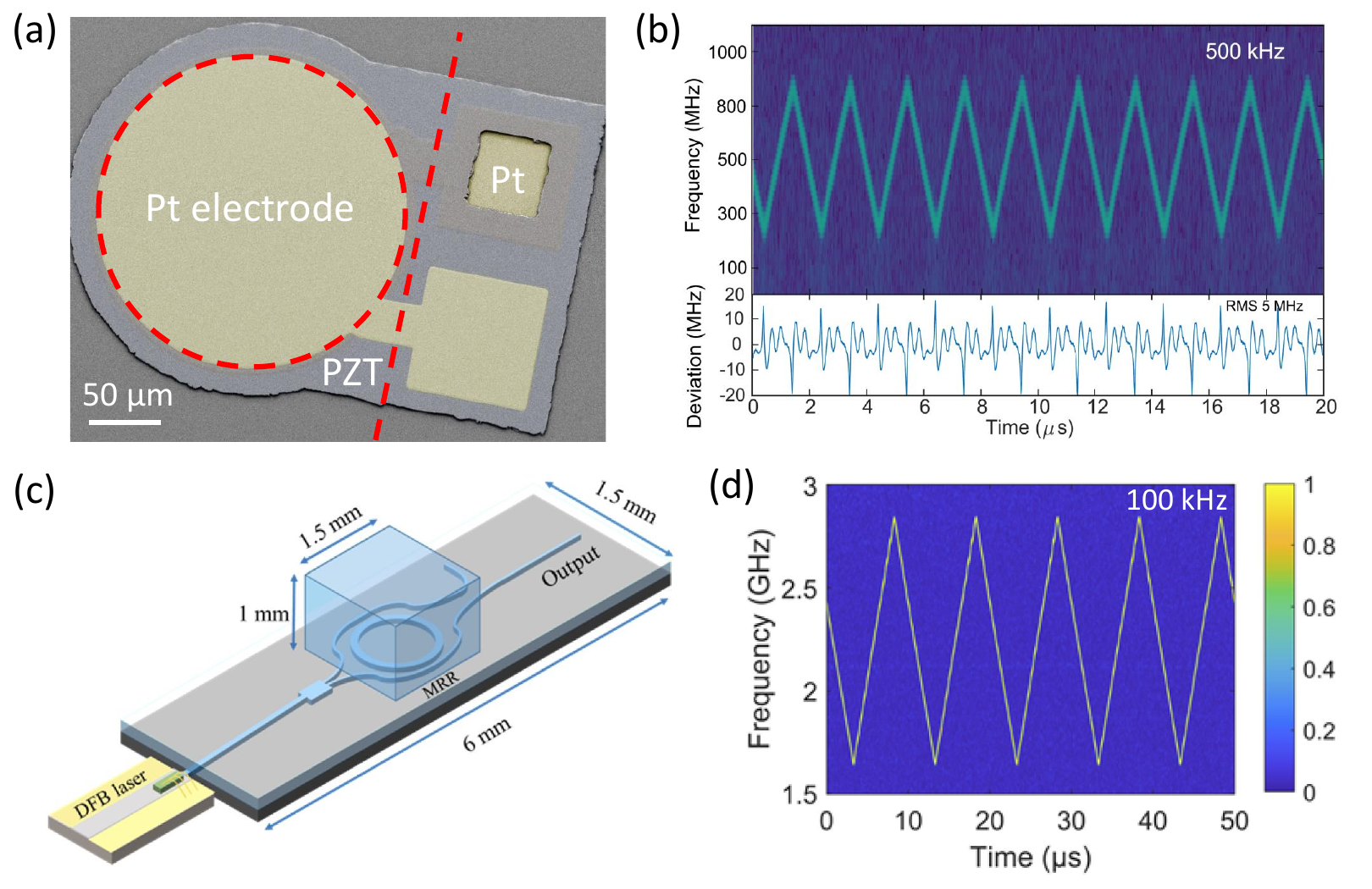}
\caption{
Hybrid DFB laser system tuned with PZT actuator.
(a) False-color SEM of the PZT actuator. 
The Si$_3$N$_4$ waveguides are indicated as red dashed lines.
(b) Time-frequency spectrogram of the heterodyne beat-note of the SIL laser under 500 kHz triangular chirp repetition rate by applying 1.1 V$_{\text{pp}}$ to the PZT actuator in (a).
(a, b) reprinted from Ref. \cite{lihachev2022}. 
The lower panel shows the deviation from an ideal linear chirping.
(c) DFB SIL laser actuated by a bulk PZT chip covering the Si$_3$N$_4$ microring resonator.
(d) Spectrogram showing the laser frequency chirping by applying 100 kHz 40 V$_{\text{pp}}$ triangular wave to the PZT chip. 
The maximum frequency excursion is 1.2 GHz and the chirp nonlinearity is 12 MHz.
(c, d) reprinted from Ref. \cite{tang2022}. 
}
\label{Ch4_DFB_PZT}
\end{figure}

In addition to AlN, other piezoelectric materials with larger piezoelectric coefficient, such as lead zirconium titanate (PZT), can be used to reduce the required voltage for a given frequency excursion. 
The same PZT actuator shown in Fig.\ref{Ch3_PZTSiN} was used by the same group to tune the frequency of a SIL DFB laser, as demonstrated in Fig. \ref{Ch4_DFB_PZT}(a,b). 
A 525 MHz frequency excursion at the repetition rate of 500 kHz is achieved by applying only 1.1 V$_{\text{pp}}$ triangular wave (see Fig. \ref{Ch4_DFB_PZT}(b)).
The laser tuning was linearized using an iterative algorithm to correct for the intrinsic nonlinearity and hysteresis of the PZT actuator \cite{lihachev2022}. 
As a result, the measured RMS nonlinearity is 5 MHz, and the relative nonlinearity, defined as the ratio of the nonlinearity to the total frequency excursion, is less than 1$\%$. 
The small sweeping voltage is also compatible with CMOS integrated circuits (e.g., application-specific integrated circuit (ASIC)), paving the way for a fully photonic-electronic integrated tunable hybrid laser systems \cite{lukashchuk2023}. 

\begin{figure}[t]
\centering
\includegraphics[width=13.3 cm]{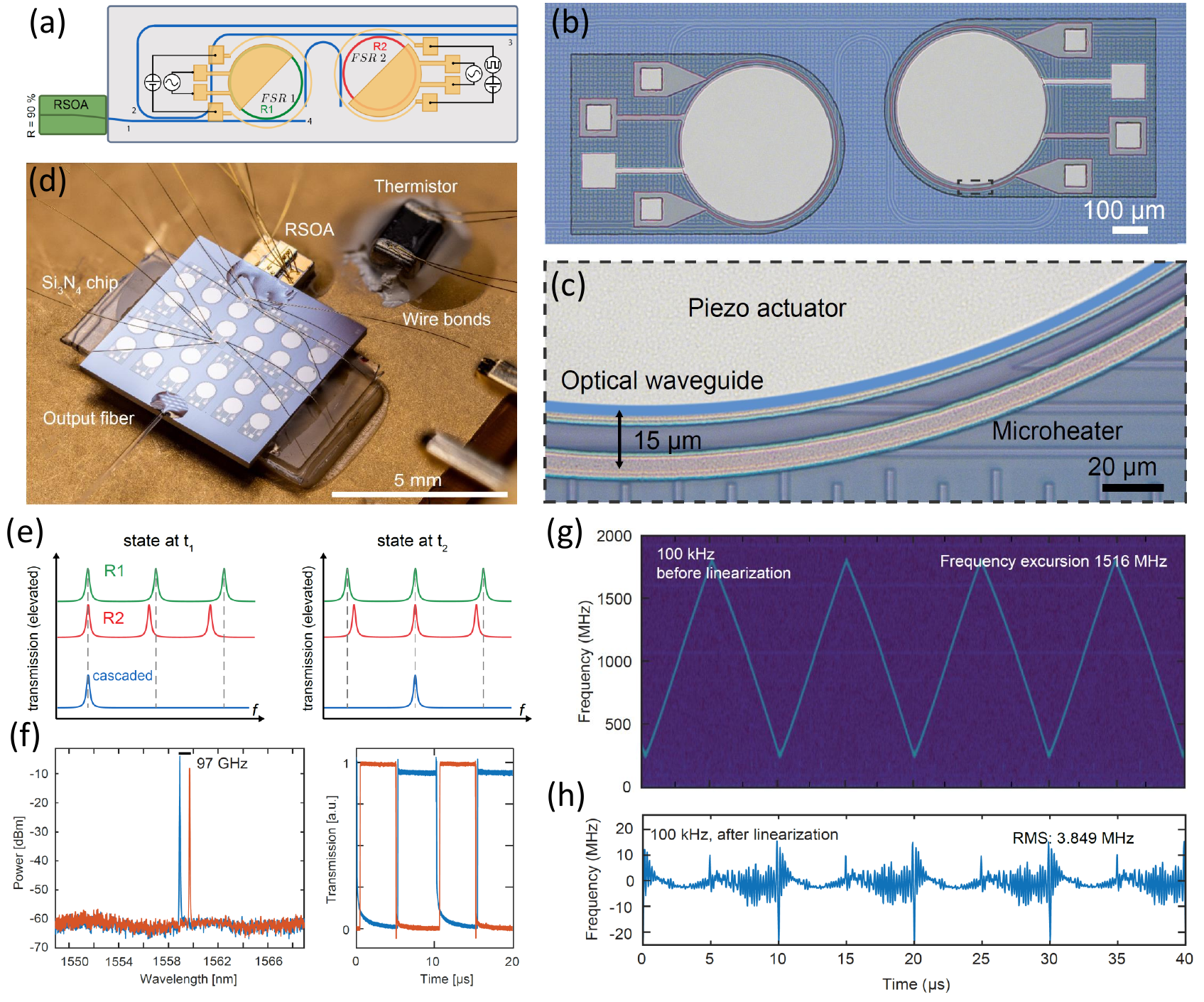}
\caption{
PZT tunable external cavity laser with Vernier filter. 
(a) Schematic of the laser system where a reflective semiconductor optical amplifier (RSOA) is edge coupled to a Si$_3$N$_4$ Vernier filter made from two cascaded microring resonators. 
(b) Optical microscope image of PZT actuators. 
(c) Zoom-in of the black dashed box in (b), showing the microheater surrounding the Si$_3$N$_4$ microring (blue line). 
(d) Photo of the hybrid packaged laser system. 
(e) Schematic of the principle of wavelength switching via the Vernier effect, where the total reflection (blue curve) changes from one resonance to the other. 
(f) Experimental results of the wavelength switching.
Left: Output laser spectra showing wavelength switched by one FSR.
Right: Time domain dynamics of the switching with 100 kHz repetition rate. 
(g) Time-frequency spectrogram of the frequency tuning at 100 kHz before linearization.
(h) Deviation from linear tuning after linearization with RMS of 3.849 MHz. 
Reprinted from Ref. \cite{lihachev2023}. 
}
\label{Ch4_vernier}
\end{figure}

Another group has demonstrated the tuning with a piece of bulky PZT chip that is glued on top of the Si$_3$N$_4$ circuits \cite{tang2022}, as shown in Fig. \ref{Ch4_DFB_PZT}(c).
The PZT chip has a size of 1.5$\times$1.5 mm$^2$ and thickness of 1 mm. 
The advantage of this approach is the lower cost compared to thin-film PZT, where the bulk PZT is widely available from off-the-shelf. 
Also, bulk PZT can generate a stronger piezoelectric force. 
However, the disadvantage is the incompatibility with Si$_3$N$_4$ resonators with sub-mm diameter and wafer-scale production.
The bulk PZT chip also supports a rich series of mechanical modes with frequency as low as 300 kHz.
To alleviate this problem, a low-pass filter with 50 kHz cutoff frequency is applied to the input electrical signal.
To maintain linear tuning, the authors also adopted iterative learning to regulate the applied voltage until a small nonlinearity is reached \cite{zhang2019laser}.
The experimental result of the tuning is illustrated in Fig. \ref{Ch4_DFB_PZT}(d), where a 40 V$_{\text{pp}}$ 100 kHz triangular wave is applied. 
A frequency excursion of 1.2 GHz is obtained with a nonlinearity of 12 MHz.
The higher required voltage is due to the much thicker PZT chip. 
A narrow intrinsic linewidth of 22 Hz is achieved in the SIL state with a maximum output power of 8.9 mW. 

\subsubsection{Wavelength Switchable RSOA-Based External Cavity Laser}
Although self-injection locked DFB laser has produced a narrow linewidth on par with the fiber laser, it requires careful tuning of the setpoint.
Also, the manufacture of DFB diode requires making gratings on III-V dies which increases the cost. 
In contrast, the external cavity laser (ECL) based on reflective semiconductor optical amplifier (RSOA) has enabled turn-key operation and low cost.
Moreover, the external cavity increases the photon lifetime in the laser cavity which narrows the linewidth compared to the DFB laser \cite{xiang2019, guo2022chip}. 
The coupling with photonic feedback circuits provides frequency-selective reflection and makes sub-kHz linewidth within reach \cite{boller2019, tran2019}.
A variety of PIC platforms have been demonstrated such as Si \cite{tran2019, kita2015}, Si$_3$N$_4$ \cite{boller2019}.
However, these lasers are tuned by microheaters with limited speed.
A recent work has coupled the RSOA with a LiNbO$_3$ photonic circuit and demonstrated tuning up to 500 MHz \cite{li2022integrated}, but an intrinsic linewidth of 11 kHz is obtained, which is primarily limited by the charge  noise \cite{zhang2023fundamental}. 
Therefore, the piezoelectric tuning provides a solution for both low noise and fast frequency tuning. 

The piezoelectric tuning of RSOA-based ECL has been demonstrated using PZT actuators as shown in Fig. \ref{Ch4_vernier} \cite{lihachev2023}. 
An RSOA is edge coupled to a Si$_3$N$_4$ chip consisting of a Vernier filter made from two microring resonators, which have slightly different FSR (FSR$_1$ = 96.7 GHz and FSR$_2$ = 97.9 GHz). 
The Vernier filter serves a mirror of the laser cavity and only reflects light at the frequency when the resonances of the two resonators align. 
This selective reflection provides us a way to tune the laser frequency by programming the resonances of the two resonators.
The fabricated device is illustrated in Fig. \ref{Ch4_vernier}(b), where each of the microring is covered by a PZT actuator. 
Different from previous DFB laser, a microheater is co-integrated with PZT which is 15 $\mu$m away from the microring (see Fig. \ref{Ch4_vernier}(c)).
To keep a small distance between the heater and microring, the optical microring is placed right beneath the edge of the top electrode, which would compromise the piezoelectric tuning efficiency.
Nevertheless, an efficiency of 166 MHz/V is obtained.
As shown in Fig. \ref{Ch4_vernier}(d), the laser system is hybrid packaged, including RSOA, Si$_3$N$_4$ chip, and temperature control.
The electrical signal is applied through wirebonding of all actuators and heaters in a custom butterfly package, and the laser is output through optical fiber fixed by epoxy at the edge.
Maximum of 6 mW laser is output with an intrinsic linewidth of 400 Hz.
The packaging will improve the long-term laser stability and reduce laser frequency noise below 1 kHz offsets by eliminating acoustic instabilities \cite{lihachev2023}. 

There are two frequency tuning schemes supported by the Vernier filter. 
The first one is wavelength switching by tuning the resonance alignment of the two resonators from one pair to another that is one FSR away, as illustrated in Fig. \ref{Ch4_vernier}(e). 
This frequency tuning is discrete but can cover a large range (determined by the FSR) by only tuning a small amount of the resonant frequency of one microring (determined by the FSR difference), which is preferable for piezoelectric actuators.
In the experiment, a heater is necessary to initially align the two resonators which are generally far off from each other due to the fabrication variability. 
Following that, the resonance of one resonator can be tuned by its piezoelectric actuator relative to the other. 
The wavelength switching direction is determined by which microring is actuated. 
Figure \ref{Ch4_vernier}(f) shows the emission wavelength under two switching states that are separated by an FSR.
The switching speed is tested by applying 8 V$_{\text{pp}}$ 100 kHz square wave and a 7 ns rise time is achieved. 
The second scheme is the continuous frequency tuning by actuating the two microrings concurrently. 
As illustrated in Fig. \ref{Ch4_vernier}(g), a frequency excursion of 1.5 GHz is generated under 100 kHz triangular wave with 11 V$_{\text{pp}}$. 
To suppress the nonlinearity, two algorithms are combined to correct the input voltage, including the frequency response inversion \cite{feneyrou2017frequency} and interative signal correction. 
After linearization, the RMS nonlinearity changes from 15 MHz to 3.8 MHz, as shown in Fig. \ref{Ch4_vernier}(h). 

\begin{figure}[t]
\centering
\includegraphics[width=13.3 cm]{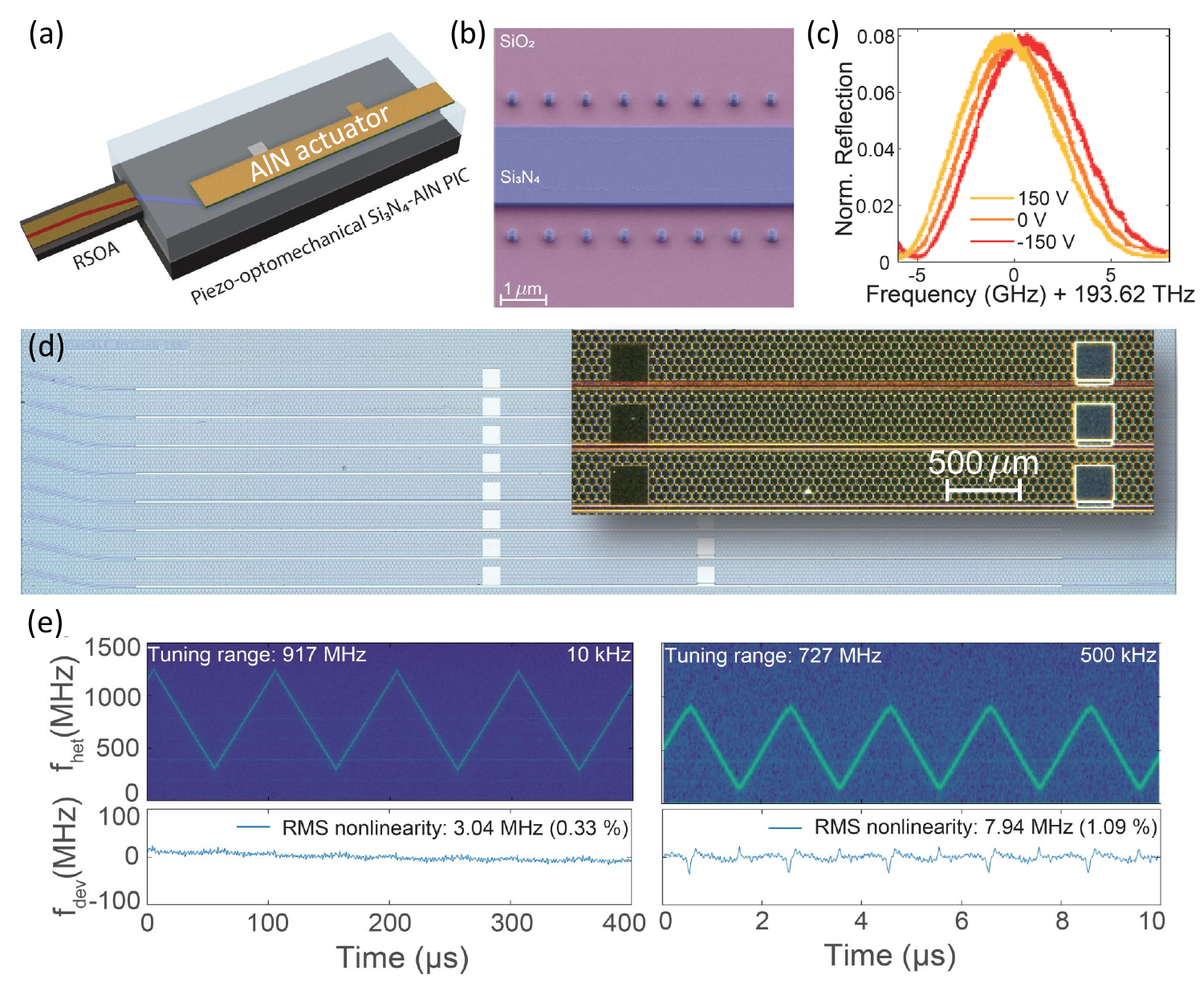}
\caption{
Piezoelectric tunable RSOA-based E-DBR laser.
(a) Schematic of the hybrid integrated E-DBR laser which is tuned by the AlN actuator. 
(b) False-color SEM of the Si$_3$N$_4$ Bragg gating formed by circular posts on both sides. 
(c) Piezoelectric tuning of the reflection of the Bragg grating.
(d) Optical microscope image of the photonic chip with 8-mm long Si$_3$N$_4$ E-DBR and AlN actuator. 
Inset: zoom-in of the AlN actuator. 
(e) Time-frequency spectrogram of the laser frequency tuning by applying 10 kHz (left) and 500 kHz (right) triangular wave with 250 V$_{\text{pp}}$. 
The lower panels show the deviation from ideal linear tuning.
Reprinted from Ref. \cite{siddharth2023}. 
}
\label{Ch4_DBR}
\end{figure}

\subsubsection{High Power and Tunable Extended-DBR Laser}
In addition to the Vernier filter, the extended distributed Bragg reflector (E-DBR) made from Si$_3$N$_4$ PIC has also been widely studied due to its narrow bandwidth, high side-lobe supperssion, and ease of design and fabrication \cite{xiang2019, xiang2021}. 
Recently, the piezoelectric tuning of Si$_3$N$_4$ E-DBR and the consequent laser frequency has been demonstrated \cite{siddharth2023}, as shown in Fig. \ref{Ch4_DBR}. 
An 8-mm long Si$_3$N$_4$ E-DBR is formed by placing periodic circular posts on both sides of the Si$_3$N$_4$ waveguide which weakly perturbs the evanescent field (see Fig. \ref{Ch4_DBR}(b)).
As shown in Fig. \ref{Ch4_DBR}(d), a straight AlN actuator covers the whole E-DBR. 
A novel fabrication procedure for the AlN actuator (`pullback process') is introduced in this work to eliminate the bottom metal beneath the positive signal pad (made from top metal), which reduces the total capacitance and eases the wirebonding.
The voltage tuning of the reflection from the Bragg grating is shown in Fig. \ref{Ch4_DBR}(c) where the central frequency is shifted.
By edge coupling to an RSOA gain chip, a single frequency lasing is generated with maximum power of 25 mW and an intrinsic linewidth of 3.45 kHz.
The laser intrinsic linewidth can be further narrowed down to 3.8 Hz by SIL to a high-$Q$ Si$_3$N$_4$ spiral resonator with 800 MHz FSR.
The linear frequency tuning of the RSOA-based E-DBR laser is shown in Fig. \ref{Ch4_DBR}(e).
Thanks to the linearity of the AlN actuator, a small relative nonlinearity of 0.33-1$\%$ is obtained without any predistortion correction \cite{siddharth2023}.  

\begin{table}[t]
    %\centering
    \caption{
    Comparison between the hybrid or heterogeneous integrated lasers that are not tunable and tunable with thermal-optic, electro-optic (EO), and stress-optic effects. 
    PDH: Pound–Drever–Hall locking.
    LN: LiNbO$_3$. NA: not available. 
    The photonic circuits (except for the EO devices) are all based on the Si$_3$N$_4$ PIC. 
    Note 1: Nonlinearity is the value after linearization algorithm being implemented.  
    }
    
    \begin{tabular}{ p{0.8cm}|p{0.8cm}|p{1.8cm}|p{1.5cm}|p{1.2cm}|p{1.2cm}|p{1.2cm}|p{1.2cm}|p{1.2cm}|p{1.2cm}
    }
 \hline
 Ref.  &  Year &Laser scheme & Tuning scheme & Linewidth (Hz)& Max. power (mW) & Tuning range & V$_{\text{pp}}$ (V) & Speed (kHz) & Nonlinearity \\
 \hline
\cite{Jin:21}   & 2021 & DFB SIL & - & 1.2 & 30 & -  & -  & - & - \\
\cite{xiang2021}   & 2021 & E-DBR & - & 780 & >10 & -  & -  & - & - \\
\cite{xiang2021}   & 2021 & E-DBR SIL & - & 3 & - & -  & -  & - & - \\
\cite{guo2022chip}   & 2022 & E-DBR SIL + PDH& - & 1.1 & - & -  & -  & - & - \\
\hline
\cite{fan2020hybrid}   & 2020 & Vernier & Thermal & 40 & 23 & 74 nm & -  & - & - \\
\cite{tang2021}   & 2021 & DFB SIL & Thermal & 49.9 & 1 & 5.6 GHz & 6  & 1 & 0.004$\%$$^1$ \\
\hline
\cite{li2022integrated}   & 2022 & Vernier & EO (LN) & 11,300 & 5.5  & 2 GHz & 6 & 600,000 & 10$\%$ \\
\cite{Snigirev:23}   & 2023 & DFB SIL & EO (LN+Si$_3$N$_4$) & 3,000 & 0.15  & 0.6 GHz & 25 & 10,000 & 1.7$\%$ \\
\hline
\cite{lihachev2022}   & 2022 & DFB SIL & Stress (AlN) & 900 & 1.5  & 1.1 GHz & 150 & 800 & 0.9$\%$ \\
\cite{lihachev2022}   & 2022 & DFB SIL & Stress (PZT) & 900 & 1.5  & 525 MHz & 1.1 & 500 & 1$\%$ \\
\cite{tang2022}   & 2022 & DFB SIL & Stress (PZT) & 22 & 8.9  & 1.2 GHz & 40 & 100 & 1$\%$$^1$ \\
\cite{lihachev2023}   & 2023 & Vernier & Stress (PZT) & 400 & 6  & 1.5 GHz & 1.1 & 100 & 0.25$\%$$^1$ \\
\cite{siddharth2023}   & 2023 & E-DBR & Stress (AlN) & 3,450 & 25  & 900 MHz & 250 & 500 & <1$\%$ \\
\hline

\end{tabular}
    \label{table_laser}
\end{table}

Table \ref{table_laser} summarizes the performances of the piezoelectric tuned hybrid lasers and compares them with the state-of-the-art hybrid or heterogeneous integrated lasers with fixed frequency, as well as lasers that are tuned by the heater or electro-optic effect. 
We can see most of the lasers with sub-kHz linewidth are achieved by self-injection locking a DFB laser or an ECL with E-DBR to an ultra-high $Q$ Si$_3$N$_4$ resonator. 
The frequency noise is primarily limited by the thermal-TRN of the Si$_3$N$_4$ \cite{huang2019}. 
A recent work further suppresses noise beyond TRN by locking an E-DBR SIL laser to a Fabry-Pérot cavity via the PDH technique \cite{guo2022chip}. 
While the PDH signal was generated by an off-chip electro-optic modulator, the required phase modulation may also be generated via a piezoelectric actuator on chip signal has been generated on-chip by the piezoelectric actuator that is compatible with the Si$_3$N$_4$ PICs \cite{Liu:20a}. 
The thermo-optic tuning has achieved the largest tuning range at the expense of high power consumption and slow speed, and the tuning range decreases as the speed increases.
On the other hand, electro-optic tuning based on LiNbO$_3$ PICs presents the highest speed up to 600 MHz.
However, the laser linewidth is above kHz even with SIL.
A recent study reveals that the charge noise of the LiNbO$_3$ ultimately sets the noise floor of the hybrid laser \cite{zhang2023fundamental}.
As for the piezoelectric tuning, lasers with sub-kHz linewidth, over 1 GHz tuning range, and hundreds of kHz modulation speed have been demonstrated. 
So far, these experiments are implemented by edge coupling the DFB laser or gain chip to the Si$_3$N$_4$ photonic chip.
In the furture, the development heterogeneous integration of the laser with piezoelectric actuators and the photonic wirebonding technique will be of great interest for building a more compact and stable chip-scale laser system \cite{xiang2023, lindenmann2014}.

%%%%%%%%%%%%%%%%%%%%%%%%%%%%%%%%%%%%%%%%%%%%%%
%%%%%%%%%%%%%%%%%%%%%%%%%%%%%%%%%%%%%%%%%%%%%%
\subsection{Monolithic Piezoelectric Control of Soliton Microcombs}

In this section, using integrated piezoelectric actuators to control soliton microcombs is demonstrated, which allows deterministic soliton initiation, switching, tuning, long-term stabilization, and phase locking with 0.6 MHz bandwidth.
The same device shown in Fig. \ref{Ch3_device} is employed for the soliton tuning, where a disk-shaped AlN actuator is placed on top of the Si$_3$N$_4$ microring resonator.
The AlN actuators feature linear and bi-directional tuning, free from hysteresis, and operating at ultralow electrical power. 
Also, the optical $Q$ is unchanged upon tuning, which is important for maintaining stable soliton states. 
More details of the resonance tuning can be found in the corresponding section. 

\subsubsection{Voltage Initiation and Tuning of Soliton Microcombs}

\begin{figure*}[t]
\centering
\includegraphics[width=13.5 cm]{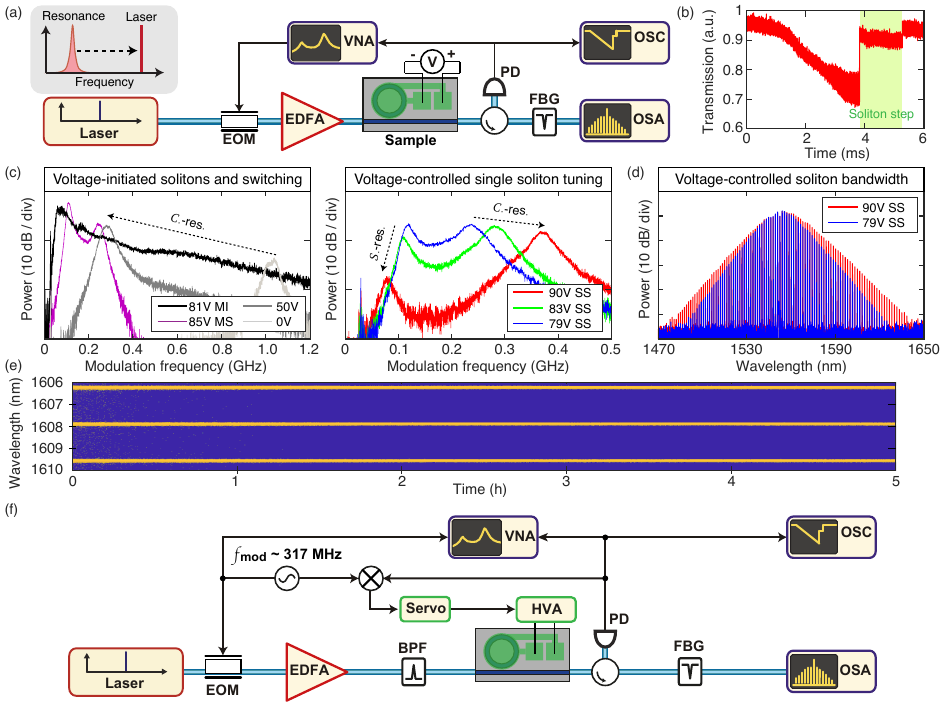}
\caption{
Experimental demonstration of soliton control via AlN actuation.
(a) Experimental setup. 
%OSA: optical spectrum analyser. PD: photodiode
OSC: oscilloscope. FBG: Fibre Bragg grating. 
Inset: the soliton is initiated by tuning the resonance relative to the fixed-frequency laser.
(b) A typical soliton step with millisecond length.
(c) Soliton initiation and state switching via AlN actuation. 
Left: Initially, the resonance is 1 GHz detuned from the laser (0 V). 
As the resonance is tuned to the laser, modulation instability (MI, 81 V) and a multi-soliton state (MS, 85 V) are generated. 
Right: the voltage is decreased to switch to a single soliton state (SS, 79 V). 
The soliton detuning and the bandwidth are further increased via increasing the voltage (90 V).
\textit{S}-res.: Soliton resonance. \textit{C}-res.: Cavity resonance. 
(d) Spectrum of the soliton under different applied voltages. 
(e) Soliton stabilization over 5 hour by locking the resonance to the laser.
(f) Experimental setup for long-term stabilization of the soliton microcomb. 
%OSC: oscilloscope. FBG: fiber Bragg grating. 
HVA: high voltage amplifier. 
BPF: bandpass filter. 
Reprinted from Ref. \cite{Liu:20a}. 
}
\label{Fig:2-1}
\end{figure*}

A soliton is initiated by coupling a CW laser into the microresonator and piezoelectric tuning the resonance to the laser \cite{Joshi:16} via varying the voltage, using the setup shown in Fig. \ref{Fig:2-1}(a).
The laser is initially blue-detuned by 1~GHz from the microresonator resonance with 15 mW power coupled into the waveguide.
A typical soliton step of millisecond duration is shown in Fig. \ref{Fig:2-1}(b).
Though not required for soliton initiation, the resonance-laser detuning is monitored using an EOM and a vector network analyser (VNA) \cite{guo2017}. 
As shown in Fig. \ref{Fig:2-1}(c), the resonance is gradually tuned towards the laser, and a modulation instability (MI) state is generated at 81 V, followed by a multi-soliton state (MS, 85 V).
Next, the AlN voltage is reduced such that the backward tuning enables switching \cite{guo2017} to the single soliton state (SS, 79 V). 
The voltage is increased again (90 V) to increase the soliton bandwidth. 
Figure \ref{Fig:2-1}(d) shows different soliton spectrum with different applied voltages.

The piezoelectric control of soliton via AlN actuators enables the implementation of feedback and elimination of detuning fluctuations over a long term.
The experimental setup to stabilize the soliton microcomb over 5 hours is shown in Fig. \ref{Fig:2-1}(f). 
A feedback loop is employed to lock the soliton detuning at 317 MHz. 
The VNA is used to monitor the soliton detuning.
The final soliton loss after 5 hours is caused by the drift of the fibre-chip coupling using suspended lensed fibres, which can be mitigated via gluing the fibre to the chip \cite{raja2020chip}. 
Figure \ref{Fig:2-1}(e) shows the evolution of three soliton comb lines over 5 hours. 

As seen from the previous chapter, the resonance tuning via stress-optic effect achieves 1 GHz range with more than 60 V applied voltage. 
Assuming that initially the laser and the resonance have the same frequency, for soliton initiation, soliton detuning of $\sim20\kappa/2\pi$ is needed to access the soliton regime. 
This means that, the resonance linewidth $\kappa/2\pi$ must be below 50 MHz in the case of critical coupling, i.e. the intrinsic loss $\kappa_0/2\pi$ must be below 25 MHz.   
Therefore, it is worth to emphasize that, the key reason why the piezoelectric AlN actuator can be used to initiate solitons with Si$_3$N$_4$ integrated photonics is due to the ultralow loss (i.e. high microresonator $Q$) of the Si$_3$N$_4$ waveguides. 
%%%%%%%%%%%%%%%%%%%%%%%%%%%%%%%%%%%%%%%%%%%%%%%%%%%%%%%%%%%%%%%%%%%%%%%%%%%%%%%%%
\subsubsection{Fast Soliton Microcomb Actuation and Locking}
\begin{figure*}[b]
\centering
\includegraphics[width=1 \textwidth]{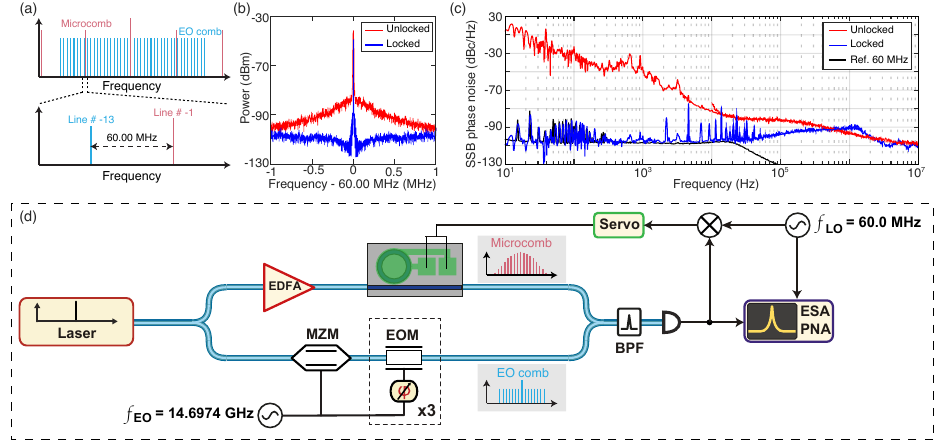}
\caption{
High-speed piezoelectric actuation for soliton repetition rate stabilization using an EO comb. 
(a) Schematic of soliton repetition rate stabilization using an EO comb with 14.6974 GHz line spacing. 
The $-1^{\text{st}}$ line of the microcomb and the $-13^{\text{th}}$ line of the EO comb are locked, referenced to a 60.0 MHz microwave signal. 
(b) Measured beatnote signal of the microcomb and the EO comb, in the cases of locked and unlocked (free-running) states. 
Resolution bandwidth (RBW) is 1 kHz. 
(c) Measured phase noise of the beat signal, in comparison to the 60.0 MHz microwave signal. 
The locking bandwidth of the AlN actuator is 0.6 MHz. 
(d) Experimental setup to characterize the \emph{in-loop} phase noise of the beat signal between the $-1^{\text{st}}$ line of the microcomb and the $-13^{\text{th}}$ line of the EO comb.
Reprinted from Ref. \cite{Liu:20a}. 
}
\label{Fig:3-2}
\end{figure*}

The advantage of piezoelectric actuator over heater is its fast actuation speed, which can be utilized to stabilize the soliton repetition rate with a high bandwidth. 
As studied in the previous chapter, the actuation speed of the AlN actuator is primarily limited by the mechanical modes of the photonic chip (100 kHz to 10 MHz) and the HBAR modes of Si substrate (10 MHz to 100 MHz). 
Strategies to mitigate the chip modes have been presented in Fig. \ref{Ch3_strategy}, where a flat frequency response has been achieved up to 10 MHz. 
The stabilization of the  soliton repetition rate via AlN actuator is shown in Fig. \ref{Fig:3-2}. 
The soliton repetition rate ($\nu_\text{rep}=191$ GHz) is indirectly measured from the the beat signal between the $-1^{\text{st}}$ line of the microcomb and the $-13^{\text{th}}$ line of an EO comb as illustrated in Fig. \ref{Fig:3-2}(a). 

Figure \ref{Fig:3-2}(d) shows the experimental setup to stabilize the soliton repetition rate referenced to the EO comb, which is generated using the scheme described in Refs. \cite{obrzud2017, anderson2021} with a comb line spacing of 14.6974 GHz. 
The EO comb and soliton microcomb are pumped by the same laser (Toptica CTL). 
The measured beat signal between the microcomb and the EO comb is compared to a reference signal of 60.0 MHz. 
The error signal is applied directly on the AlN actuator, such that the actuation on the microresonator stabilizes the soliton repetition rate to the EO comb line spacing.
Figure \ref{Fig:3-2}(b, c) compare the beat signal and the phase noise in the cases of free-running and locked states, in comparison with the phase noise of the 60.0 MHz reference microwave.
A 600 kHz locking bandwidth is determined from the merging point of the phase noise of beat signal and 60 MHz reference.
Note that the 600 kHz bandwidth is larger than conventional lasers where the piezo response is typically limited to few kilohertz, similar to integrated heaters. 

\begin{figure*}[b]
\centering
\includegraphics[width=1 \textwidth]{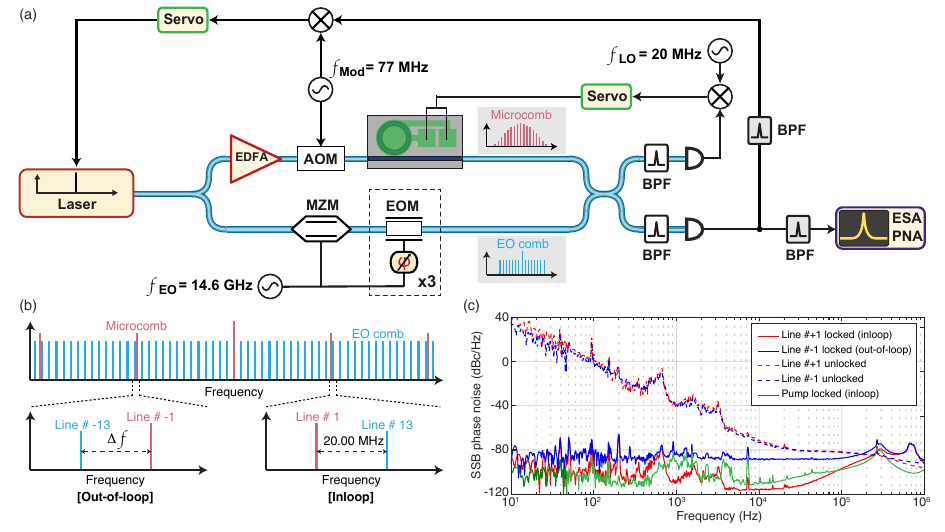}
\caption{Out-of-loop beat signal characterization using a modified setup.
(a) Modified experimental setup to characterize the \emph{out-of-loop} phase noise of the beat signal between the $-1^{\text{st}}$ line of the microcomb and the $-13^{\text{th}}$ line of the EO comb. 
(b) Schematic of referencing the microcomb to the EO comb. 
The beatnote (in-loop) between the $+1^{\text{st}}$ line of the microcomb and $+13^{\text{th}}$ line of the EO comb is detected and referenced to a 20.0 MHz microwave signal.
The beatnote (out-of-loop) between the $-1^{\text{st}}$ line of the microcomb and $-13^{\text{th}}$ line of the EO comb is characterized.
(c) Comparison of SSB phase noises measured in different cases.
Reprinted from Ref. \cite{liu2020silicon}. 
}
\label{Fig:3-3}
\end{figure*}

The above experiment only demonstrates the stabilization of one comb line that is involved in the feedback loop.
To show the stability of the entire comb, the out-of-loop comb needs to be characterized. 
To measure the \emph{out-of-loop} beat signal and its phase noise, a modified setup is used as shown in Fig. \ref{Fig:3-3}(a). 
The pump laser's frequency to generate the soliton microcomb is shifted by 77.0 MHz via a fibre-coupled AOM. 
The reason to shift the microcomb's pump frequency is to cancel out the drift and noise caused by the imbalanced paths in delayed self-homodyne measurement, by detecting the beatnote (77.0 MHz shift) between the pump lines of the microcomb and the EO comb. 
By down-mixing the 77.0 MHz heterodyne beatnote signal using the same microwave source that drives the AOM, the feedback signal is applied to the laser current such that the pump laser's frequency is stabilized and the noise in the delayed self-homodyne measurement is removed.

Then, the beatnote between the $+1^{\text{st}}$ line of the microcomb and $+13^{\text{th}}$ line of the EO comb is detected and referenced to a 20.0 MHz microwave signal, in order to stabilize the microcomb repetition rate. 
The entire schematic of referencing the microcomb to the EO comb is shown in Fig. \ref{Fig:3-3}(b).
Figure \ref{Fig:3-3}(c) compares the single-sideband (SSB) phase noise for in-loop (red) and out-of-loop (blue) beat signals when in the locked (solid) and free-running (dashed) states. 
The solid green line is the phase noise between the EO comb's pump and the microcomb's pump (shifted by 77 MHz), when the pump laser's frequency is locked,  
A reduction in the phase noise of the out-of-loop beat signal is observed with the AlN actuation, which demonstrates the stability of the repetition rate. 
The locking bandwidth in this case is $>300$ kHz. 

\begin{figure*}[b]
\centering
\includegraphics[width=8 cm]{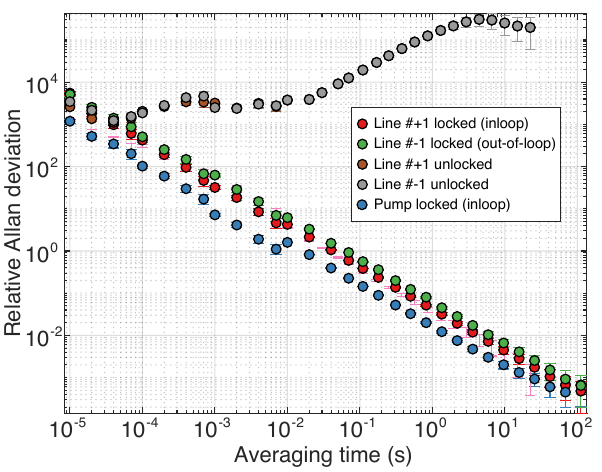}
\caption{Comparison of measured relative Allan deviation for the microcomb and EO comb beat signals in different cases, to evaluate the long-term stability of the locked system.
Reprinted from Ref. \cite{liu2020silicon}. 
}
\label{Fig:3-4}
\end{figure*}

To further evaluate the long-term stability of the locked system, frequency counting measurements of the relative Allan deviations are performed, as shown in Fig. \ref{Fig:3-4}.
The 77 MHz microwave source (used to lock the pump laser's frequency) is referenced to the 20 MHz microwave oscillator (used to down-mix the in-loop beat signal to derive the error signal). 
Similarly, the microwave source driving the EOMs (14.6974 GHz) for EO comb generation is also referenced to the same 20 MHz microwave oscillator. 
The relative Allan deviation of the beat signals between the free-running microcomb and EO comb do not converging, while the beat signals between the locked pump lines of the microcomb and the EO comb re at the level of 10$^{-2}$ with 1 s averaging time.
After locking the soliton repetition rate to the EO comb by actuating on AlN, the \emph{in-loop} and the \emph{out-of-loop} beat signals show similar frequency stability. 

\begin{figure*}[b]
\centering
\includegraphics[width=1 \textwidth]{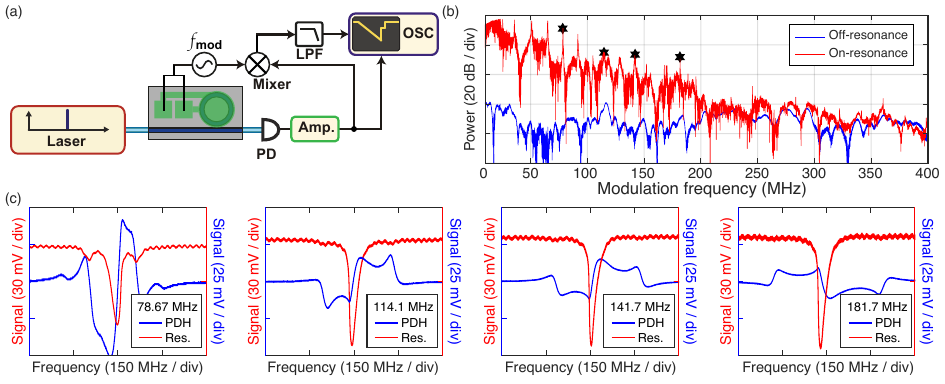}
\caption{
On-chip generation of PDH error signals using the HBAR modes induced by the AlN actuation.
(a) Experimental setup.
LPF: low-pass filter. 
Amp.: RF power amplifier. 
(b) The measured $S_{21}(\omega)$ response of the AlN actuator in the linear frequency scale. 
Both cases, when the laser is on- and off-resonance, are measured.
(c) The PDH error signals modulated at the HBAR frequencies marked with stars in (b).
Reprinted from Ref. \cite{Liu:20a}. 
}
\label{Fig:3-1}
\end{figure*}

In addition to the quasi-DC tuning, the HBAR modes of the AlN actuator provide a novel functionality for error signal generation using the PDH technique. 
Figure \ref{Fig:3-1}(a) shows the experimental setup to generate PDH error signals using HBAR modes. 
The measured $S_{21}(\omega)$ response of the AlN actuation, up to 400 MHz, is plotted in the linear frequency scale in Fig. \ref{Fig:3-1}(b), showing both cases when the laser is on- and off-resonance. 
Different modulation frequencies corresponding to different HBAR modes are investigated, which are marked with stars in Fig. \ref{Fig:3-1}(b). 
The PDH error signals modulated at these HBAR frequencies are shown in Fig. \ref{Fig:3-1}(c), as well as the corresponding microresonator's resonance.
A microwave source providing $\sim$8 dBm RF power is used to modulate the Si$_3$N$_4$ microresonator via AlN actuation. 
The decrease in error signal contrast at higher HBAR frequency is caused by the finite bandwidth of the optical resonance. 
This PDH signal can be used to stabilize the laser frequency referenced to an on-chip high optical Q Si$_3$N$_4$ microresonator. 

%%%%%%%%%%%%%%%%%%%%%%%%%%%%%%%%%%%%%%%%%%%%%%%%%%%%%%%%%%%%%%%%%%%%%%%%%%%%%%%%%
\subsubsection{Soliton Microcomb Based Parallel LiDAR Engine}
\begin{figure*}[t]
\centering
\includegraphics[width=1 \textwidth]{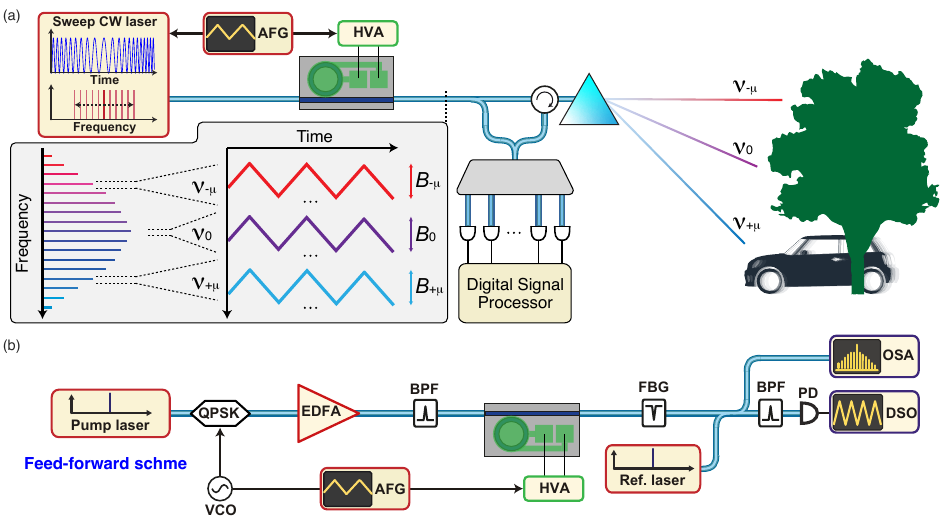}
\caption{Schematic and experimental setup of hybrid AlN-Si$_3$N$_4$ soliton LiDAR engine.
(a) Schematic of soliton-based parallel FMCW LiDAR. 
A chirped pump translates modulation to other comb lines, with the same modulation rate and frequency excursion.
(b) Experimental setup for synchronous scan of the laser frequency and the microresonator resonance, using the feed-forward scheme. 
VCO: voltage-controlled oscillator. 
AFG: arbitrary waveform generator. 
QPSK: quadrature phase shift keying. 
DSO: digital storage oscilloscope.
Reprinted from Ref. \cite{Liu:20a}. 
}
\label{Fig:4}
\end{figure*}

The AlN actuator can be a key component for a soliton-based, parallel, FMCW LiDAR \cite{riemensberger:2020}. 
This scheme conventionally relies on scanning the pump laser's frequency $\nu_\text{L}$ over the soliton existence range $\Delta\nu_\text{s}=\nu_2-\nu_1$, i.e. $\nu_\text{c}-\nu_\text{L}\in[\nu_1, \nu_2]$, with $\nu_\text{c}$ being the resonance frequency, $\nu_1$ and $\nu_2$ the boundary of soliton existence detuning range. 
As a result, the pump frequency chirp is transferred to all soliton comb lines, as shown in Fig. \ref{Fig:4}(a). 
The diffractive optics subsequently separates the comb lines spatially, which allows high-speed and parallel acquisition of both velocity and position in each pixel. 
However, the varying soliton detuning has several limitations. 
First, the soliton detuning variation leads to changes in soliton spectrum bandwidth and power, which limits the number of usable optical channels.
Second, the Raman self-frequency shift \cite{karpov2016, yi2017} causes a varied soliton repetition rate. 
Third, GHz frequency excursion of the pump laser ($\Delta\nu_\text{L}$) requires GHz-wide soliton existence range ($\Delta\nu_\text{s}$), which leads to elevated pump power to several Watts. 

%%%%%%%%%%%%%%%%%%%%%%%%%%%%%%%%%%%%%%%%%%%%%%%%%%%%%%%%%%%%%%%%%%%%%%

These limitations can be overcome by using the AlN actuation on the microresonator to synchronously modulate the microresonator resonance $\nu_\text{c}$ with the pump frequency $\nu_\text{L}$, while maintaining a constant soliton detuning $\nu_\text{c}-\nu_\text{L}$. 
This scheme is called ``feed-forward scan'' \cite{Liu:20a}. 
Figure \ref{Fig:4}(b) shows the experimental setup to synchronously scan the microresonator and the pump laser, where a triangular signal of frequency $f_\text{mod}$ is applied on both the pump laser and the AlN actuator.
A single-sideband modulator driven by a voltage-controlled oscillator (VCO) is used to fast scan the laser frequency.
The ramp signal sent to the AlN actuator is amplified by the HVA with $\times50$ voltage amplification and 3-dB bandwidth of $\sim5$ MHz. 
The synchronization of the laser frequency and the microresonator resonance is performed by adjusting the amplitude and the phase of the ramp signal applied on the VCO.
A PDH lock can further improve the synchronization by locking the resonance to the laser with a constant detuning \cite{stone2018}.
A reference laser is used to probe the chirp of different comb lines (the pump line, $\pm 10^{th}$ comb lines etc).  
A fast oscilloscope of 2 GHz bandwidth and 5 GSamples/s is used to capture the heterodyne beatnote detected on the fast photodiode, for further off-line data processing such as fast Fourier transform and fitting triangular signal. 

%%%%%%%%%%%%%%%%%%%%%%%%%%%%%%%%%%%%%%%%%%%%%%%%%%%%%%%%%%%%%%%%%%%%%%
\begin{figure*}[t]
\centering
\includegraphics[width=1 \textwidth]{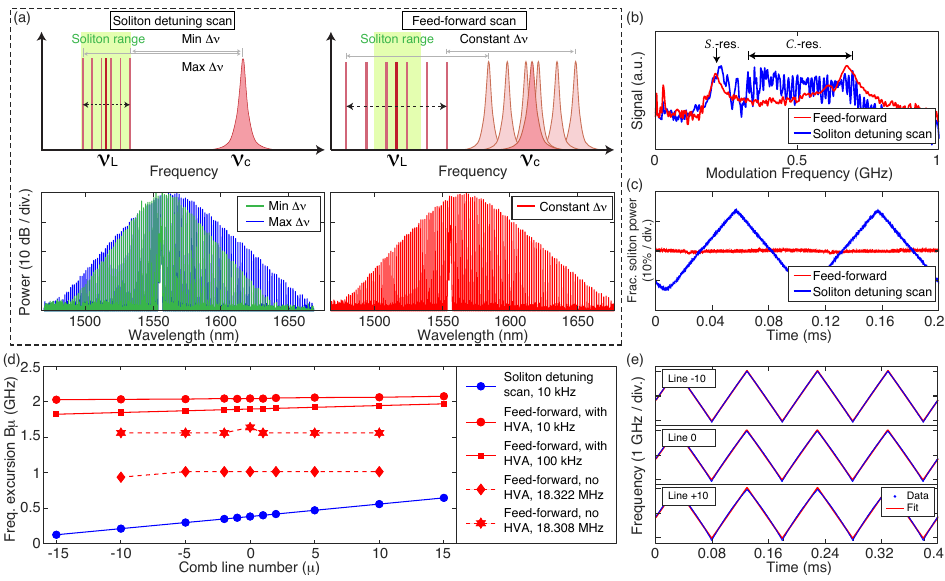}
\caption{Results of hybrid AlN-Si$_3$N$_4$ soliton LiDAR engine.
(a) Comparison of two schemes for soliton-based FMCW LiDAR. 
Left: The scheme employing soliton detuning scan, with a fixed microresonator resonance $\nu_\text{c}$, while the pump $\nu_\text{L}$ chirping within the soliton existence range (green shaded region). 
The soliton spectrum varies with the soliton detuning, as shown in the bottom panel. 
Right: Feed-forward scheme employing synchronous modulation of the pump frequency and the resonance, such that the soliton detuning $\Delta\nu$ is constant, resulting in a constant soliton spectrum. 
Note that in this scheme, the pump frequency excursion can be significantly larger than the soliton existence range.
(b) Comparison of both schemes for the soliton response on VNA. 
The soliton detuning scan shows a varying \textit{C}-resonance, while the resonance in the feed-forward scheme is fixed. 
(c) Comparison of both schemes for the integrated soliton pulse power. 
The soliton detuning scan shows a varying soliton pulse power, while the feed-forward scheme shows a constant power.
(d) Comparison of the frequency excursions of different soliton comb lines, using different approaches.
(e) Time-frequency diagram of different comb lines chirping at 10 kHz modulation frequency, with frequency excursion of 2.04 GHz.
Reprinted from Ref. \cite{liu2020silicon}. 
}
\label{Fig:4-1}
\end{figure*}
%%%%%%%%%%%%%%%%%%%%%%%%%%%%%%%%%%%%%%%%%%%%%%%%%%%%%%%%%%%%%%%%%%%%%%

The comparison between the two schemes, i.e. (a) scanning the pump laser within the soliton existence range, and (b) feed-forward scheme, is experimentally studied in Figure~\ref{Fig:4-1}. 
Figure \ref{Fig:4-1}(a-c) illustrate the differences in soliton spectrum, soliton detuning, and integrated soliton pulse power in both schemes, showing the advantages of the feed-forward scheme using the AlN actuator. 
First, the pump frequency chirp $\Delta\nu_\text{L}$ is no longer limited by the soliton existence range $\Delta\nu_\text{s}$, which allows operating the soliton with tens of milliwatt power, compatible with the state-of-art integrated laser \cite{HuangD:19}.
Second, the soliton spectrum, repetition rate, and power are maintained constant and not affected by the pump frequency chirp. 
Figure~\ref{Fig:4-1}(d) compares the frequency excursions of different soliton comb line generated with different approaches. 
In the scheme of soliton detuning scan at frequency $f_\text{mod}=10$ kHz (blue), the frequency excursion depends on the comb line number, resulting from the soliton repetition rate variation due to the change of soliton detuning. 
The feed-forward scheme with modulation frequency $f_\text{mod}=10$ kHz and 100 kHz (red, solid lines) allows frequency excursion of $\sim$2 GHz, which is more than 5 times larger than the soliton existence range ($\sim$400 MHz).
Moreover, the variation of frequency excursion among comb lines is largely suppressed. 
Figure~\ref{Fig:4-1}(e) shows the time-frequency diagram of different comb lines chirped at 10~kHz modulation frequency, with frequency excursion of 2.04 GHz. 

Although the HBAR and mechanical modes of the chip are unwanted in linear and fast tuning, they enable modulation with higher frequency and efficiency (red dashed lines in Fig. \ref{Fig:4-1}(d)).
By applying modulation frequency ($f_\text{mod}=18.308$ MHz) that coincides with the fundamental HBAR mode and a contour mode, frequency excursion of 1.6 GHz is generated without HVA.
The voltage applied on the AlN is only 7.48 V, which is compatible with CMOS circuits. 
Additionally, this modulation corresponds to an equivalent tuning speed of 58.6 PHz/s. 
The corresponding resonance tuning efficiency is $\delta\nu/\delta V=219$ MHz/V, which is 14 times larger than the DC tuning of $\delta\nu/\delta V=15.7$ MHz/V.
Applying modulation with $f_\text{mod}=18.322$ MHz, which coincides with the fundamental HBAR mode but not the contour mode, the tuning efficiency is reduced to $\delta\nu/\delta V=133$ MHz/V. 
Therefore, it will be of great interest in the future to optimize the design of the AlN actuator that is compatible with high frequency triangular wave modulation. 
For instance, the higher-order HBAR modes can be designed to match with the higher harmonics of the triangular wave, such that the modulation signal can be linearly transduced onto the optical microresonator. 

In conclusion, integrated piezoelectric actuators that are linear, fast, non-absorptive, bidirectional, and consume ultralow electric power endow integrated soliton microcombs with capability in power-critical applications, e.g., in space, data centers and portable atomic clocks, in extreme environments such as cryogenic temperatures, or in emerging applications for coherent LiDAR \cite{riemensberger:2020}, frequency synthesizers \cite{Spencer:18} and RF photonics \cite{Torres-Company:14, Wu:18}. 
Therefore, this capability is important for tuning and stabilization of soliton microcombs, and may allow soliton injection locking for stabilization and microwave synthesis \cite{obrzud2017, weng2020frequency}.
While polycrystalline AlN is used in current works, the operation voltage can be reduced by more than 2 times using scandium-doped AlN \cite{fichtner2019alscn}. 
In the future, the co-integration of CMOS electronic circuitry on a close-by die paves the way for packaged soliton microcombs with rapid electronic actuation \cite{lukashchuk2023}.

\section{Radio-Frequency Acousto-Optic Modulation}

While piezoelectric actuators have found wide applications in tuning of integrated photonic devices, fast actuation at microwave frequencies will excite acoustic waves that enable efficient acousto-optic modulation. 
At the acoustic resonance, integrated AOMs have been demonstrated with efficiency comparable to or even higher than state-of-the-art electro-optic modulators, albeit at the expense of narrow bandwidth \cite{Shao:19, Tadesse:14, Tian:20, jiang2020efficient, mirhosseini2020}. 
Nevertheless, it has been applied not only in classical applications, but also in quantum regime working as quantum transducers between microwave and light \cite{safavi2019, chu2020perspective}. 
In the following, we will review recent progress on chip-scale AOMs that can be classified into three categories in terms of the acoustic mode being utilized: surface acoustic wave, bulk acoustic wave, and optomechanical crystal.

%The figure of merit for characterizing a classical optical modulator is $V_{\pi}$, the applied voltage that induces a $\pi$ phase shift of the light in the optical cavity. 

%%%%%%%%%%%%%%%%%%%%%%%%%%%%%%%%%%%%%%%%%%%%%%
%%%%%%%%%%%%%%%%%%%%%%%%%%%%%%%%%%%%%%%%%%%%%%
\subsection{Surface Acoustic Wave Based AOM}

The surface acoustic wave excited by IDT has been one of the traditional ways of launching acoustic waves on a chip \cite{slobodnik1976}, and has become the key technology across a wide range of research areas, such as microfluidics \cite{ding2013}, biosensors \cite{lange2008}, radio-frequency filters and delay lines \cite{campbell1989, campbell2012surface}, and quantum acoustics \cite{aref2016}. 
As nano-fabrication advances, the ability to create sub-micron electrode fingers has extended the frequency of the acoustic wave to beyond 10 GHz \cite{tadesse2015, li2015}. 
By interacting with an optical cavity, the stress and strain field of the acoustic wave can be used to modulate the optical refractive index, achieving the chip-scale AOM.

\begin{figure}[t]
\centering
\includegraphics[width=13.3 cm]{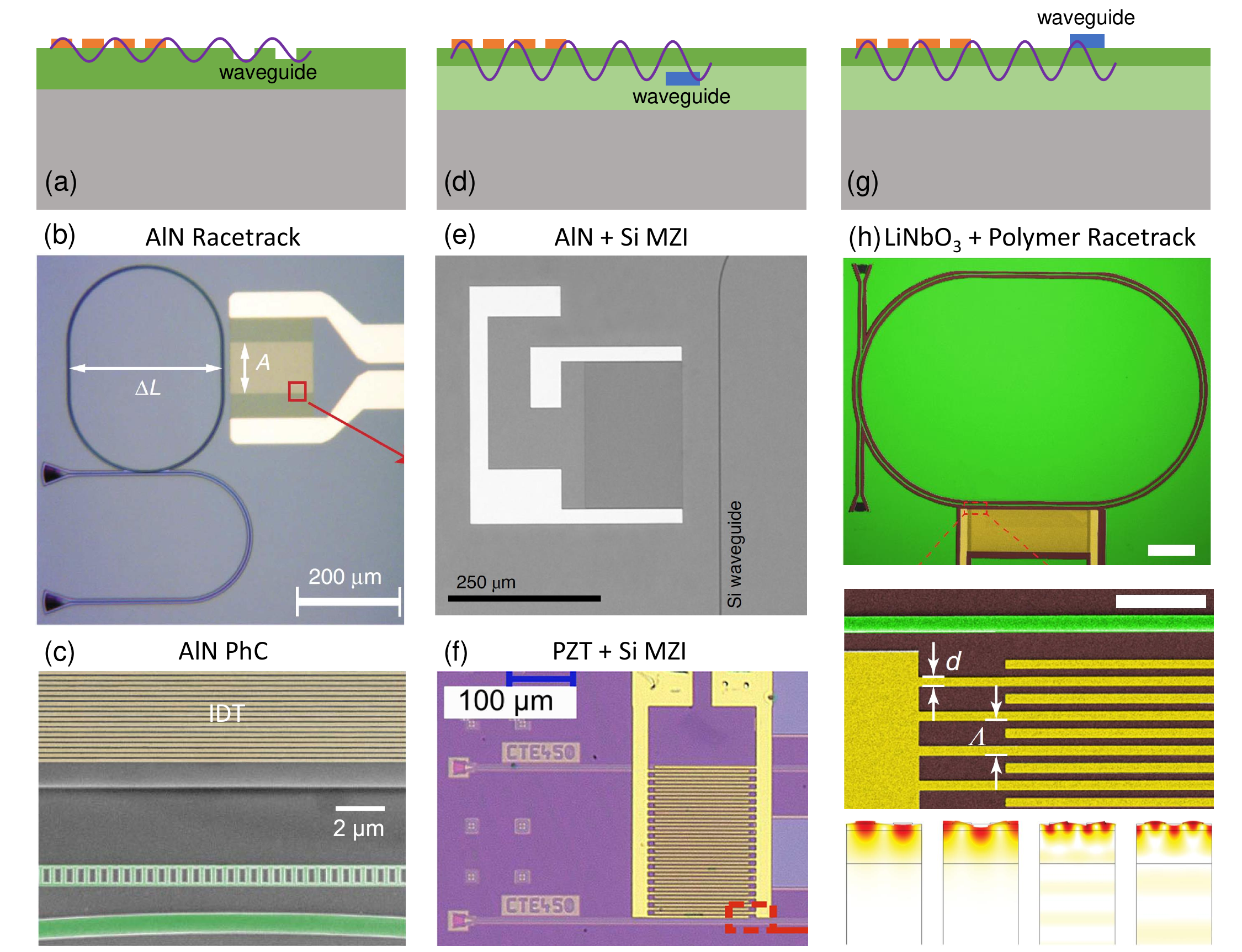}
\caption{
SAW AOM demonstrated with unreleased structures. 
(a) Schematic of the SAW AOM with the optical waveguide made from the piezoelectric material. 
Examples with the structure in (a) have been demonstrated for (b) an AlN racetrack resonator and (c) an AlN photonic crystal (PhC) resonator. 
(d) and (g) are schematics of the SAW AOMs with a hybrid optical waveguide made from a different optical material. 
While the waveguide is embedded in a cladding in (d), the waveguide is bonded on top of the piezoelectric layer in (g). 
(e) and (f) are examples for the structure in (d) where a Si MZI is modulated by (e) AlN and (f) PZT SAW. 
(h) is the example of (g), where a SAW excited on LiNbO$_3$ modulates the waveguide formed by polymer. 
The middle inset is the zoom-in near the IDT (yellow) and the polymer waveguide (green). 
The lower inset shows the simulated fundamental and higher order SAW Rayleigh modes. 
(b) reprinted from Ref. \cite{Tadesse:14}. (c) reprinted from Ref. \cite{li2015}. (e) reprinted from Ref. \cite{Kittlaus:21}. (f) reprinted from Ref. \cite{ansari2022}. (h) reprinted from Ref. \cite{yu2020}. 
}
\label{Ch5_SAW_unrelease}
\end{figure}

One advantage of SAW is that the fabrication is relatively simple and easy, where the IDT electrodes can be made by directly depositing and patterning metal on top of the released or unreleased piezoelectric films. 
For the unreleased SAW AOM demonstrated to date, there are mainly three typical structures, as shown in Fig. \ref{Ch5_SAW_unrelease}. 
Since the SAW propagates in plane, the optical waveguide needs to be brought as close as possible to the plane of SAW. 
The most straightforward and widely adopted way is to etch the piezoelectric film to form the optical waveguide, as shown in Fig. \ref{Ch5_SAW_unrelease}(a). 
It has been realized on LiNbO$_3$ \cite{sarabalis2020, cai2019}, AlN \cite{Tadesse:14, li2015}, and AlGaAs \cite{crespo2013} platforms. 
Among them the optical resonator is mostly a racetrack with the long straight side along the IDT fingers such that the SAW propagates perpendicular to the waveguide [see Fig. \ref{Ch5_SAW_unrelease}(b)]. 
In the design shown in Fig. \ref{Ch5_SAW_unrelease}(b), only one straight segment is close to the IDT while the other side is $\Delta L$ away. 
When $\Delta L$ is longer than the attenuation length of the SAW, the acoustic energy reaching the other side of the racetrack will be largely suppressed. 
One solution is to place the IDT inside the racetrack which, however, poses challenges on the electrode routing without compromising the optical $Q$. 
The other strategy is to shorten $\Delta L$ without introducing significant bending loss of the bending part. 
One should also notice that $\Delta L$ should be an integer number of the acoustic wave's wavelength such that modulation on both sides of the racetrack can be constructively interfered \cite{Tadesse:14}. 
To better confine acoustic energy inside the racetrack, acoustic reflectors can be placed on both sides of the device to form an acoustic cavity \cite{cai2019}. 
On the other hand, the AOM of an 1-D AlN photonic crystal resonator has been demonstrated with the crystal along the IDT [see Fig. \ref{Ch5_SAW_unrelease}(c)] \cite{li2015}. 
In this case, the optical mode is confined in a small volume and can have a good overlap with the acoustic wave. 

The drawback of using piezoelectric materials as the optical waveguide is that it requires etching of the piezoelectric film which is still challenging for most piezoelectric materials.
In recent decade, promising progresses have been made in achieving high-Q optical integrated resonators, where intrinsic Q as high as 3.7 million and 11 million have been demonstrated in AlN \cite{liu2022fundamental} and LiNbO$_3$ \cite{wu2018long} microring resonators, respectively. 
Additionally, the waveguide cannot be cladded, which is prone to perturbation from the environment. 
One should also notice that the piezoelectric material also possesses electro-optic effect, such that the strain can induce electric field across the waveguide via the piezoelectricity which in turn changes its refractive index through the electro-optic effect \cite{Shao:19}. 
This may counteract the optomechanical interaction if not designed correctly. 
To mitigate these problems, a hybrid optical waveguide can be implemented. 
As mentioned above, the Si and Si$_3$N$_4$ photonic integrated circuits have evolved into a mature platform which is compatible with foundry processes \cite{Ye:23}. 
Moreover, ultra-high-Q Si$_3$N$_4$ integrated resonators have been demonstrated by multiple groups \cite{ji2021methods}, and intrinsic Q as high as 442 million has been achieved \cite{Puckett:21}. 
As shown in Fig. \ref{Ch5_SAW_unrelease}(d), the waveguide is entirely embedded in the optical cladding (which is usually SiO$_2$), and the piezoelectric film is deposited on top. 
Depending on the thickness of the top cladding, the piezoelectric layer can be either fully separated from or partially overlapped with the optical mode. 
However, since the SAW is primarily confined at the surface, it decays exponentially into the substrate, as can be seen in Fig. \ref{Ch5_SAW_unrelease}(h). 
In that sense, the waveguide needs to be brought as close as possible to the surface to have a good acousto-optic overlap. 
The SAW AOM of the Si MZI has been demonstrated on both AlN \cite{Kittlaus:21} and PZT \cite{ansari2022} platforms, as shown in Fig. \ref{Ch5_SAW_unrelease}(e-f). 
The modulation of a hybrid LiNbO$_3$ on Si$_3$N$_4$ waveguide has also been achieved recently \cite{ghosh2021}. 

Another hybrid waveguide that can be designed is to place the waveguide above the piezoelectric film, as shown in Fig. \ref{Ch5_SAW_unrelease}(g). 
Waveguide material may have a refractive index either higher or lower than the piezoelectric film. 
While the former case confines the light via total internal reflection \cite{yang2022}, the latter forms the so-called bound states in the continuum (BIC) \cite{yu2020}. 
The fabrication of hybrid waveguides in this way is not trivial, as waveguide etching can rough the top surface of the piezoelectric layer, reducing the SAW quality. 
Despite of that, efficient AOM has been demonstrated on hybrid chalcogenide-LiNbO$_3$ racetrack resonator \cite{yang2022}. 
For the work shown in Fig. \ref{Ch5_SAW_unrelease}(h), the waveguides are manufactured by patterning an electron-beam resist onto the piezoelectric layer without etching \cite{yu2020}. 
Another solution is to use the pick-and-place method, which will be shown later. 

\begin{figure}[t]
\centering
\includegraphics[width=13.3 cm]{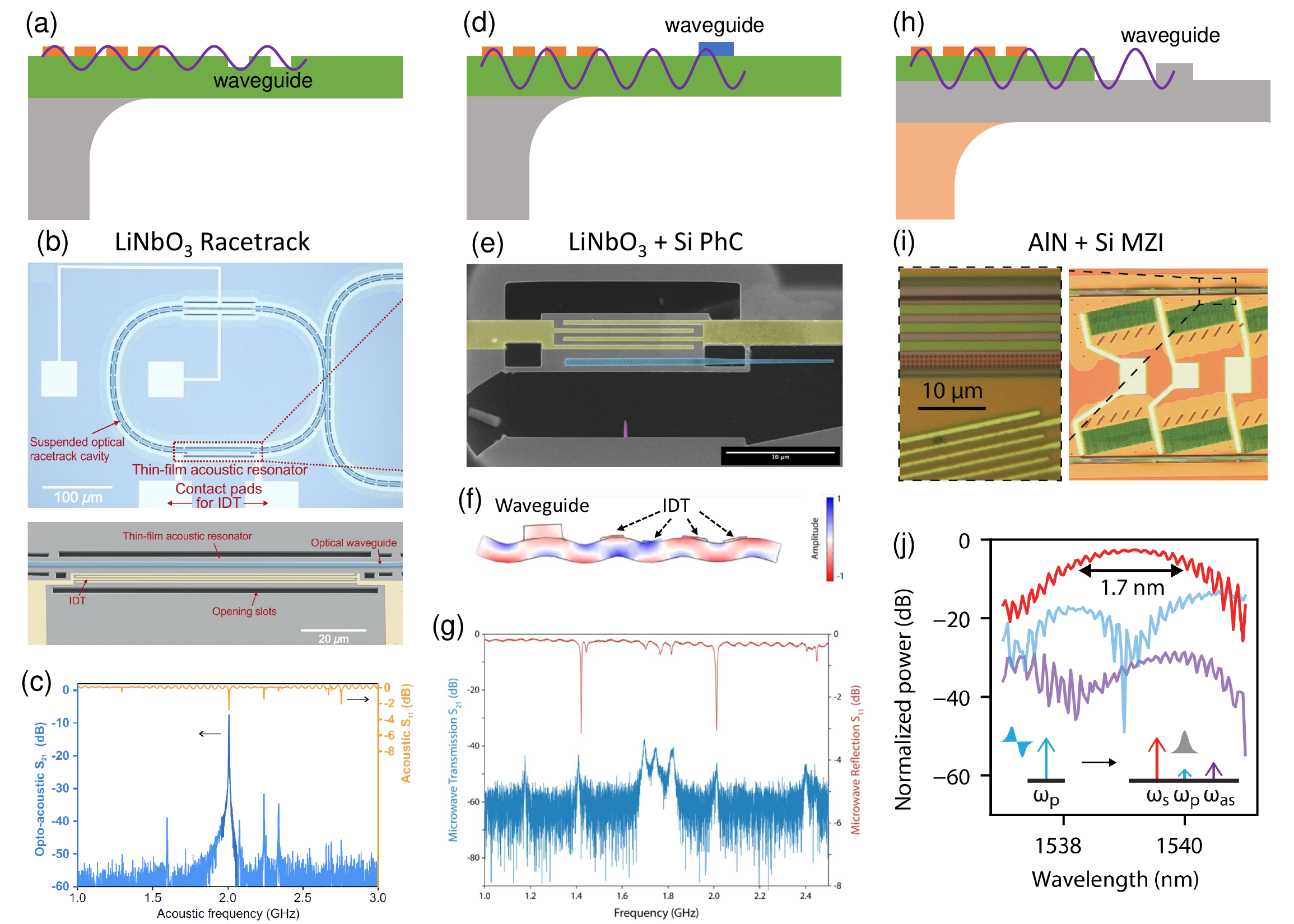}
\caption{
SAW AOM demonstrated on released structures. 
(a) Schematic of the AOM with the optical waveguide made from piezoelectric materials. 
(b) AOM of a LiNbO$_3$ optical racetrack resonator with the structure as in (a). 
The lower inset is the false colored SEM of the region in the red dashed box. 
(c) Characterization of the device in (b), showing the acousto-optic S$_{21}$ response (blue) and the microwave reflection spectrum S$_{11}$ (yellow). 
(d) Schematic of the hybrid optical waveguide that is on top of the piezoelectric layer. 
(e) False-colored SEM of the device that modulates the Si PhC (blue) via the SAW on the released LiNbO$_3$ membrane. 
The Si waveguide is fabricated using the pick-and-place technique. 
(f) Simulated Lamb wave excited by the IDT. 
(g) Acousto-optic response (blue) and microwave reflection (red) of the device in (e). 
(h) Schematic of the ridge waveguide that is separated from the piezoelectric material. 
(i) Optical image of the device that modulates the Si ridge waveguide using AlN SAW. 
(j) Modulation spectra showing a single sideband modulation with bandwidth of 1.7 nm. 
Red and purple curves are the optical power in the Stokes and anti-Stokes sidebands, respectively. 
(b-c) reprinted from Ref. \cite{Shao:19}. (e-g) reprinted from Ref. \cite{marinkov2021}. (i-j) reprinted from Ref. \cite{zhou2022intermodal}.
}
\label{Ch5_SAW_release}
\end{figure}

Although unreleased structures are straightforward to deal with, there is a trade-off of choosing the substrate that confines both the acoustic and optical modes tightly. 
For most of the examples shown above, the device sits on a SiO$_2$ layer for the sake of confining the optical mode.
However, SiO$_2$ has an acoustic velocity close to or smaller than the piezoelectric layer, which leads to a low acoustic index contrast, such that the acoustic wave partially leaks into the substrate \cite{wang2020high, zhang2020}.
Therefore, to efficiently excite the SAW and maintain a high $Q$, the substrate should not only have a low optical refractive index, but a high acoustic velocity (equivalently a low acoustic index). 
Although it is non-trivial to find a material that meets these requirements, sapphire has attracted much attention due to its low refractive index of 1.75 and high longitudinal and transverse acoustic velocities of 10685 m/s and 5796 m/s, respectively \cite{fu2019phononic}. 
Tight acoustic mode confinement in a phononic waveguide has been demonstrated with a variety of piezoelectric materials, such as LiNbO$_3$ \cite{mayor2021}, AlN \cite{streque2020}, and GaN \cite{fu2019phononic, xu2022high}.
With further investigation of the optical waveguide design in these platforms, an optomechanical integrated circuit can be realized, which confines both mechanical and optical modes in the same waveguide \cite{sarabalis2021}. 

An alternative way to confine the acoustic mode is by releasing the piezoelectric layer at the cost of a complicated fabrication process.
As shown in Fig. \ref{Ch5_SAW_release}, the piezoelectric thin film is suspended by isotropic etching of the substrate through opening holes.
Similar as the unreleased case, both single and hybrid optical waveguides have been explored to date, with the former being the most widely adopted and demonstrated on AlN \cite{ghosh2016, tadesse2015}, LiNbO$_3$ \cite{hassanien2021, Shao:19}, and GaAs \cite{khurana2022}. 
An example of the AOM of a LiNbO$_3$ racetrack resonator is shown in Fig. \ref{Ch5_SAW_release}(b), where a ridge waveguide is formed by partially etching the LiNbO$_3$.
Two pairs of IDT fingers are deposited on the remaining LiNbO$_3$ slabs.
The opening slots on both sides serve as acoustic reflectors that form an acoustic cavity. 
The waveguide should be placed at the antinode of the acoustic wave in the cavity to maximize efficiency.
The modulation of a hybrid Si-on-LiNbO$_3$ 1-D photonic crystal resonator is illustrated in Fig. \ref{Ch5_SAW_release}(e) \cite{marinkov2021}.
The Si PhC is fabricated on a separate chip and attached to the piezoelectric layer using the pick-and-place technique \cite{kim2017hybrid, elshaari2018}. 
The Lamb acoustic wave is excited by the IDT as simulated in Fig. \ref{Ch5_SAW_release}(f), which is the most commonly used mode in released SAW AOMs.
In contrast to the racetrack resonator where only a small portion of the resonator is modulated, the PhC confines the optical mode in a small volume where the whole resonator is inside the Lamb wave. 
The AOM of a Si ridge waveguide has also been demonstrated as shown in Fig. \ref{Ch5_SAW_release}(h-j) \cite{zhou2022intermodal}. 
The optical waveguide is spatially separated from the piezoelectric actuator and is modulated by the SAW propagating in the Si slab.
These different structures provide us the possibilities for exploring new materials and novel applications, aiming at high modulation efficiency, low cost, and small size.

The AOM devices are generally characterized by measuring the microwave reflection S$_{11}$ and acousto-optic transduction S$_{21}$ as shown in Fig. \ref{Ch5_SAW_release}(c) and (g). 
The S$_{11}$ takes the ratio of the magnitude between the input and the reflected microwave from the IDT, where the acoustic resonances can be found as dips in the spectrum. 
The impedance of the acoustic resonator $Z_{11}$ can be inferred from the S$_{11}$ as:
\begin{equation}
    S_{11}=\frac{Z_{11}-Z_0}{Z_{11}+Z_0}
\end{equation}
where $Z_0$ is the impedance of the microwave cable, which is 50 $\Omega$ in most cases.
The S$_{21}$ is measured by biasing the laser near the optical resonance and detecting the light intensity and phase modulation through a photodetector.
S$_{21}$ is then the magnitude ratio of the microwave from the photodetector and the input microwave drive. Both S$_{11}$ and S$_{21}$ can be easily measured using a commercial vector network analyzer at the same time. 
This measurement can help us to obtain most of the key parameters, such as the frequency and linewidth of the acoustic resonance, the electromechanical coupling strength, as well as the microwave to optical transduction efficiency. 

One of the important figures of merit for traditional electro-optic modulators is $V_{\pi}$, the voltage that induces $\pi$ phase shift of the light in the waveguide.
Similarly, an effective $V_{\pi}$ can be defined for the AOM of an optical resonator. 
To do that, a modulation index $\beta$ can be defined as \cite{Zhao:23, jiang2020efficient}:
\begin{eqnarray}
\label{beta}
    \beta & = & \frac{2g_0\sqrt{n_{\text{phon}}}}{\Omega_m} \\
    a(t) & = & \alpha(t) e^{-i\beta \text{sin}(\Omega_mt) }
\end{eqnarray}
where $a(t)$ represents the intra-cavity electric field and $n_{\text{phon}}$ is the phonon number in the acoustic mode. 
It can be seen $\beta$ denotes the amplitude of the sinusoidal modulation of the light's phase inside the cavity. 
$V_{\pi}$ is thus the voltage when $\beta=\pi$ and can be measured by fitting to the transmission spectrum of the optical mode under the driving of AOM, where the optical resonance would split under large modulation index $\beta$ \cite{Shao:19, Zhao:23, jiang2020efficient}. 

The $V_{\pi}$ can be also related with the key parameters of the AOM as \cite{Zhao:23}:
\begin{equation}
\label{Vpi}
    V_{\pi} = \frac{\pi \gamma_m \Omega_m}{4g_0}\sqrt{\frac{2Z_0 \hbar\Omega_m}{\gamma_{m,\text{ex}}}}
\end{equation}
where $\gamma_{m,\text{ex}}$ is the external coupling rate from the microwave channel to the acoustic mode, $\gamma_m$ is the total loss rate, and $g_0$ is the single photon optomechanical coupling rate. 
$\gamma_m$ and $\Omega_m$ can be found from the $S_{11}$ and $S_{21}$ measurements.
There are mainly two ways to calibrate the $g_0$: one is by measuring the change in mechanical linewidth induced by the optomechanical back action \cite{Zhao:23}; the other is by homodyne detection of the frequency fluctuation of the optical cavity induced by mechanical thermal noise \cite{gorodetksy2010}.
On the other hand, $\gamma_{m,\text{ex}}$ can be measured by fitting the $S_{11}$ spectrum \cite{Shao:19}, or measuring the injected intra-cavity phonon number by calibrating to the thermal occupation or from the modulation index by fitting the optical transmission spectrum (see Eq. \ref{beta}) \cite{Zhao:23}. 

\begin{table}[t]
    %\centering
 \caption{
 Comparison among the state-of-the-art integrated acousto-optic modulation based on surface acoustic waves. 
 MZI: Mach-Zehnder interferometer.
 PhC: photonic crystal.
 WG: waveguide. 
 $\gamma_{m,\text{ex}}/\gamma_m$ represents external coupling efficiency of the SAW. 
 The optical loss of the optical resonator is characterized by its linewidth $\kappa_o$ in unit of GHz, while that for MZI (or WG) is by its propagation loss $\alpha$ in unit of dB/cm. 
 The modulation efficiency is characterized by $V_{\pi}$ for optical resonators in unit of V, and by $V_{\pi}L$ for MZI (or WG) in unit of V$\cdot$cm.}
    
\begin{tabular}{ p{0.8cm}|p{0.8cm}|p{1.3cm}|p{1.2cm}|p{1.8cm}|p{1.2cm}|p{1.2cm}|p{1.2cm}|p{1.2cm}|p{1.5cm} }
 \hline
 Ref.  &  Year &Structure & Released & Material platform & $\Omega_m/2\pi$ (GHz)& $\gamma_m/2\pi$ (MHz) & $\gamma_{m,\text{ex}}/\gamma_m$ &Optical loss ($\kappa_o/2\pi$ or $\alpha$) &$V_{\pi}$ or $V_{\pi} L$\\
 \hline
\cite{Shao:19}   & 2019 & Racetrack & Yes & LiNbO$_3$ & 2 & 1.28 & 0.17  & 0.095 GHz  & 0.113 V \\
\cite{marinkov2021}   & 2021 & PhC & Yes & LiNbO$_3$+Si& 1.7 & 12  & 0.015 & 6.6 GHz  & 0.05 V \\
\cite{yang2022}   & 2022 & Racetrack & No & LiNbO$_3$+ChG & 0.843 & 2.63  & 0.48 & 0.387 GHz & 1.74 V \\
\cite{khurana2022}   & 2022 & Ring & Yes & GaAs & 1.94 & 6.07  & 0.013
 &9.74 GHz & 0.21 V \\
\cite{crespo2013}   & 2013 & MZI & No & AlGaAs & 0.52 &  -- & --
 &--  & 0.065 V$\cdot$cm \\
 \cite{cai2019}   & 2019 & MZI & No & LiNbO$_3$ & 0.112 &  0.062 & 0.118 &0.7 dB/cm  & 5 V$\cdot$cm \\
 \cite{hassanien2021}   & 2021 & PhC WG & Yes & LiNbO$_3$ & 1.16 &  0.48 & 0.146 &330 dB/cm  & 0.019 V$\cdot$cm \\
\cite{Kittlaus:21}   & 2021 & MZI & No & AlN+Si & 3.11 &  20 & --
 &0.15 dB/cm  & 1.8 V$\cdot$cm \\
 \cite{ansari2022}  & 2022 & MZI & No & PZT+Si & 2 &  -- & --
 &--  & 3.6 V$\cdot$cm \\
\hline
\end{tabular}
    \label{table_AOM}
\end{table}

The magnitude of $S_{21}$ can be related with $V_{\pi}$ as \cite{Shao:19, yang2022, tianthesis, khurana2022}:
\begin{equation}
\label{S21Vpi}
    |S_{21}|=\frac{\kappa_{o,\text{ex}}}{\kappa_o}\frac{\pi R_{\text{PD}}I_{\text{rec}}}{V_{\pi}}
\end{equation}
where $\kappa_{o,\text{ex}}$ and $\kappa_o$ are the external and total loss of the optical mode, $R_{\text{RD}}$ is the responsivity of the photodetector in unit of V/W, $I_{\text{rec}}$ is the DC optical power reaching the photodetector.
From Eq. \ref{S21Vpi}, $V_{\pi}$ can also be estimated from $S_{21}$ and the smaller the $V_{\pi}$ the larger the $S_{21}$. 
Note that Eq. \ref{S21Vpi} is derived by assuming a resolved sideband where the mechanical frequency is much larger than the optical linewidth.
The non-resolved sideband case is derived from Refs. \cite{tianthesis,khurana2022}. 
Also, note that when studying the literature and in real applications, we should pay attention to the definition of $S_{21}$ which can be either the ratio of amplitude \cite{yang2022} or power \cite{Shao:19} of the input and output microwave signals.
Also, when converting from log scale (in dB) to linear scale, there is a factor of 2 between power and amplitude.
While the above analysis are done for AOM of optical resonator, the $V_{\pi}$ of the AOM of MZI can be obtained from $S_{21}$ in a similar fashion \cite{Shao:19, hassanien2021}:
\begin{equation}
\label{S21VpiMZI}
     |S_{21}|=\frac{\pi R_{\text{PD}}I_{\text{rec}}}{V_{\pi}}
\end{equation}

The performance of the state-of-the-art SAW AOM is summarized and compared in Table. \ref{table_AOM}.
One can see that released AOMs generally have higher efficiency (smaller $V_{\pi}$) than unreleased structures.
However, the unreleased AOM has the advantages of easier fabrication, better thermal conductivity, and high-power handling capability.
Although the photonic crystal structure demonstrates the smallest $V_{\pi}$ due to its tight optical mode confinement, the optical loss is still higher than in other platforms.
On the material side, the SAWs made from LiNbO$_3$ show a higher SAW excitation efficiency due to its high electromechanical coupling coefficient. 
In the future, $V_{\pi}$ can be further reduced with the help of novel structures and new materials, targeting at higher microwave to acoustic wave conversion efficiency, smaller acoustic losses, and a larger optomechanical coupling rate (which benefits from a smaller mechanical and optical mode volume, better mode overlap, and larger photoelastic coefficient). 

\begin{figure}[t]
\centering
\includegraphics[width=12 cm]{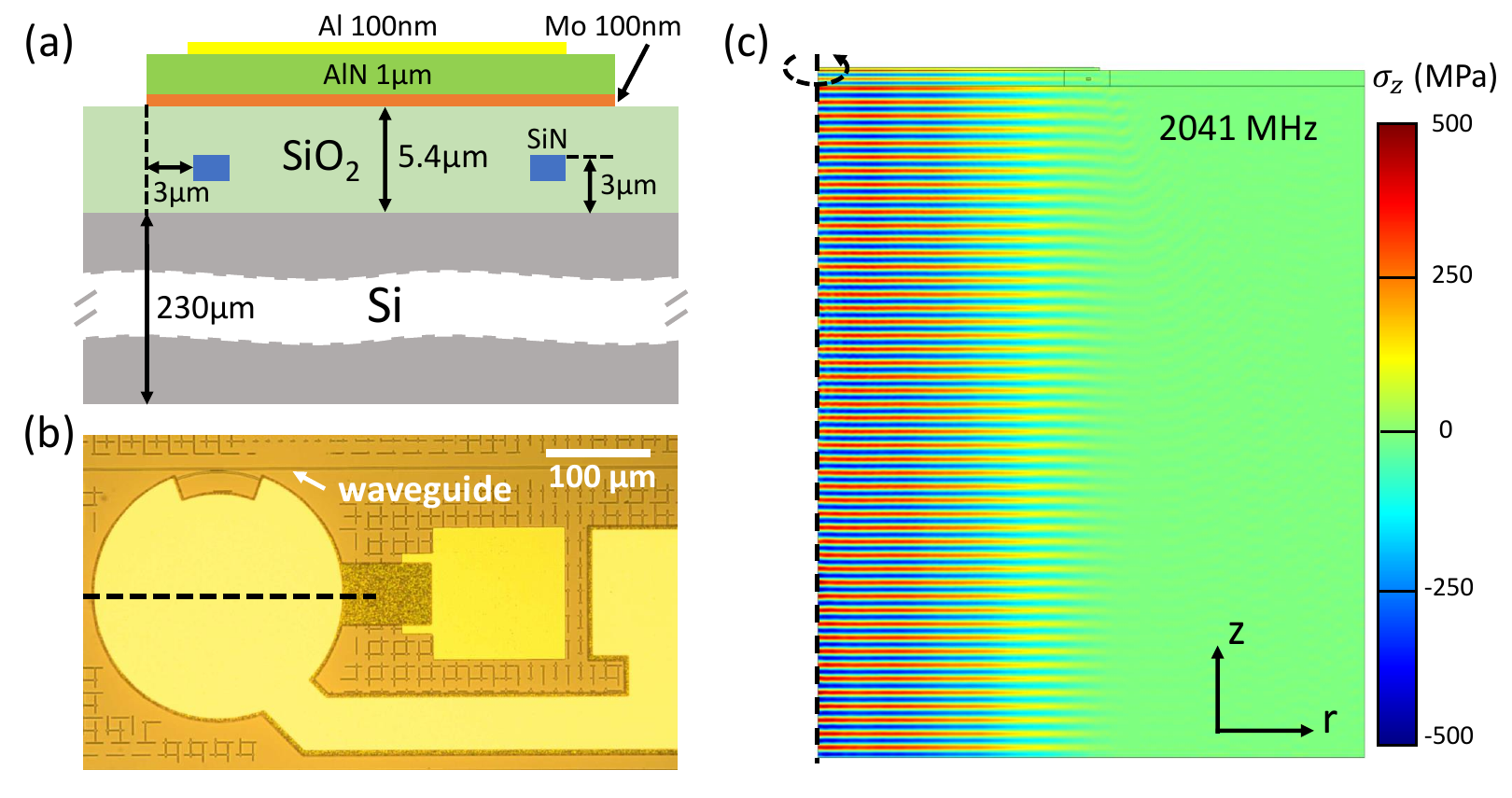}
\caption{
BAW AOM on an unreleased structure. 
(a) Cross-section of the device along the black dashed line in (b).
(b) Optical image of the disk AlN actuator with optical microring under its edge. 
(c) Numerical simulation of the vertical stress distribution of one HBAR mode at 2.041 GHz. 
Reprinted from Ref. \cite{Tian:20}.
}
\label{Ch5_BAW_unrelease}
\end{figure}

\subsection{Bulk Acoustic Wave Based AOM}
Apart from the surface acoustic wave, a bulk acoustic wave that transmits vertically in the substrate has also been utilized in modulating the Si$_3$N$_4$ microring resonator \cite{Tian:20}.
As shown in Fig. \ref{Ch5_BAW_unrelease}(a, b), the device shares the same structure as the one shown in Fig. \ref{Ch3_device}, where a disk-shaped actuator is designed to tune the optical resonance of a Si$_3$N$_4$ resonator. 
If we drive the actuator at the microwave frequency in stead, an acoustic wave will be excited vertically and confined in an acoustic Fabry-Pérot cavity, forming the so-called HBAR mode. 
The mode distribution of one typical HBAR mode is illustrated in Fig. \ref{Ch5_BAW_unrelease}(c). 
We can see that the acoustic wave is confined laterally beneath the actuator and is distributed uniformly throughout the substrate. 
Since the HBAR shown here is a longitudinal wave, the stress is dominated by the vertical component. On the other hand, transverse HBAR with large shear stress has also been demonstrated on LiNbO$_3$ \cite{xie2023sub, holzgrafe2020}.
An efficient AOM can be achieved by placing the optical waveguide on the antinode of the stress distribution. 

\begin{figure}[t]
\centering
\includegraphics[width=13 cm]{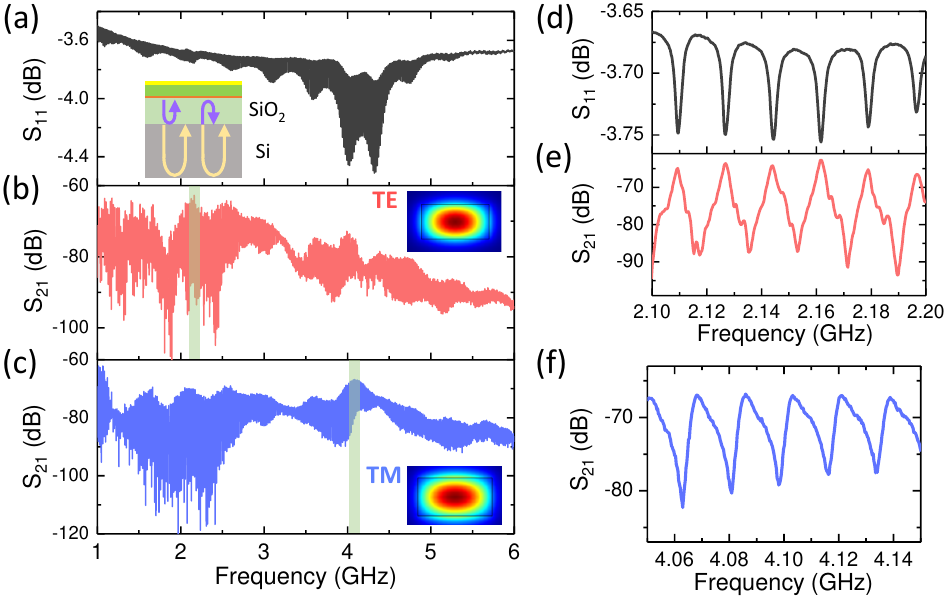}
\caption{
Experimental results of the unreleased HBAR AOM.
(a) Electromechanical S$_{11}$ response. 
The inset schematic illustrates the reflection of acoustic waves at interfaces. 
Optomechanical S$_{21}$ spectra showing the modulation of (b) TE$_{00}$ and (c) TM$_{00}$ optical modes. 
Zoom-in of (d) S$_{11}$ and (e) S$_{21}$ of TE$_{00}$ mode in the green shaded region of (b) around 2 GHz.
Periodic acoustic resonances can be observed with spacing of 17.5 MHz.
(f) Zoom-in of TM$_{00}$ mode's S$_{21}$ around 4 GHz (green shaded region) in (c). 
Reprinted from Ref. \cite{Tian:20}.
}
\label{Ch5_BAW_unreleaseS21}
\end{figure}

The HBAR modes can be found from its electromechanical S$_{11}$ response as presented in Fig. \ref{Ch5_BAW_unreleaseS21}(a) and (d). 
Similarly to an optical Fabry-Pérot cavity, there exists a series of equally spaced mechanical resonances, covering multiple octaves in the microwave frequencies.
The mode spacing of 17.5 MHz is primarily determined by the thickness of the substrate. 
The resonances have a linewidth of $\sim$3 MHz, corresponding to a mechanical $Q$ of $\sim$1000. 
In addition, the envelope of these resonances is periodically modulated.
This is due to the presence of an acoustic resonance in the SiO$_2$ layer.
The coupling between the SiO$_2$ and Si cavities further modifies the mechanical dispersion and perturbs the mode spacing. 
This behavior can be captured by the Mason model, as detailed in Ref. \cite{Tian:20}. 

Each of the acoustic resonances can modulate the optical resonator as manifested in the optomechanical S$_{21}$ measurement in Fig. \ref{Ch5_BAW_unreleaseS21}(b, c) for TE$_{00}$ and TM$_{00}$ optical modes.
Different from S$_{11}$, where the resonance dips are less than 1 dB, the S$_{21}$ shows clear peaks with high contrast ($>20$ dB) between resonance and anti-resonance due to the high sensitivity and large signal to noise ratio (SNR) of the optical readout.
Maximum S$_{21}$ of -65 dB can be obtained for both TE$_{00}$ and TM$_{00}$ modes which leads to a $V_{\pi}$ of $\sim$200 V according to Eq.\ref{S21VpiMZI}. 
This is much higher than the SAW AOM. 

\begin{figure}[t]
\centering
\includegraphics[width=13 cm]{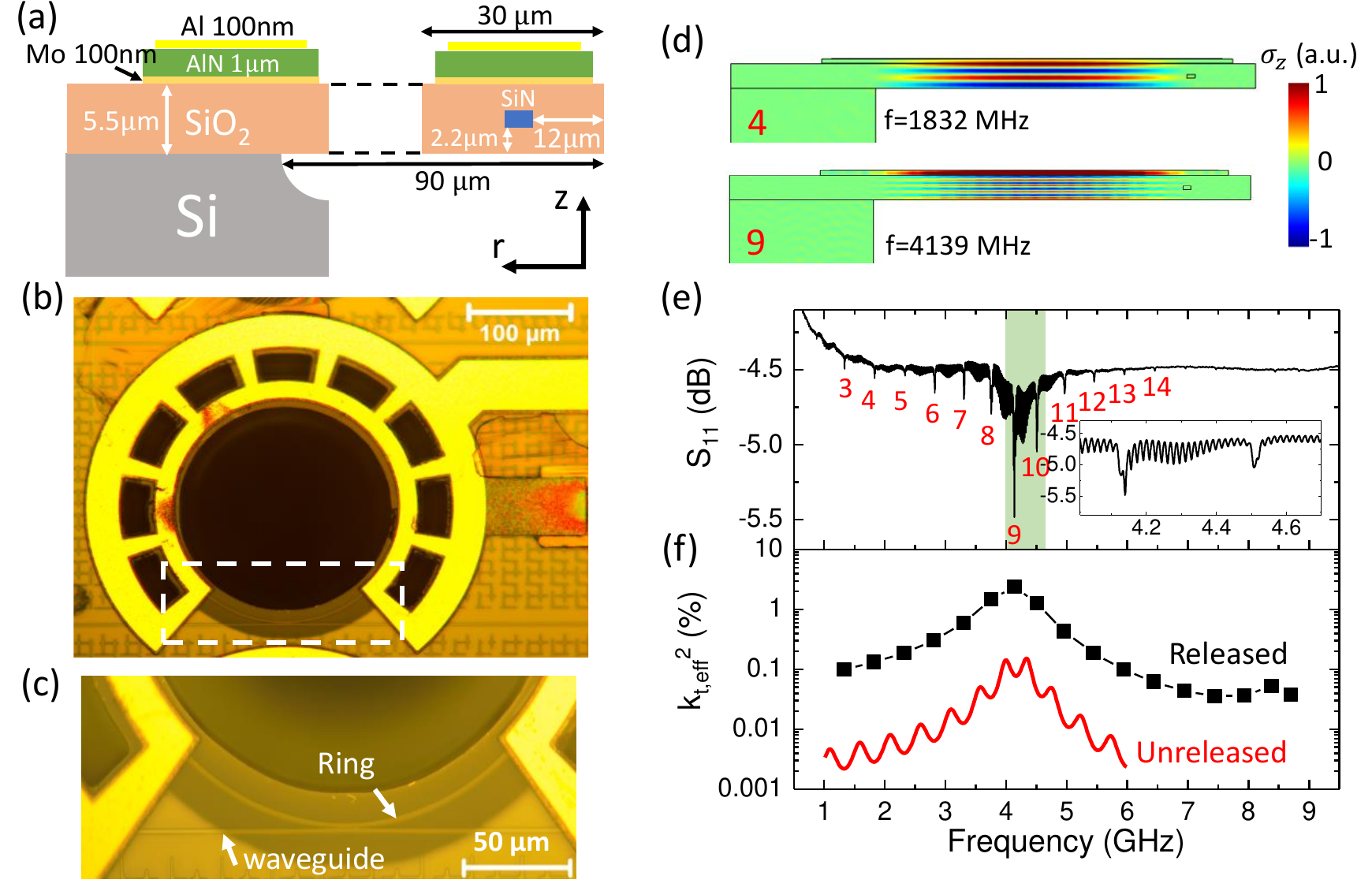}
\caption{
BAW AOM on a released structure. 
(a) Cross-section of the released HBAR. 
(b) Optical image of the fabricated device.
(c) Zoom-in of the white dashed box in (b), showing the microring to bus waveguide coupling region.
(d) Numerical simulation of the vertical stress distribution for two typical modes.
The scale bar is the stress normalized by the maximum. 
(e) Electromechanical S$_{11}$ spectrum. 
The mode order for each HBAR in the oxide cavity is labeled.
The inset shows the zoom-in of the green shaded region.
(f) Calculated electromechanical coupling efficiency $k^2_{t,\text{eff}}$ for released (black square dot) and unreleased (red curve) HBAR modes. 
Reprinted from Ref. \cite{Tian:21}.
}
\label{Ch5_BAW_release}
\end{figure}

The modulation efficiency can be further improved by releasing the SiO$_2$ membrane to tightly confine the acoustic wave in smaller volume \cite{Tian:21}.
As shown in Fig. \ref{Ch5_BAW_release}(a-c), the region where the Si$_3$N$_4$ resides is undercut and supported by a few cantilever beams connected to the substrate. 
By getting rid of the Si, a series of HBAR modes can be excited in the SiO$_2$ cavity, as can be found from the S$_{11}$ response in Fig. \ref{Ch5_BAW_release}(e).
Due to the thinner SiO$_2$ layer, the mode spacing is increased to around 450 MHz. 
The mode order is labeled for each resonance, which corresponds to the number of wavelengths in one round-trip inside the cavity. 
Two examples of HBAR modes on the order of 4 and 9 are simulated in Fig. \ref{Ch5_BAW_release}(d).
The electromechanical response around 4 GHz (green shaded region) is enhanced by the fundamental acoustic resonance in the AlN cavity, where the acoustic wavelength is twice of the AlN thickness.
Besides the deep resonances, there exist small resonances with a much smaller FSR of 17 MHz (see the inset in Fig. \ref{Ch5_BAW_release}(e)). 
This is due to the HBAR resonances from the Si substrate since part of the contact pads and traces are not released.
The electromechanical coupling efficiency $k^2_{t,\text{eff}}$ is calculated based on the Mason model \cite{tianthesis} for released and unreleased structures, as shown in Fig. \ref{Ch5_BAW_release}(f).
The efficiency is maximized around the AlN resonance at 4 GHz and improved by more than one order of magnitude upon releasing. 

\begin{figure}[t]
\centering
\includegraphics[width=13 cm]{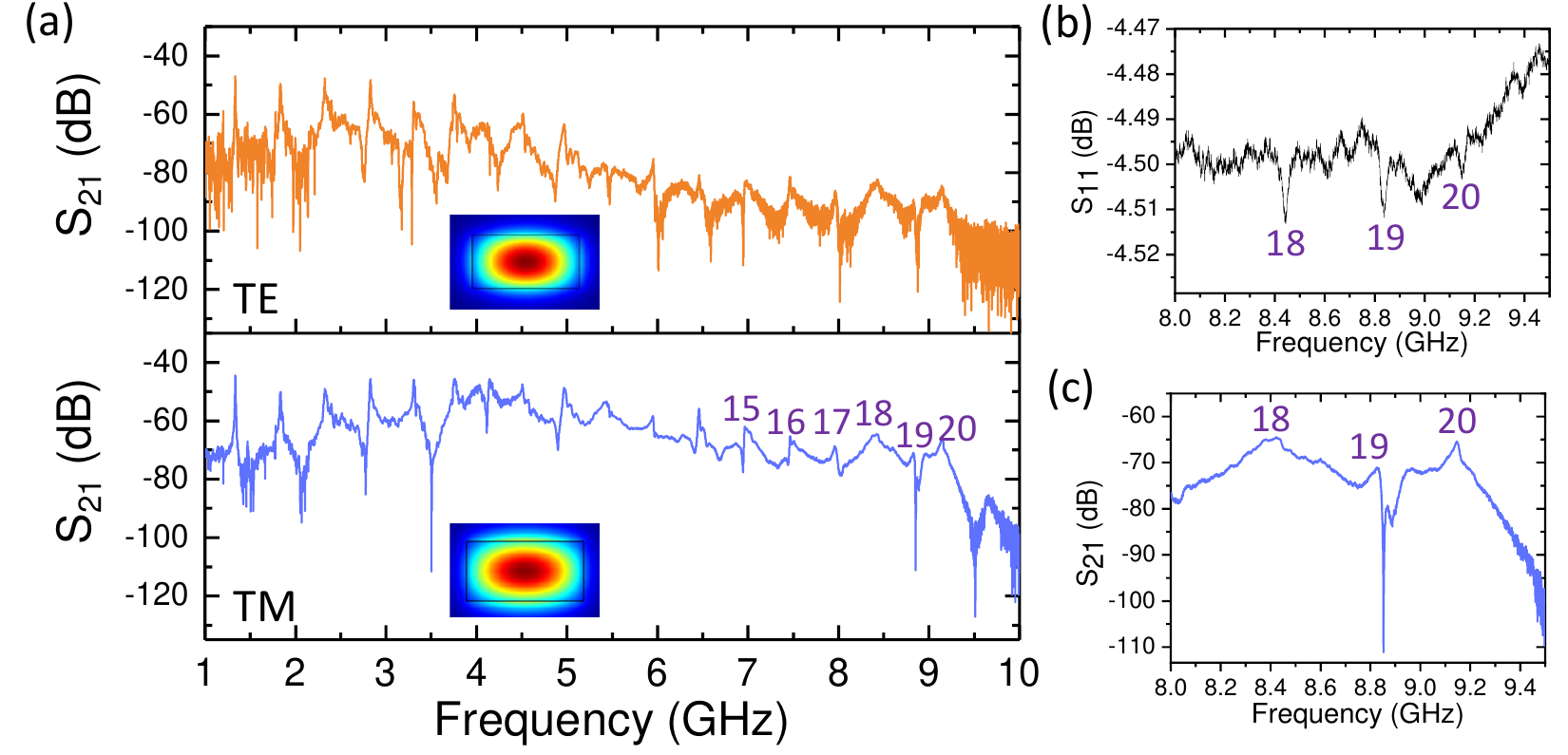}
\caption{
Experimental results of the released HBAR AOM. 
Optomechanical S$_{21}$ for TE$_{00}$ (top) and TM$_{00}$ (bottom) modes. 
Zoom-in of (b) S$_{11}$ and (c) S$_{21}$ of TM$_{00}$ mode for higher order modes around 9 GHz.
Reprinted from Ref. \cite{Tian:21}.
}
\label{Ch5_BAW_releaseS21}
\end{figure}

The AOM is characterized by the optomechanical S$_{21}$ response in Fig. \ref{Ch5_BAW_releaseS21}(a).
Each of the SiO$_2$ HBAR modes manifests itself as a peak in S$_{21}$. 
The Si HBAR modes do not show up since they mainly reside at the outer periphery that is far away from the inner Si$_3$N$_4$ microring.
Modulation up to 9.2 GHz can be achieved that enters the microwave X-band, as shown in Fig. \ref{Ch5_BAW_releaseS21}(c).
A maximum of -45 dB of S$_{21}$ is reached at 1.5 GHz and $V_{\pi}$ is estimated to be 18 V, which is much smaller than the unreleased HBAR AOM.
The improvement mainly comes from the enhancement of the electromechanical and optomechanical interactions as a result of the better confinement of the HBAR mode.

\begin{figure}[h]
\centering
\includegraphics[width=13.5 cm]{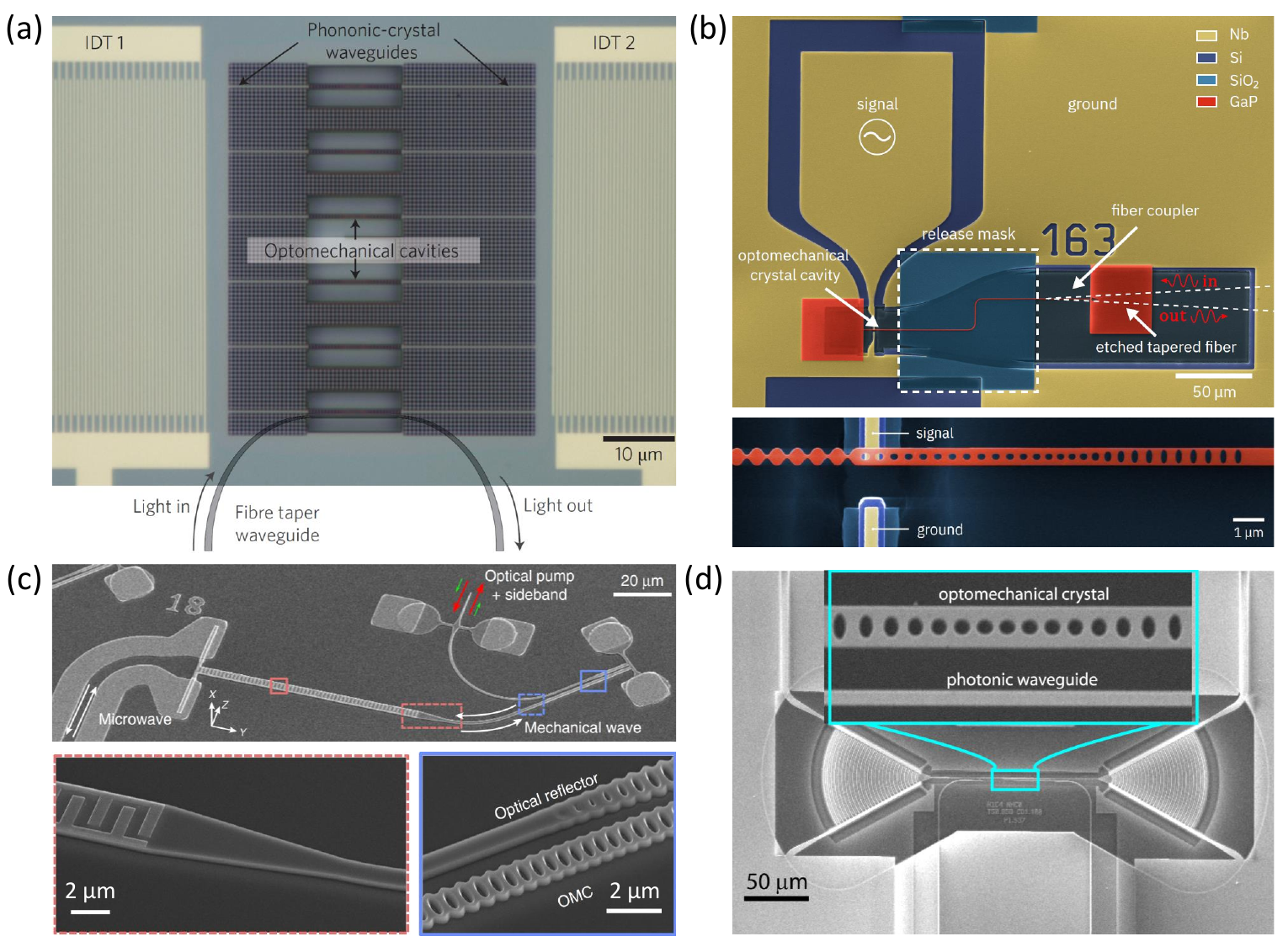}
\caption{
Piezo-optomechanical crystal based on single piezoelectric material. 
(a) GaAs OMC mechanically excited by the Lamb wave from IDT. Reprinted from Ref. \cite{balram2016}.
(b) GaP OMC excited by a pair of electrodes. Reprinted from Ref. \cite{honl2022}.
The lower inset shows the zoom-in of the OMC above the electrodes.
(c) LiNbO$_3$ OMC coupled through a mechanical waveguide to an IDT. 
The lower left inset shows the IDT, while the lower right inset is the OMC. Reprinted from Ref. \cite{jiang2020efficient}.
(d) AlN OMC excited by a focusing IDT. Reprinted from Ref. \cite{vainsencher2016}.
}
\label{Ch5_OMC_single}
\end{figure}

Although the BAW AOM demonstrated so far has lower efficiency than the SAW AOM, it possesses advantages that are preferred in specific applications. 
First, since the HBAR mode is distributed uniformly in the vertical direction of the substrate, the optical waveguide can be buried deep inside the cladding to maintain a high optical $Q$ and immunity to the environment perturbation. 
Also, the out-of-plane transmission of the acoustic wave makes it possible for independent optimization of the in-plane optical circuits.
Secondly, the acoustic wave is confined laterally by the actuator, which guarantees a compact size and a low cross-talk between adjacent actuators \cite{Tian:2021b}.
Lastly, since the acoustic resonance is mainly dictated by the thickness of each layer, the lateral degree of freedom is released such that larger critical dimensions of the actuator can be designed, in contrast to the sub-micron electrodes of SAW IDT.
This not only eases the fabrication, but also increases the power handling capability.
On the other hand, since the thickness of each layer can be well controlled over the whole wafer, HBAR actuators with the same frequency can be massively produced at the wafer scale, which would improve the yield and lower the cost.
In the future, the AOM efficiency can be improved by engineering the mechanical and optical mode, such as coupling HBAR with a photonic crystal.
In addition, materials with larger piezoelectric coefficient, such as LiNbO$_3$ and BaTiO$_3$, will be of great interest for excitation of HBAR modes \cite{xie2023sub}. 

\begin{figure}[h]
\centering
\includegraphics[width=13.5 cm]{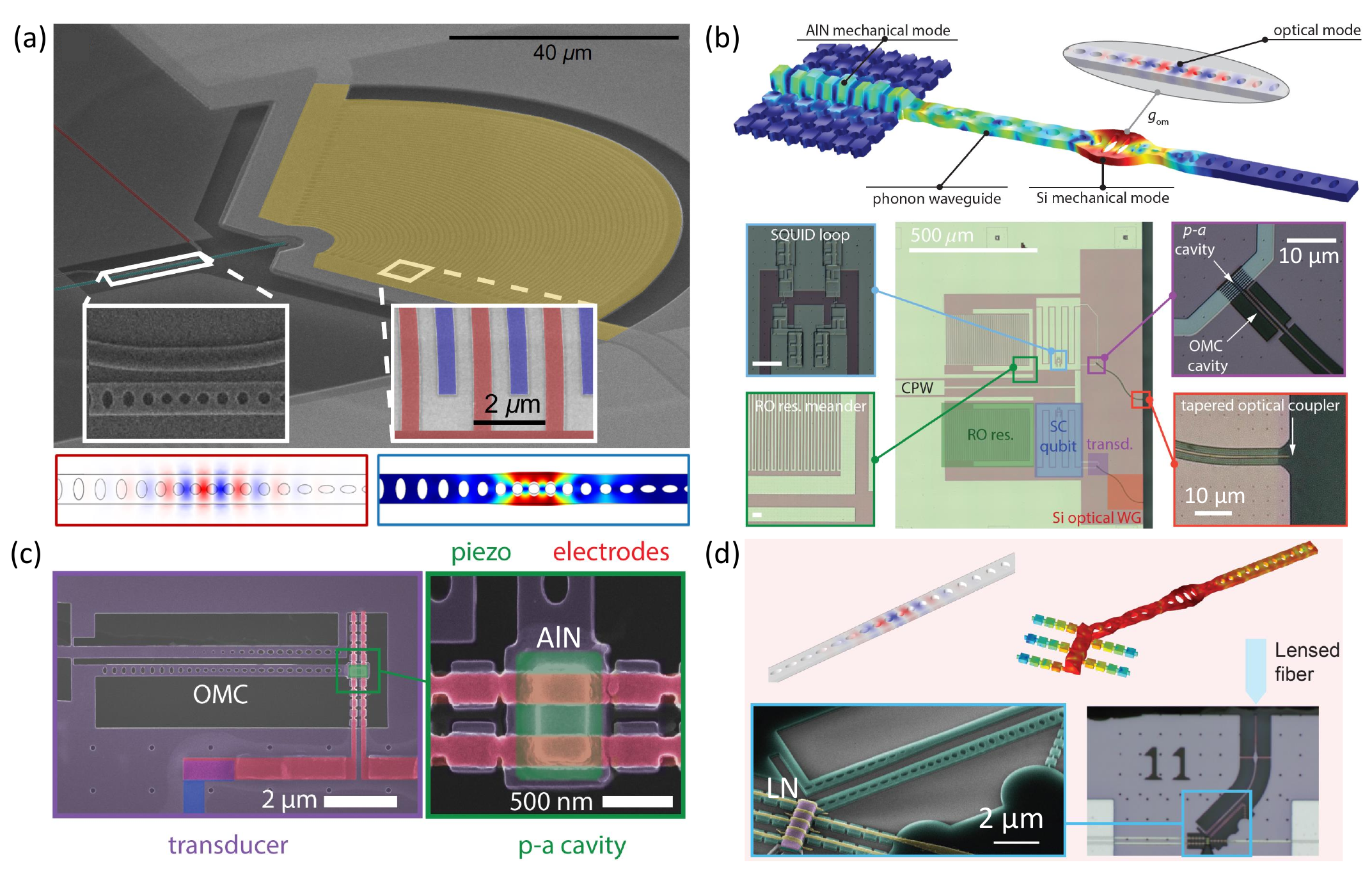}
\caption{
Hybrid piezo-optomechanical system with Piezo-on-Si structure.
(a) Si OMC excited by a focusing AlN IDT.
The lower insets are the optical (left) and mechanical (right) modes.
Reprinted from Ref. \cite{peairs2020}.
Si OMC coupled with (b) an AlN piezoelectric resonator, (c) a half-wavelength AlN piezo-acoustic cavity (p-a cavity), and (d) a LiNbO$_3$ (LN) piezoelectric resonator (purple) through a phonon waveguide. 
The insets in (b) and (d) show the optical mode and the hybridized mechanical mode. (b) Reprinted from Ref. \cite{mirhosseini2020}. (c) Reprinted from Ref. \cite{meesala2023}. (d) Reprinted from Ref. \cite{jiang2023}.
}
\label{Ch5_OMC_hybrid}
\end{figure}

\subsection{Piezo-Optomechanical Crystal}

Since its first demonstration by Eichenfield et al. \cite{eichenfield2009} in 2009, the optomechanical crystal (OMC) has provided a way to tightly confine mechanical and optical modes together in a small volume by co-design of the photonic and phononic band structure.
This has enabled us to achieve the highest optomechanical coupling rate on the order of 1 MHz. 
OMC has been realized in most common optical materials including Si \cite{chan2012}, Si$_3$N$_4$ \cite{grutter2014si}, diamond \cite{burek2016} and SiC \cite{lu2020}, as well as piezoelectric materials, such as AlN \cite{fan2013}, LiNbO$_3$ \cite{jiang2019lithium}, GaP \cite{schneider2019, stockill2019}, and GaAs \cite{Balram:14}. 
By depositing electrodes, the OMC made from piezoelectric material can be directly coupled to the external microwave channel for optical modulation, as shown in Fig. \ref{Ch5_OMC_single}. 
Either a single pair of electrodes or an IDT can be designed to excite the mechanical vibration. 
However, to maintain a high optical $Q$, the electrodes need to be placed a few microns away from the OMC.
A phononic crystal waveguide is required to deliver the acoustic energy to the OMC (see Fig. \ref{Ch5_OMC_single}(a)), which must be tailored to support a large acoustic wavenumber but in the same time serve as an optical mirror for the optical mode \cite{Zhao:23}.
On the other hand, due to the small size of the OMC mechanical mode, there is a large mismatch with the acoustic mode of the IDT, making the coupling efficiency low.
One solution is to confine the IDT mode in a wavelength-scale mechanical waveguide, as illustrated in Fig. \ref{Ch5_OMC_single}(c) \cite{jiang2020efficient, dahmani2020}. 
Another way is to design an IDT that focuses the acoustic wave onto the phonon waveguide, as demonstrated in Fig. \ref{Ch5_OMC_single}(d) \cite{vainsencher2016, decrescent2022, kandel2023}. 

\begin{table}[t]
    %\centering
 \caption{
 Comparison between the state-of-the-art AOMs of OMC. 
 WG: waveguide.
 WL: wavelength. 
 $V_{\pi}$ is calculated using Eq. \ref{Vpi} based on the reported experimental results. 
 }
    
    \begin{tabular}{ p{0.8cm}|p{0.8cm}|p{1.8cm}|p{1.5cm}|p{1.2cm}|p{1.2cm}|p{1.2cm}|p{1.2cm}|p{1.2cm}|p{1.2cm}}
 \hline
 Ref.  &  Year &Piezo coupling scheme & Material platform & $\Omega_m/2\pi$ (GHz)& $\gamma_m/2\pi$ (MHz) & $\gamma_{m,\text{ex}}/\gamma_m$ &Optical linewidth (GHz) & $g_0/2\pi$ (kHz) &$V_{\pi}$ (V) \\
 \hline
\cite{vainsencher2016}   & 2016 & Focusing SAW IDT & AlN & 3.78 & 5 & $2\times 10^{-4}$  & 15  & 115 & 0.162 \\
\cite{balram2016}   & 2016 & SAW IDT & GaAs & 2.42 & 1.6 & $3\times 10^{-10}$ & 5.2  & 1100 & 0.625 \\
\cite{forsch2020}   & 2020 & SAW IDT & GaAs & 2.7 & 0.2  & $3.6\times 10^{-10}$ & 5.8 & 1300 & 1.29 \\
\cite{jiang2019lithium}   & 2019 & Electrodes & LiNbO$_3$ & 2 & 0.5  & $1.8\times 10^{-8}$ & 0.78 & 120 & 1.99 \\
\cite{jiang2020efficient}   & 2020 & WL-scale IDT & LiNbO$_3$ & 1.85 & 1.9  & 0.001 & 1.2 & 80 & 0.024 \\
\cite{honl2022}   & 2022 & Electrodes & GaP & 3.31 & 2.8  & $1.36\times 10^{-9}$ & 3 & 280 & 15.6 \\
\cite{peairs2020}   & 2022 & Focusing SAW IDT & AlN on Si & 4.74 & 6.2  & $\sim$0.002 & 10.3 & 734 & 0.013 \\
\hline
\end{tabular}
    \label{table_OMC}
\end{table}

Apart from piezo-optomechanical systems made from single material, a hybrid Piezo-on-Si platform has been widely studied, showing the advantages of high mechanical and optical $Q$ and large optomechanical coupling strength.
This primarily benefits from the low acoustic losses, large refractive index and photoelastic coefficients of Si \cite{chiappina2023}.
AlN and LiNbO$_3$ have been the two main piezoelectric materials demonstrated so far, as shown in Fig. \ref{Ch5_OMC_hybrid}. 
Different acoustic modes can be excited by the IDT, such as Lamb wave (Fig. \ref{Ch5_OMC_hybrid}(a)) and wavelength-scale acoustic cavity (Fig. \ref{Ch5_OMC_hybrid}(b)). 
The acoustic wave is then guided by the phononic crystal waveguide to the OMC. 
The coupling between the OMC and piezoelectric modes leads to a hybridized mechanical mode when they are designed to have the same frequency.
This hybridized mode is partially coupled with the optical field and the electric field of IDT for transduction between optical and microwave photons.
The acoustic energy participated in the piezoelectric cavity does not contribute to the optomechanical interaction, which could lower the optomechanical coupling rate $g_{\text{om}}$. 
Therefore, it is preferable to have a piezo-acoustic cavity with a small mode volume. 
Recently, a half-wavelength scale acoustic resonator has been designed \cite{chiappina2023} and demonstrated \cite{meesala2023}, as shown in Fig. \ref{Ch5_OMC_hybrid}(c).
A Si-like hybrid mode is formed, which supports a large $g_0$ and a strong coupling to a high impedance microwave resonator. 

The performances of the state-of-the-art AOMs achieved in piezo-optomechanical crystal have been summarized in Table \ref{table_OMC}. 
Despite of the large $g_0$, the efficiency is largely limited by the external coupling rate from microwave to OMC, especially for the excitation schemes using SAW IDT and simple electrodes.
In that sense, the wavelength-scale IDT has been more popular due to its better mode matching with OMC, and a $V_{\pi}$ much smaller than the SAW AOM has been demonstrated \cite{jiang2020efficient}. 
On the other hand, GaAs exhibits the largest $g_0$ such that a much improved $V_{\pi}$ can be foreseen with optimized IDT design in the future. 

In summary, this chapter reviews the integrated AOMs realized using various acoustic modes, which have their own advantages in different aspects. 
While the SAW has a high microwave-to-acoustic-wave conversion efficiency ($\gamma_{m,\text{ex}}/\gamma_m$), the HBAR supports high optical $Q$ by cladding the optical waveguide, and the OMC is superior in optomechanical coupling $g_0$ through sub-wavelength mode confinement. 
Therefore, depending on the requirement of the application, the appropriate scheme should be chosen, taking into account the material compatibility, efficiency, and fabrication cost. 
A few of the most widely explored applications will be outlined in the next chapter, and the comparison between different acoustic modes will be made. 

%\subsection{Comparison of different structures}

\section{Applications of On-chip Acousto-Optic Modulator}

While the on-chip AOM features a narrow modulation bandwidth that is primarily limited by the mechanical resonance, there are many advanced applications that only require single frequency or narrow band microwave drive. 
In this chapter, we will review three of the most widely explored applications that will benefit from the high modulation efficiency of the AOM. 
Despite of being made from AOMs with similar structures, they realize unique functions, ranging from frequency converter, optical isolator, to frequency shifter. 
This manifests the universality of the AOM and the possibility of performing multiple functions on the same chip.

\subsection{Microwave to Optical Converter}

Superconducting circuits and qubits have experienced a rapid progress in recent decades, featuring a full set of microwave quantum gates with high fidelity \cite{chow2012}.
Although the prototype of quantum processing unit (QPU) has been demonstrated \cite{arute2019}, the requirement of working at cryogenic temperature limits its scalability, due to the finite cooling power of current dilution refrigerators and the limited space imposed by the need for coaxial cables to drive and readout of SC qubits. 
These challenges prevent the realization of fault-tolerant quantum computing in a foreseeable future, which requires millions of qubits \cite{preskill2018}.
A potential solution is to interconnect multiple quantum processors hosted in separate dilution refrigerators.
Although the proof-of-concept experiment has demonstrated the connection via coaxial cables \cite{storz2023}, they must be kept at cryogenic temperatures to maintain a low thermal noise, which is expensive and not scalable.
However, optical photons possess frequencies four orders of magnitude higher than the microwave, leading to negligible thermal noise at room temperature.
Indeed, the realization of the optical quantum key distribution (QKD) \cite{lo2014} has demonstrated its capability to maintain quantum superposition and entanglement at room temperature.
Moreover, optical fiber telecommunication systems have become one of the major infrastructures which can be readily upgraded for building large-scale quantum processor and quantum network. 
Therefore, conversion between microwave and optical photons with high quantum efficiency has become an active research area recently, and promising progress has been made. 

For an ideal microwave to optical converter in quantum transduction, it should have high quantum efficiency with low added noise and wide transduction bandwidth. 
Although classical optical modulator based on electro-optic effect has been well studied, due to the large frequency and energy difference between microwave and optical photons, it is nontrivial to convert directly with high quantum efficiency through the weak nonlinearity of most optical materials.
Nevertheless, electro-optic conversion has been explored in various platforms, such as integrated AlN \cite{Fan:18, fu2021} and LiNbO$_3$ \cite{mckenna2020, holzgrafe2020, xu2021} photonic circuits, as well as bulk LiNbO$_3$ disk \cite{hease2020, rueda2016, sahu2022}.
Quantum transduction via an intermediary mode, such as mechanical mode \cite{andrews2014, bochmann2013}, magnon mode \cite{hisatomi2016, zhu2020}, and atomic transition \cite{bartholomew2020} has attracted much attention.
Among them, the mechanical mode has been more popular due to its high efficiency, ease of fabrication, and compatibility with SC qubits. 

There are mainly three ways of coupling mechanics with microwave signals, including radiation pressure, electrostatic and piezoelectric. 
The highest conversion efficiency has been achieved in electromechanically coupled mechanical membrane via radiation pressure, from 10$\%$ \cite{andrews2014} to 47$\%$ \cite{higginbotham2018}, approaching the 50$\%$ efficiency required for having finite quantum capacity in quantum communication \cite{wolf2007}.
However, since the membrane operates at MHz frequencies, it only supports a limited bandwidth, and the added noise is high even at cryogenic temperature.
Progresses have been made recently in electrostatically exciting Si OMC which supports GHz mechanical modes \cite{Zhao:23}. 
Despite being applicable to most materials, the electrostatic coupling is still weaker than the piezoelectric force. 
In that sense, piezo-optomechanical transducer has drawn much attention in coherent conversion between microwave and optics.
The variety of mechanical modes as presented in the previous chapter for AOM have also been explored in quantum transducer, from SAW \cite{Shao:19, khurana2022}, BAW \cite{han2020, valle2021, blesin2023}, to OMC \cite{jiang2019lithium, jiang2020efficient, mirhosseini2020}. 

The same mode coupling scheme as presented in Fig. \ref{mode} for AOM also holds for the quantum transduction, where the mechanical and optical modes are coupled through the optomechanical coupling rate $g_{\text{om}}$.
The Hamiltonian that describes the interaction is \cite{aspelmeyer2014}:
\begin{equation}
    H_{\text{int}}=\hbar g_0 \hat{a}^{\dagger}\hat{a}(\hat{b}+\hat{b}^{\dagger})
\end{equation}
where $\hat{a}$ and $\hat{b}$ are the quantum operators for optical and mechanical modes, $g_0$ is the single photon optomechanical rate.
The above equation is usually linearized by substituting $\hat{a}$ by $\Bar{\alpha}+\delta \hat{a}$, where $\Bar{\alpha}$ is the average amplitude of the pump optical field, and $\delta \hat{a}$ is the fluctuation.
Under rotating-wave approximation, the interaction Hamiltonian can be deduced as:
\begin{eqnarray}
    H_{\text{int}}=\hbar g_{\text{om}}(\delta \hat{a}^{\dagger}\hat{b}+\delta \hat{a}\hat{b}^{\dagger})+\hbar g_{\text{om}}(\delta \hat{a}^{\dagger}\hat{b}^{\dagger}+\delta \hat{a}\hat{b})
\end{eqnarray}
where $g_{\text{om}}=g_0\sqrt{n_{\text{cav}}}$ with $n_{\text{cav}}$ the intra-cavity pump photon number. 
The first and second terms in the above equation correspond to the beam-splitter and two-mode squeezing interaction, respectively.
Depending on the detuning of the pump laser (assuming resolved sideband $\omega_m \gg \kappa_o$), one interaction will dominate the other, with red detuning for beam-splitter and blue detuning for two-mode squeezing \cite{aspelmeyer2014}.

\begin{figure}[b]
\centering
\includegraphics[width=13.5 cm]{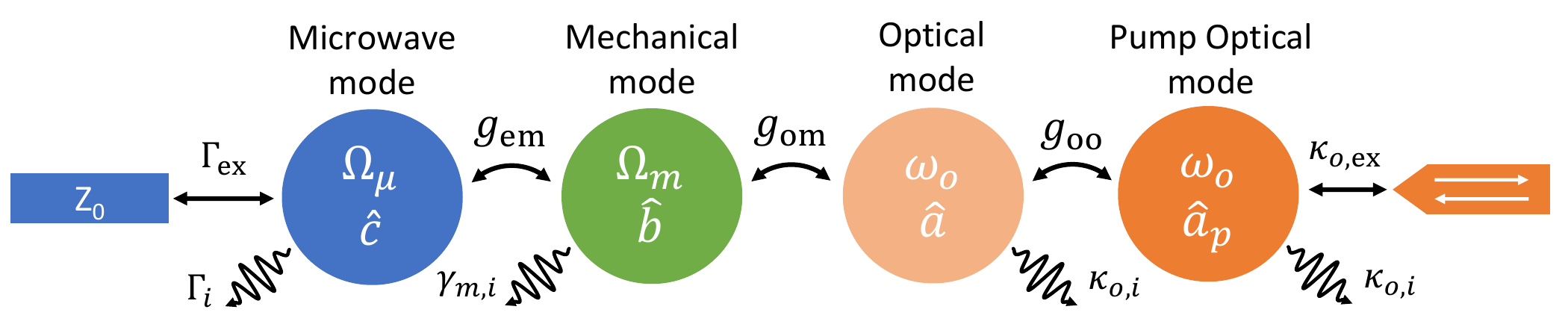}
\caption{
Mode coupling scheme of the microwave to optical transduction by including a microwave mode to boost the mechanical excitation efficiency and a second optical mode to increase the intra-cavity photon number. 
}
\label{Ch6_mode}
\end{figure}

The microwave-to-optical conversion relies on the beam-splitter interaction where the annihilation of an input microwave photon leads to the generation of an output optical photon, and vice versa. 
The quantum efficiency is defined as the ratio of the output optical photon flux to input microwave photon flux in up-conversion, and vice versa in down-conversion. 
Practically, the total efficiency consists of the coupling efficiency from the off-chip (microwave or optical) sources / detectors to the on-chip waveguides and the on-chip conversion efficiency $\eta_{\text{oc}}$. 
The $\eta_{\text{oc}}$ can serve as a figure of merit for comparing different works and can be derived from the Heisenberg-Langevin equations of motion (see Eqs. \ref{eqmotion1}-\ref{eqmotion2}) as \cite{blesin2023}:
\begin{equation}
    \eta_{\text{oc}} = \eta_{\text{ext}}\eta_{\text{int}} = \frac{\kappa_{o,\text{ex}}}{\kappa_o}\frac{\gamma_{m,\text{ex}}}{\gamma_m}\frac{4C_{\text{om}}}{(1+C_{\text{om}})^2}
\end{equation}
where $\eta_{\text{ext}}=\kappa_{o,\text{ex}}\gamma_{m,\text{ex}}/(\kappa_o\gamma_m)$ is the external coupling efficiency from the waveguides to the optical and mechanical modes. 
The internal efficiency $\eta_{\text{int}}$ is the efficiency between mechanical and optical modes and is determined by the optomechanical cooperativity $C_{\text{om}}=4g_{\text{om}}^2/(\kappa_o\gamma_m)$, which leads to its maximum of 1 when $C_{\text{om}}$ is equal to 1.
The external efficiency will become larger when $\kappa_{o,\text{ex}}$ and $\gamma_{m,\text{ex}}$ increase.
However, this would lower the $C$ and the internal efficiency. 
Therefore, an optimal efficiency can be achieved when both of the mechanical and optical modes are critically coupled to their waveguides \cite{blesin2023}.

\begin{figure}[t]
\centering
\includegraphics[width=12 cm]{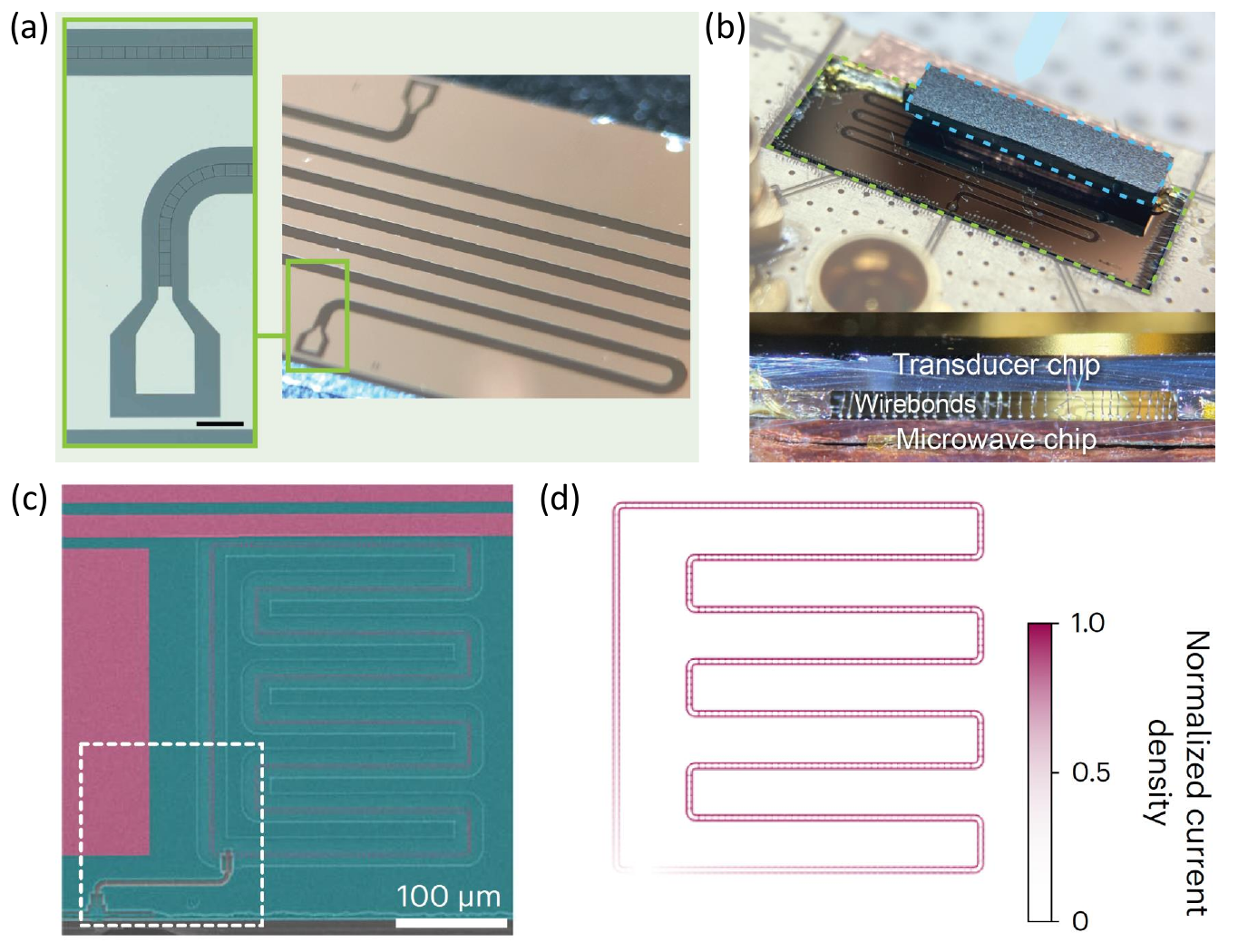}
\caption{
High-impedance microwave resonator with ladder geometry. 
(a) Optical images of an NbTiN microwave resonator fabricated on a chip separate from the transducer chip.
The resonator is formed in a coplanar waveguide made from nanowires patterned in a ladder shape. The scale bar is 200 $\mu$m.
(b) Assembly of the transducer chip (top) with the microwave chip (bottom) via wirebonds.
(c) False-colored SEM of a MoRe microwave resonator made on the same chip with the transducer. 
(d) Finite element simulation of the current density of the microwave resonator. 
(a-b) Reprinted from Ref. \cite{jiang2023}. (c-d) Reprinted from Ref. \cite{weaver2023}. 
}
\label{Ch6_microwave}
\end{figure}

As discussed in the previous chapter, the direct coupling rate from microwave transmission line to the mechanical mode is generally small due to impedance mismatch, especially for the high impedance, sub-micron OMC mode. 
To further improve the excitation efficiency, a new scheme has been proposed that involves a microwave resonator, as illustrated in Fig. \ref{Ch6_mode}.
It helps by recycling the microwave reflected by the piezoelectric transducer.
The microwave and mechanical modes are coupled through $g_{\text{em}}$, which can be calculated as \cite{wollack2021loss}:
\begin{eqnarray}
    g_{\text{em}} & \simeq & \frac{1}{2}\sqrt{\Omega_{\mu}\Omega_m}\sqrt{\frac{C_m}{C_{\mu}+C_m+C_0}}\\
    & \approx & \frac{1}{2}\sqrt{\frac{\gamma_{m,\text{ex}}}{Z_0}}\sqrt{\frac{1}{C_{\mu}+C_m+C_0}} 
    \label{gem}
\end{eqnarray}
where $\Omega_{\mu}$ and $C_{\mu}$ are the frequency and capacitance of the microwave resonator, respectively.
When $\Omega_{\mu}$ equals $\Omega_m$, $g_{\text{em}}$ can be related with the external coupling rate of the mechanical mode to the transmission line $\gamma_{m,\text{ex}}$ by Eq. \ref{gem}.
It can be seen for large $g_{\text{em}}$, the microwave resonator is desirable to have small capacitance $C_{\mu}$ which corresponds to a high impedance $Z_{\mu}$ ($=\sqrt{L_{\mu}/C_{\mu}}$). 
The development of high-impedance and compact superconducting resonator has benefited from the study of thin film superconductors with high kinetic inductance and has been applied in microwave to optical transduction, such as TiN \cite{Zhao:23}, NbTiN \cite{jiang2023}, NbN \cite{meesala2023}, and MoRe \cite{weaver2023}. 
Moreover, the design of these resonators with ladder geometry allows tuning via external magnetic flux \cite{xu2019}, as shown in Fig. \ref{Ch6_microwave}. 

The Hamiltonian that describes the interaction between the microwave and mechanical resonators is $\hbar g_{\text{em}}( \hat{c}^{\dagger}\hat{b}+ \hat{c}\hat{b}^{\dagger})$. 
The conversion efficiency by including the microwave resonator can be obtained similarly from the equations of motion as \cite{Zhao:23, jiang2020efficient, han2021microwave}:
\begin{equation}
\label{eq_CemCom}
    \eta_{\text{oc}} = \eta_{\text{ext}}\eta_{\text{int}} = \frac{\kappa_{o,\text{ex}}}{\kappa_o}\frac{\Gamma_{\text{ex}}}{\Gamma}\frac{4C_{\text{om}}C_{\text{em}}}{(1+C_{\text{om}}+C_{\text{em}})^2}
\end{equation}
where $\Gamma$ and $\Gamma_{\text{ex}}$ are the total and external coupling rate of the microwave resonator, and $C_{\text{em}}=4g_{\text{em}}^2/(\Gamma\gamma_m)$ is the electromechanical cooperativity.
The internal efficiency $\eta_{\text{int}}$ reaches maximum when $C_{\text{om}}=C_{\text{em}}$, and approaches 1 as $C_{\text{om}}$ and $C_{\text{em}}$ increase. 
Interestingly, the previous two-mode case with only optical and mechanical modes can be considered as a special case of Eq. \ref{eq_CemCom} by treating the mechanical mode coupled to an overcoupled broad microwave resonator such that $C_{\text{em}}\ll 1$ and $C_{\text{em}}=\gamma_{m,\text{ex}}/{\gamma_m}$ \cite{Blesin:21, jiang2020efficient, Zhao:23}. 

\begin{figure}[t]
\centering
\includegraphics[width=13.5 cm]{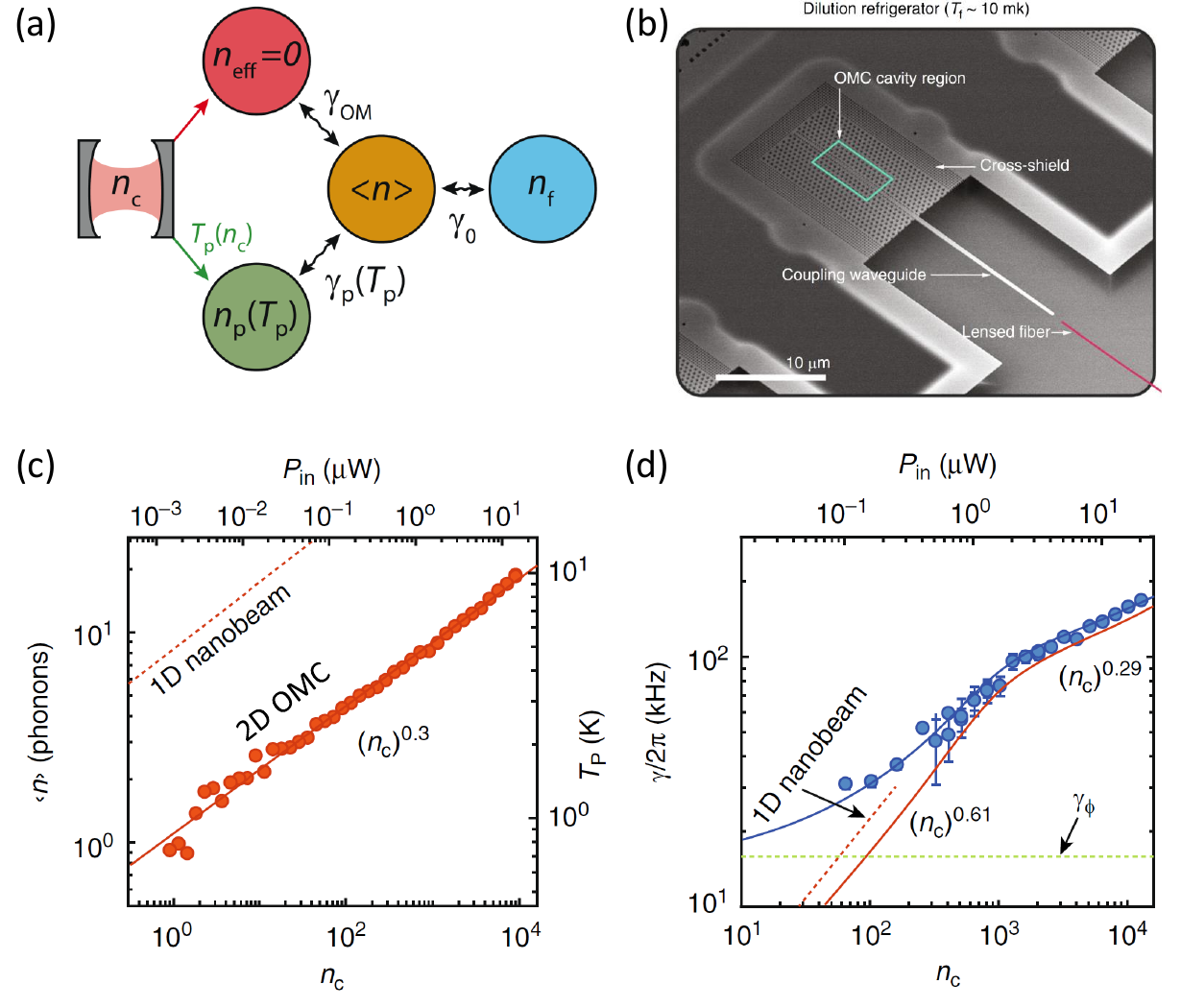}
\caption{
Effects of optical absorption induced hot bath.
(a) Schematic of the model describing different baths the mechanical mode is coupled to. 
Reprinted from Ref. \cite{meenehan2014}.
(b) SEM of a two-dimensional optomechanical crystal with better thermal conductance.
(c) Experimental measurement of $n_p$ versus intra-cavity photon number $n_c$ for 1D OMC (red dashed line) and 2D OMC (red dots). 
Solid red line is the fitting leads to power-law scaling of $n_c^{0.3}$. 
(d) Measurement of mechanical linewidth versus $n_c$ for 2D OMC (blue dots) with theory fitting (blue solid line).
Red solid line is the fitting for $\gamma_p$. 
Dashed red line is $\gamma_p$ for 1D OMC.  
(b-d) Reprinted from Ref. \cite{ren2020two}. 
}
\label{Ch6_bath}
\end{figure}

Since the conversion efficiency is proportional to the intracavity pump photon number, and thus the on-chip optical pump power, high efficiency can be achieved by increasing the pump power.
However, the coherent quantum transduction requires the added noise to be smaller than one.
Therefore, the maximum power is ultimately limited by the added noise resulting from optical absorption, which has two main negative effects on the transducer.
One is the generation of quasi-particles on the superconductor, and the other is the increased thermal occupancy of the mechanical mode.
The quasi-particle is generated from broken Cooper pairs which can come either directly from the light absorption or through thermal heating \cite{weaver2023, mirhosseini2020}.
It has been experimentally examined that the quasi-particle could lower the lifetime of SC qubits \cite{mirhosseini2020} and shift the frequency and increase the intrinsic loss of the superconducting resonator \cite{weaver2023}. 
The quasi-particle can be effectively mitigated by reducing the time of optical illumination through optical pulses \cite{weaver2023}. 
Also, the use of superconductors with shorter quasi-particle lifetime could accelerate the recovery time of the superconductor, such as Nb and NbTiN which have quasi-particle lifetime on the order of nanoseconds \cite{johnson1991}. 

The optical-absorption-induced hot bath becomes more prominent at sub-Kelvin temperature, which has been experimentally studied in Ref. \cite{meenehan2014}.
The effect of the hot bath on the mechanical resonator can be described by the model in Fig. \ref{Ch6_bath}(a), where the mechanical mode couples not only to the refrigerator bath $n_f$ at a rate of $\gamma_0$, but also to the hot bath $n_p$ from the optical absorption with rate $\gamma_p$. 
Both $n_p$ and $\gamma_p$ depend on the temperature of the hot bath $T_p$ which is a function of the intra-cavity photon number $n_c$.
The thermal occupancy of the mechanical mode can be described as:
\begin{equation}
     \left<n \right>(n_c) = \frac{\gamma_0n_f+\gamma_p(n_c)n_p(n_c)}{\gamma_0+\gamma_p(n_c)+\gamma_{\text{om}}(n_c)}
     \label{heating}
\end{equation}
The dependence of $n_p$ and $\gamma_p$ on $n_c$ for an 1D OMC is measured as shown in Fig. \ref{Ch6_bath}(c-d). 
It can be seen that the optical absorption not only increases the thermal occupancy but also the dissipation of the mechanical resonator.

A few strategies are under investigation to mitigate the optical absorption. 
The most widely adopted method is the use of optical pulses, where high optical peak power can be achieved while the average power remains low \cite{meenehan2015 ,ramp2019}.
However, the repetition rate of the pulses is ultimately limited by the power handling capability of the transducer, which limits the transduction bandwidth.
By increasing the thermal conductance of the optical cavity to the surrounding cold environment, the heating from the optical absorption can be effectively dissipated.
For instance, a recent study has shown that the two-dimensional optomechanical crystal has thermal conductance 40 times larger than the 1D OMC, and the corresponding occupancy of the hot bath is 7 times smaller (see Fig. \ref{Ch6_bath}(b-c)) \cite{ren2020two}. 
The thermal conductance can be further increased by utilizing unreleased OMC structure which has been demonstrated in both 1D \cite{kolvik2023} and 2D OMC \cite{liu2022optomechanical}.
Bulk acoustic wave has also demonstrated good optical power handling at cryogenic temperature \cite{doeleman2023}. 

For most of the optomechanical system in quantum transduction, it works in the resolved sideband regime where the mechanical frequency is much larger than the optical linewidth.
This inevitably leads to low pump efficiency when the laser is red or blue detuned by $\Omega_m$ where most of the pump light is reflected by the optical cavity.
Moreover, research indicates that the reflected light can constructively interfere with the incoming pump light, resulting in light absorption and heating within the coupling waveguide comparable to that in the cavity \cite{ren2020two}.
In this case, $n_c$ in the phenomenological model in Eq. \ref{heating} needs to be replaced by $n_c+\beta P_{\text{in}}$ in order to include the heating from coupling waveguide, where $P_{\text{in}}$ is the input optical power \cite{sonar2024}. 
In the demonstration in Ref. \cite{ren2020two}, the coupling waveguide is butt coupled with the OMC such that they are mechanically connected and the heating from the waveguide can be directly conducted to the OMC. 
The prefactor $\beta$ is experimentally measured to be 15 $\mu$W$^{-1}$ which largely increases the thermal occupation \cite{ren2020two}. 
To circumvent this problem, it has been proposed to mechanically detach the waveguide from the OMC though side coupling \cite{sonar2024}. 
$\beta$ equals to zero has been demonstrated with good theoretical fit with the experimental results \cite{sonar2024}. 

A second method to mitigate the waveguide heating is to include a second optical mode that aligns with the pump frequency, as shown in Fig. \ref{Ch6_mode}.
In this way, the same intra-cavity photon number can be achieved with much smaller input optical power $P_{\text{in}}$, which largely suppresses the heating contribution from the waveguide. 
This approach is initially proposed and demonstrated in electro-optic converters \cite{javerzac2016, soltani2017, mckenna2020, holzgrafe2020, xu2021, fu2021}. 
Recently, it has been implemented in HBAR \cite{blesin2023} and OMC \cite{primo2023} schemes, where a second optical cavity is evanescently coupled to the cavity with optomechanical interaction.
The two cavities form hybrid symmetric and asymmetric modes with spacing equal to the mechanical frequency.
Depending on the application, either the lower or higher frequency mode can be pumped to drive the quantum transduction.

\begin{table}[t]
    %\centering
    \caption{
    Comparison of piezo-optomechanical microwave to optical converters based on different mechanical modes. 
    The last two columns show the comparison with other schemes using electro-optic (EO) and electromechanical (EM) coupling. For EO transduction, $C_{\text{om}}$ represents cooperativity between optical and microwave resonators. 
    }
    
    \begin{tabular}{ p{1.4cm}|p{0.9cm}|p{1.2cm}|p{1.5cm}|p{1.2cm}|p{1.2cm}|p{1.2cm}|p{1.2cm}|p{1.2cm}|p{1.2cm}}
 \hline
 Scheme  &  SAW & BAW & \multicolumn{5}{c|}{OMC} & EO & EM \\
 \hline
Ref.   & \cite{Shao:19} & \cite{blesin2023} & \cite{jiang2020efficient} & \cite{stockill2022} & \cite{meesala2023} & \cite{jiang2023}  & \cite{weaver2023}  & \cite{xu2021} & \cite{Zhao:23} \\
Year   & 2019 & 2023 & 2020 & 2022 & 2023 & 2023 & 2023  & 2021 & 2023 \\
Piezo Material  & LiNbO$_3$ & AlN & LiNbO$_3$ & GaP & AlN & LiNbO$_3$ & LiNbO$_3$ & - & - \\
Optical Material & LiNbO$_3$ & Si$_3$N$_4$ & LiNbO$_3$ & GaP & Si  & Si & Si & LiNbO$_3$ & Si \\
$\Omega_m/2\pi$ (GHz)  & 2 & 3.48 & 1.85 & 2.8 & 5  & 3.596 & 5.043 & 7.836 & 4.93 \\
$\gamma_m/2\pi$ (MHz)   & 1.28 & 13 & 1.9 & 0.067 & 0.15  & 1.1 & 2.63 & - & 3.3 \\
$\gamma_{m,\text{ex}}/\gamma_m$   & 0.17 & 0.11 & 0.001 & $3.6\times 10^{-6}$ & -  & - & - & - & $2.58\times 10^{-7}$ \\
$\kappa_o/2\pi$ (GHz)  & 0.095 & 0.17 & 1.2 & 4.17 & 1.3  & 1.12 & 4.99 & 0.28 & 1.39 \\
$g_0/2\pi$ (kHz)  & 1.1 & 0.042  & 80 & 700 & 270  & 410 & 530 & 0.75 & 577 \\
$\Gamma/2\pi$ (MHz)  & - & - & - & - & 1.75  & 3.06 & 3.44 & 9.06 & - \\
$\Gamma_{\text{ex}}/\Gamma$   & - & - & - & - & 0.69  & 0.993 & 0.64 & 0.36 & - \\
$g_{\text{em}}/2\pi$ (kHz)  & - & -  & - & $\sim$68 & 800  & 420 & 7400 & - & - \\
Pump mode   & CW & CW \& Pulse  & CW & CW \& Pulse & Pulse  & CW \& Pulse & CW \& Pulse & CW \& Pulse & CW \\
Optical power   & 1 mW & 126 mW  & 3.3 $\mu$W & 0.5 $\mu$W & 6.2  & 10.2 $\mu$W & 1 $\mu$W & 20 mW & - \\
$C_{\text{om}}$   & $1.7\times 10^{-4}$ & $5\times 10^{-4}$  & $6.5\times 10^{-3}$ & 1.04 & $1.2\times 10^{-3}$  & 0.3 & 0.13 & 0.041 & 0.268 \\
$C_{\text{em}}$   & - & -  & - & - & 9.75  & 0.2 & 24.2 & - & - \\
$\eta_{\text{int}}$   & $6.7\times 10^{-4}$ & $2\times 10^{-3}$  & $2.6\times 10^{-2}$ & 0.032 & -  & 0.11 & 0.019 & 0.152 & 0.67 \\
$\eta_{\text{oc}}$   & $1.7\times 10^{-5}$ & $7.9\times 10^{-5}$  & $1.1\times 10^{-5}$ & $6.8\times 10^{-8}$ & -  & 0.049 & 0.009 & 0.01 & $1.72\times 10^{-7}$ \\
\hline
\end{tabular}
    \label{table_converter}
\end{table}

The comparison of microwave to optical converters realized in different schemes is summarized in Table \ref{table_converter}. 
It can be seen that the OMC has an optomechanical coupling strength $g_0$ orders of magnitude larger than SAW and BAW due to the tight mode confinement, which leads to large cooperativity $C_{\text{om}}$ under modest optical power.
However, the sub-micron mechanical mode makes it hard to be driven directly through a large microwave feed line, resulting in a small $\gamma_{m,\text{ex}}$.
The development of high-impedance microwave resonators in recent works \cite{meesala2023, jiang2023, weaver2023} helps boost the drive efficiency. 
On the other hand, the advantages of SAW and BAW include lower optical losses and better optical power handling capability. 
Also, the electromechanical extraction efficiency and phonon injection efficiency for SAW and BAW are much higher.
Even without additional microwave resonators, critically coupled HBAR has been demonstrated by engineering the acoustic impedance \cite{gokhale2020c}. 
Nevertheless, OMC is the most promising platform so far that would meet most of the requirements for quantum transduction.
The highest efficiency of 5$\%$ has been achieved with LiNbO$_3$ on Si OMC with only 10 $\mu$W optical pump power \cite{jiang2023}. 
The added noise referred to the input is measured to be $99\pm 10$ for optical-to-microwave conversion and 181 for microwave-to-optical conversion \cite{jiang2023}. 

\begin{figure}[t]
\centering
\includegraphics[width=12 cm]{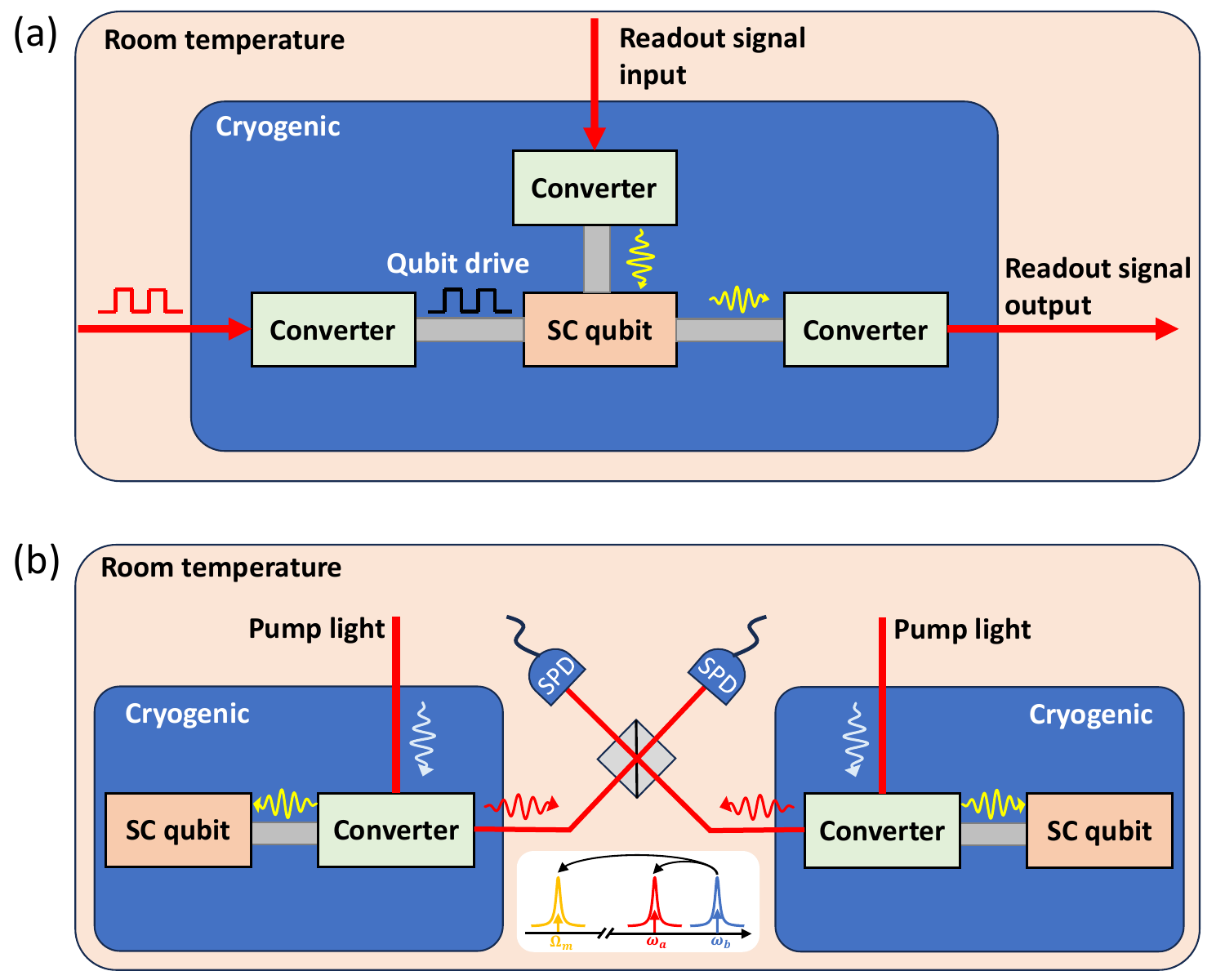}
\caption{
Potential applications of microwave-to-optical converter in near future.
(a) Schematic of photonic links, where the converters are used to coherently drive and readout SC qubits, with the potential of replacing coaxial cables.
(b) Schematic of the optically heralded remote entanglement generation between distant SC qubits.
The inset shows the generation of microwave-optical photon pair by a high frequency pump light.
SPD: single photon detector.  
}
\label{Ch6_scheme}
\end{figure}

Although the quantum transduction efficiency of current converters is still far below the requirement for building the quantum internet \cite{kimble2008, wolf2007}, the superconducting quantum processor can already benefit from the development of this technology in the near furture.
One potential application is the classical optical control and readout of SC qubits.
Conventionally, each qubit is connected via a coaxial cable to the control electronics at room temperature.
However, this poses challenges on the scaling of the number of qubits which is limited to a thousand qubits based on current technology \cite{lecocq2021}.
Apart from the space constraint, the heat load from coaxial cables would ultimately overwhelm the cooling power of the dilution refrigerator \cite{krinner2019}.
Optical fibers made from silica, on the other hand, show negligible thermal conductivity at cryogenic temperatures. 
Also, the ultra-wide optical bandwidth allows signal multiplexing in a single fiber.
In that sense, replacing the coaxial cable with the photonic link could become an attracting solution to large-scale quantum computer in the future.
As shown in Fig. \ref{Ch6_scheme}(a), the converter down-converts the input light into microwave signals to drive and readout the qubit, and the readout signal is then up-converted into light for qubit measurement.
It has recently been demonstrated that the optically generated microwave via photodiodes can faithfully serve as both drive and readout signal for an SC qubit \cite{lecocq2021}.
Another work realized the optical readout of SC devices using off-the-shelf electro-optic modulators \cite{youssefi2021}.
Based on these proofs of the concept, the conversion efficiency and added noise of the photonic link can be further improved using the-state-of-the-art microwave to optical converters.
While most recent works on this topic have been realized on electro-optical converters \cite{shen2024, arnold2023all, warner2023}, the optical multi-shot readout of SC qubit has been demonstrated on piezo-optomechanical transducer with high fidelity \cite{van2023high}. 
In the future, the investigation of heat load from these converters and the demonstration of optical multiplexing could enhance the scalability of this technology as a photonic link for SC circuits. 

Aside from the classical control and readout of qubits, it has been proposed that current converter technology can be readily applied for the optically heralded remote entanglement between distant SC systems \cite{krastanov2021} and the quantum teleportation based transduction between microwave and optical photons \cite{wu2021}, which has less strict requirement on the conversion efficiency.
Following the well-known DLCZ protocol \cite{duan2001long}, the diagram of the proposed architecture is illustrated in Fig. \ref{Ch6_scheme}(b).
In stead of red detuned optical pump, the light is pumped at the blue side, which generates an entangled microwave-optical photon pair through an optomechanical parametric down conversion process \cite{zhong2020proposal, zhong2020entanglement}.
Interestingly, the transducer preferably works in a low optomechanical cooperativity regime such that the probability of multi-photon events is kept low \cite{warner2023}. 
In this way, by passing the optical photons generated from separate SC systems through a beamsplitter to erase the path information, the detection of single optical photons in both outputs of the beam splitter heralds the entanglement between microwave photons located in each SC circuits.
The entangled microwave states can then be transferred to corresponding SC qubits in each node to build the quantum network.
Different entanglement schemes have been proposed and the entanglement fidelity has been theoretically analyzed based on practical experimental parameters, such as time-bin \cite{zhong2020proposal} and frequency-bin \cite{zhong2020entanglement} encoded entanglement. 
To have a high entanglement fidelity, the system prefers large electromechanical cooperativity $C_{\text{em}}$ and microwave extraction efficiency to suppress the microwave photon loss during the transduction \cite{zhong2020proposal, zhong2020entanglement, wu2021}. 
Large $C_{\text{em}}$ is relatively easier to reach than $C_{\text{om}}$ since the piezo-electromechanical coupling rate $g_{\text{em}}$ is generally on the similar order of magnitude with the microwave and mechanical resonators' linewidths (see Table \ref{table_converter}). 
On the other hand, entanglement purification protocol and quantum error correction codes can be utilized to improve the entanglement fidelity at the presence of microwave loss \cite{krastanov2021, bennett1996mixed}. 
Ref. \cite{zhong2020proposal} proposes to improve the fidelity by postselecting the quantum states via a joint parity measurement of the SC qubits in a time-bin encoded entanglement. 
Preliminary experiments have demonstrated the generation of correlated microwave-optical photon pairs and optically heralded non-classical microwave photons in piezo-optomechanical transducers \cite{jiang2023, meesala2023}.
Further improvement of the converter's performance and reduction of the losses in the system could increase the entanglement generation rate and fidelity, paving the way to the quantum network in a foreseeable future.

\subsection{Magnetic-Free Optical Isolator}

Integrated photonics has witnessed major progresses in recent decades, where most of the important building blocks have been successfully integrated on chip, from semiconductor lasers \cite{Xiang:21, Jin:21}, modulators \cite{WangC:18}, low-loss waveguides \cite{Liu:20b}, to photodetectors \cite{Kang:09}.
Despite of these achievements, integration of optical non-reciprocal devices, such as optical isolators and circulators, remains as a challenging task, although they are indispensable components for compact signal routing and protecting lasers from reflection. 
Traditionally, these devices are realized using magneto-optical materials based on the Faraday effect, where non-reciprocal polarization rotation is induced by external magnetic field. 
Recently, achievements have been made on the integration of cerium-substituted yttrium iron garnet (Ce:YIG) on Si photonic circuits via wafer bonding \cite{Huang:16} and on Si$_3$N$_4$ by pulsed laser deposition \cite{Yan:20}. 
However, the optical loss remains high and a bulky external magnet is generally required to generate the magnetic field. 

Magnetic-free optical non-reciprocity has gained much research efforts, and many different schemes have been proposed and demonstrated, including optical nonlinearity \cite{DelBino:18, YangKY:20, white2023}, optomechanically induced transparency (OMIT) \cite{Ruesink:16, Shen:16}, stimulated Brillouin scattering (SBS) \cite{Kim:17, Dong:15, Kang:11, Kim:15}, and dynamic modulation \cite{williamson2020, Sounas:17, Yu:09, yu2009optical}. 
They all show pros and cons in practical applications.
For example, while the optical nonlinearity can work passively and all-optically, the isolation depends on the input optical power and is subject to the dynamic reciprocity \cite{Shi:15}. 
The OMIT and SBS are mostly realized in a stand-alone sphere or microtoroid that is difficult to integrate with PICs.
Moreover, the isolation bandwidth is limited by the mechanical linewidth which is generally between kilohertz to megahertz range (see Table \ref{table_isolator}). 
In contrast, the dynamic modulation scheme benefits from currently developed electro-optic and acousto-optic modulators which can be readily incorporated into existing photonic systems. 

The dynamic modulation breaks the Lorentz reciprocity and time-reversal symmetry through the spatio-temporal modulation of optical waveguides which induces non-reciprocal interband or intraband photonic transitions.
The intraband transition-based optical isolator has been demonstrated in an electro-optic phase modulator by exciting traveling microwave that satisfies the phase matching condition only in one direction \cite{yu2023integrated}.
The intraband transition is hard to implement on AOM since it is nontrivial to match the group velocity of the optical mode with the acoustic wave.
On the contrary, the interband transition between different optical modes has been widely studied \cite{Yu:09, yu2009optical, Shi:18} and is well suited for acoustic waves.
Depending on the specific arrangement of the optical and acoustic modes and the structure design, different schemes of optical isolator based on dynamic modulation have been proposed, which are summarized in Ref. \cite{williamson2020} in more detail.
Here, we review two main schemes that have been successfully demonstrated in experiments. 

\begin{figure}[t]
\centering
\includegraphics[width=13.5 cm]{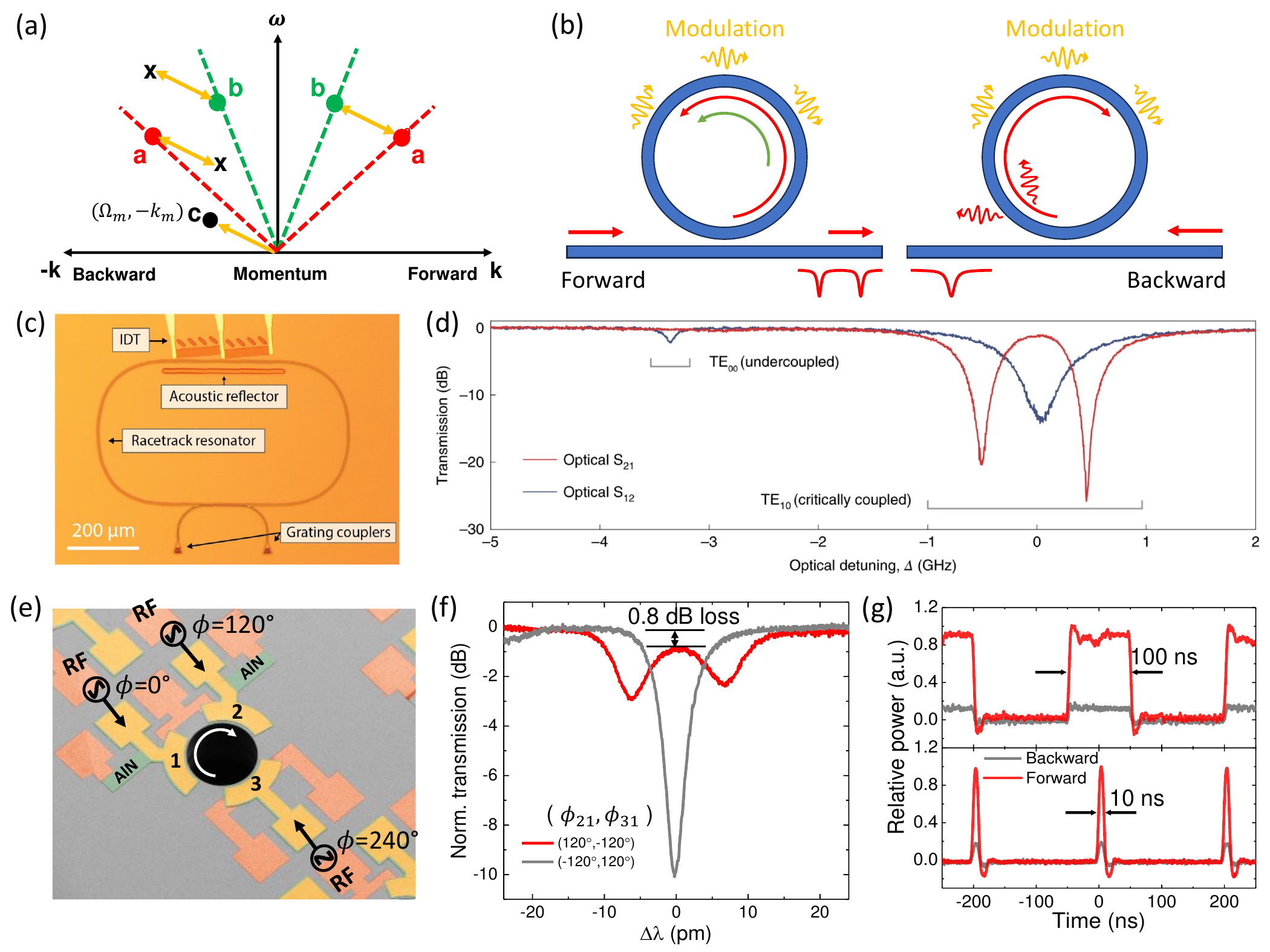}
\caption{
Optical isolator based on a spatio-temporal modulated optical resonator. 
(a) Frequency-momentum diagram illustrating the indirect interband transition between two optical modes $a$ and $b$ induced by an acoustic wave $c$. 
(b) Schematic of the optical transmission in forward and backward directions.
In the forward direction, the strong coupling between the two optical modes induces Rabi mode splitting, whereas in the backward direction the transmission remains as original single resonance.
(c) Optical isolator realized in a LiNbO$_3$ optical resonator modulated by an unreleased SAW.
(d) Optical transmission spectrum in the forward (red) and backward (blue) directions.
(c-d) Reprinted from Ref. \cite{sohn2021}.
(e) False-colored SEM of an optical isolator realized in a Si$_3$N$_4$ microring resonator modulated by three HBAR driven at rotational phases.
(f) Optical transmission when the actuators are reconfigured into phase matching (red) and phase mismatch (gray) conditions.
(g) Transmission of optical pulses in the forward (red) and backward (gray) directions for pulse width of 100 ns (upper) and 10 ns (lower).
(e-g) Reprinted from Ref. \cite{Tian:2021b}.
}
\label{Ch6_resonator}
\end{figure}

The first scheme is based on an optical resonator, and the working principle is illustrated in Fig. \ref{Ch6_resonator}(a-b). 
The optical resonator supports a pair of two closely spaced optical modes as shown in the frequency-momentum ($\omega-k$) space diagram in Fig. \ref{Ch6_resonator}(a), where the sign of the momentum indicates the propagation direction in the optical waveguide. 
Under the modulation of an acoustic wave that carries a non-zero momentum, the light in one mode can be scattered to the other if the energy and momentum conservation conditions are satisfied, where the frequency and momentum of the acoustic wave equal to the corresponding difference between the two optical modes.
This is also known as the phase matching condition.
In that sense, the momentum of the acoustic wave prescribes the direction to which the light can be scattered.
From Fig. \ref{Ch6_resonator}(a), it can be seen the phase matching can only be fulfilled in the forward direction, where the two optical modes can be coupled strongly.
The same equations of motion developed for the microwave-to-optical converter can be used to describe this three wave mixing process.
Instead of an optical pump, the acoustic drive serves as the pump to induce the optical coupling.
In the strong coupling regime where the coupling rate is much larger than the loss rate, the optical modes start showing Rabi-splitting \cite{Shi:18}, and a transparency window appears at the original optical resonance position.
In the backward direction, the two optical modes are not coupled due to phase mismatch and show a single resonance.
If the optical resonator is critically coupled to the waveguide, most of the light is dissipated into the environment without transmission. 
With this non-reciprocal transmission, an optical isolator can be built. 

Initial experimental attempts have been made on a released AlN optical resonator modulated by SAW, where a fundamental TE$_{00}$ and a second order TE$_{10}$ are coupled \cite{Sohn:18, sohn2019}.
However, limited by the modulation efficiency and power handling capability, only non-reciprocal modulation sideband generation is demonstrated without entering into the strong coupling regime. 
Recently, the same group achieved strong coupling using an unreleased LiNbO$_3$ SAW which has a larger piezoelectric response and better thermal conductance (see Fig. \ref{Ch6_resonator}(c)) \cite{sohn2021}. 
The finger of the IDT has a non-zero angle with the optical waveguide in order to generate momentum parallel to the waveguide.
Non-reciprocal transmission of a TE$_{10}$ mode at 1556 nm has been realized, as shown in Fig. \ref{Ch6_resonator}(d). 
One of the most important metrics for an optical isolator is the isolation ratio defined as the transmission ratio between forward and backward directions.
A maximum of 12.75 dB isolation ratio is achieved at optical resonance with insertion loss of 1.13 dB \cite{sohn2021}.
The isolation ratio is primarily determined by how critically the optical resonator is coupled to the bus waveguide, which is sensitive to the fabrication disorder.
We should also note that, in the forward direction, there will be non-zero modulation sideband generated in the other optical mode with a frequency separated by the acoustic frequency.
This sideband can be filtered by either a frequency filter or a mode multiplexer. 
Not only for telecommunication band, the same design is also suitable for visible light and similar result has been demonstrated \cite{sohn2021}.

Apart from the SAW AOM, HBAR has also been utilized in demonstrating an optical isolator on the Si$_3$N$_4$ photonic circuit \cite{tian2023}, as shown in Fig. \ref{Ch6_resonator}(e).
Differently from the SAW, HBAR excites the out-of-plane acoustic waves, which carry momentum perpendicular to the optical waveguide. 
To generate the in-plane momentum, it is proposed to include three discrete HBAR modulators that are driven by relative phases rotating in a circle.
In this way, an effective rotational wave can be synthesized along the optical microring resonator.
The resonator supports a pair of TE$_{00}$ and TM$_{00}$ modes with orthogonal polarization, which are then coupled through the modulation. 
Despite of the entirely different material systems and the choice of optical and mechanical modes, similar optical isolation performance is achieved with 10 dB isolation and 0.8 dB insertion loss, which demonstrates the universality of the dynamic modulation approach. 
Furthermore, the actuator phases can be reconfigured to switch the rotation direction such that the isolation direction can be controlled in situ, as illustrated in Fig. \ref{Ch6_resonator}(f). 
The modulation sideband in the forward direction can be easily separated by a polarizing beamsplitter.
The isolation of optical pulses is also demonstrated in Fig. \ref{Ch6_resonator}(g).
Despite of the compact size, the isolation bandwidth is limited by the optical linewidth to hundreds of megahertz to gigahertz. 
Although bandwidth can be increased by adding extra optical losses, the required mode splitting also increases, which leads to larger RF power.
Nevertheless, the narrow bandwidth is sufficient in protecting laser systems and suitable for rejecting back-reflection noise for microwave-to-optical quantum transducers.

\begin{figure}[t]
\centering
\includegraphics[width=13.5 cm]{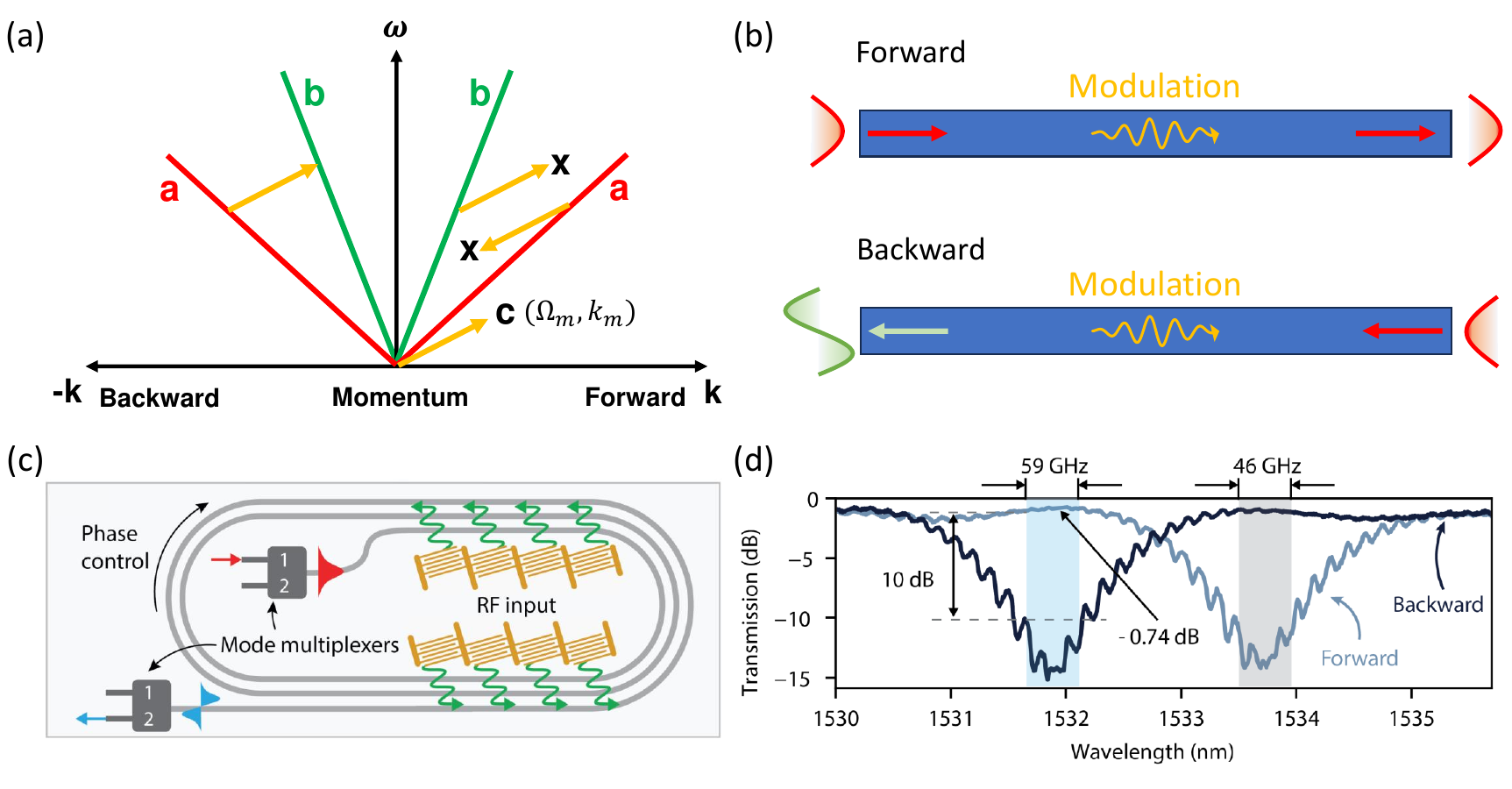}
\caption{
Optical isolator based on a spatio-temporal modulated optical waveguide.
(a) Frequency-momentum diagram showing the phase matching condition between two optical waveguide modes ($a$ and $b$) and the acoustic wave $c$. 
(b) Schematics illustrating the mode transmission in forward and backward directions.
In the forward direction, the light transmits without mode conversion, while it completely converts to another mode due to phase matched acoustic modulation.
(c) Schematic of the experimental realization of a waveguide-based optical isolator through SAW excited by a series of cascaded IDTs. 
The mode multiplexers at the input and output ports combine and separate the two waveguide modes.
(d) Optical transmission in the forward and backward directions showing a wide isolation bandwidth (shaded area). 
(c-d) Reprinted from Ref. \cite{zhou2022intermodal}.
}
\label{Ch6_waveguide}
\end{figure}

The broadband optical isolator can be realized in a dynamically modulated optical waveguide. 
Similar to the case of the resonator, indirect interband transition can be achieved between two optical waveguide modes (e.g. TE$_{00}$ and TE$_{10}$) under a proper spatio-temporal modulation \cite{Yu:09, williamson2020}. 
As can be seen from Fig. \ref{Ch6_waveguide}(a), the phase matching condition can only be satisfied in the backward direction, such that under sufficient modulation strength, one mode can be completely converted into the other mode (see Fig. \ref{Ch6_waveguide}(b)). 
By separating and rejecting the converted mode, the backward transmission can be blocked. 
On the contrary, in the forward direction the original mode remains intact and gets fully transmitted.
The proof of concept is initially demonstrated on an electrically modulated doped Si waveguide \cite{Lira:12}. 
However, the insertion loss is as high as 70 dB.
Modulation through SAW has been theoretically proposed \cite{dostart2020} and experimentally studied on an unreleased SI waveguide modulated by AlN SAW \cite{Kittlaus:21}. 
Due to the inefficient AOM, only non-reciprocal modulation sideband is demonstrated.
Recently, the same group improved the modulation efficiency by releasing the Si waveguide and cascading more IDTs as shown in Fig. \ref{Ch6_waveguide}(c) \cite{zhou2022intermodal}.
Full mode conversion was achieved under 22 dBm RF power, which led to an optical isolation of 14 dB and 0.74 dB insertion loss, as shown in Fig. \ref{Ch6_waveguide}(d).
The isolation is mainly limited by the unwanted interband crosstalk and evanescent coupling between waveguides. 
It is interesting to notice that there is a second band that shows isolation in the opposite direction, where the phase is matched in the forward direction. 
The maximum bandwidth for isolation bigger than 10 dB is 59 GHz which is suitable for isolating high speed signals.
The bandwidth is primarily limited by the difference of group velocity of the two optical modes, which can be further improved through the dispersion engineering of waveguide bands. 

Borrowing the concept from the solid state physics, the aforementioned two schemes are both based on indirect interband transition where the momentum is involved in the scattering process. 
Another scheme exploring the direct interband transition has also been proposed where the phase of modulation induces an effective gauge potential for photons \cite{fang2012photonic, Fang:12}. 
In analogy to the Aharonov-Bohm effect for electrons \cite{aharonov1959}, the effective gauge potential can be used to construct an effective magnetic field for photons and break the reciprocity by creating a direction dependent phase \cite{Tzuang:14}.
Although the concept has been demonstrated using an electrically modulated doped Si waveguide, the isolation is small and the insertion loss is unacceptable \cite{Tzuang:14}. 
While the implementation using AOM has been realized based on off-the-shelf bulk AOM devices \cite{li2014photonic}, it will be of great interest to integrate the system on chip using SAW or BAW AOM in the future for a new type of magnetic-free chip-scale optical isolator with broad bandwidth. 

\begin{table}[t]
\centering 
 \caption{
 Comparison of optical non-reciprocal devices realized using different schemes: MO (magneto-optic), NL (nonlinear optics), OM (optomechanical), SBS (stimulated Brillouin scattering), AB (Aharonov-Bohm effect), SM (synthetic magnetic field), and ST (spatio-temporal modulation).
WG: waveguide. 
IL: Insertion loss.  
$a$: Non-reciprocal sideband modulation. 
$b$: Optical pump power. 
}
\begin{tabular}{p{1.1cm}|p{1.1cm}|p{1.1cm} |p{1.2cm}|p{1.87cm}|p{1.3cm}|p{1.1cm}|p{1.5cm}|p{1.2cm}|p{1cm}  }
\hline
Ref. & Year & Scheme & Structure & Material & Isolation & IL (dB) & Bandwidth & Power & CMOS\\
\hline
\cite{Huang:16}&2016 & MO & Ring & Si+Ce:YIG & 32 dB & 2.3 & 15 GHz& 10 mW & No\\
\cite{Yan:20}&2020 & MO & Ring & Si$_3$N$_4$ +Ce:YIG & 28 dB & 1 & 15 GHz&  & No\\
\cite{DelBino:18}&2018 & NL & Toroid & SiO$_2$ & 24 dB & 7 & 1 MHz& No drive & No\\
\cite{YangKY:20}&2020 & NL & Ring & Si & 20 dB & 1.3 & 20 GHz& No drive & Yes\\
\cite{Ruesink:16}&2016 & OM & Toroid &  SiO$_2$ & 10 dB & 14  & 250 kHz & 17 $\upmu$W$^b$ & No\\
\cite{Ruesink:18}&2018 & OM & Toroid &  SiO$_2$ & 10 dB & 3  & 60 kHz & 60 $\upmu$W$^b$ & No\\
\cite{Shen:18}&2018 & OM & Sphere &  SiO$_2$ & 17 dB & 2 & 200 kHz & 7.8 mW$^b$ & No\\
\cite{Kim:17}&2017 & SBS & Sphere &  SiO$_2$ & 11 dB & 0.14 & 400 kHz & 235 $\upmu$W$^b$ & No\\
\cite{Tzuang:14}&2014 & AB & WG &  Doped Si & 2.4 dB & & 20 nm& 34 dBm & Yes\\
\cite{Kim:21} &2021& SM & Ring &  AlN & 3 dB & 9 & 4 GHz& 16 dBm & Yes\\
\cite{Lira:12}&2012 & ST & WG &  Doped Si & 3 dB & 70 & 200 GHz& 25 dBm & Yes\\
\cite{Kittlaus:18}&2018 & ST$^a$ & WG &   Si & 39 dB & NA & 125 GHz& 90 mW$^b$ & Yes\\
\cite{Sohn:18}&2018 & ST$^a$ & Ring &  AlN & 15 dB & NA& 1 GHz& 18 dBm & Yes\\
\cite{Kittlaus:21}&2021 & ST$^a$ & WG &  AlN+Si & 16 dB & NA& 100 GHz& 21 dBm & Yes\\

\cite{Tian:2021b}&2021 & ST & Ring &  AlN+Si$_3$N$_4$ & 10 dB & 0.8 & 0.7 GHz& 25 dBm & Yes\\
\cite{sohn2021}&2021 & ST & Ring &  LiNbO$_3$ & 12.75 dB & 1.13 & 0.199 GHz& 29 dBm & Yes\\
\cite{zhou2022intermodal}&2022 & ST & WG &  AlN+Si & 14.37 dB & 0.74 & 59 GHz& 22 dBm & Yes\\
\cite{yu2023integrated}&2023 & ST & WG &  LiNbO$_3$ & 48 dB & 0.5 & 13 GHz& 21 dBm & Yes\\
\hline
\end{tabular}
\label{table_isolator}
\end{table}

Table \ref{table_isolator} summarizes and compares the performances of optical isolator realized on different platforms.
It can be seen that the scheme of spatio-temporal modulation has become one of the most promising methods in terms of compatibility with current photonic circuits, and the development of integrated AOM technology facilitates the recent demonstration of optical isolator on Si$_3$N$_4$ \cite{tian2023}, Si \cite{zhou2022intermodal}, and LiNbO$_3$ \cite{sohn2021} photonics. 
While the ring-based structure is more compact, the waveguide structure supports wider bandwidth, which can be chosen according to specific application.
Thus far, the widest bandwidth is achieved on the SAW modulated Si waveguides \cite{zhou2022intermodal}, and the highest isolation is demonstrated on the LiNbO$_3$ electro-optic phase modulator \cite{yu2023integrated}. 
The required RF power is on the order of 100 mW. 
These performances can be further improved by engineering the dispersion of optical waveguides, studying different pair of modes (e.g. TE$_{00}$ and TM$_{00}$ \cite{wang2020non}), and exploiting new piezoelectric material. 

\begin{figure}[b]
\centering
\includegraphics[width=13.5 cm]{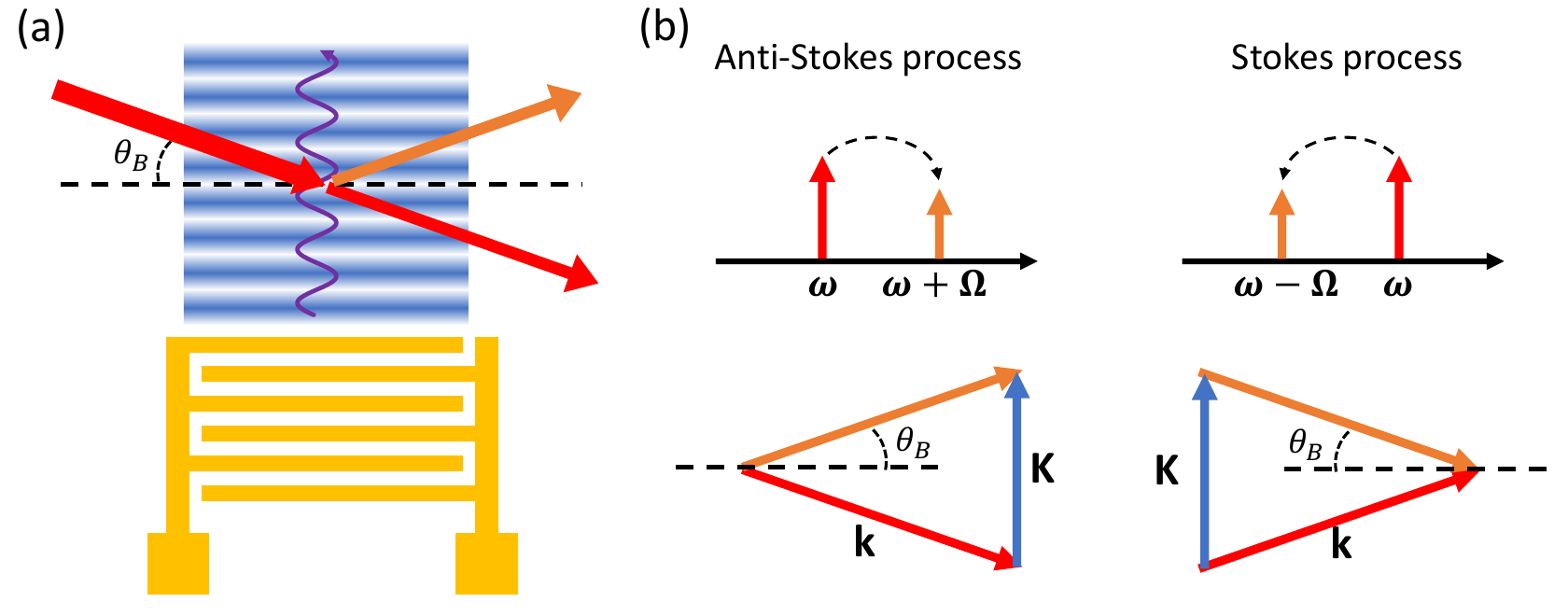}
\caption{
Working principle of traditional AOM based on Bragg diffraction. 
(a) Schematic of the AOM where the incident light is diffracted by the traveling SAW excited by IDT.
The Bragg angle is labeled. 
(b) Phase matching condition for anti-Stokes process (left) and Stokes process (right), showing the frequency shifting (top) and momentum conservation (bottom). 
}
\label{Ch6_Bragg}
\end{figure}

\subsection{Integrated Acousto-Optic Frequency Shifter}

Traditional AOM has long been an indispensable component in laser systems and applications. 
It provides a convenient way to control the intensity, frequency, and direction of the light.
The working principle of traditional AOM is illustrated in Fig. \ref{Ch6_Bragg}(a), where an incident light undergoes a Bragg diffraction after interaction with a traveling (or resonant) acoustic wave excited by an IDT.
For most widely used AOM, it works at the Bragg regime, where the light is mostly scattered into the first diffraction order. 
The frequency of the deflected light is shifted by the acoustic frequency.
For traveling acoustic wave, it naturally carries momentum, such that, to satisfy the phase matching condition, the light is only scattered into a single sideband.
As shown in Fig. \ref{Ch6_Bragg}(b), the light can be scattered to the Stokes (lower frequency) or anti-Stokes (higher frequency) sideband, depending on its incident direction relative to the acoustic wave. 
The diffraction efficiency is maximized when the light is incident at the Bragg angle that is dictated by the phase matching condition as \cite{shao:2020}:
\begin{equation}
    \text{sin}(\theta_B)=\frac{K}{2k}=\frac{\Omega v_{\text{opt}}}{2\omega v_{\text{ac}}}
\end{equation}
where $K$ ($k$), $\Omega$ ($\omega$), and $v_{\text{ac}}$ ($v_{\text{opt}}$) are the momentum, angular frequency, and velocity of acoustic (optical) wave, respectively. 
The intensity of the diffracted light can be modulated by controlling the diffraction efficiency which is related with the power and frequency of the RF drive \cite{yu2021}. 
This capability is conventionally utilized in the Q-switch laser \cite{barmenkov2014} and in the generation of optical pulses with ultra-high on-off contrast \cite{ramp2019}. 
The frequency shifting of the acoustically scattered light naturally leads to the desirable single sideband suppressed carrier (SSB-SC) modulation, which is complicated to implement using electro-optical modulators \cite{loayssa2001}.
This unique feature of AOM has been translated into an indispensable optical device, acousto-optic frequency shifter (AOFS), which is widely used in the heterodyne detection \cite{Sohn:18, patel2021}.

\begin{figure}[b]
\centering
\includegraphics[width=13.5 cm]{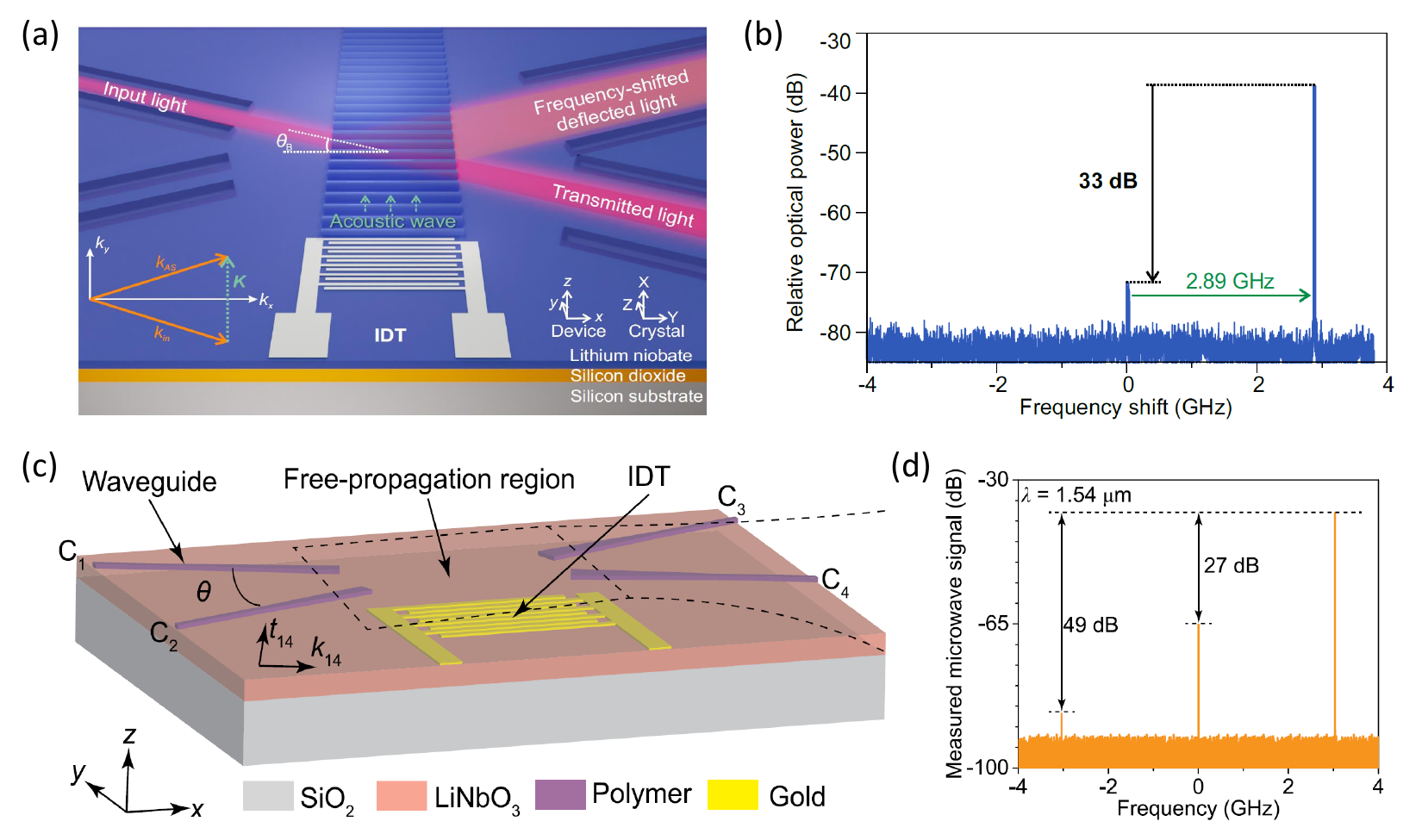}
\caption{
Experimental demonstrations of integrated AOFS based on Bragg diffraction. 
(a, c) Schematic of AOFS using SAW excited on thin film LiNbO$_3$.
The input and output light is confined in etched LiNbO$_3$ rib waveguides for (a) and in BIC mode formed by polymer for (c).
(b, d) Optical spectrum of the deflected light for device in (a) and (c), respectively. 
(a-b) Reprinted from Ref. \cite{shao:2020}. (c-d) Reprinted from Ref. \cite{yu2021}.
}
\label{Ch6_AOFS}
\end{figure}

Despite of the ubiquitous application of AOM, current commercially available AOMs are fabricated on a stand-alone bulk piezoelectric substrate and the acoustic frequency is limited to a few hundreds of megahertz. 
The required RF drive power is at Watts scale due to the large acoustic and optical mode volume. 
The implementation of Bragg diffraction on integrated planar photonics has been demonstrated on a released AlN membrane for the first time \cite{li2019}.
Despite the low scattering efficiency, asymmetric modulation with 11 GHz frequency shift is achieved with only 100 mW RF power.
Recently, the efficiency is improved by exciting SAW on an unreleased LiNbO$_3$ thin film, which has higher piezoelectric response and power handling capability \cite{shao:2020, yu2021}, as shown in Fig. \ref{Ch6_AOFS}.
The input and output light is guided through either an etched LiNbO$_3$ rib waveguide \cite{shao:2020} or a BIC waveguide mode formed by patterned polymer on LiNbO$_3$ film \cite{yu2021}.
The light from the waveguide transits into a two-dimensional Gaussian beam confined in the slab formed by the LiNbO$_3$ thin film, which interacts with the SAW in the same region.
An ideal AOFS would require most of the input light to be shifted to the desired frequency with zero power remaining in the original frequency or scattered to other sidebands. 
This behavior can be characterized by the modulation efficiency (the fraction of input light that is scattered to the target sideband), and carrier and sideband suppression ratios. 
The optical spectrum of the shifted light is shown in Fig. \ref{Ch6_AOFS}(b, d). 
It can be found that most optical power is in the anti-Stokes sideband with around 3 GHz higher frequency, and the center carrier and the Stokes sideband is 30 dB and $>$40 dB smaller, respectively. 
The efficiency is measured to be around 3$\%$ per 1 W RF drive.
Frequency shifting to the Stokes sideband can be realized by changing the input port \cite{shao:2020, yu2021}. 
The optical bandwidth is proportional to $\lambda \Lambda/W$, where $\lambda$, $\Lambda$, and $W$ are the optical wavelength, period of IDT fingers, and width of the IDT, respectively \cite{shao:2020}. 
Optical bandwidths of 14 nm and 35 nm are achieved for the works in Ref. \cite{shao:2020} and \cite{yu2021}, respectively. 
These integrated AOFS have compact size, low power consumption, and gigahertz operation, which can be used for on-chip heterodyne detection, microwave signal processing, and optical signal routing. 

\begin{figure}[t]
\centering
\includegraphics[width=13.5 cm]{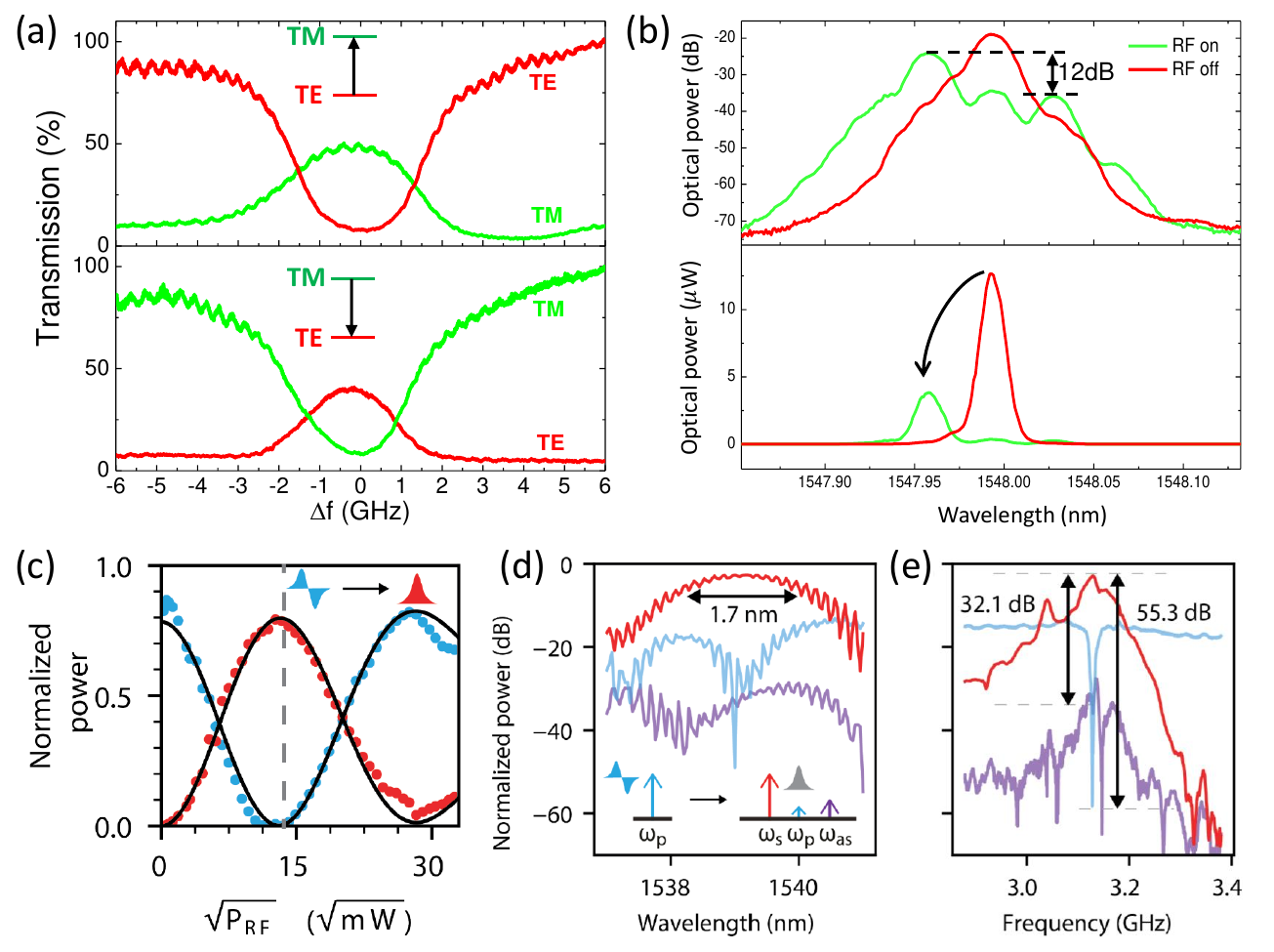}
\caption{
Experimental realizations of integrated AOFS through interband transition. 
(a, b) AOFS realized using optical microring resonator, the same device presented in Fig. \ref{Ch6_resonator}(e). 
(a) Optical transmission of TE$_{00}$ and TM$_{00}$ modes when TE$_{00}$ (TM$_{00}$) is converted to TM$_{00}$ (TE$_{00}$) in upper (lower) panel. 
The x-axis is the laser frequency relative to the corresponding optical resonance.
(b) Optical spectrum of output light in TM$_{00}$ mode when RF drive is turned on (green) and light in TE$_{00}$ mode when RF is off (red).
The upper panel is in dB scale while the lower is linear. 
(c-e) AOFS using integrated Si waveguide, the same device presented in Fig. \ref{Ch6_waveguide}(c). 
(c) Energy exchange between symmetric (red) and anti-symmetric (blue) modes as a function of RF power.
(d) Optical spectrum of output symmetric mode when input anti-symmetric mode into the waveguide. 
The light is mainly converted into Stokes sideband (red), while some power remains in carrier (blue) and is scattered to anti-Stokes sideband (purple). 
(e) Optical power in different frequency components when sweeping the RF drive frequency around the acoustic resonance. 
(c-e) Reprinted from Ref. \cite{zhou2022intermodal}.
}
\label{Ch6_AOFS2}
\end{figure}

To further improve the modulation efficiency, the acoustic and optical interaction length should be increased. 
However, this is hard to implement in the scheme presented in Fig. \ref{Ch6_AOFS}.
This is because longer interaction length would require wider acoustic width and IDT aperture, which leads to broad deflected light beam with lower collection efficiency by integrated waveguide.
Also, wider IDT inevitably compromises the optical bandwidth as mentioned before. 
This issue can be circumvented by confining light in optical waveguides which also helps with higher interaction strength due to better mode confinement \cite{wiederhecker2019}. 
AOFS based on this scheme has been demonstrated in both microring and straight waveguide structures which all enhance the interaction length \cite{Tian:2021b, zhou2022intermodal}.
It is interesting to find that the AOFS and the isolator discussed before share a similar working principle, where one optical mode is completely converted into another mode through an indirect interband transition \cite{yu2009optical, Yu:09}.
Therefore, the same device can function as both an AOFS and an optical isolator. 

For the same device in Fig. \ref{Ch6_resonator}(e), where a pair of TE$_{00}$ and TM$_{00}$ modes of a Si$_3$N$_4$ microring resonator are coupled by HBAR, AOFS can be realized by designing the optical resonator in overcoupled regime. 
The overcoupling is required to have high extraction efficiency of the frequency shifted light.
The strong coupling between the two modes leads to complete optical conversion when cooperativity equals one.
The optical transmission when injecting TE$_{00}$ light into the waveguide and under 20 dBm RF drive is shown in the upper panel of Fig. \ref{Ch6_AOFS2}(a).
At the optical resonance, 50\% of the input light is converted into TM$_{00}$ mode, which has a higher frequency. 
Since the device is not in the strong overcoupled regime, the rest 50\% is dissipated as the intrinsic loss of the optical resonator.
The conversion is also bidirectional, where the input TM$_{00}$ mode can be converted into TE$_{00}$ mode (see the lower panel of Fig. \ref{Ch6_AOFS2}(a)). 
The asymmetry of the modulation sidebands can be seen from the output optical spectrum as shown in Fig. \ref{Ch6_AOFS2}(b).
A single frequency laser at TE$_{00}$ mode is input and modulated with a fixed frequency microwave drive.
The spectrum is broad because of the limited resolution of the spectrum analyzer.
The main peak of the output TM$_{00}$ light shifts to a lower wavelength. 
The light remaining in the carrier and other sideband is more than 10 dB smaller.
The frequency shift can be seen more clearly on the linear scale.
The choice of TE$_{00}$ and TM$_{00}$ mode pair rotates the polarization in addition to the frequency shift, which also makes it convenient to separate the converted light from the carrier using commercial polarization beam-splitter. 

In addition to the microring resonator, AOFS is also performed on a straight Si waveguide, the same device shown in Fig. \ref{Ch6_waveguide}(c). 
A pair of symmetric and anti-symmetric waveguide modes is coupled by a SAW excited from the AlN IDT.
As can be seen in Fig. \ref{Ch6_AOFS2}(c), a Rabi-like oscillation between the two modes is observed by sweeping the RF drive power, demonstrating complete mode conversion.
The optical spectrum of the output light at different sidebands is illustrated in Fig. \ref{Ch6_AOFS2}(d). 
The Stokes sideband dominates and is more than 30 dB larger than the carrier and anti-Stokes sideband. 
The Si waveguide supports a broad optical bandwidth of 1.7 nm, in stark contrast to the microring resonator. 
The RF bandwidth is limited by the bandwidth of the IDT to a few MHz (see Fig. \ref{Ch6_AOFS2}(e)), which can be improved in the future by adopting chirped IDT design \cite{fall2017}. 

While the integrated optical waveguides help to extend the interaction length between photon and phonon on chip, the confining of acoustic wave could further improve the propagation length and interaction strength. 
In that sense, the development of acoustic waveguide and phononic circuits has drawn much attention recently, and major progress has been made in both released \cite{dahmani2020} and unreleased \cite{mayor2021, fu2019phononic} waveguiding.
Furthermore, the co-propagation of phonon and photon in the same waveguide has been demonstrated successfully, paving the path for furture optomechanical circuits \cite{chen2023, zhang2024}. 

Apart from the classical application in heterodyne detection, AOFS will find application in future frequency-domain optical quantum information processing and quantum computing based on frequency-bin optical qubits \cite{Lukens:17, lu2018quantum}. 
As demonstrated in Fig. \ref{Ch6_AOFS2}(c), the conversion between the two optical modes can be controlled by the RF power such that the output power is equally split into the two modes.
Thus, a frequency domain beam splitter can be built which is the building block for optical quantum computing \cite{o2007, hu:2020}.
The frequency beam splitter is conventionally constructed by cascading electro-optical modulators and optical pulse shaper \cite{lu2018electro, lu2020fully}, which is bulky and power consuming, making it difficult to scale.
Therefore, it can be foreseen that the integrated AOFS will become an indispensable component for future large-scale optical quantum computing.
So far, the experiments of the aforementioned AOFS are done in the classical domain, it will be necessary to explore the quantum behavior of it, such as the operation with non-classical light, quantum coherence and interference \cite{Kobayashi:16}, and characterization of the quantum fidelity \cite{lu2018electro}. 

\begin{figure}[b]
\centering
\includegraphics[width=13.5 cm]{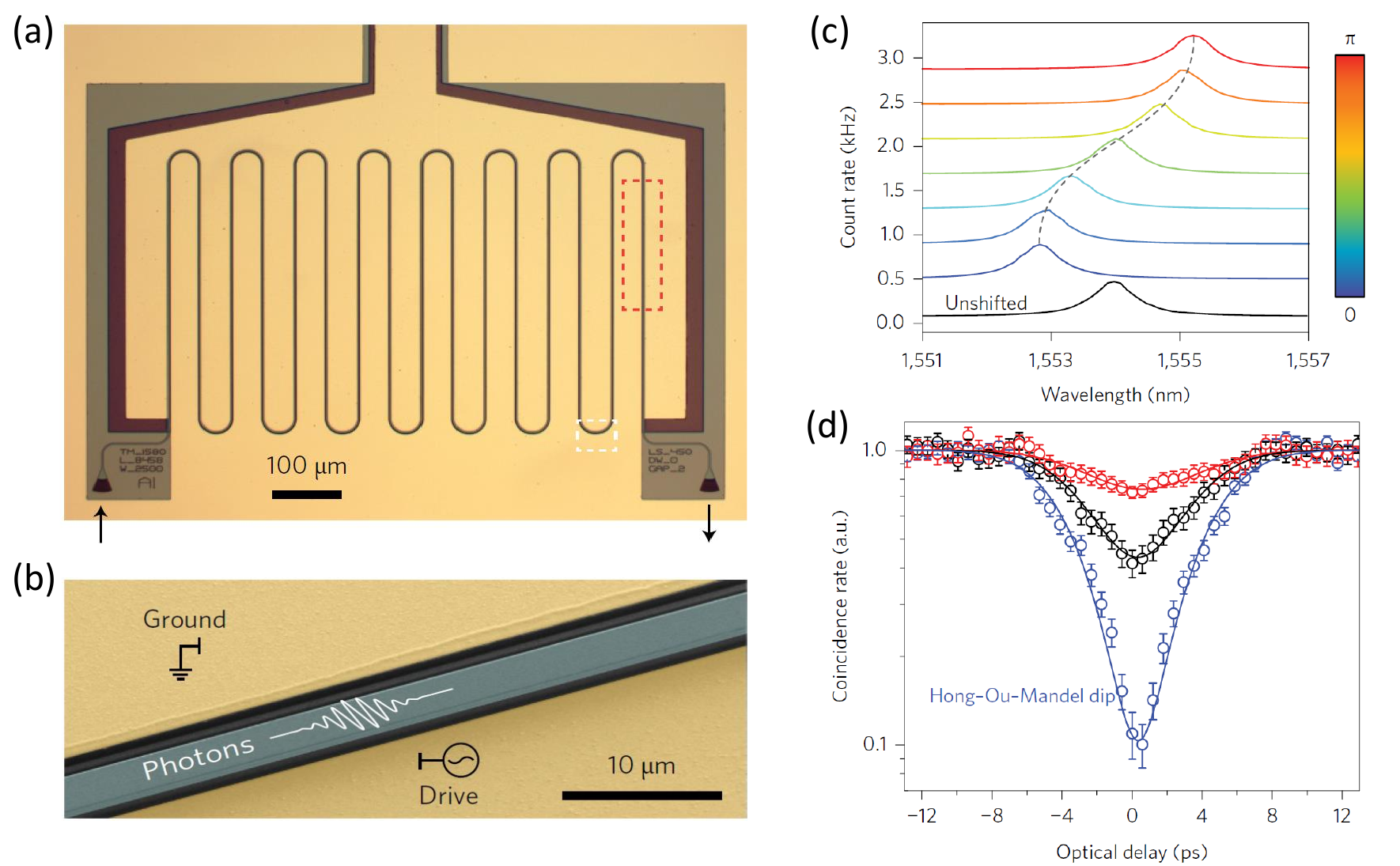}
\caption{Single photon frequency shifting based on spectral shearing. (a) Optical image of a suspended serpentine AlN waveguide that is modulated optomechanically by the mechanical thickness mode of the AlN waveguide. (b) False-colour SEM (red dashed box in (a)) showing the AlN waveguide (grey) and two electrodes (yellow) made from gold. (c) Single-photon spectrum under different RF drive phases from 0 to $\pi$. The black curve is the original spectrum with RF drive off. Each curve is shifted vertically for clarity. (d) Frequency domain Hong-Ou-Mandel interference between two photons with different colors, demonstrating the preservation of quantum coherence after frequency shifting. The coincidence rate between two single photon detectors is measured for unshifted photon (black), blue-shifted photon (blue), and red-shifted photon (red). 
Reprinted from Ref. \cite{fan2016}.
}
\label{ch6_AOFS3}
\end{figure}

Besides the frequency shifts caused by Brillouin scattering between optical and acoustic modes, frequency shifts can also occur through linear phase modulation of the optical carrier $\phi(t)=-\delta \omega t$, resulting in a photon frequency shift by $\delta \omega$. 
This phenomenon is a key aspect of the Fourier transform relationship between frequency and time, commonly referred to as spectral shearing in the literature \cite{zhu2022spectral}. 
Optical frequency shifters utilizing spectral shearing have been demonstrated with both bulk and integrated electro-optic modulators \cite{zhu2022spectral, wright2017} as well as piezo-optomechanical phase modulators \cite{fan2016, fan2019}. 
Figure \ref{ch6_AOFS3} (a, b) illustrate an optomechanical frequency shifter, where a suspended AlN waveguide is stretched piezoelectrically which then changes the effective refractive index of the waveguide and the phase experienced by the photon \cite{fan2016}.
The mechanical thickness mode of the AlN waveguide is activated, enhancing modulation efficiency through mechanical resonance. 
In most cases, the phase is modulated sinusoidally at RF frequencies. 
To achieve linear phase modulation, it is essential that optical pulses are significantly shorter than the sinusoidal wave period, ensuring that the phase during the pulse is linearly modulated. 
Consequently, this approach is not suitable for long optical pulses or CW light, as it would lead to spectral distortion and the generation of multiple sidebands \cite{fan2016}.
It is noteworthy that the frequency shift does not depend directly on the modulation frequency itself, but on the rate of change of the phase over time, which is influenced by the modulation depth, modulation frequency, and the relative phase between the RF drive and the photon arrival time.
The dependence of the frequency shift on the RF drive phase is experimentally measured, as shown in Fig. \ref{ch6_AOFS3}(c).
Over 300 GHz frequency shift has been achieved with near-unity intrinsic efficiency under 2 W RF drive \cite{fan2016}. 
The preservation of quantum coherence after frequency shifting is further demonstrated by conducting the Hong-Ou-Mandel interference between two photons with different colors, as shown in Fig. \ref{ch6_AOFS3}(d).
In the experiment, one red photon is blue-shifted to have the same frequency as the other. 
The large dip of the coincidence rate between two single photon detectors indicates indistinguishability of the two photons after the frequency shift (blue curve in  Fig. \ref{ch6_AOFS3}(d)).
If the red photon is unshifted (black) or red-shifted (red), the visibility becomes low. 
This optomechanical frequency shifter is broadband and does not require pump light, which makes it readily be applicable in optical quantum communication and information processing. 
In the future, the study of new mechanical modes (e.g., SAW and BAW), photonic crystal waveguide, and optical materials with larger photoelastic coefficients could further improve the modulation efficiency and reduce the device footprint by stronger optomechanical interaction strength. 

\section{Future Developments}
In the previous chapters, we have reviewed the design of integrated piezoelectric actuators and their applications in tunable laser and microcombs, acousto-optic modulators, and non-reciprocal devices. In this chapter, we will give our perspectives on potential developments of piezoelectric actuators in future integrated photonic and quantum systems. 

\begin{figure}[t]
\centering
\includegraphics[width=9 cm]{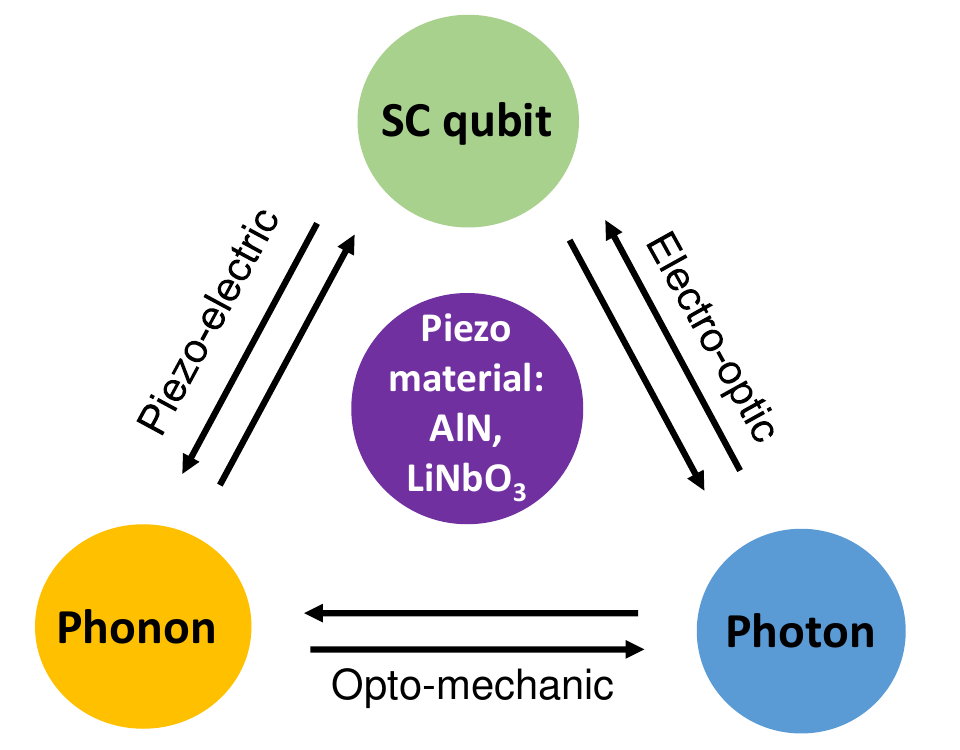}
\caption{
Schematic of the interaction between photon, phonon, and SC qubit enabled by piezoelectric materials. The transduction between each node is bidirectional. Reprinted from Ref. \cite{tianthesis}.
}
\label{Ch7_scheme}
\end{figure}

\subsection{Hybrid Integrated Quantum Systems}
Hybrid integrated quantum systems have attracted extensive efforts recently, as they take advantages of distinct quantum objects for high-performance quantum applications \cite{kurizki2015, clerk2020hybrid}. 
For example, the SC qubit has become one of the mature platforms for quantum computing. 
However, the operation at cryogenic temperature limits its scalability. 
On the other hand, optical photon is good at quantum information communication with low quantum noise, which makes it possible for scalable quantum network \cite{duan2001long}. In contrast to the electromagnetic wave, the low speed of acoustic wave endows it with sub-micron wavelength at microwave frequencies, which allows compact device size.
Moreover, the long coherence time (low losses) of the mechanical resonator makes it suitable to store quantum information. 
A hardware-efficient Quantum Random Access Memory (QRAM) has been theoretically proposed \cite{hann2019hardware}. 
Therefore, hybrid quantum system built from these different modules paves the way for a multi-functional and scalable quantum processor.
However, achieving the transduction between different domains with high quantum efficiency and low added noise is still challenging. 

\begin{figure}[t]
\centering
\includegraphics[width=13 cm]{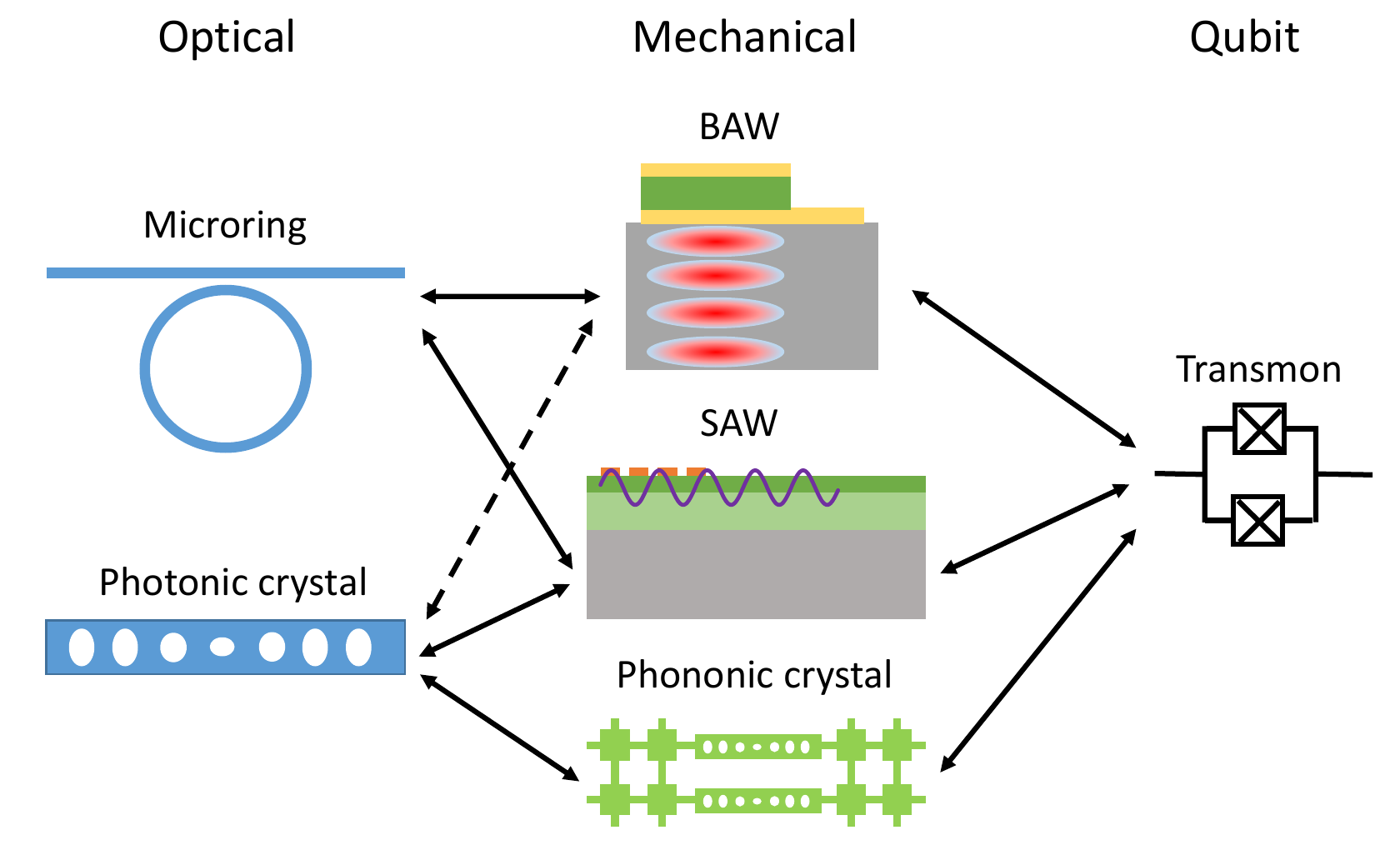}
\caption{
Diagram showing the coupling between optical, mechanical, and SC qubit devices with different optical and mechanical structures.
The black link shows the work demonstrated in the literature, while the black dashed link hasn't been achieved yet. Reprinted from Ref. \cite{tianthesis}.
}
\label{Ch7_combo}
\end{figure}

Figure \ref{Ch7_scheme} illustrates a diagram in which the piezoelectric material plays an important role in mutually coupling SC qubit, optical photon, and phonon. 
The coupling between SC qubit and mechanics through the piezoelectric effect has been initially demonstrated by Andrew N Cleland \textit{et al}. in 2010 using an AlN BAW resonator \cite{o2010quantum}. 
As nanofabrication technologies advance, major progress has been made in coupling SC qubit with various mechanical structures, such as SAW \cite{bienfait2019, sletten2019, dumur2021}, HBAR \cite{chu2017, chu2018}, and phononic crystal \cite{arrangoiz2019, arrangoiz2018, wollack2021}, as shown in Fig. \ref{Ch7_combo}. 
It has grown as a new field of the so-called circuit quantum acousto-dynamics (cQAD). 
The cQAD has recently witnessed promising advances, from phonon-number-resolved readout by SC qubit \cite{arrangoiz2019} to phonon-mediated remote communication and entanglement between SC qubits \cite{dumur2021, bienfait2019}. 
The further development of cQAD will lead to indispensable building blocks in the future SC quantum computing system, featuring a long coherence time and a compact size. 

On the other hand, piezoelectric materials also exhibit an electro-optic effect which allows for direct coupling between SC qubit and optical photon. 
Although the quantum transduction efficiency from microwave to optics is still low compared with the transduction mediated by mechanics, classic optical control and readout of SC qubit has been demonstrated recently \cite{warner2023, arnold2023all, shen2024}, which can serve as the photonic link for SC circuits between cryogenic and room temperatures.
The optomechanical interaction between photon and phonon is the main focus of this review, and its quantum application was discussed in the previous chapter. 
As shown in Fig. \ref{Ch7_combo}, the coupling between various optical and mechanical modes has been demonstrated except for the modulation of photonic crystal using BAW resonator. 
It is worth exploring this combination in the future since it supports higher optical $Q$ and thermal conductance than the OMC, and smaller mode volume than the optical ring.

From the above discussion, one can envision an architecture of the hybrid quantum system where different functional modules based on distinct quantum objects are coordinated together to produce a high-level quantum processor that is capable of computing (SC qubit), storing (phonon) and communicating (photon) quantum information with high fidelity and scalability.
Additionally, piezoelectric materials (which are also electro-optic materials) serve to electrically connect these modules, allowing hybrid integration through wire bonding or flip-chip techniques \cite{malik2023}. 
The materials required for each module are compatible with currently standard CMOS processes, such as Si, Si$_3$N$_4$, and AlN, which enables chip-scale integration and massive production in the future. 

\begin{figure}[t]
\centering
\includegraphics[width=13 cm]{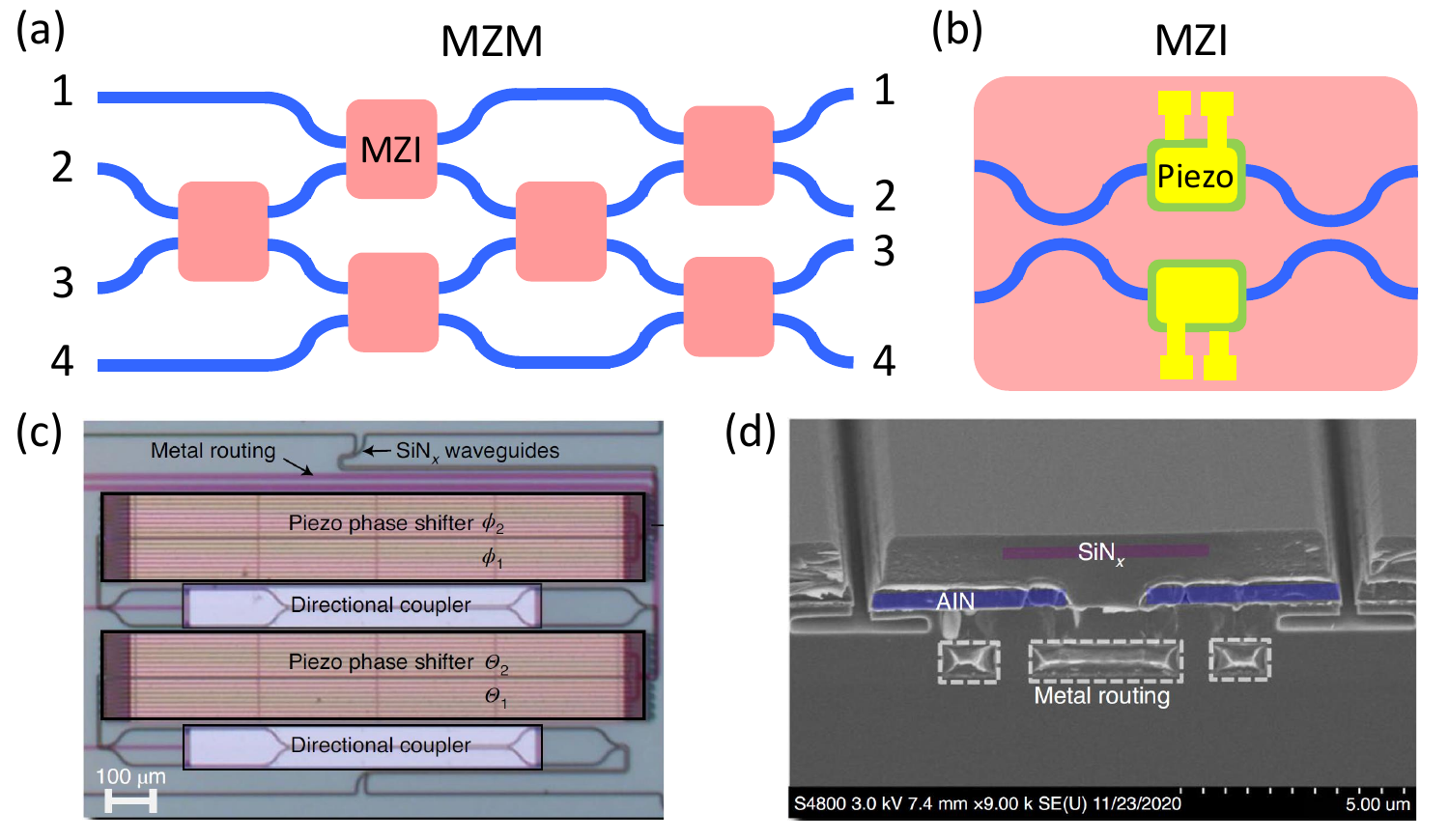}
\caption{
Programmable photonic circuits enabled by piezoelectric actuators. 
(a) Schematic of the Mach-Zehnder Mesh (MZM) for arbitrary linear optical transformation, which consists of cascaded MZI.
(b) Example of one MZI with the optical phases of each arm controlled by the piezoelectric phase shifter. 
(c) Experimental realization of an MZI with four piezoelectric phase shifters. 
(d) False colored SEM image of the cross-section of the released AlN piezoelectric actuator. 
(c-d) Reprinted from Ref. \cite{dong2022high}.
}
\label{Ch7_MZI}
\end{figure}

\subsection{Piezoelectric Programmable Photonic Circuits}
Programmable photonic circuits \cite{bogaerts2020, harris2018, harris2016large} have evolved as the core processor in today's advanced classical and quantum optical computing applications, such as optical neural networks (ONN) for deep learning \cite{shen2017deep} and the photonic quantum simulator \cite{Arrazola:21}.
The main building block of the circuit is a mesh of Mach-Zehnder Interferometers as shown in Fig. \ref{Ch7_MZI}(a). 
In each MZI unit, the light output can be programmed by controlling the relative phases between the top and bottom arms (see  Fig. \ref{Ch7_MZI}(b)). 
By encoding the optical information in each input channel (1, 2, 3, ...), the circuit is capable of implementing a unitary linear-optical transformation at the output, which is one of the fundamental operations in complex computation tasks such as optical quantum information processing and machine learning. 

The key component of the programmable MZI is the optical phase shifter. 
For better performance of the MZI, an ideal phase shifter would require a high-speed response (large bandwidth), low power consumption, and compact size. 
The latter two metrics together determine the scalability of the MZM. 
To achieve the unitary transformation of $N$ input channels, total of $N(N-1)/2$ MZI will be needed \cite{harris2018}.
This $O(N^2)$ scaling places a restrict requirement on the power consumption and size of the designed phase shifter.
Conventionally, the MZI is thermally tuned to build the programmable circuit, from Si ONN \cite{shen2017deep} to the implementation of quantum algorithms on a Si$_3$N$_4$ photonic circuits \cite{Arrazola:21}. 
Although the thermal phase shifter is easy to design and fabricate, its actuation speed is slow (millisecond) and power consumption is large (milliwatt).
Moreover, the thermal tuning is incompatible with cryogenic experiments, such as co-integration with SC circuits, single photon detector, or trapped ion.
Modulation of the Si waveguide through free carrier injection has been demonstrated with low power and high speed \cite{baehr2012}. 
However, the light propagation loss is too high for large-scale integration. 
Although the electrostatic phase shifter provides a solution for low power and compact size, its bandwidth is limited by the mechanical resonance to 10-100 kHz \cite{gyger2020, QuackMEMS, edinger2021silicon, kim2023}.
Therefore, a lower power, low optical loss, high-speed, and compact-size phase shifter will become essential for future programmable photonic circuits.

Piezoelectric actuators could become a good candidate for programming MZI in terms of power and speed.
As shown in Fig.\ref{Ch7_MZI}(c, d), the piezoelectric tuning of a Si$_3$N$_4$ MZI using AlN actuators has recently been demonstrated with a dynamic and static power consumption of 200 $\mu$W and 6 nW, respectively, allowing operation at a cryogenic temperature of 5 K \cite{dong2022high}. 
The actuator shows a 3-dB bandwidth of up to 120 MHz. 
Implementing a $4\times4$ unitary transformation is achieved using 6 MZI.
The main factor that currently limits the scalability of the piezoelectric programmed is the relatively high voltage length product $V_{\pi}L$, which trades off between drive voltage and device length.
For the above work, the $V_{\pi}L$ is 50 V$\cdot$cm. 
To have a low voltage that is compatible with CMOS circuits, the length of the phase shifter must be on the order of centimeters.
Although folding the waveguide shrinks the footprint of an MZI into 1.3$\times$0.6 mm$^2$ \cite{dong2022high}, the long propagation length and bending loss lead to a high insertion loss which also prevents large-scale integration.
Therefore, it is essential to optimize the design of the piezoelectric actuator to have smaller $V_{\pi}L$.
If we recall from Eq. \ref{eq1.18}, the change of refractive index due to the photoelastic effect is proportional to $n^3$ of the waveguide.
Thus, Si waveguide with higher refractive index is worth of reconsidering, in addition to its mature CMOS-compatible fabrication. 
On the other hand, the choice of materials with higher piezoelectric coefficients, such as PZT and AlScN, could further improve the tuning efficiency by five to ten times.
With these improvements, the building of scalable and high-speed programmable photonic circuits will be within reach in the near future.

\subsection{Synthetic Frequency Dimensions in Photonics}
Synthetic dimension has drawn much research interest in recent decades, as it provides a way to study high-dimensional physics beyond the intuitive three spatial dimensions. 
By exploring the coupling between degrees of freedom other than the physical space, a high-dimensional lattice can be synthesized in a low-dimensional structure. 
This allows the investigation of rich topological phenomena that are otherwise hard or impossible to implement in real space.
For instance, the magnitude and phase of the coupling between sites in the synthesized lattice can be controlled easily, such that the long-range hopping can be induced, where novel topological and condensed matter phenomena can be studied \cite{yuan2021synthetic}.
Various platforms have been proposed and developed to realize synthetic dimensions, such as cold atoms \cite{boada2012, price2015four}, superconducting circuits \cite{tsomokos2010, hung2021quantum}, optomechanics \cite{fang2017, schmidt2015}, and photonics \cite{luo2015quantum, dutt2019experimental}. 

Using optical photons as the media to build the synthetic dimension has been attractive in recent research, benefiting from the versatile internal degrees of freedom of photons, such as frequency, polarization, orbital angular momentum (OAM), waveguide modes, and arrival time of pulses. 
These degrees have long been utilized in signal multiplexing for high-speed optical telecommunication.
Here, optical lattices synthesized in these new dimensions paves the way to simulating high-dimensional lattices for studying fundamental physics and exploring nontrivial topological effects \cite{senanian2023, luo2015quantum}, such as the four-dimensional quantum Hall effect \cite{ozawa2016}, photonic gauge potential \cite{yuan2016photonic, hey2018}, quantum walks \cite{chen2018, chalabi2019, hu2020}, Bloch oscillation \cite{yuan2016bloch}, and photonic topological insulator \cite{lustig2019}. 
The exploration of these effects in turn helps the development of novel photonic devices that provide new ways to manipulate the frequency and propagation of photons. 
For example, the synthetic quantum Hall lattice enables unidirectional frequency translation that is topologically protected \cite{dutt2020single}, and nonreciprocal optical transmission has been demonstrated \cite{kim2021chip}. 

\begin{figure}[t]
\centering
\includegraphics[width=13 cm]{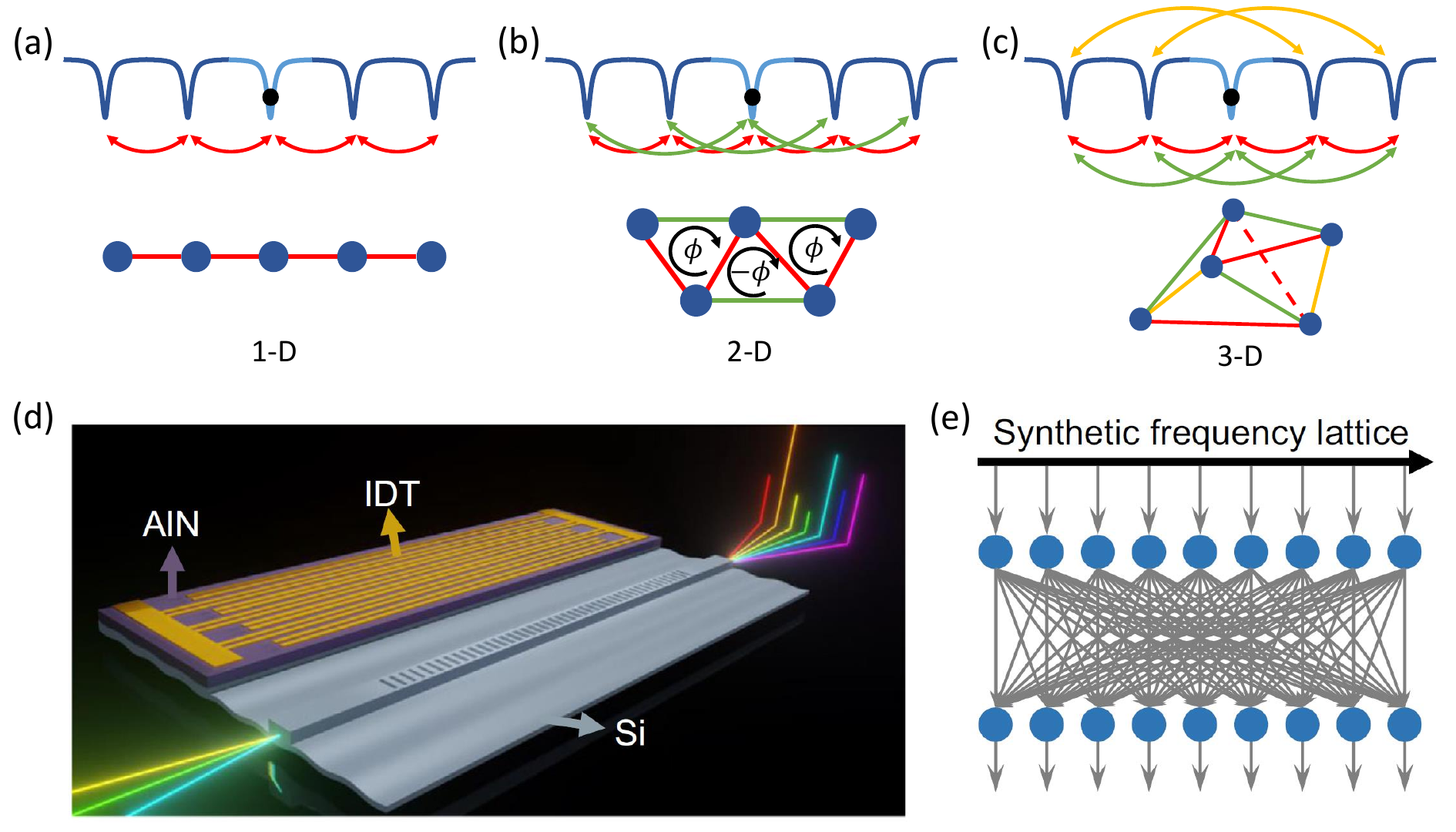}
\caption{
Synthetic frequency dimension formed in dynamically modulated optical resonator. 
Schematics of the synthesized optical frequency lattice in (a) 1D, (b) 2D, and (c) 3D by inducing the coupling of adjacent modes (red), the next-nearest neighbour (green), and the next-next-nearest neighbour (yellow).
(a-c) Reprinted from Ref. \cite{tianthesis}.
(d) Schematic of the experimental realization of the optical computing in frequency dimension using acousto-optically modulated photonic crystal resonator. 
(e) Diagram illustrating the matrix vector multiplications in frequency lattice by programming the mutual coupling among lattice sites through optical modulation. 
(d-e) Reprinted from Ref. \cite{zhao2022}.
}
\label{Ch7_synthetic}
\end{figure}

Synthetic dimension in frequency is of particular interest because of the ultra-wide optical bandwidth which provides a large computational space. 
In addition, the optical phase modulator has been well developed, making it convenient to couple different frequency components.
An optical ring resonator consists of a series of equidistant optical resonances, which naturally forms an optical lattice in frequency. 
Phase modulation of the resonator at the frequency equal to the FSR of the resonator introduces coupling between each two adjacent modes, as shown in Fig. \ref{Ch7_synthetic}(a).
The dynamics of the coupling can be well described by the tight-binding model, which was originally applied in solid-state physics to model the band structure of the electron in a crystal lattice. 
The corresponding Hamiltonian can be written as \cite{yuan2021synthetic}:
\begin{equation}
    H_{\text{TB}}=\sum_k \omega_k a_k^{\dag}a_k + \sum_k g a_{(k+1)}^{\dag}a_k e^{i\phi}+ h.c.
\end{equation}
where $\omega_k$ is the frequency of each optical mode which serves as an on-site potential.
$g$ and $\phi$ are the coupling amplitude and phase between two adjacent modes, which are determined by the microwave drive.
$\hbar$ is omitted for simplicity. 
Therefore, it can be seen the dynamics of the photon in the synthesized frequency lattice mimics the hopping of electron in an atom lattice, indicating the potential of optical simulation. 
Since the reciprocal variable of frequency is time, the band structure can be easily obtained by measuring the time resolved spectrum of the modulated resonator \cite{dutt2019experimental}. 

Beyond 1D lattice, higher dimensional frequency lattice can be constructed by adding microwave drives at multiple FSRs, as shown in Fig. \ref{Ch7_synthetic}(b, c). 
These drives induce long-range hopping terms in the Hamiltonian as $\sum_k g_N a_{(k+N)}^{\dag}a_k e^{i\phi_N}$, which allows us to fold the 1D lattice into more complex structure \cite{yuan2018, senanian2023}, such as 2D, 3D, or beyond. 
Very rich topological phenomena can be studied by engineering the magnitude and phase of each coupling term $g_N e^{i\phi_N}$. 
For instance, the modulation phase $\phi_N$ serves as an effective gauge potential for the photon in the frequency dimension \cite{yuan2021synthetic}. 
An effective magnetic field can be generated, depending on the phase difference between coupling terms, as shown in Fig. \ref{Ch7_synthetic}(b). 
Additionally, if the drive is detuned from the FSR, an effective electric force will be formed, making it possible to study the Bloch oscillation of photon in the frequency lattice \cite{yuan2016bloch}. 
The frequency dimension can further combine with the spatial dimension by coupling multiple optical resonators, enabling unique control of the photon in frequency and real space \cite{ozawa2016, hey2018, kim2021chip}. 

The experimental realization of the synthetic frequency dimension in photonics has witnessed promising progress recently. 
The initial proves of the concept were demonstrated using fiber optics and off-the-shelf electro-optic modulators, including synthetic gauge potential \cite{dutt2019experimental}, synthetic Hall ladder \cite{dutt2020single}, effective electric field \cite{li2021dynamic}, and high-dimensional lattices \cite{senanian2023}. 
Although this approach is simple to implement, the large size of the fiber optical ring resonator limits its scalability and application in making novel optical devices.
A few recent works have achieved the synthetic frequency dimension in integrated photonic platforms, such as thin film lithium niobate microresonator that is electro-optically modulated \cite{hu2020, herrmann2022}, and Si photonic resonator modulated by a p-n junction \cite{balvcytis2022}. 
Since the synthetic frequency dimension only requires narrow band frequency modulation, acousto-optic modulator could play an important role, benefiting from its high modulation efficiency at mechanical resonances.
A synthetic Hall effect has been demonstrated by modulating two coupled AlN photonic resonators through SAW, and nonreciprocal optical transmission is achieved by engineering the modulation phases \cite{kim2021chip}. 
The main challenge for AOM is to match the acoustic resonance with the FSR, especially for on-chip optical resonators with FSR typically from 10 GHz to THz range.
Low-FSR Si$_3$N$_4$ ring resonators have been demonstrated with FSR of 2.5 GHz and optical $Q$ over 5 million \cite{lihachev2022, Puckett:21}, which allows for on-chip synthetic frequency dimension enabled by AOM. 

In addition to studying the topological effects, the synthetic frequency dimension has been utilized to implement an arbitrary linear transformation by engineering the coupling between frequency modes \cite{buddhiraju2021}.
Similar to the MZM presented in the last section, the input optical states can be arbitrarily mapped to the output but in the frequency space.
The advantages of frequency domain implementation include the compact device size and scalability, since all optical channels transmit in the same optical waveguide. 
Recently, matrix vector multiplication has been demonstrated by acousto-optic modulating a single Si photonic crystal resonator \cite{zhao2022}, as shown in Fig. \ref{Ch7_synthetic}(d, e). 
In the future, linear optical quantum information processing in the frequency dimension can be achieved by incorporating non-classical light. 
Although experimental realizations using bulk EOM and pulse shaper have been achieved \cite{lu2020fully, lu2018electro}, chip-scale photonic circuit and AOM could enable large-scale optical computing with high-fidelity frequency-bin logic gates \cite{buddhiraju2021, zhao2022}. 
Additionally, a non-unitary transformation can be realized by introducing the modulation of gain or loss of the optical resonator, which is generally the case in neural networks \cite{buddhiraju2021}.
Finally, combining with the optical nonlinearity of the photonic waveguide, such as Si$_3$N$4$, Si, and LiNbO$_3$, novel dynamics of the soliton generation has been theoretically predicted \cite{tusnin2020}. 
The nonlinearity is also required for implementing deep learning algorithms, a step for fully optical neural networks \cite{shen2017deep}. 
Therefore, we believe the integrated AOM will play an important role in both classical and quantum optical computing in the synthetic frequency dimension through efficient modulation and frequency conversion. 

\begin{figure}[t]
\centering
\includegraphics[width=13.5 cm]{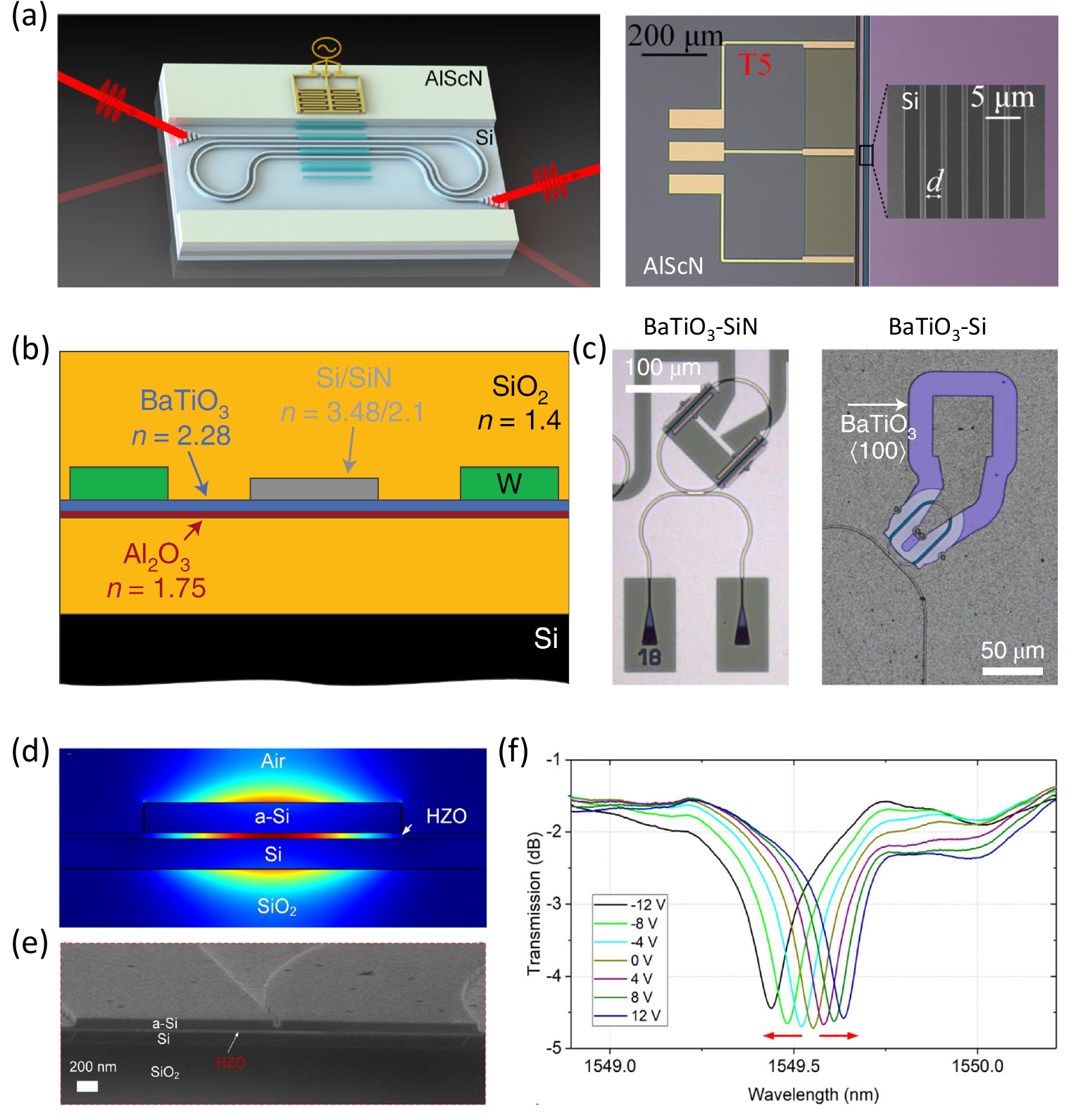}
\caption{Application of novel piezoelectric materials in photonic integrated devices. (a) AOM of Si waveguide on a hybrid AlScN-Si platform. Left: 3D schematic of the device, where the Si waveguide is modulated by the SAW excited from IDT made on AlScN. Right: optical image of the fabricated device. Reprinted from Ref. \cite{huang2024}. (b) Schematic of the cross-section and (c) optical images of integrated EOM made from hybrid BTO-Si and BTO-Si$_3$N$_4$ optical racetrack resonators. Reprinted from Ref. \cite{abel2020}. (d-f) Piezoelectric tuning of HZO-Si horizontal slot waveguide. (d) FEM simulation of the TM$_{00}$ mode of the slot waveguide. (e) SEM of the cross-section of the fabricated slot waveguide. 20 nm HZO is sandwiched between two Si layers. (f) Tuning of an optical resonance under different voltages. Reprinted from Ref. \cite{shen2023}. 
}
\label{Ch7_material}
\end{figure}

\subsection{Novel Piezoelectric Materials}
As a crucial element in piezoelectric actuators, investigating new materials with improved piezoelectric properties will significantly boost the efficiency, reduce the size, and lower the power consumption of these devices. 
Benefiting from the advancement of high-quality thin film deposition on semiconductor substrates, a number of good candidates have received intensive research efforts in the past decade, such as AlScN \cite{chen2022, fichtner2019alscn}, BaTiO$_3$ (BTO) \cite{zgonik1994}, hafnium zirconium oxide (HZO) \cite{shen2023, guberna2023}, and potassium–sodium niobate (KNN) \cite{wu2015potassium}. 
Among them, AlScN has witnessed promising progress in recent years, where a variety of thin-film deposition techniques have been explored, from reactive sputter deposition, molecular beam epitaxy (MBE), metal organic chemical vapor deposition (MOCVD), and pulsed laser deposition (PLD).
A review of these deposition methods is provided in Ref. \cite{chen2022}.
Compared to AlN, dry etching of AlScN is more challenging due to the low volatility of ScCl$_3$, especially for high Sc doping concentration \cite{airola2022high}. 
Nevertheless, inductively coupled plasma reactive ion etching (ICP-RIE) of AlScN has been studied in Ref. \cite{yan2023}. 
Alternatively, AlScN can be etched physically by Ar$^+$ ion milling \cite{pearton1994}. 
Wet etching of AlScN has also been studied with various wet chemical solutions, such as alkaline solutions tetramethyl ammonium hydroxide (TMAH) and potassium hydroxide (KOH) and acid solution phosphoric acid (H$_3$PO$_4$) \cite{airola2022high, shifat2023}.
The integration of CMOS-compatible AlScN PICs at the wafer scale has been achieved \cite{wang2024cmos, bian2023integrated}, together with enhanced second-order nonlinearity \cite{yoshioka2021, yang2024unveiling} and electro-optic modulation \cite{xu2024silicon, yoshioka2024cmos}. 
Efficient AOMs based on AlScN have been demonstrated with various structure designs \cite{huang2022, huang2024, bian2024, erdil2024}. 
An example of the AOM is shown in Fig. \ref{Ch7_material}(a), where a SAW is excited by the IDT fabricated on the AlScN thin film \cite{huang2024}.
A Si waveguide is modulated by the SAW and a voltage-length product $V_{\pi}L$ of 2.34 V$\cdot$cm is reported, which is still comparable to that of AlN (see Table \ref{table_AOM}). 

In the past decade, BTO has attracted considerable research interest because of its substantial Pockels and piezoelectric coefficients (see Table \ref{tablepiezo}) and the successful deposition of high-quality thin films on silicon substrates \cite{abel2013, karvounis2020}.
CMOS-compatible wet etching via buffered hydrogen fluoride (HF) \cite{lim2020} and dry etching through fluorine-based \cite{li2014} and Ar-based \cite{chen2021} plasma have made major progresses.
Further optimization of the etching conditions would allow for photonic and optomechanical circuits made from patterned BTO. 
Integrated electro-optic modulators made from BTO have achieved over 20 GHz modulation bandwidth in both room \cite{abel2019} and cryogenic temperatures \cite{abel2020}, as shown in Fig. \ref{Ch7_material}(b, c).
Nonetheless, BTO's significant piezoelectric response has not been sufficiently investigated for applications in piezoelectric actuation and AOM.
Alongside its ferroelectric characteristics, BTO will pave the way for new tunable and reconfigurable electro-optic and acousto-optic devices \cite{abel2022}. 

Since its first report of ferroelectricity in 2011 \cite{muller2011}, HZO has received intensive research interest by serving as ferroelectric transistors and memories. 
Atomic layer deposition (ALD) of HZO thin films, being compatible with CMOS process, has become a commonly used method \cite{bang2008}. 
However, the thickness of these films is typically limited to several tens of nanometers. 
The recently developed atomic layer etching (ALE) technique enables precise control of the etching of HZO thin films \cite{hoffmann2022atomic}. 
Beyond microelectronic devices, high Q-factor electromechanical resonator based on an ultra-thin HZO film (10 nm) has been demonstrated \cite{ghatge2019}. 
Even though it is still in its early stages, significant progress has been made in applying HZO in integrated photonic circuits. 
For example, a nonvolatile optical phase shifter utilizing a hybrid Si$_3$N$_4$-HZO optical waveguide has been realized, paving the way for low-power programmable photonic circuits in the future \cite{taki2024}.
Recently, an ultra-low power piezoelectric tuning of a hybrid Si-HZO optical ring resonator has been demonstrated with 1 GHz/V tuning efficiency and 1 pW/MHz power consumption \cite{shen2023}, as illustrated in Fig. \ref{Ch7_material}(d-f). 
The comparison with AlN and PZT is included in Table \ref{table_compare}, where the tuning efficiency of HZO is much larger than that of AlN and comparable to PZT.
However, the optical waveguide loss is still high (48 dB/cm), primarily due to the roughness of the HZO film and the surface of amorphous Si film \cite{shen2023}.

Due to the demand for eco-friendly, lead-free piezoelectric materials, KNN has emerged as a viable substitute for conventional piezoelectric materials like PZT.
Superior piezoelectric response of KNN ceramics with $d_{33}\approx570$ pC/N \cite{xu2016superior} and KNN single crystal with $d_{33}\approx9000$ pC/N \cite{hu2020ultra} have been reported.
KNN thin-films are generally synthesized via RF magnetron sputtering \cite{chen2024composition}, sol-gel \cite{akmal2018}, and chemical solution deposition techniques \cite{kupec2012lead}. 
A post-annealing process at around 700 $^{\circ}$C is required for the KNN film to crystallize \cite{chen2024composition}. 
The precise regulation of the composition of KNN thin films through optimized processes is responsible for the enhanced piezoelectric response of KNN \cite{gao2021mechanism}. 
Various etching methods have been explored including wet etching with alkaline solution \cite{zhao2024design} and drying plasma etching \cite{kurokawa2012micro}. 
Owing to their piezoelectric characteristics that are on par with PZT, KNN thin films are being investigated for various applications \cite{zhang2019potassium}, such as sensors, actuators \cite{zhao2024design}, and devices for energy harvesting. 
Future research on further improving the performance and reliability of KNN films will facilitate its application in photonic circuits for piezoelectric tuning and AOM.

\section{Conclusion}
In this review, the role of integrated piezoelectric actuators in controlling and modulating photonic devices via optomechanical interactions, including photoelastic and moving boundary effects, was explored.
Compared to electro-optic modulation, piezoelectric actuation works with optical materials devoid of the Pockels effect, like Si and Si$_3$N$_4$.
This review thoroughly examines the piezoelectric tuning of optical ring resonators and waveguides, focusing on piezoelectric materials (mainly AlN and PZT) and the design of mechanical structures (either released or unreleased). 
Achievements include low power consumption (tens of nanowatts) and rapid actuation speeds (sub-microsecond), offering benefits over traditional thermal tuning methods.
Although the tuning range of piezoelectric actuators remains less than that of integrated heaters, their quick response facilitates a broad range of applications in integrated lasers and frequency comb sources, including frequency-agile narrow-linewidth lasers, the initiation and control of Kerr microcombs, and the stabilization of frequency comb lines. 

Beyond quasi-DC tuning, the piezoelectric actuator also facilitates acousto-optic modulation at microwave frequencies, exhibiting higher efficiency compared to EOMs at the expense of narrow bandwidth. 
We explored various integrated AOMs utilizing distinct acoustic modes, ranging from surface and bulk acoustic waves to optomechanical crystals, each offering unique benefits. 
For instance, surface acoustic waves provide high modulation efficiency, whereas bulk acoustic wave resonators are noted for their small size and compatibility with thick optical cladding to achieve high optical $Q$.
Moreover, optomechanical crystals have shown the greatest optomechanical coupling strength, enhancing their utility in quantum transduction.
These miniature AOMs are extensively used in a range of classical and quantum optical devices, including optical isolators, microwave to optical converters, and optical frequency shifters, leading the way in controlling light propagation and frequency on a chip-scale. 
This is crucial for the development of future hybrid integrated systems for optical classical and quantum information processing.

Equipped with DC tuning and AOM, the concept of a multifunctional photonic processor is conceivable, as depicted in Fig. \ref{ch1_chip} at the outset of this review.
Ranging from tunable laser sources and programmable beam splitters to optical modulators, the Piezo-on-Photonic platform facilitates the integration of these elements on a single chip, albeit requiring moderate adjustments in device design and fabrication processes to suit various materials, etching selectivity, and release compatibility.
Looking ahead, piezoelectric actuators are poised to significantly contribute to the development of programmable photonic circuits across spatial and frequency domains, enabling sophisticated optical simulation and computation.

In the future, advances in nanofabrication technologies, driven by breakthroughs in chemical and material science, could enhance the performance of piezoelectric actuators and the creation of cutting-edge functional devices. 
For instance, the epitaxial growth of single crystalline piezoelectric thin films at the wafer scale that is CMOS compatible, such as GaN \cite{gokhale2020a} and AlN \cite{kim2021low}, will attract more research interest due to their lower acoustic and optical losses, negligible leakage current, and better piezoelectric response.
Furthermore, investigating novel piezoelectric materials with higher piezoelectric coefficient, such as AlScN \cite{chen2022}, BTO \cite{zgonik1994}, HZO \cite{shen2023, guberna2023}, and KNN \cite{wu2015potassium}, could enhance the tuning and modulation efficiency, and thus reduce the required drive voltage $V_{\pi}$ and shrink the device size. 

Finally, exploring innovative fabrication techniques for the monolithic integration of III-V gain components on the same piezoelectric-actuated photonic chip could provide tunable lasers with reduced noise and enable large-scale production \cite{xiang2023, tran2022}. 
Consequently, we anticipate that piezoelectric actuation will evolve into an essential component in future photonic circuit designs, offering distinctive tuning and modulation capabilities with high efficiency, minimal power requirements, and cost-effectiveness.

\section*{Funding}
U.S. National Science Foundation’s RAISE TAQS program (PHY 18-39164); 
Air Force Office of Scientific Research (FA8655-20-1-7009); 
EU H2020 research and innovation programme (732894 HOT); 
Swiss National Science Foundation (SNSF) (176563 and 211728 BRIDGE);
Defense Advanced Research Projects Agency (DARPA) (W911NF212024 NINJA LASER);
Microsystems Technology Office (MTO) (HR0011-15-C-055 DODOS).

\section*{Acknowledgement}
We thank Jiahao Sun for valuable discussion and critical reading of this review. 
We thank Bin Dong, Michael Zervas, Erwan Lucas, Arslan S. Raja, Tianyi Liu, Miles H. Anderson, Wenle Weng, Ben Yu, Mert Torunbalci, Wil Kao, Diego Visani, Jijun He for valuable discussion during the fabrication and experimental realization of piezoelectric actuated photonic devices.
We also thank the collaborations from Christian Koos, Yung Chen, and Huanfa Peng.
We acknowledge the support from the Defense Advanced Research Projects Agency (DARPA), Microsystems Technology Office (MTO), Air Force Office of Scientific Research, U.S. National Science Foundation, the EU H2020 research and innovation program, and the Swiss National Science Foundation (SNSF). 
H.T. acknowledges the support from NSF QISE-Net under grant no. DMR 17-47426.
We thank the support from the EPFL centre of MicroNanoTechnology (CMi) and the Birck Nanotechnology Center at Purdue University for the fabrication of integrated Si$_3$N$_4$ photonic chips and AlN actuators.
We also thank Plasma-Therm LLC. and Advanced Modular Systems Inc. for AlN deposition, and Radiant Technologies Inc. for fabricating PZT actuators. 

\section*{Disclosures}
TJK: LiGenTec SA (I, P), DEEPLIGHT SA (I, P); AB: DEEPLIGHT SA (I, E, P); AV: DEEPLIGHT SA (I, P)

\section*{Data Availability}
No data were generated or analyzed in the presented research.

%%%%%%%%%%%%%%%%%%%%%%% References %%%%%%%%%%%%%%%%%%%%%%%%%

\newpage
%%%%%%%%%% If using BibTeX:
\bibliography{sample}

\newpage
\begin{wrapfigure}{l}{0.3\textwidth}
\includegraphics[width=0.3\textwidth]{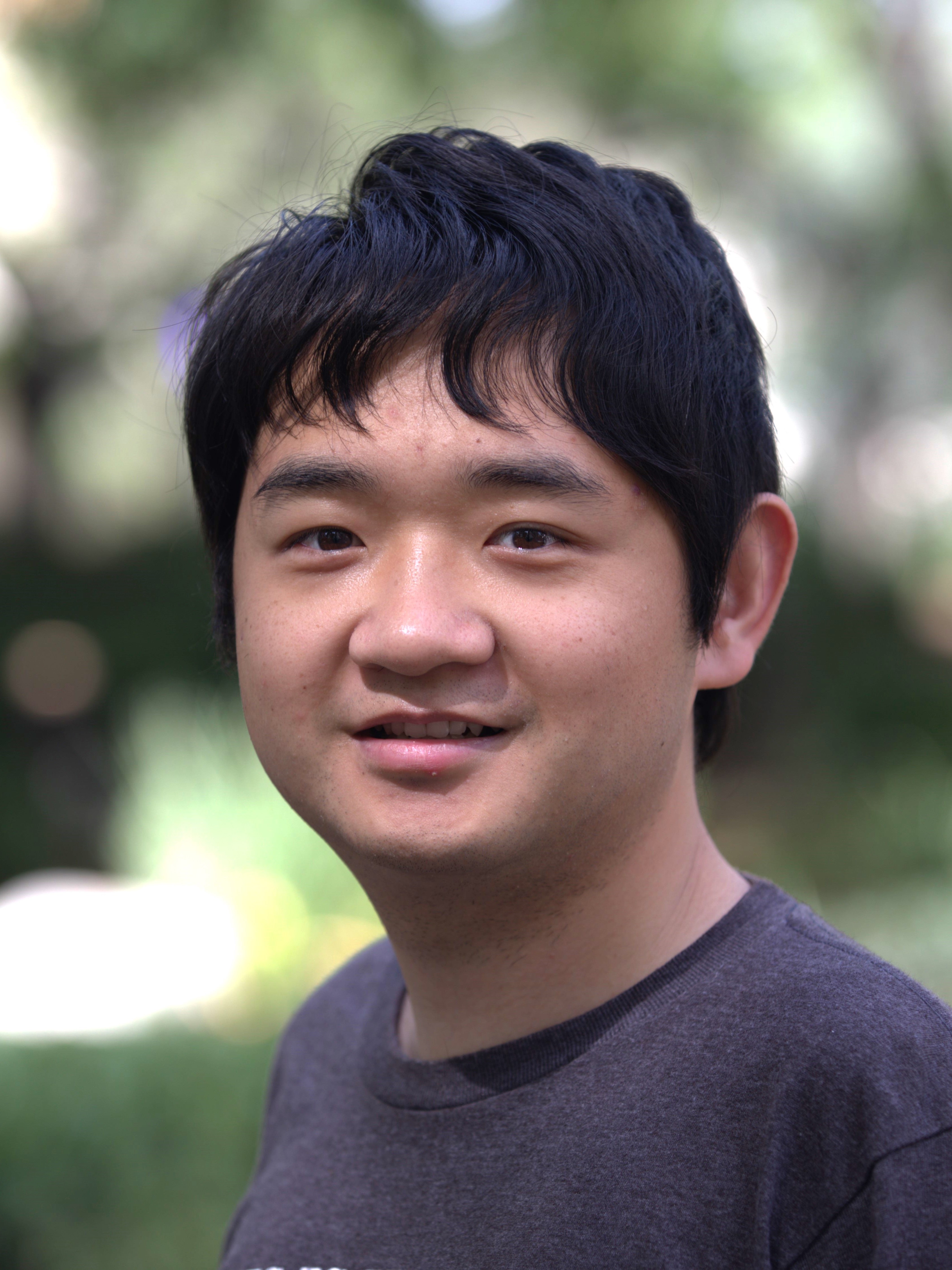} 
\end{wrapfigure}\par
\textbf{Hao Tian} is an IQIM postdoctoral fellow at Caltech. He received Ph.D. in Electrical and Computer Engineering from Purdue University in May 2022, where he worked on piezoelectric actuation on Silicon Nitride photonic circuits, and its applications in integrated optical isolator, tunable frequency combs and semiconductor lasers. He currently works on quantum transduction between mechanics, photonics, and superconducting qubits for building hybrid quantum systems. He is the recipient of Outstanding Student Paper Award in IEEE MEMS conference in both 2021 and 2023. His Ph.D. thesis on piezoelectric control of Silicon Nitride photonics received the 2022 Outstanding Graduate Student Research Award in Purdue University. 
\\

\begin{wrapfigure}{l}{0.3\textwidth}
\includegraphics[width=0.3\textwidth]{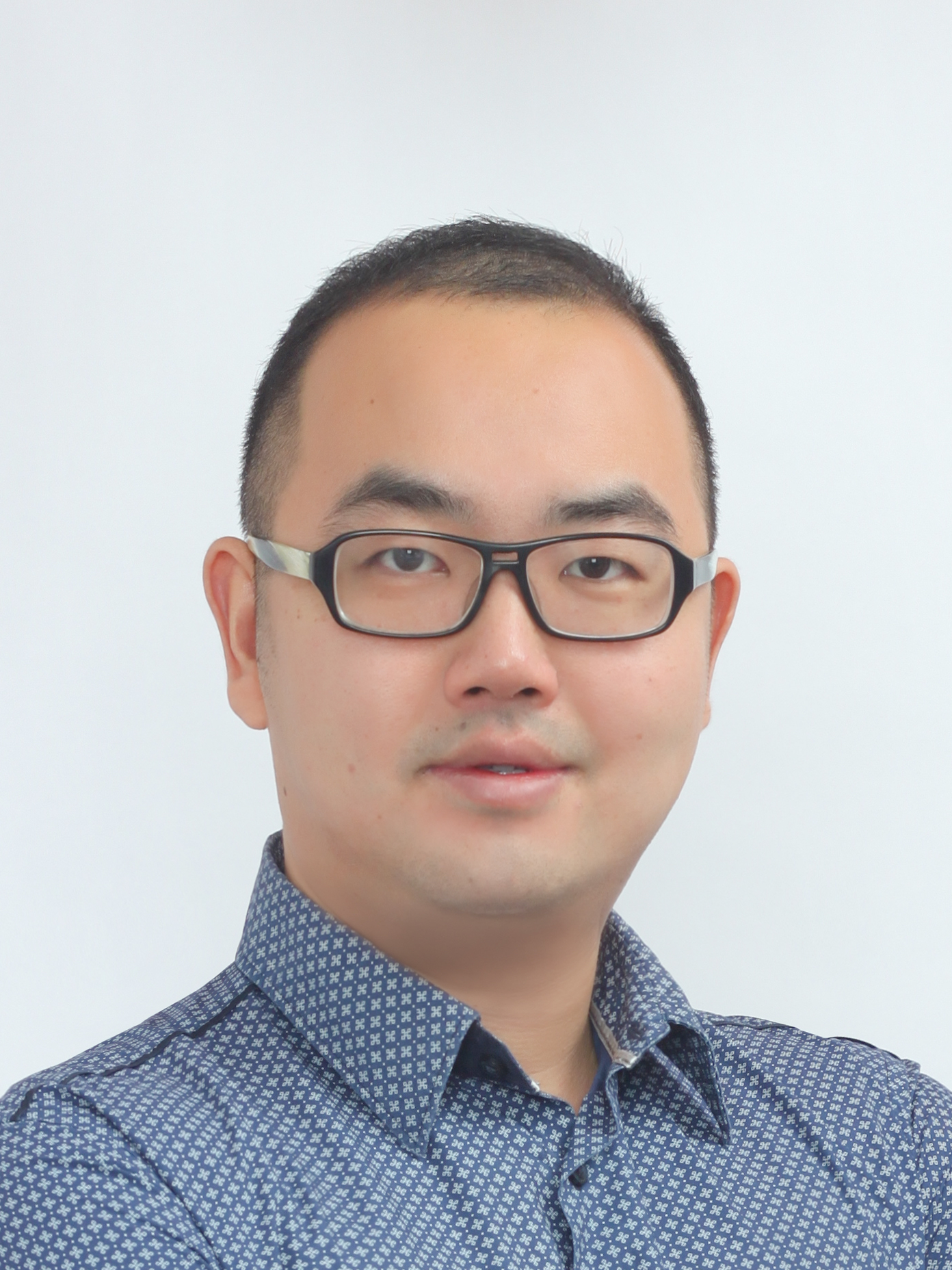} 
\end{wrapfigure}\par
\textbf{Junqiu Liu} is currently a principal investigator at Shenzhen International Quantum Academy and a tenure-track assistant professor at Hefei National Laboratory, University of Science and Technology of China (USTC). He received PhD from EPFL in 2020, MSc (with highest distinction) from the University of Erlangen-Nuremberg in 2016, and BSc from USTC in 2012. His research is in the domains of integrated photonics, nonlinear optics, quantum optics, MEMS, microwave photonics and lase spectroscopy. His PhD thesis entitled “Silicon Nitride Nonlinear Integrated Photonics” was awarded 2021 EPFL Doctorate Award, “For ground breaking experiments in the field of chip-scale frequency combs and the extraordinary record of scientific accomplishments”. He has published more than 70 papers in peer-reviewed journals and has received numerous awards for young scholars.
\\

\begin{wrapfigure}[14]{l}{0.3\textwidth}
\includegraphics[width=0.3\textwidth]{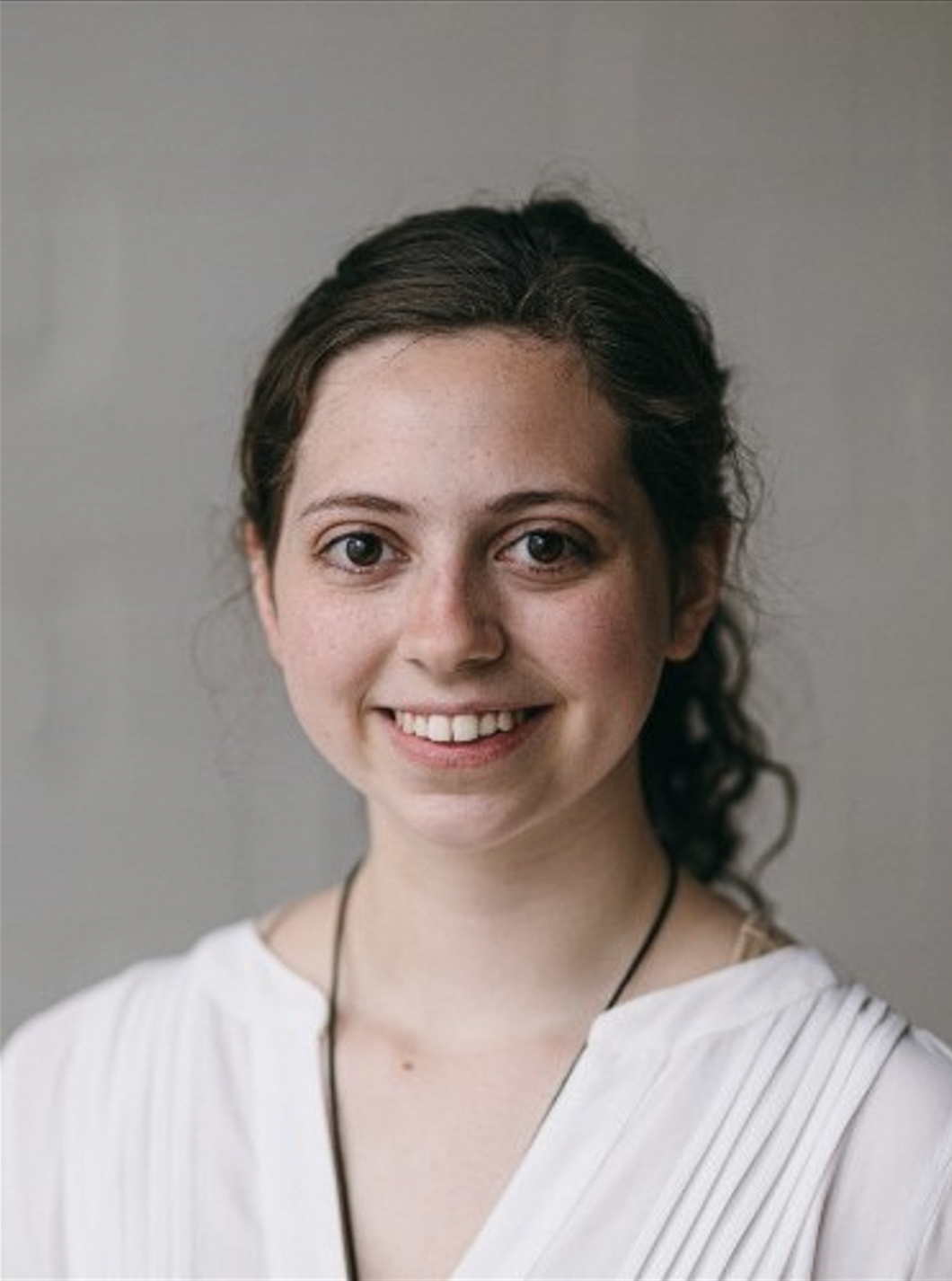}
\end{wrapfigure}\par
\textbf{Alaina Attanasio} is currently a third year graduate student pursuing her doctorate at Purdue University. She received her B.S in Electrical and Computer Engineering at University of Rochester in May 2021. She has been focusing on designing and testing MEMS optomechanical accelerometers and photonic HBAR resonators for a wide variety of applications including tunable lasers, microwave-to-optical transducers, and quantum converters. She co-authored "Programmable Silicon Nitride Photonic Integrated Circuits" which won the Best Paper Award at the IEEE MEMS 2023 conference. 
\\
\\

\newpage
\begin{wrapfigure}[13]{l}{0.3\textwidth}
\includegraphics[width=0.3\textwidth]{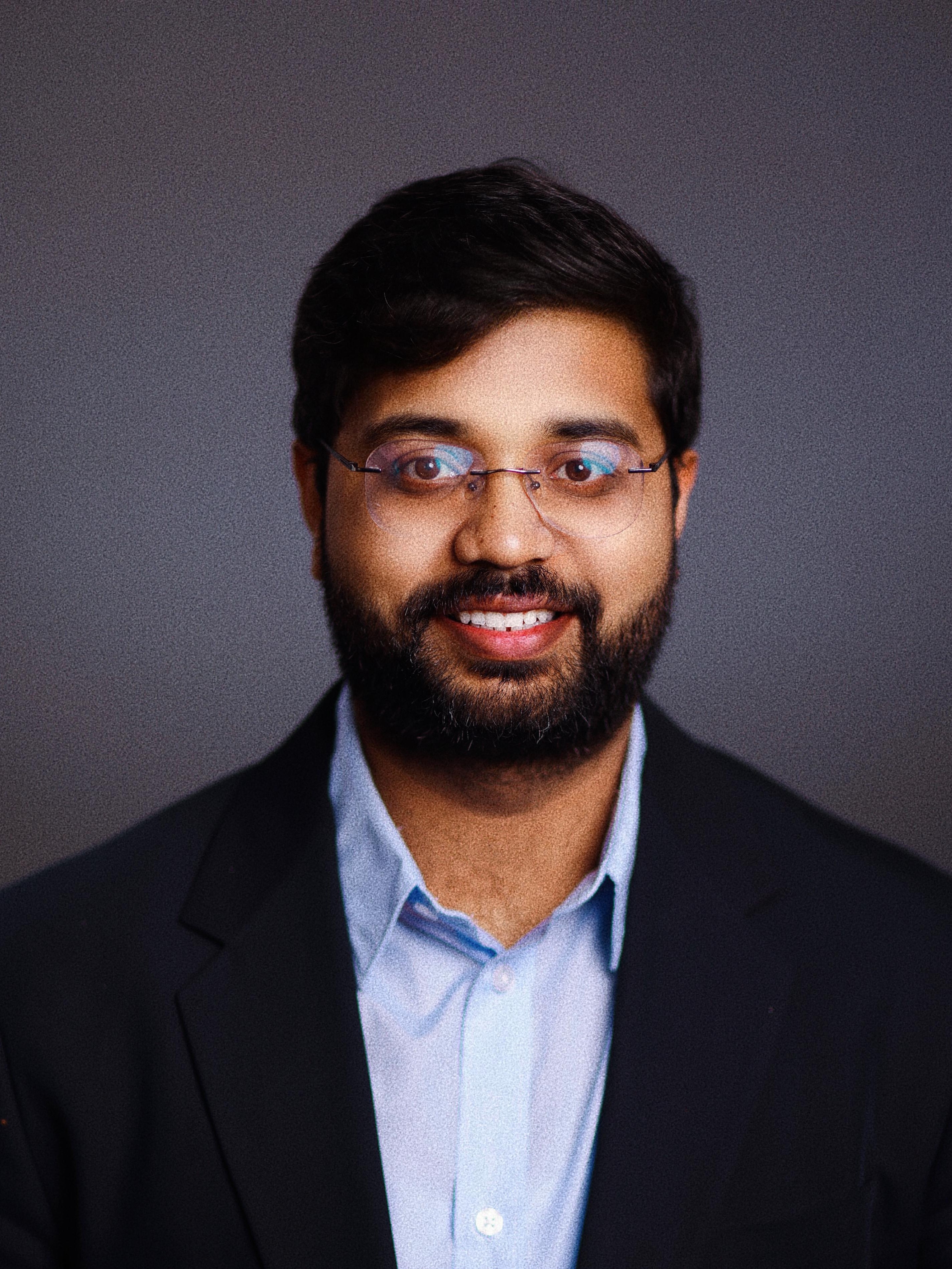}    
\end{wrapfigure}\par
\textbf{Anat Siddharth} received his Bachelor of Technology in Engineering Physics from the Indian Institute of Technology, Guwahati, and a Master of Technology in Applied Optics from the Indian Institute of Technology, Delhi. He is currently pursuing his PhD in the research group of Prof. Tobias J. Kippenberg. He is an ESA OSIP fellow whose work focuses on low-noise, frequency-agile photonic integrated lasers. Formerly, he worked on nonlinear photonics at IIT Delhi and served as an engineer at FINISAR Malaysia. 
\\
\\

\begin{wrapfigure}[14]{l}{0.3\textwidth}
\includegraphics[width=0.3\textwidth]{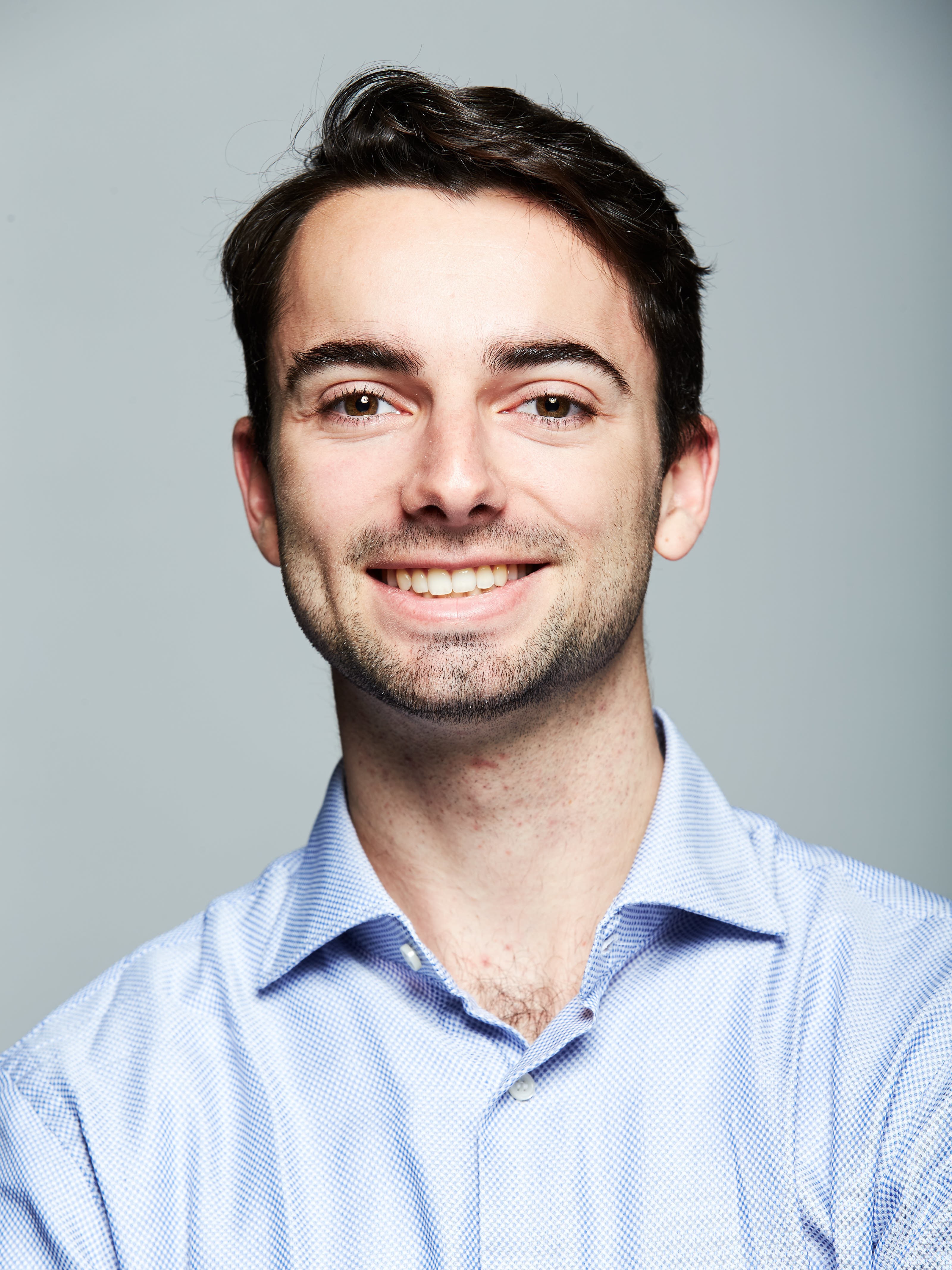}
\end{wrapfigure}
\textbf{Terence Blésin} is a Ph.D. candidate in Physics at the Swiss Technology Institute in Lausanne (EPFL). He received his M.Sc. in Applied Physics and Physico-informatics from Keio University, his M.Sc. in Engineering Physics from the Université Libre de Bruxelles (ULB) and his B.Eng. from the Université Libre de Bruxelles (ULB). He is a recipient of the JASSO scholarship and T.I.M.E. double degree. He is currently working on microwave-optical transduction using integrated photonics in Prof. Tobias Kippenberg’s group.
\\
\\
\\

\begin{wrapfigure}[14]{l}{0.3\textwidth}
\includegraphics[width=0.3\textwidth]{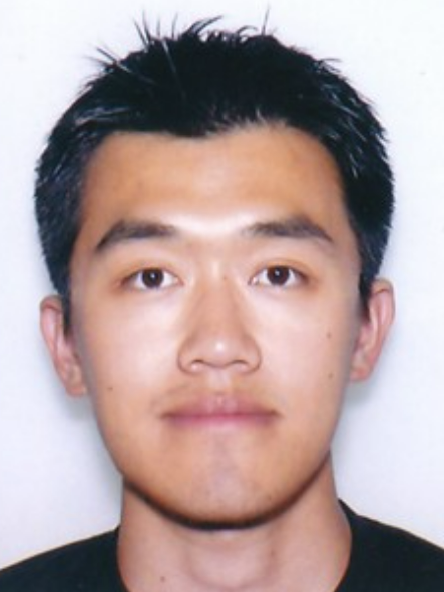}
\end{wrapfigure}
\textbf{Rui Ning Wang} is head of production and R\&D at Luxtelligence S.A., a start-up in Switzerland offering microfabrication services on thin-film lithium Niobate. Prior to that, he worked for 5 years in the group of Prof. Tobias J. Kippenberg at EPFL as a process engineer in microfabrication on silicon nitride and lithium niobate integrated photonic platforms. Rui studied for his M.Sc. in Material Science at EPFL in Switzerland. He then obtained a Ph.D. degree in Physics, working at the Paul-Drude Institut in Berlin on molecular beam epitaxy of chalcogenide superlattices.
\\
\\
\newpage
\begin{wrapfigure}{l}{0.3\textwidth}
\includegraphics[width=0.3\textwidth]{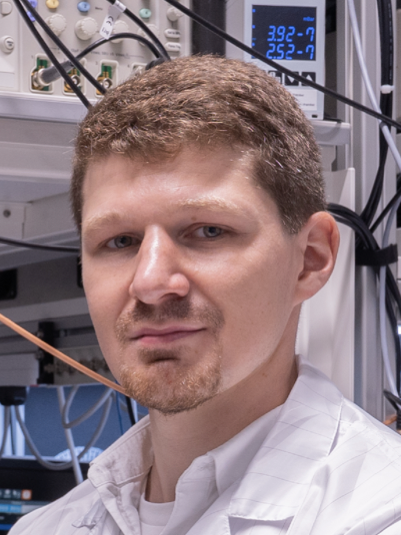}
\end{wrapfigure}
\textbf{Andrey Voloshin} is an expert in photonics, a researcher at the Swiss Federal Institute of Technology, Lausanne (EPFL), and a co-founder of Deeplight SA. He obtained his Ph.D. from Moscow State University in 2016. In 2017, Andrey demonstrated the first fully integrated optical frequency comb based on self-injection locking to optical microresonators. Recently, at Deeplight and EPFL, he was awarded the Marie Skłodowska-Curie Individual Fellowship and led the development of photonic chip-based ultra-low noise lasers with high-speed actuation enabled by monolithic integration with MEMS actuators.
\\
\\
\\

\begin{wrapfigure}[14]{l}{0.3\textwidth}
\includegraphics[width=0.3\textwidth]{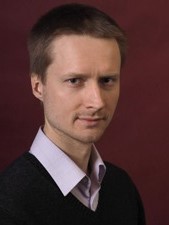}
\end{wrapfigure}
\textbf{Grigory Lihachev} is a postdoctoral fellow in the group of Prof. Tobias Kippenberg at EPFL working on projects in nonlinear integrated photonics. Grigory worked on soliton microcombs in the group of Michael Gorodetsky during his Ph.D. studies at Moscow State University.
\\
\\
\\
\\
\\
\\
\\
\\

\begin{wrapfigure}{l}{0.3\textwidth}
\includegraphics[width=0.3\textwidth]{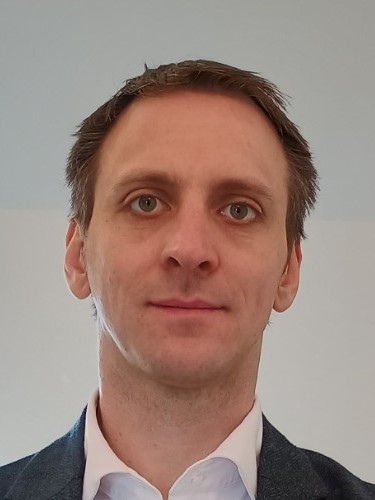}
\end{wrapfigure}
\textbf{Johann Riemensberger} is an Onsager Fellow and Associate Professor at the Norwegian University of Science and Technology. He received a PhD in Physics from Technical University of Munich, where he worked on attosecond science. His Ph.D. thesis received the PhD thesis prize of the “Bund der Freunde der TUM”. He joined the Swiss Federal Institute and Technology as a postdoc where he received a Marie Skłodowska-Curie Individual Fellowship from the European Commission and an Ambizione Fellowship of the Swiss National Science Foundation demonstrating novel approaches and applications of nonlinear integrated photonic circuits in amplification, low-noise laser, and for coherent laser ranging. 
\\
\newpage
\begin{wrapfigure}[14]{l}{0.3\textwidth}
\includegraphics[width=0.3\textwidth]{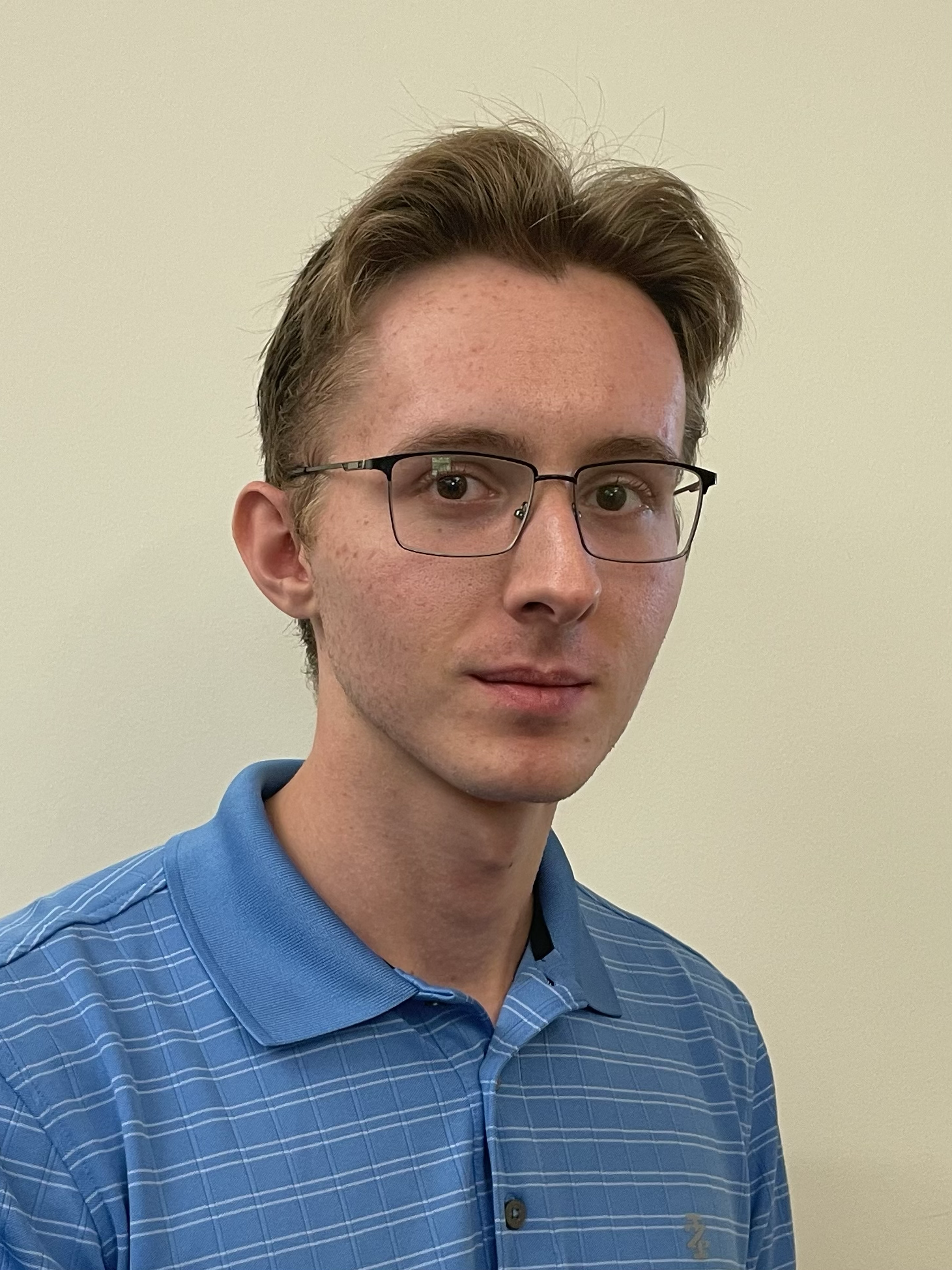}
\end{wrapfigure}
\textbf{Scott E. Kenning} is an Electrical and Computer Engineering PhD student in the OxideMEMS lab at Purdue University. He received his B.S. in Electrical Engineering from Purdue in 2022. His interests involve piezoelectric actuation of integrated photonics, nonlinear optics, and integrated optical readout of micro-mechanical structures. During undergraduate studies, he worked on silicon integrated photonics and free-space measurement of mechanical membrane structures, shaping his interests. He is supported by the NSF GRFP, awarded in 2022, to pursue such topics. 
\\
\\

\begin{wrapfigure}{l}{0.3\textwidth}
\includegraphics[width=0.3\textwidth]{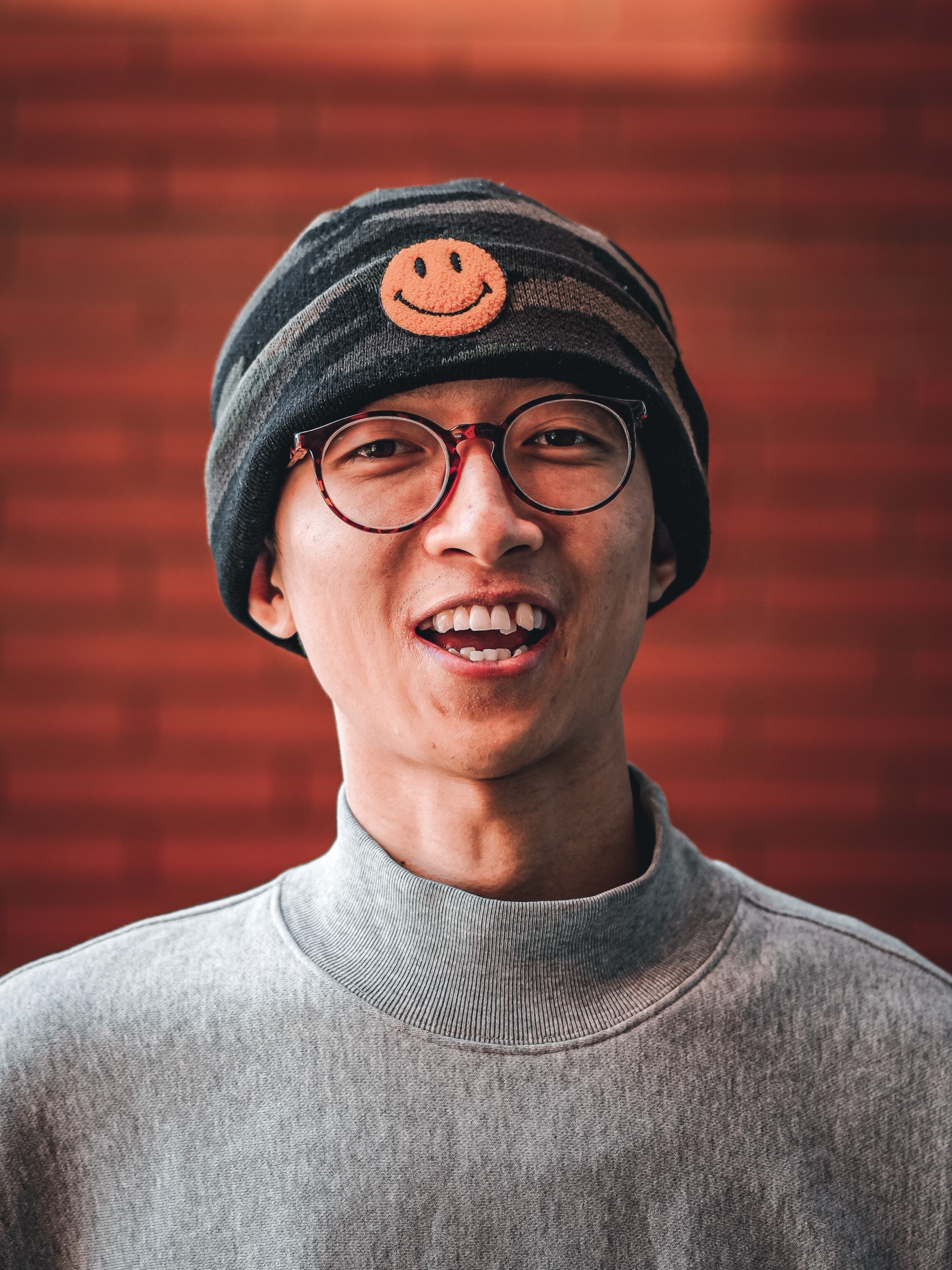}
\end{wrapfigure}
\textbf{Yu Tian} is a Ph.D. candidate in Electrical and Computer Engineering at Purdue University, specializing in the development of MEMS-based actuators for photonic integrated circuits. His research focuses on enhancing the efficiency and precision of these systems. He is also passionate about implementing optomechanical systems for both classical and quantum applications. Yu holds an M.S. degree in Mechanical Engineering from SUNY Binghamton, where he conducted evaluations and optimizations across various MEMS scenarios, with a particular emphasis on nonlinear dynamics.
\\
\\
\\

\begin{wrapfigure}[14]{l}{0.3\textwidth}
\includegraphics[width=0.3\textwidth]{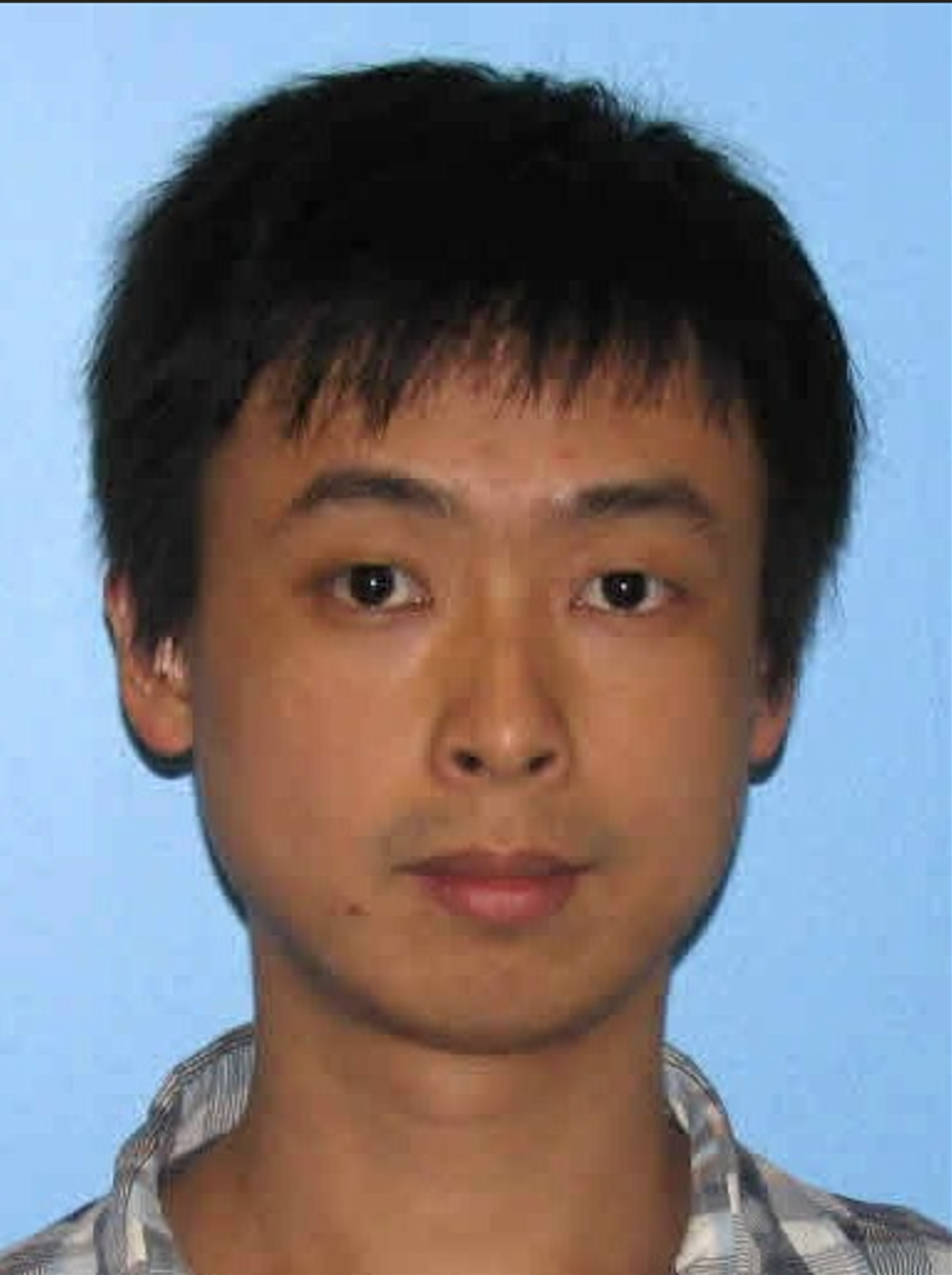}
\end{wrapfigure}
\textbf{Tzu-Han Chang} is a postdoctoral fellow in Sunil Bhave's group at Purdue University, focusing on optomechanics, chip-based lasers, photonic packaging, and fabrication. During his Ph.D. studies at Purdue, he worked on nanophotonics and atomic physics, developing an atom-photonic hybrid system in Chen-Lung Hung's group. Prior to that, he earned his Master's and Bachelor's degrees from National Taiwan University.
\\
\\
\\
\\
\\
\\
\\
\newpage
\begin{wrapfigure}{l}{0.3\textwidth}
\includegraphics[width=0.3\textwidth]{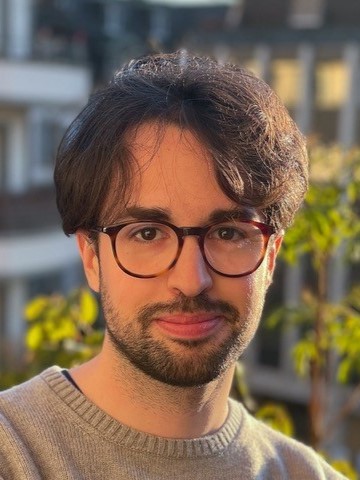}
\end{wrapfigure}
\textbf{Andrea Bancora} is co-founder and Senior Photonics Engineer at Deeplight SA, a startup in Lausanne, Switzerland, specialized in ultra-low noise photonic circuit-based laser sources. He earned his M.Sc. in Physics from the University of Milan in 2021. Andrea worked on superconducting microwave quantum optomechanics during his thesis in Tobias Kippenberg’s lab, LPQM, before venturing in the photonics domain as a Research Assistant in the same group. He conducted applied research on hybrid-integrated silicon nitride-based laser, also joining Deeplight in 2022. Since 2024 he is leading the technological development and industrialization of the laser sources.
\\
\\

\begin{wrapfigure}[14]{l}{0.3\textwidth}
\includegraphics[width=0.3\textwidth]{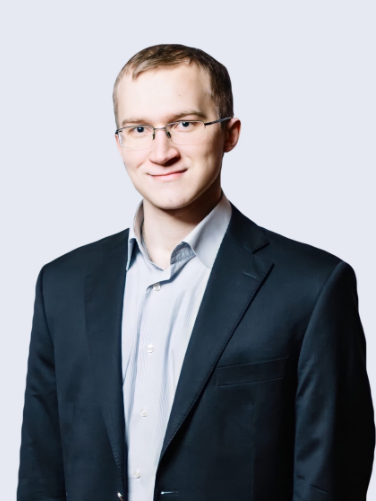} 
\end{wrapfigure}
\textbf{Viacheslav Snigirev} is a Ph.D. student at EPFL. He received his M.Sc. degree from Lomonosov Moscow State University in June 2020, where he worked on metaphotonic structures design and characterization. He is currently working on ferroelectric-based integrated photonics with a wide selection of materials (including lithium niobate, lithium tantalate and barium titanate) and a broad range of application (including frequency-agile lasers, electro-optic quantum transduction, single-shot optical q-bit read-out, and high speed electro-optic modulators). He is a recipient of ECIO 2022 Best Poster Award. 
\\
\\
\\
\begin{wrapfigure}[14]{l}{0.3\textwidth}
\includegraphics[width=0.3\textwidth]{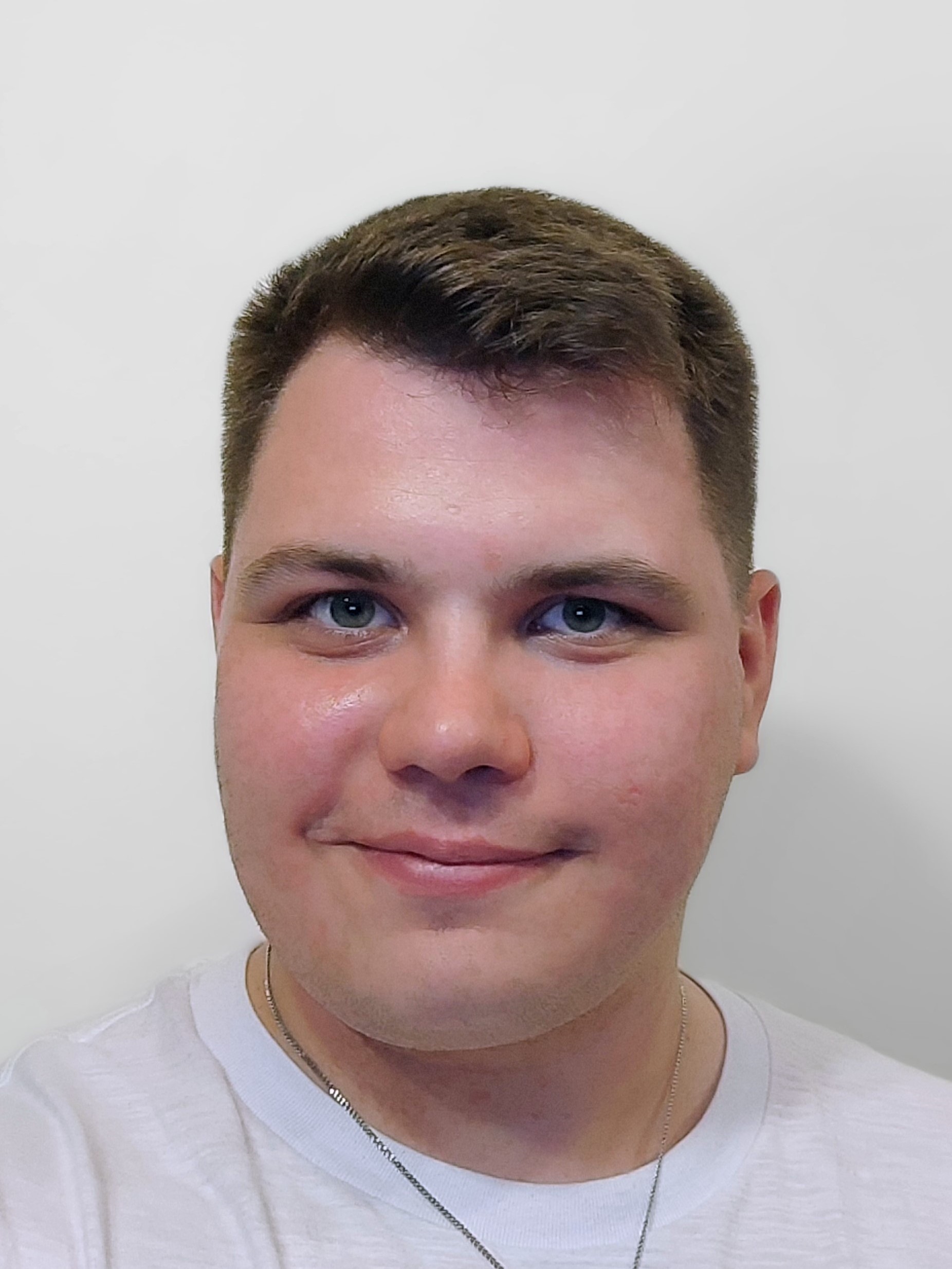}
\end{wrapfigure}\par
\textbf{Vladimir Shadymov} received the M.Sc. degree in Applied Mathematics and Physics from the Moscow Institute of Physics and Technology (MIPT), Russia, in collaboration with the Laboratory of Photonic and Quantum Measurements (LPQM) at École Polytechnique Fédérale de Lausanne (EPFL), Switzerland, in 2022. His thesis was titled "Frequency-Agile Hybrid Integrated Lasers with High Linearity for the Visible and Near-Infrared Spectral Range."
He is an expert in the field of photonic integrated circuits (PICs), with more than 3 years of experience. He specializes in the development and commercialization of PICs whose functionality is extended with other integrated technologies for various applications, including space, automotive, sensing, and telecom. From 2021 to 2024, he played a significant role in commercializing frequency-agile hybrid integrated lasers based on silicon nitride PICs monolithically integrated with MEMS for wavelengths ranging from visible to near-infrared at DeepLight SA, a spin-off from LPQM, EPFL.
He is currently a Photonic Integrated Circuit Design Engineer at Polariton Technologies AG, where he focuses on the development of plasmonic hybrid modulators.
In addition to his professional roles, Vladimir has contributed to the academic field with publications, including articles in journals such as Nature Communications. He has been actively involved in several US, Swiss and European research projects, enhancing the integration and reliability of photonic technologies.
\\

\begin{wrapfigure}[14]{l}{0.3\textwidth}
\includegraphics[width=0.3\textwidth]{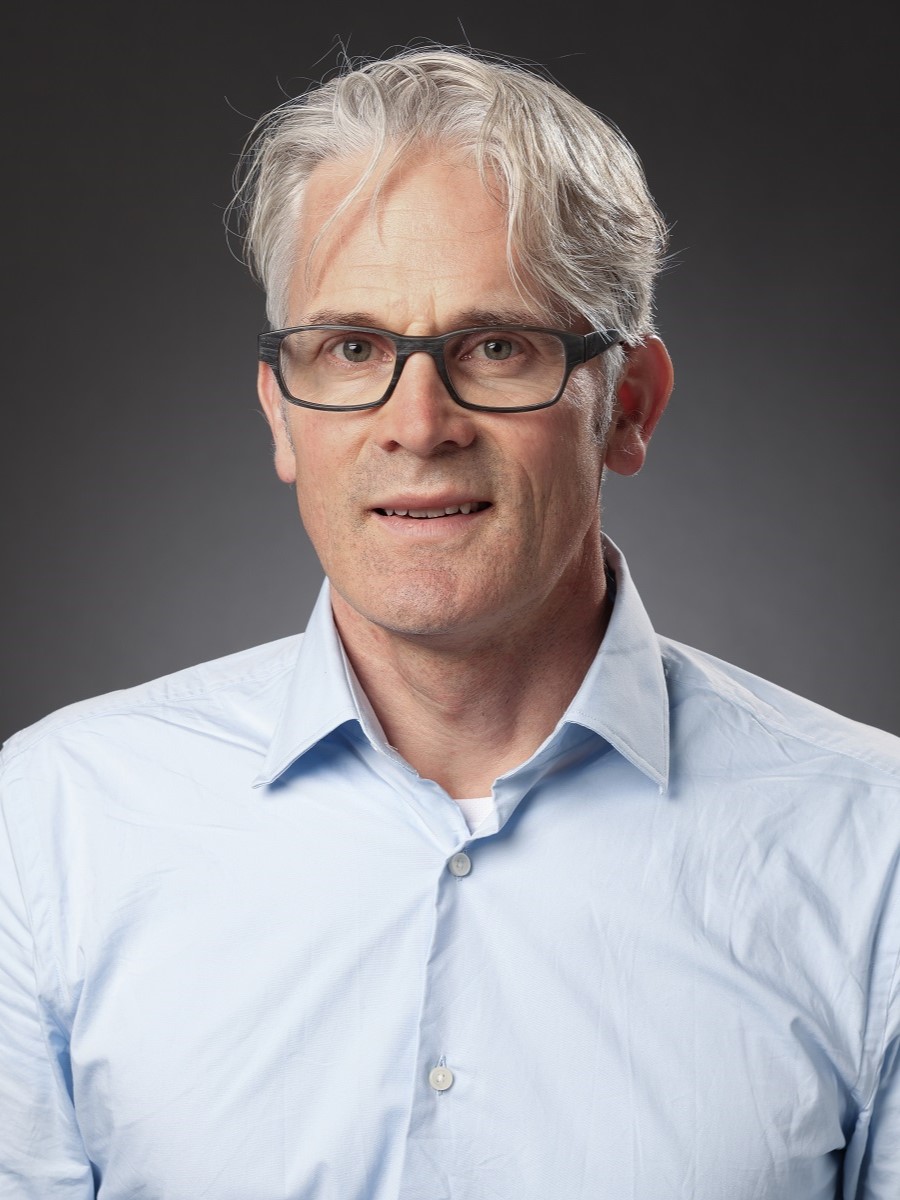}
\end{wrapfigure}
\textbf{Tobias J. Kippenberg} has been Full Professor in the Institute of Physics and Electrical Engineering at EPFL in Switzerland since 2013. He joined EPFL in 2008 as Tenure Track Assistant Professor. Prior to EPFL, he was Independent Max Planck Junior Research group leader at the Max Planck Institute of Quantum Optics in Garching, Germany. While at the MPQ he demonstrated radiation pressure cooling of optical micro-resonators, and developed techniques with which mechanical oscillators can be cooled, measured and manipulated in the quantum regime that are now part of the research field of Cavity Quantum Optomechanics. Moreover, his group discovered the generation of optical frequency combs using high Q micro-resonators, a principle known now as micro-combs or Kerr combs. This discovery unlocked record data transmission rate which led to the development of new concepts in telecommunications in collaborations with industry.
For his early contributions in these two research fields, he was recipient of the EFTF Award for Young Scientists (2011), The Helmholtz Prize in Metrology (2009), the EPS Fresnel Prize (2009), ICO Award (2014), Swiss Latsis Prize (2015), the Wilhelmy Klung Research Prize in Physics (2015), the 2018 ZEISS Research Award, and the R. Wood Award (2021). Moreover,
he is 1st prize recipient of the "8th European Union Contest for Young Scientists" in 1996. He has been ranked in the top 1\% as a highly cited physicist by Clarivate Analytics since 2014. He is co-founder of the startups LIGENTEC SA, an integrated photonics foundry, DEEPLIGHT SA, a supplier of advanced laser sources, LUXTELLIGENCE SA, a thin-film lithium niobate foundry, and EDWATEC SA, a company developing rare-earth ion doped photonic integrated circuits for photonic integrated circuit-based Erbium amplifiers and lasers.
Prof. Kippenberg has authored more than 260 papers. He is an elected member of the German National Academy of Sciences Leopoldina, the United States National Academy of Sciences (NAE), and the Swiss Academy of Engineering Sciences (SATW).
\\

\begin{wrapfigure}{l}{0.3\textwidth}
\includegraphics[width=0.3\textwidth]{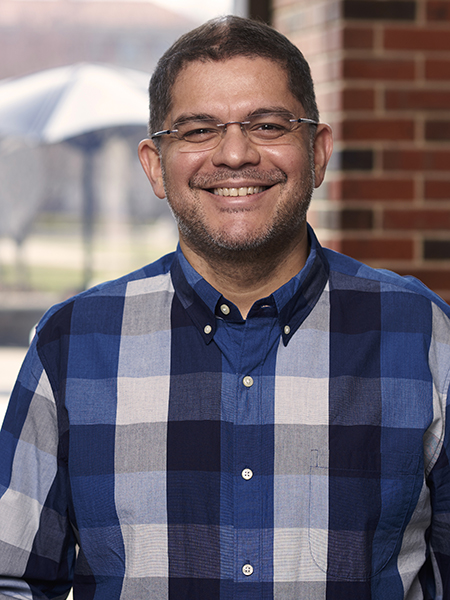}
\end{wrapfigure}
\textbf{Sunil Bhave} received the B.S. and Ph.D. degrees from Berkeley in EECS in 1998 and 2004 respectively. In April 2015, he joined the Elmore Family School of Electrical and Computer Engineering at Purdue University. Sunil’s research explores inter-domain coupling in Opto-mechanical, Spin-Acoustic and Atom-MEMS devices and strives to leverage understanding of these coupled systems to design and fabricate inertial sensors, clocks, frequency combs and computing and microwave sub-systems.
Sunil received the NSF CAREER Award in 2007, the DARPA Young Faculty Award in 2008, the IEEE Ultrasonics Society’s Young Investigator Award in 2014, the Google Faculty Research Award in 2020 and IEEE UFFC’s 2023 Water G. Cady Award. His students have received Best Paper Awards at IEEE MEMS 2023 and 2021, IEEE Frequency Control Symposium 2021 and 2020, IEEE Photonics 2012, Ultrasonics 2009 and IEDM 2007. Before joining Purdue, Sunil was a professor at Cornell for 10 years and worked at Analog Devices for 5 years.

\end{document}